\documentclass[12pt]{article}
\pdfoutput=1

\usepackage{fullpage}
\usepackage{graphicx}
\usepackage{cite}
\usepackage{authblk}
\usepackage{bbm}
\usepackage[backgroundcolor=green!5, linecolor=blue!60]{todonotes}
\usepackage{mathrsfs}
\usepackage{dsfont}
\usepackage{feynmp}
\usepackage{feynmp-auto}
\usepackage{tikz}
\usetikzlibrary{arrows,decorations.markings,shapes.misc,math}

\usepackage{amsmath}
\usepackage{amssymb}
\usepackage{subcaption}

\usepackage{empheq}

\newcommand{\bbone} { {\mathds 1}}
\newcommand{\be}{\begin{equation}}
\newcommand{\ee}{\end{equation}}

\newcommand{\tDisc} {\text{Disc}_{\text{tot}}}

\definecolor{darkyellow}{rgb}{0.5, 0.5, 0.0}
\definecolor{darkpurple}{rgb}{0.5, 0.2, 0.8}
\definecolor{darkblue}{rgb}{0.0, 0.0, 0.8}
\definecolor{darkgreen}{rgb}{0.0, 0.4, 0.0}
\definecolor{darkred}{rgb}{0.5, 0.0, 0.0}
\usepackage{hyperref}
\hypersetup{
    linktocpage,
     colorlinks,
     citecolor=darkgreen,
     linkcolor= darkgreen,
     urlcolor=darkgreen
}

\newcommand{\circleleft}{
\begin{tikzpicture}
 \draw[
        decoration={markings, mark=at position 0.3 with   {\arrow[scale=1.4]{>}}},
        postaction={decorate}
        ] (-0.01,0.1) -- (-0.011,0.1);
\draw[] (0.05,0) circle (0.1);
\end{tikzpicture}
}
\newcommand{\circleright}{
\begin{tikzpicture}
 \draw[
        decoration={markings, mark=at position 0.3 with   {\arrow[scale=1.4]{<}}},
        postaction={decorate}
        ] (0.012,0.1) -- (0.012-0.001,0.1);
\draw[] (0.05,0) circle (0.1);
\end{tikzpicture}
}

\tikzset{->-/.style={decoration={
  markings,
  mark=at position 1 with {\arrow[scale=1.3]{>}}},postaction={decorate}}}

  \tikzset{-<-/.style={decoration={
  markings,
  mark=at position .4 with {\arrow[scale=1.3]{<}}},postaction={decorate}}}

\newcommand{\linebubR}{
\begin{tikzpicture}
\tikzmath{\rad = 0.1; \ang =150; \dy = \rad*sin(\ang);}
 \draw[-<-] (0,-\dy) to (0.2, -\dy);
 \draw[-] (0.2, \dy) arc (\ang:-\ang:\rad);
 \draw[-<-] (0.2, \dy) to (0, \dy);
\end{tikzpicture}
}

\newcommand{\linebub}{
\begin{tikzpicture}
\tikzmath{\rad = 0.1; \ang =150; \dy = \rad*sin(\ang);}
 \draw[->-] (0,-\dy) to (0.2, -\dy);
 \draw[-] (0.2, \dy) arc (\ang:-\ang:\rad);
 \draw[->-] (0.2, \dy) to (0, \dy);
\end{tikzpicture}
}

\newcommand{\linebubUp}{
\begin{tikzpicture}
\tikzmath{\rad = 0.1; \ang =150; \dx = \rad*sin(\ang);}
 \draw[->-] (-\dx,0.05) to (-\dx,0.2);
 \draw[-] (-\dx,0.2) arc (90+\ang:90-\ang:\rad);
 \draw[-] (\dx,0.2) to (\dx,0.05);
  \draw[
        decoration={markings, mark=at position 0.3 with   {\arrow[scale=1.4]{>}}},
        postaction={decorate}
        ] (0.05,0.39) -- (0.05+0.001,0.39);
\end{tikzpicture}
}

\newcommand{\linebubDown}{
\begin{tikzpicture}
\tikzmath{\rad = 0.1; \ang =150; \dx = \rad*sin(\ang);}
 \draw[->-] (-\dx,-0.05) to (-\dx,-0.2);
 \draw[-] (\dx,-0.2) arc (270+\ang:270-\ang:\rad);
 \draw[-] (\dx,-0.2) to (\dx,-0.1);
  \draw[
        decoration={markings, mark=at position 0.3 with   {\arrow[scale=1.4]{>}}},
        postaction={decorate}
        ] (0.05,-0.39) -- (0.05+0.001,-0.39);
\end{tikzpicture}
}

\newcommand{\linebubDownR}{
\begin{tikzpicture}
\tikzmath{\rad = 0.1; \ang =150; \dx = \rad*sin(\ang);}
 \draw[-<-] (-\dx,-0.05) to (-\dx,-0.2);
 \draw[-] (\dx,-0.2) arc (270+\ang:270-\ang:\rad);
 \draw[-] (\dx,-0.2) to (\dx,-0.05);
  \draw[
        decoration={markings, mark=at position 0.3 with   {\arrow[scale=1.4]{>}}},
        postaction={decorate}
        ] (-0.05,-0.38) -- (-0.05-0.001,-0.38);
\end{tikzpicture}
}

\renewcommand{\circlearrowleft}{\circleleft}
\renewcommand{\circlearrowright}{\circleright}

\newcommand{\blue}[1]{ {\color{darkblue}{#1}}}
\newcommand{\red}[1]{ {\color{darkred}{#1}}}
\newcommand{\green}[1]{ {\color{darkgreen}{#1}}}
\newcommand{\purple}[1]{ {\color{darkpurple}{#1}}}

\newcommand{\cA}{\mathcal A}
\newcommand{\cV}{\mathcal V}

\newcommand{\cF}{\mathcal F}
\newcommand{\cU}{\mathcal U}
\newcommand{\cL}{\mathcal L}
\newcommand{\cM}{\mathcal M}
\newcommand{\cN}{\mathcal N}
\newcommand{\sM}{\mathscr M}
\newcommand{\cMb}{\overline \cM}

\newcommand{\cO}{\mathcal O}

\newcommand{\cT}{\mathcal T}

\newcommand{\Li}{\text{Li}}

\newcommand{\eps}{\varepsilon}
\newcommand{\im}{\text{Im }}
\newcommand{\re}{\text{Re }}

\newcommand{\zb}{\bar{z}}
\newcommand{\disc}{\text{Disc}}
\newcommand{\cut}{\text{Cut}}

\newcommand{\set}{\mathcal{S}}

\newcommand{\fwbox}[2]{\text{\makebox[#1][c]{$\hspace{-150pt}\displaystyle#2\hspace{-150pt}$}}}

\numberwithin{equation}{section}

\newcommand*\circled[1]{\tikz[baseline=(char.base)]{
            \node[shape=circle,draw,inner sep=2pt] (char) {#1};}}

\tikzset{cross/.style={cross out, draw=black, fill=none, minimum size=2*(#1-\pgflinewidth), inner sep=0pt, outer sep=0pt}, cross/.default={2pt}}

\newcommand{\toptM}{M}
\newcommand{\toptMb}{\overline{M}}

\newcommand{\equivD}{\equiv}

\newcommand\bin[2]{ \begin{pmatrix} #1 \\ #2  \end{pmatrix} }
\newcommand\stirling[2]{ \begin{Bmatrix} #1 \\ #2  \end{Bmatrix} }
\newcommand\bins[2]{ (^{#1}_{#2})}
\newcommand\stirlings[2]{\{^{#1}_{#2}\}}

\newcommand{\mc}[1]{\mathcal{#1}}
\newcommand{\eq}[1]{\begin{equation}#1\end{equation}}

\title{Sequential Discontinuities of Feynman Integrals\\ and the Monodromy Group}
\author[1,2]{Jacob L.\ Bourjaily}%
\author[3]{Holmfridur Hannesdottir}%
\author[1]{Andrew~J.~McLeod}%
\author[3]{Matthew D. Schwartz}%
\author[1]{Cristian~Vergu}%
\affil[1]{\small \emph{Niels Bohr International Academy and Discovery Center, Niels Bohr Institute\\University of Copenhagen, Blegdamsvej 17, DK-2100, Copenhagen \O, Denmark}}
\affil[2]{\small \emph{Institute for Gravitation and the Cosmos, Department of Physics\\Pennsylvania State University, University Park, PA 16892, USA}}
\affil[3]{\small \emph{Department of Physics, Harvard University, Cambridge, MA 02138, USA}}

\begin{document}

\begin{fmffile}{feyngraph}
\unitlength = 0.4mm
\maketitle
\thispagestyle{empty}

\begin{abstract}
We generalize the relation between discontinuities of scattering amplitudes and cut diagrams to cover sequential discontinuities (discontinuities of discontinuities) in arbitrary momentum channels. The new relations are derived using time-ordered perturbation theory, and hold at phase-space points where all cut momentum channels are simultaneously accessible. As part of this analysis, we explain how to compute sequential discontinuities as monodromies and explore the use of the monodromy group in characterizing the analytic properties of Feynman integrals.  We carry out a number of cross-checks of our new formulas in 
polylogarithmic examples, in some cases to all loop orders.
\end{abstract}

\hypersetup{pageanchor=false}
\newpage
\enlargethispage{20pt}
\tableofcontents
\newpage
\hypersetup{pageanchor=true}
\pagenumbering{arabic}

\section{Introduction}

Feynman integrals---integrals over Feynman propagators appearing in perturbative quantum field theory calculations---are primarily useful for making observable predictions about particle physics experiments. Famously, they have been used to make some of the most precise predictions in the history of science~\cite{Tanabashi:2018oca}. However, these integrals have also increasingly become recognized as interesting mathematical objects in their own right, exhibiting a variety of geometric, analytic, and number-theoretic properties. 

One of the aspects of Feynman integrals that has become better understood in recent years is the class of transcendental functions they evaluate to in integer dimensions. In particular, at low loop order and low particle multiplicity, they can often be expressed in terms of generalized polylogarithms~\cite{Chen,G91b,Goncharov:1998kja}. These functions are under good theoretical and numerical control, due in part to the symbol and coaction~\cite{Goncharov:2005sla,Goncharov:2010jf,2011arXiv1101.4497D,Brown:2011ik,Duhr:2011zq}, which provide a systematic way to understand their analytic structure and to exploit identities among them. In particular, arbitrarily complicated polylogarithms can be broken down into simpler building blocks such as logarithms and Riemann zeta values, at the cost of losing only integration boundary data. 

Knowing the analytic structure of polylogarithms has proven especially useful in the computation of Feynman integrals and scattering amplitudes, as the branch cut structure of these quantities is constrained by physical principles such as locality and causality. For example, in the Euclidean region where all Mandelstam invariants are negative, Feynman integrals can only have logarithmic branch points at the vanishing loci of sums of external momenta. This places strong constraints on the symbol and coaction of the polylogarithms these integrals produce.\footnote{This constraint on the symbol can also be extended to Feynman integrals that evaluate to elliptic polylogarithms~\cite{brown2011multiple,Broedel:2017kkb,Broedel:2018iwv}; however, no coaction has been worked out for the types of worse-than-elliptic integrals that appear in Feynman integrals in integer dimensions (see for instance~\cite{Bloch:2016izu,Bourjaily:2018ycu,Bourjaily:2018yfy,Bogner:2019lfa}).}

That Feynman integrals have branch cut singularities has been known since the early days of quantum field theory. In a seminal paper by Landau~\cite{landau1959}, these branch cuts were shown to be associated with regions of external momenta where the poles in Feynman propagators coalesce around the integration contour, so that the contour is pinched between the singularities (see also~\cite{coleman1965singularities,Collins:2020euz}). Cutkosky subsequently gave a general formula relating the discontinuity across these branch cuts to ``cut graphs'' in which some Feynman propagators are replaced by delta functions\cite{cutkosky}. 't Hooft and Veltman later gave a simple diagrammatic derivation of Cutkosky's cutting rules\cite{tHooft:1972qbu,tHooft:1973wag}. However, these works are mostly confined to the study of a {\it single} discontinuity of Feynman integrals. 

In this paper, we are interested in studying the cutting rules for  discontinuities of discontinuities: is there a way to compute sequential discontinuities of Feynman integrals with cut diagrams, as there is for single discontinuities? Cutkosky and his contemporaries touched on this topic, but computing sequential discontinuities is significantly more complicated than computing a single discontinuity. 
Some progress on the study of sequential discontinuities was made in~\cite{Abreu:2014cla}, where a formula relating sequential discontinuities in different channels to a sum over cuts was conjectured. Drawing inspiration from this work, we make use of time-ordered perturbation theory (TOPT) to derive more general relations between the sequential discontinuities of Feynman integrals and cut integrals. In particular, our method clarifies the role of the $\pm i\eps$ prescription in the cut integrals, and emphasizes the importance of considering monodromies around branch points rather than discontinuities across branch cuts. These new results apply to sequential discontinuities in any channels, including discontinuities in the same channel. 

Sequential cuts of Feynman integrals can also be computed using the multivariate residue calculus of Leray~\cite{BSMF_1959__87__81_0}. This has been worked out explicitly at one loop~\cite{Abreu:2017ptx}. While this approach is both general and mathematically rigorous, it quickly becomes computationally onerous. More Hodge-theoretic approaches were also considered in~\cite{Bloch:2010gk,Bloch:2015efx}. In this paper, our goal was thus to come up with a prescription for computing sequential discontinuities that was more computationally tractable than these approaches.

One set of constraints on sequential discontinuities are the Steinmann relations. As originally studied by Steinmann~\cite{steinmann1}, these relations follow from causality and express linear relations between vacuum expectation values of certain types of operator products called $R$-products (as defined in~\cite{Lehmann:1957zz}). Steinmann originally studied these relations for the case of four local gauge-invariant operators; they were subsequently generalized to higher multiplicity~\cite{steinmann2,Ruelle:1961,Araki1960,doi:10.1063/1.1703695}. Later, it was shown that the Steinmann relations imply scattering amplitudes cannot have double discontinuities in partially overlapping momentum channels~\cite{Lassalle:1974jm}. 
The Steinmann relations have also been studied directly from the point of view of \(S\)-matrix theory, without reference to local fields and their commutators; for a review, see~\cite{Cahill:1973qp}. 

Steinmann-type constraints have proven extremely useful for the modern amplitude bootstrap program, which attempts to determine the functional forms of Feynman integrals or scattering amplitudes from their general properties (such as symmetries, analytic properties, and factorization in certain kinematic limits). So far, these methods have been mostly applied to processes in  planar \(\mathcal{N} = 4\) supersymmetric Yang-Mills theory~\cite{Dixon:2011pw,Dixon:2013eka,Dixon:2014voa,Drummond:2014ffa,Dixon:2014iba,Dixon:2015iva,Dixon:2016apl,Caron-Huot:2016owq,Dixon:2016nkn,Drummond:2018caf,Caron-Huot:2019vjl,Caron-Huot:2018dsv}, where there is rich theoretical data available and integrability-based computations provide crucial consistency checks~\cite{Basso:2013vsa,Basso:2013aha,Belitsky:2014sla,Basso:2015uxa}. However, analytic constraints and bootstrap techniques are expected to extend to non-supersymmetric quantities as well (see for instance~\cite{Li:2016ctv,Almelid:2017qju,Basso:2017jwq}). 

Scattering amplitudes in Yang-Mills theories necessarily involve massless particles, so the Steinmann relations, originally derived in field theories with a mass gap, do not necessarily apply. Indeed, massless particles engender infrared divergences in these theories. In planar $\cN=4$, instead of studying the amplitude itself, one typically studies finite Feynman integrals (see for instance~\cite{Drummond:2010cz,DelDuca:2011wh,Bourjaily:2018aeq,Caron-Huot:2018dsv,Henn:2018cdp,McLeod:2020dxg}) or remainder functions, defined as ratios of amplitudes or ratios of amplitudes to the exponentiation of lower-order amplitudes. It is to these types of remainder functions that Steinmann-type constraints are often applied~\cite{Caron-Huot:2016owq,Dixon:2016nkn,Golden:2018gtk,Golden:2019kks}.\footnote{The Steinmann relations were first used to analyze these amplitudes in the multi-Regge limit, where it was also pointed out that normalizing by the BDS ansatz did not preserve these relations~\cite{Bartels:2008ce}.} While there has been some progress in systematically
extracting the infrared-finite content of the $S$-matrix (for example, through the construction of an infrared-finite S-matrix~\cite{Hannesdottir:2019rqq,Hannesdottir:2019opa}), there remains some uncertainty over how and when constraints like Steinmann relations should hold. One goal of this paper is to pry away some of the strong assumptions used in the axiomatic field theory approach. Thus, rather than full scattering amplitudes in mass-gapped theories, we study Feynman integrals directly.

More broadly, in this paper we set out to provide some clarity on how to think about and compute sequential discontinuities of Feynman integrals, and to study the types of constraints these sequential discontinuities satisfy. We treat this problem both at the level of cut integrals and at the level of polylogarithmic functions. In particular, we make use of time-ordered perturbation theory (TOPT) to prove new relations between the sequential discontinuities of Feynman integrals and their cuts. We also describe how these discontinuities can be computed systematically from polylogarithmic representations of these integrals with the use of variation matrices and the monodromy group, both of which we describe in some detail. 

The main practical results of this paper take the form of relations between discontinuities of Feynman integrals and cuts of those integrals. For example, we show that the $m^\text{th}$ discontinuity of the Feynman integral $\cM$ in a momentum channel corresponding to the Mandelstam invariant $s$ satisfies the relation
\begin{equation}
          \big[\disc_s^m \cM \big]_{R^s}= m!
      \sum_{k=m}^\infty \stirling{k}{m}
      \left(-1\right)^{m-k}
      \big[\cM_{\text{$k$-cuts}}\big]_{R^s_+} \, ,
      \label{summeq}
\end{equation}
where 
$\stirlings{k}{m} = \frac{1}{m!} \sum_{\ell=1}^m (-1)^{m-\ell} \bins{m}{\ell}  \ell^k$ are the Stirling numbers of the second kind. 
On the left side of the equation, we compute $m$ discontinuities in the $s$ channel by taking $m$ monodromies
around a branch point in s. We write this as
\begin{equation}
\disc_s^m \cM = (\bbone - \sM_{\linebub_s})^m \cM
\end{equation}
These monodromies are taken by analytically continuing along a closed contour that goes between the region $R^s$, which we define to be the region in which all Mandelstam invariants are real and negative, except for $s$ which is real and positive, and the Euclidean region $R^\star$, where all invariants are negative. On the right-hand side, $\cM_{\text{$k$-cuts}}$  denotes the sum over all ways of cutting the Feynman integral $k$ times, with positive energy flowing across all cuts. These cuts must be computed in the region $R_+^s$, where the + subscript indicates that all the Feynman propagators in these cut diagrams should be assigned $+i\eps$. A careful treatment of the $\pm i\eps$ in the cut diagrams is essential to have a sensible (and correct) formula relating discontinuities and cuts. Eq.~\eqref{summeq} is derived in Section~\ref{sec:sequential}. We also derive similar relations between cuts and discontinuities in different channels.

One thing that our analysis makes clear is that sequential discontinuities can only be nonzero when there exists at least one TOPT diagram that depends on the energies corresponding to each cut momentum channel. When one of these energies is not present, the cut in this channel vanishes. Since the energies that appear in TOPT diagrams always take the form of sums of external energies $\sum_{i \in J} E_i$, where the sets of summed-over external particle indices $J$ that appear in a given diagram are strict subsets or supersets of each other, TOPT graphs never have sequential discontinuities in partially-overlapping momentum channels. This amounts to a new proof of the Steinmann relations in perturbation theory. We emphasize that the relations we derive between sequential discontinuities and cuts hold for individual Feynman integrals, and as such the Steinmann relations must also be obeyed by individual Feynman integrals. 

This is a long paper, partly because we wanted to give a pedagogical introduction to various subjects relevant for the main results in a uniform language. We begin in Sections~\ref{sec:cutting} and~\ref{sec:discon} by reviewing first the cutting rules and then the discontinuities of integrals in both covariant perturbation theory and TOPT. These sections essentially review what is needed to understand and prove the relation between single discontinuities and cuts, as in the optical theorem. We proceed in Section~\ref{sec:math} to introduce the main mathematical tools we use for computing sequential discontinuities. Here, we discuss the maximal analytic continuation of polylogarithmic functions and introduce the formalism of variation matrices. We then show how the discontinuities of polylogarithms can be computed using the action of the monodromy group. Our treatment of these topics draws heavily  from~\cite{MR1265552,MR3469645}, but is intended to be introductory since these topics have not featured prominently in the physics literature.
In Section~\ref{sec:sequential} we use these tools to prove our main results for sequential discontinuities and cuts of Feynman integrals. A corollary is a  new integral-by-integral proof of the Steinmann relations. In Section~\ref{sec:examples} we work through some explicit examples that illustrate these new relations between the cuts and discontinuities of Feynman integrals, including bubble, triangle, and box diagrams up to $L$-loop order. A summary and discussion of some possible implications of our work and future directions are given in Section~\ref{sec:conclusions}.

We also include in this paper a number of appendices with some technical details not needed for the main results of the paper. Appendix~\ref{subsec:coaction} discusses the relation between the variation matrix and the coproduct. Appendix~\ref{sec:fundamental_group} discusses the relationship between the monodromy group associated with a polylogarithm and the fundamental group of the manifold on which it is defined, and explicitly works out the relation between these groups in the case of the triangle and box ladder integrals. Appendix~\ref{app:singlevalue} shows how single-valued functions can be easily constructed in the variation matrix formalism. In Appendix~\ref{sec:symmetry}, we provide details on how the permutation symmetry of the one-loop triangle integral acts on its rational and transcendental parts. Appendix~\ref{app:Phi2} presents the variation matrix for the transcendental function $\Phi_2(z,\zb)$ appearing in the two-loop ladder triangle and box diagrams. Finally, Appendices~\ref{sec:A_3looptri} and~\ref{sec:massless3pt} give some details of the calculation of cuts of the three-loop and $L$-loop triangle diagrams.

\section{Cutting rules: a review \label{sec:cutting}}

The branch points and branch cuts of Feynman integrals have been studied since the early days of \(S\)-matrix theory.  Landau described how to compute the location of these branch hypersurfaces\cite{landau1959}, and later Cutkosky described how to compute discontinuities across these hypersurfaces, using Feynman integrals with cut propagators~\cite{cutkosky}. In this section we review the cutting rules and the relationship between cuts, discontinuities, and the imaginary part of a scattering amplitude. 

\subsection{Cutkosky, 't Hooft and Veltman}
We begin with the generalized optical theorem, which states that the imaginary part of a scattering amplitude ${\cal A}$ is given by a sum over intermediate states $X$,
\be
\text{Im} {\cal A}(A\to B)  = i \sum_X \int d \Pi_X (2\pi)^4 \delta^4(p_A -  p_X) {\cal A} (A \to X) {\cal A}^\star (X \leftarrow B)\,.
\label{optical}
\ee
This optical theorem is non-perturbative and follows from the unitarity of the $S$-matrix. By expanding each side order-by-order in any coupling, the theorem implies a constraint on the sum of all Feynman diagrams contributing to $\cA$ at any order. However, it does not provide any constraints on individual diagrams. Some nontrivial checks on the optical theorem, including examples where disconnected diagrams play a crucial role, can be found in~\cite{Frye_2019}.

One can derive stronger results than the optical theorem by directly studying individual Feynman integrals. These integrals are Lorentz-invariant integrals over Feynman propagators, and take the form
\begin{equation}
 \cM(p) = \int \prod_{\ell} \frac{d^d k_\ell}{i (2\pi)^d} \prod_{j} \frac{1}{[q_j(k,p)]^2 - m_j^2 + i \eps}\,. \label{Mdef}
\end{equation}
In our notation, the integer $\ell$ indexes $L$ loop momenta $k_\ell$, and $j$ indexes the internal lines. The variables $k$ and $p$ denote the collective set of loop and external momenta, respectively, while $q_j(k,p)$ and $m_j$ denote the momentum and mass of the $j^{\text{th}}$ internal line. We do not include factors of $i$ in the numerators of the propagators, but include a factor of $1/i$ per loop integral in anticipation of the $i$'s generated by the $k_\ell^0$ integrals.  Throughout this paper, we take incoming particles to have positive energy.

Feynman integrals are defined in terms of external four-momenta $p^\mu$, but since they are Lorentz invariant they depend only on invariants of the form $s_I =P_I^2$, where $P_I^\mu\equivD \sum_{i\in I}p_i^\mu$ denotes a sum of external momenta.  These invariants  cannot all be independent. For instance, in four dimensions a Feynman integral $\cM(p)$ depends on $n$ external momenta and hence (at most) $4n$ independent quantities, while there are $2^n$ invariants $s_I$. The number of independent invariants is further reduced by momentum conservation and the on-shell condition for each external particle. Thus, the $s_I$ are highly interdependent. The constraints on the $s_I$ are easiest to derive using their expression in terms of four-momenta. 

The integral $\cM$ may become singular as $i \eps \to 0$ in the propagators. For physical momenta the Mandelstam invariants $s_I$ are real, but we can analytically continue $\cM$ to be a function of complex $s_I$. Then the singularities as $i\eps \to 0$ can be thought of as the endpoints of branch cuts on a Riemann surface (more generally a hypersurface of maximal analytic continuation) associated to $\cM$. In 1959, Landau derived a set of equations whose solutions indicate the regions of momenta where these singularities may reside, collectively known as the {\it Landau surface} \cite{landau1959}. The Landau surface may be disconnected, but each connected component corresponds to some set of propagators becoming singular: $[q_j(k,p)]^2 = m_j^2$. 

\paragraph{Cutkosky}~\\
Shortly after Landau's paper, Cutkosky gave a prescription for computing the discontinuity across one region of the Landau surface~\cite{cutkosky}. If the singularity is associated with the region $\cL_{J}$ where the propagators $j\in J$ go on-shell, then the discontinuity is given by
\begin{equation}
     \disc_{\cL_J} \cM = \int \prod_{\ell} \frac{d^d k_\ell}{i(2\pi)^d}
     \left[\prod_{j\in J} (-2\pi i)\delta(q_j^2 - m_j^2)\Theta(q_j^0)\right]
     \prod_{k\notin J} \frac{1}{q_k^2 - m_k^2 }\,. \label{Cutcut}
\end{equation}
Cutkosky also considered the singularities of $ \disc_{\cL_J} \cM$. He argued that the discontinuity across a region of the Landau surface associated with a set of propagators $K$ (that are in the complement of $J$) going on shell is given by
\begin{equation}
    \disc_{\cL_K} \disc_{\cL_J}\cM = \disc_{\cL_{J\cup K}} \cM\,.
\end{equation}
This is the type of sequential discontinuity we focus on in this paper.

Unfortunately, Cutkosky's results are phrased entirely in terms of discontinuities across regions of the Landau surface where particular propagators go on-shell.
However, it is generally not possible to isolate a region corresponding to the singularity locus of (just) a given set of propagators in the space of independent invariants. For example, a string of bubbles depends only on a single external kinematic invariant $p^2$, but the Landau equations identify a different branch hypersurface when the propagators in different bubbles are cut. Thus, Cutkosky's formula gives no constraint for sequential discontinuities in the same channel, a central focus of this paper.

\paragraph{'t Hooft and Veltman}~\\
A simplified treatment of cuts and discontinuities was provided in the 1970's by 't Hooft and Veltman \cite{tHooft:1972qbu,tHooft:1973wag}. Their approach sidestepped the Landau equations and analytic continuation entirely, to provide a constraint on $\cM$ directly. They start with the Feynman graph associated with the Feynman integral $\cM$, and consider all possible colorings of the vertices of this graph as either black or white. The following rules are then assigned to the edges between these colored vertices:
\begin{equation}
\hspace{-14pt}\raisebox{-14.5pt}{\fwbox{0pt}{\begin{tikzpicture}
\newcommand{\midarrow}{\tikz \draw[-triangle 45] (0,0) -- +(.1,0);};
\draw [fill=black] (3.5,0) circle[radius = 2 pt] -- (4.5,0) circle[radius = 2 pt] node[label=right:{$\!\!\equivD\!{\blue{\dfrac{1}{p^2 -m^2+ i \eps}}}$}]{};
\draw [darkblue, thick] (3.5,0)  -- (4.5,0);
\draw [fill=white] (7.92,0) circle[radius = 2 pt];
\draw [darkred,thick] (8,0) -- (9,0);
\draw [fill=white] (9.08,0) circle[radius = 2 pt] node[label=right:
{$\!\!\equivD\! {\red {\dfrac{1}{p^2 -m^2- i \eps}}}$}]{};
\draw [fill=black] (12.5,0) circle[radius = 2 pt];
\draw [darkgreen,thick] (12.6,0) -- node {\midarrow} (14.0,0);
\draw [fill=white] (14.1,0) circle[radius = 2 pt] node[label=right:{$\!\!\equivD\!
{\green{-2\pi i \delta(p^2-m^2)\Theta(p_0)}}$}]{};
\end{tikzpicture}}}
\end{equation}
The graph with all black vertices is the original time-ordered Feynman integral
$\cM$, with all $+i\eps$ propagators, while the graph with all white vertices corresponds to $-\cMb$, where $\cMb$ is defined by
\begin{equation} 
 \cMb(p) = \int \prod_{\ell} \frac{d^d k_\ell}{-i(2\pi)^d} \prod_{j} \frac{1}{[q_j(k,p)]^2 - m_j^2 - i \eps}\,. \label{Mdef2}
\end{equation}
Propagators connecting black and white vertices are said to be cut, meaning these lines are on-shell and positive energy flows from black to white. Using the position-space version of these rules, 't Hooft and Veltman showed that the sum over all possible assignments of white and black vertices is zero. This implies what we call the {\it covariant} cutting rules
\begin{multline}     \label{Cutcov}
    \cM - \cMb = \sum_{\bullet,\circ} (-1)^{L_\circ}\!\!\!
     \int \prod_{\ell} \frac{d^d k_\ell}{i(2\pi)^d}
    \prod_{\bullet - \bullet}{\blue{ \frac{1}{q_j^2 - m_j^2 + i\eps}}} 
     \\\times \prod_{\bullet - \circ}{\green{(-2\pi i)\delta(q_j^2 - m_j^2)\Theta(q_0)}}
    \prod_{\circ - \circ} {\red{\frac{1}{q_j^2 - m_j^2 - i\eps}}} \,, 
\end{multline}
where the sum is over all diagrams with mixed black and white vertices and $L_\circ$ is the number of loops connecting exclusively  white vertices. 
 
There are a few important aspects of this equation to note. First, the covariant cutting rules (like Cutkosky's rules) do not require unitarity. Eq.~\eqref{Cutcov} is derived algebraically, as a constraint among integrals over propagators and delta functions. In a unitary theory, $\cMb$ is related to the complex-conjugated integral $\cM^\star$ (where the numerators and vertices are complex conjugated in addition to $+i\eps \to -i\eps$), and the numerators of cut propagators correspond to a sum over physical spins. Then the sum over cuts gives the total scattering cross section, and the generalized optical theorem in Eq.~\eqref{optical} results.

Second, even in a non-unitary theory the covariant cutting rules relate an integral with all $+i\eps$ propagators to an integral with all $- i \eps$ propagators. Since the Feynman integrals we consider have all the other sources of imaginary parts stripped out, the cutting rules directly compute $\text{Im} \cM$. Although we would like to view $\cM$ as an analytic function, so that $\text{Im} \cM$ is related to the discontinuity of $\cM$ around a branch point, this has to be done with some care. The covariant cutting rules directly let us compute only $\cM - \cMb$.

Third, if we compare to Cutkosky's formula in Eq.~\eqref{Cutcut} we note that  the covariant cutting rules involve mixed $+i\eps$ and $-i\eps$ propagators, while Eq.~\eqref{Cutcut} is agnostic to the pole positions of the propagators. This does not make the two equations inconsistent, since left-hand-side of Eq.~\eqref{Cutcut} is the discontinuity across a Landau surface defined by the cut propagators while the left-hand side of Eq.~\eqref{Cutcov} is the imaginary part of $\cM$. It does however make it difficult to explicitly verify Cutkosky's equation. In contrast, Eq.~\eqref{Cutcov} can be verified in a straightforward manner in any number of examples. 

Finally, because the 't Hooft-Veltman derivation of the cutting rules builds on a single constraint among all the diagrams (the largest time equation), it is hard to break it down further to derive constraints on individual Feynman diagrams.
Although such a dissection might be possible, we find it more transparent to work in time-ordered perturbation theory where the cutting rules can be derived in a way that makes generalizations to sequential cuts and discontinuities  more straightforward.

\subsection{Time-ordered perturbation theory \label{sec:topt}}
To prove the cutting rules in time-ordered perturbation theory (TOPT) we exploit the following simple mathematical identity. If some functions ${\blue A_j},{\red B_j}$ and ${\green C_j}$ are related by 
\begin{equation}
    {\blue A_j} - {\red B_j} =  {\green C_j} \, , \label{ABD1}
\end{equation}
then
\begin{equation}
    {\blue A_1}\cdots {\blue A_n} - {\red B_1} \cdots {\red B_n} = 
    {\green C_1} {\red B_2} \cdots {\red B_n} + {\blue A_1}{\green C_2} {\red B_3} \cdots {\red B_n}
    + \cdots + {\blue A_1} \cdots {\blue A_{n-1}}{\green C_n} \label{ABD}\,.
\end{equation}
For $n=1$, there are no ${\blue A_j}$ or ${\red B_j}$ on the right hand side, and so Eq.~\eqref{ABD} reduces to Eq.~\eqref{ABD1}.

For example, if we take $\smash{{\blue A_j}={\blue {\frac{1}{p_j^2 +i \eps}} }}$, $\smash{{\red B_j}={\red {\frac{1}{p_j^2 - i \eps}} }}$
and $\smash{{\green C_j} = {\green{-2 \pi i \delta(p_j^2)}}}$, then Eq.~\eqref{ABD1} corresponds to the familiar relation
\begin{equation}
    {\blue{\frac{1}{p_j^2 +i \eps}} } - {\red{\frac{1}{p_j^2 - i \eps}}}= 
    {\green {-2 \pi i \delta(p_j^2)}}\,. \label{Fid}
\end{equation}
To be clear, this is an identity in the sense of distributions; it is the cutting equation for $\smash{\cM =  {\blue{\frac{1}{p_j^2 +i \eps}} }}$. In general, with this choice of  ${\blue A_j}, {\red B_j}$ and ${\green C_j}$,
the left hand side of Eq.~\eqref{ABD} corresponds to the difference between an integral with all $+i\eps$ propagators and one with all $-i \eps$ propagators, which is either  $\cM - \cMb$ or $\cM + \cMb$ depending on the number of loops. For an even number of loops, Eq.~\eqref{ABD} can be applied, but even then it produces some combination of propagators with $+i\eps$ propagators, 
some $-i \eps$ propagators and delta functions with no clear relation to Eq.~\eqref{Cutcov}.

To derive the cutting rules using Eq.~\eqref{ABD}, we use TOPT. Recall that any covariant Feynman diagram can be written as a sum over time-ordered diagrams.
In a time-ordered diagram, the internal lines are-on shell 
(meaning $q_0 = \omega_q = \sqrt{{\vec q}^{\,2}+m^2}$) and  three-momentum is conserved at each vertex, but energy is in general  {\it not} conserved at each vertex. The positive sign is always taken for the energy (in front of the square root), so that intermediate states are Fock-state elements of physical on-shell positive-energy particles. For example, the scalar loop can be written as\vspace{-10pt}
\begin{align} \label{bubbex}
&\begin{gathered}
\begin{tikzpicture}[baseline=-3.5]
\node at (0,0) {
\parbox{30mm} {
\resizebox{30mm}{!}{
 \fmfframe(0,00)(0,0){
 \begin{fmfgraph*}(60,40)
	\fmfleft{L1}
	\fmfright{R1}
	\fmf{plain}{L1,v1}
	\fmf{plain,foreground=(0,,0,,0.8),left=1,tension=0.5}{v1,v2,v1}
	\fmf{plain}{v2,R1}
\end{fmfgraph*}
}}}};
\node[below,scale=0.8] at (-1,0) {$p$};
\node[below,scale=0.8] at (1,0) {$p$};
\node[above,scale=0.8] at (0,0.4) {$k$};
\node[below,scale=0.8] at (0,-0.4) {$p-k$};
\end{tikzpicture}
\end{gathered}  
=
\int \frac{d^4k}{i(2\pi)^4}
{\blue{\frac{1}{k^2-m_1^2+i\eps}}}
{\blue{\frac{1}{(p-k)^2-m_2^2+i\eps}}}\\
&=  -\int \frac{d^3k}{(2\pi)^3}\frac{1}{2 \omega_k} \frac{1}{2 \omega_{p-k}}
\left[
{\blue{\frac{1}{E_p - (\omega_k + \omega_{p-k}) + i \eps}}} + 
{\blue{\frac{1}{E_p - (\omega_k + \omega_{p-k} + 2 \omega_p)+ i \eps}}} \right]\, , \nonumber
\end{align}
where $E_p=\omega_p =\sqrt{\vec{p}^2 + m^2}$ is the energy of $p^\mu$ and 
$\omega_{k} = \sqrt{\vec k^2+m_1^2}$, $\omega_{p-k} = \sqrt{(\vec p-\vec k)^2+m_2^2}$ are the energies of the virtual particles. 
Eq.~\eqref{bubbex} can be verified by performing the $k_0$ integral, which picks up two of the four poles. 
In terms of diagrams, we have
\begin{equation}
\begin{gathered}
\begin{tikzpicture}[baseline=-3.5]
\node at (0,0) {
\parbox{30mm} {
\resizebox{30mm}{!}{
 \fmfframe(0,00)(0,0){
 \begin{fmfgraph*}(60,40)
	\fmfleft{L1}
	\fmfright{R1}
	\fmf{plain}{L1,v1}
	\fmf{plain,foreground=(0,,0,,0.8),left=1,tension=0.5}{v1,v2,v1}
	\fmf{plain}{v2,R1}
\end{fmfgraph*}
}}}};
\node[below,scale=0.8] at (-1,0) {$p$};
\node[below,scale=0.8] at (1,0) {$p$};
\node[above,scale=0.8] at (0,0.4) {$k$};
\node[below,scale=0.8] at (0,-0.4) {$p-k$};
\end{tikzpicture}
\end{gathered}  
=
\begin{gathered}
\begin{tikzpicture}[baseline=-3.5]
\node at (0,0) {
\parbox{30mm} {
\resizebox{30mm}{!}{
 \fmfframe(0,00)(0,0){
 \begin{fmfgraph*}(60,40)
	\fmfleft{L1}
	\fmfright{R1}
	\fmf{plain}{L1,v1}
	\fmf{plain,foreground=(0,,0,,0.8),left=1,tension=0.5}{v1,v2,v1}
	\fmf{plain}{v2,R1}
\end{fmfgraph*}
}}}};
\node[below,scale=0.8] at (-1,0) {$\vec{p}$};
\node[below,scale=0.8] at (1,0) {$\vec{p}$};
\node[above,scale=0.8] at (0,0.4) {$\vec{k}$};
\node[below,scale=0.8] at (0,-0.4) {$\vec{p}-\vec{k}$};
\end{tikzpicture}
\end{gathered} 
+
\begin{gathered}
\begin{tikzpicture}[baseline=-3.5]
\node at (0,0) {
\parbox{20mm} {
\resizebox{15mm}{!}{
 \fmfframe(0,00)(0,0){
 \begin{fmfgraph*}(30,20)
 \fmfstraight
 \fmftop{t1,t2,t3,t4}
 \fmfbottom{b1,b2,b3,b4}
	\fmf{plain}{t1,t4}
	\fmf{plain}{b1,b4}
	\fmffreeze
	\fmf{plain,foreground=(0,,0,,0.8),left=0.3,tension=0.8}{t4,b1,t4}
\end{fmfgraph*}
}}}};
\node[above,scale=0.8] at (-0.4,0.4) {$\vec{p}$};
\node[right,scale=0.8] at (0.1,0.0) {$\vec{p}-\vec{k}$};
\node[left,scale=0.8] at (-0.5,0.2) {$\vec{k}$};
\node[below,scale=0.8] at (0.2,-0.4) {$\vec{p}$};
\end{tikzpicture}
\end{gathered} 
\label{bubblesOFPT}
\end{equation}
The intermediate state in the TOPT diagrams changes as each vertex is passed in time (where time flows to the right). In the first diagram this state includes only the $k$ and $p-k$ lines, so its energy is $\omega_k+\omega_{p-k}$; in the second diagram, the intermediate state includes also the energy of the initial and final states, and thus its energy is $\omega_k + \omega_{p-k} + 2 \omega_p$.

It is often difficult to perform the $k_0$ integrals to reduce Feynman diagrams to TOPT diagrams. Their equivalence is easiest to show from more general principles of quantum field theory, since both compute the same time-ordered products  (cf.~\cite{Sterman:1994ce,Schwartz:2013pla}). Keep in mind that although the $+i \eps$ is necessary to determine the $k_0$ integration contour, it cannot be removed after the integration is done. Indeed the $+i \eps$ originates from the fact that particles move forward in time with positive energy and is an essential part of the Lippmann-Schwinger propagator in TOPT. 

Now for each term in the TOPT decomposition we can apply the identity in Eq.~\eqref{ABD}, using the TOPT analog of Eq.~\eqref{Fid}:
\begin{equation}
    {\blue{\frac{1}{E_j - \omega_j +i \eps}} } - {\red{\frac{1}{E_j - \omega_j - i \eps}}}=  
    {\green{ -2 \pi i \delta(E_j - \omega_j)} } \, . \label{TOPTid}
\end{equation}
The sum of all TOPT diagrams with a given topology and all $+i\eps$ propagators gives the Feynman diagram $\cM$, while the sum of these diagrams with all $-i \eps$ propagators gives $\cMb$. The remaining terms have 
$\delta(E_j- \omega_j)$ factors which impose energy conservation at an intermediate time. These diagrams neatly split in two along the cut, with positive energy automatically flowing across the cut (because TOPT diagrams have positive energy at any intermediate time) and where all cut particles are on-shell (since all particles are on-shell in TOPT). By Eq.~\eqref{ABD} all the propagators before the cut have $+i \eps$ and those after the cut have $-i \eps$. Thus the cut TOPT diagram is one particular time-ordering of a white/black partition, which is one time-ordering of a cut Feynman diagram. The sum of all possible cut TOPT diagrams gives all the possible time-orderings of the black and white vertices, and therefore reproduces the full cut Feynman diagram and confirms the cutting rules.

 For example, when we apply Eq.~\eqref{ABD} to Eq.~\eqref{bubblesOFPT}, there is only one intermediate state in each diagram to cut (in contrast to the Feynman diagram, which has two intermediate propagators to cut). Cutting the first diagram gives%
\vspace{-10pt}
\begin{align}
&\hspace{-0pt} 
\begin{tikzpicture}[baseline=-3.5]
\node at (0,0) {
\parbox{30mm} {
\resizebox{30mm}{!}{
 \fmfframe(0,00)(0,0){
 \begin{fmfgraph*}(60,40)
	\fmfleft{L1}
	\fmfright{R1}
	\fmf{plain}{L1,v1}
	\fmf{plain,left=1,tension=0.5,foreground=(0,,0.5,,0.0)}{v1,v2}
	\fmf{plain,left=1,tension=0.5,foreground=(0.0,,0.5,,0.0)}{v2,v1}
    \fmf{plain}{v2,R1}
\end{fmfgraph*}
}}}};
\draw[dashed, line width = 1, dashed, darkgreen] (0,-0.8) to (0,0.8);
\end{tikzpicture}
= \int \frac{d^3k}{(2\pi)^3}\frac{1}{2 \omega_k} \frac{1}{2 \omega_{p-k}}
(-2\pi i) \delta(E_p - (\omega_k + \omega_{p-k}))\\
&=\int \frac{d^4k}{i(2\pi)^4}\int \frac{d^4 k'}{i(2\pi)^4} \left(-2\pi i\right) (2\pi)^3\delta^4(p-k-k') (-2\pi i) \delta(k^2-m_1^2)\Theta(k_0)(-2\pi i)\delta(k'{}^2-m_2^2)\Theta(k'_0)\,.\nonumber
\end{align}
So this diagram alone gives the cut of the Feynman diagram. The cut of the other diagram is zero, since energy conservation at the cut is impossible to satisfy:
\begin{equation}
    \begin{gathered}
\begin{tikzpicture}[baseline=-3.5]
\node at (0,0) {
\parbox{20mm} {
\resizebox{15mm}{!}{
 \fmfframe(0,00)(0,0){
 \begin{fmfgraph*}(30,20)
 \fmfstraight
 \fmftop{t1,t2,t3,t4}
 \fmfbottom{b1,b2,b3,b4}
	\fmf{plain}{t1,t4}
	\fmf{plain}{b1,b4}
	\fmffreeze
	\fmf{plain,foreground=(0,,0.5,,0.0),left=0.3,tension=0.8}{t4,b1,t4}
\end{fmfgraph*}
}}}};
\draw[dashed, line width = 1, dashed, darkgreen] (-0.2,-0.8) to (-0.2,0.8);
\end{tikzpicture}
\end{gathered} \hspace{-10pt}
= \int \frac{d^3k}{(2\pi)^3}\frac{1}{2 \omega_k} \frac{1}{2 \omega_{p-k}}
(-2\pi i) \delta(E_p - (2 \omega_p + \omega_k + \omega_{p-k}))
=0 \, .
\end{equation}
This is typical of TOPT graphs: when one time-ordering can be cut, the same diagram with vertices in reversed time order cannot be cut. 

More broadly, the key reason why the cutting rules can be derived diagrammatically in TOPT is that cuts in TOPT are associated with internal multiparticle states, not individual particles. So a cut, which replaces a TOPT propagator by a delta function, splits the diagram in two, ordered by time, in contrast to Feynman diagrams, where using Eq.~\eqref{ABD} just opens up a loop. 

In fact, we have  derived something stronger than the covariant cutting rules: the constraint on the amplitude holds for each time-ordered Feynman diagram separately {\it and} it holds point-by-point in phase space. Indeed, the equation that we use to prove it, Eq.~\eqref{ABD} holds at the integrand level. Let us define an individual TOPT integrand for fixed loop momenta as
\begin{equation}
\toptM \equivD
{\blue{\frac{1}{E_1 - \omega_1  + i \eps} }} \cdots {   \blue{\frac{1}{E_n - \omega_n + i \eps} }},
\qquad
\toptMb \equivD {\red {\frac{1}{E_1 - \omega_1  - i \eps} }} \cdots    {\red {\frac{1}{E_n - \omega_n - i \eps} }}\,. \label{eq:TOPT_integrand}
\end{equation}
Then, by putting in the explicit form of the TOPT propagators, Eq.~\eqref{ABD} gives
what we call the {\it time-ordered cutting rules}:
\begin{multline}
\toptM - \toptMb = 
 \sum_j
\blue{\frac{1}{E_1 - \omega_1  + i \eps} } \cdots \blue{\frac{1}{E_{j-1} - \omega_{j-1}  + i \eps} }  \green{ (-2\pi i) \delta(E_j-\omega_j) } \label{ABDT} \\
\times \red{\frac{1}{E_{j+1} - \omega_{j+1} - i \eps} } \cdots \red{\frac{1}{E_n - \omega_n - i \eps} } \,.
\end{multline}
When the loop momenta are integrated over, this equation implies the cutting rules, but this equation holds for any $E_j$ and $\omega_j$.

\section{Discontinuities \label{sec:discon}}
Having understood the cutting rules in covariant perturbation theory and in time-ordered perturbation theory, we can now proceed to connect cuts to the discontinuities of amplitudes. As discussed above, the Feynman integral $\cM$, viewed as an analytic function of Mandelstam invariants, is a multi-valued function on a complex manifold. Cutkosky showed that one can compute the discontinuity of $\cM$ across some region of its Landau surface by summing over integrals in which different sets of propagators have been cut. However, to provide practical constraints on amplitudes we need a prescription much more explicit than Cutkosky's. For example, how do we identify what region of the surface we are probing from knowledge of which Feynman propagators have been cut? And how do we actually perform the analytic continuation around the relevant branch points? 

There are two related concepts that we will discuss, and which we want to connect. The first is the total discontinuity of a Feynman integral in a particular {\it region}, which is computed by the covariant cutting rules. A region in this context is the specification of the signs of the Mandelstam invariants, and the signs of the energies (which particles are incoming and which are outgoing), if necessary. Once the signs are specified, we can compute the total discontinuity using Eq.~\eqref{Cutcov}. The second concept is the discontinuity of a Feynman integral with respect to a particular kinematic invariant $s$. More specifically, we define $\disc_s \cM$ as the difference between $\cM$ before and after analytic continuation along a path that encircles the branch point in $s$ (but no other branch points). Since Mandelstam invariants are not all independent, this has to be done with some care.

\subsection{Covariant approach \label{sec:covariantdisc}}
We begin with the total discontinuity $\cM - \cMb$, which can be computed using the covariant cutting rules in Eq.~\eqref{Cutcov}. As defined in Eqs.~\eqref{Mdef} and~\eqref{Mdef2}, $\cM$ is a Feynman integral with all $+i\eps$ propagators and $\cMb$ is the same integral with all $-i\eps$ propagators, multiplied by a factor of $(-1)^L$. At any real phase-space point, $\cM$ and $\cMb$ are complex conjugates of each other for finite values of $i\eps$. From this point of view, $\cM$ and $\cMb$ are separated by a branch cut at $i \eps = 0$, and may have a finite difference as $i\eps \to 0$ from the positive or negative direction. 
In contrast, viewed as an analytic function of the momenta, $\cM$ and $\cMb$ are evaluations of the same function at different points on a complex manifold. Thus the finite difference between $\cM$ and $\cMb$ can be thought of as the discontinuity of a single function $\cM$. We would like to understand the analytic continuation contour along which $\cM$ can be transformed into $\cMb$, as this will allow us to connect the total discontinuity computed by the covariant cutting rules to the notion of discontinuities with respect to particular Mandelstam invariants. 

The branch cut between $\cM$ and $\cMb$ starts at a branch point (more generally, a branch hypersurface) somewhere in the space of Mandelstam invariants on which $\cM$ depends. As such, the discontinuity can be computed by analytically continuing $\cM$ around this branch point to the other side of the branch cut. To do this, we can continue $\cM$ into a regime where it is analytic, and then to the region where it matches $\cMb$. 
For example, suppose $\cM = \ln(-s + i \eps)$ and $\cMb = \ln(-s - i\eps)$, and take $s>0$. Then we can continue $\cM$ along the path $s \to e^{i \alpha} s$ with $0\le \alpha \le \pi$ to the region where $s<0$. From this region we can either go back and reproduce $\cM$ using $s \to e^{-i \alpha} s$, or keep going to arrive at $\cMb$ on the other side of the branch cut using $s \to e^{i \alpha} s$. We can also continue increasing the phase of $s$ in this manner: as $\alpha$ increases, we end up on higher and higher sheets of the Riemann surface of $\ln (-s)$. A single discontinuity  corresponds to the single monodromy around the branch point at $s=0$. In equations, for the logarithm we have $\disc_s \cM = \tDisc \cM = \cM - \cMb = 2i \im \cM$ in the region where $s>0$.

A useful concept for studying the analytic properties of Feynman integrals is the Euclidean region. In this region, all Mandelstam invariants are negative and $\cM$ is analytic. 
To see that integrals are analytic in the Euclidean region, it is helpful to write a general Feynman integral in the Symanzik representation~\cite{Smirnov:2004ym}. This is done by using Feynman parameters and then integrating over the loop momenta. The result is that a Feynman amplitude as in Eq.~\eqref{Mdef} can be written as
\begin{equation}
  \cM (p)
  = \int\limits_{x_j\ge0} \prod_j d x_j\; \delta\big(1-\sum x_j\big)\; \frac{\cU^{n-2L-2}}{\cF^{n-2 L}}\,.
\end{equation}
Here, the first Symanzik polynomial $\cU$ is 
\begin{equation}
    \cU = \sum_{\cT_1} \Big[ \prod_{j \not \in \cT_1} x_j \Big] \, ,
\end{equation}
where the sum is over all 1-trees $\cT_1$, which correspond to tree diagrams that connect all vertices in the graph. The second Symanzik polynomial $\cF$ is
\begin{equation}
\cF = \sum_{\cT_2} \Big[  \prod_{j \not \in  \cT_2 } x_j \Big] ( -s_{P(\cT_2)} ) + \cU \sum_{j=1} x_j m_j^2 - i \eps,
\qquad
s_{P(\cT_2)} = \Big[ \sum_{j \not \in \cT_2} p_j \Big]^2 \, ,
\label{Fsym}
\end{equation}
where $m_j$ are the masses of the internal lines and the sum is over 2-trees $\cT_2$,  which themselves correspond to pairs of disconnected tree diagrams that involve all vertices of the original graph.  
The nice thing about this parametrization is that $\cM$ is now manifestly a function of Mandelstam invariants.

Singularities in $\cM$ can only arise when $\cF=0$. Since the integration region corresponds to $x_i \ge 0$ and in the Euclidean region $s_{P(\cT_2)} < 0$ for all $\cT_2$ and $m_j^2 \ge 0$ for all $j$, the denominator will never vanish and the result will be analytic in the external momenta. Note that the Euclidean regime is identified with a stronger requirement than that $\cM$ is analytic; it requires that all Mandelstam invariants are negative, not just those associated with 2-trees from a particular graph. We denote the Euclidean region by $R^\star$.

We denote generic regions, in which kinematic invariants can be positive or negative, by $R$. We use the more precise notation $R_+$ to indicate a region in which all positive invariants are slightly above the associated branch cut, i.e.~all propagators have $+ i \eps$. The region in which all positive invariants are instead below the associated branch cut, and all propagators have $-i\eps$, will be denoted $R_-$.\footnote{With Feynman propagators, the amplitude in this region also has a $(-1)^L$ in $R_-$ due to the additional rotation of the energies in the loop integrals. With TOPT, we simply flip $i\eps \to -i\eps$ as there are no energies in the loop integrals.} Thus, we write
\begin{equation}
    \left[\tDisc \cM\right]_R = \left[\cM - \cMb\right]_R = \cM_{R_+} - \cM_{R_-}\,.
\end{equation}
To compute the right hand side, we would like to understand how to analytically continue the amplitude between $R_+$, $R^\star$, and $R_-$. There are many ways to do this. The precise path should not affect the answer for the discontinuity. It is nevertheless important to know that the path exists, and having an explicit path can help determine which branch points are encircled. 

Since $\cM$ is Lorentz invariant, it may seem most natural to continue the invariants themselves. 
For example, we can rotate all the positive invariants to negative values via
$s_I \to e^{i \alpha} s_I$ with $0 < \alpha < \pi$ while leaving the negative invariants stationary.
This puts us in $R^\star$, where all $s_I <0$ and the amplitude is nonsingular. We can then keep going, and analytically continue all the invariants that were originally positive further by extending $0 < \alpha < 2\pi$, to end up in $R_-$. Unfortunately, since the invariants are not all independent, this procedure can be ambiguous. For example, in massless four-particle kinematics, if we want to rotate $s$ from being positive to negative while holding the other invariants fixed, we could try the above analytic continuation path. But if we rewrite our amplitude or integral to depend just on the other invariants using the relation $s=-t-u$, this rotation would seem to have no effect. Thus, one must be careful to do the rotation in a manner that respects the reparameterization invariance of the integrals. 

In this paper, we will restrict ourselves to analytic continuations in external energies that respect overall energy conservation and leave all external three-momenta fixed. In addition to avoiding the issue described above, this choice facilitates our derivation of relations between sequential discontinuities and cut integrals, and leads to unambiguous predictions. 
In addition, rotating the energies while respecting four-momentum conservation ensures that we always satisfy any Gram determinant constraints.

\begin{figure}
    \centering
    \includegraphics[width=7cm]{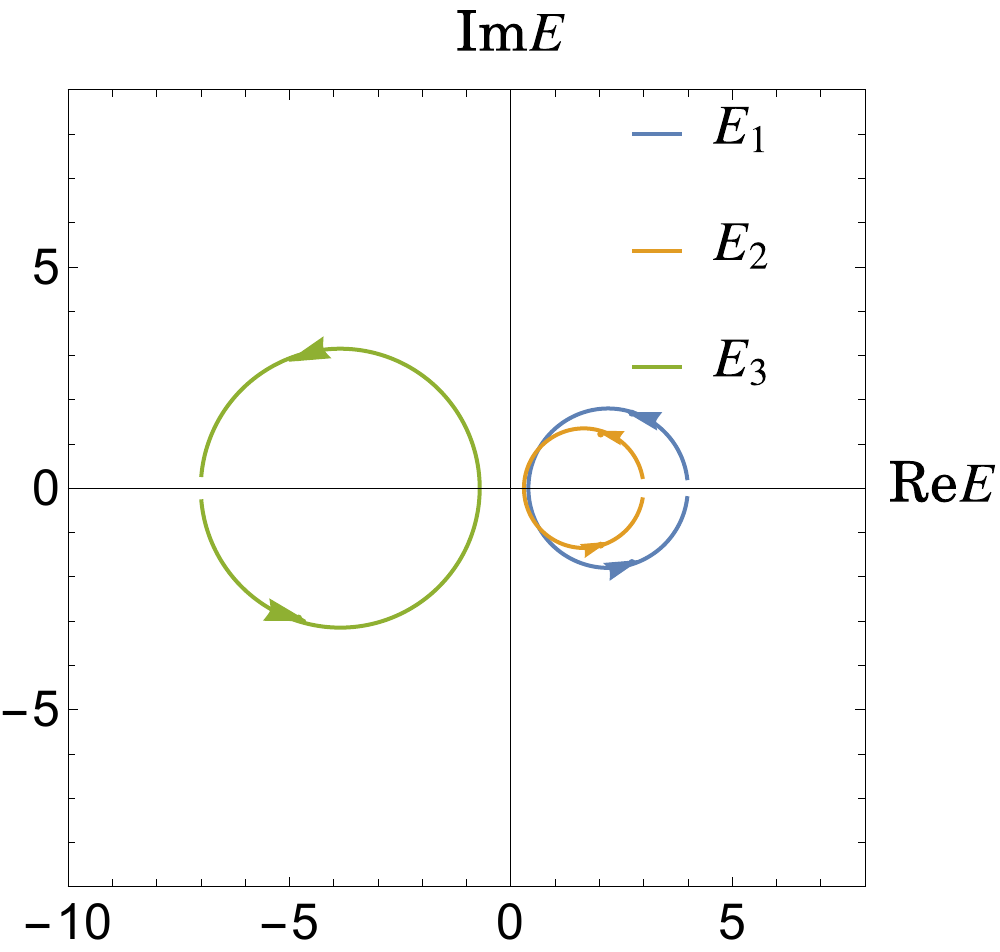}
    ~~~
    \includegraphics[width=7.2cm]{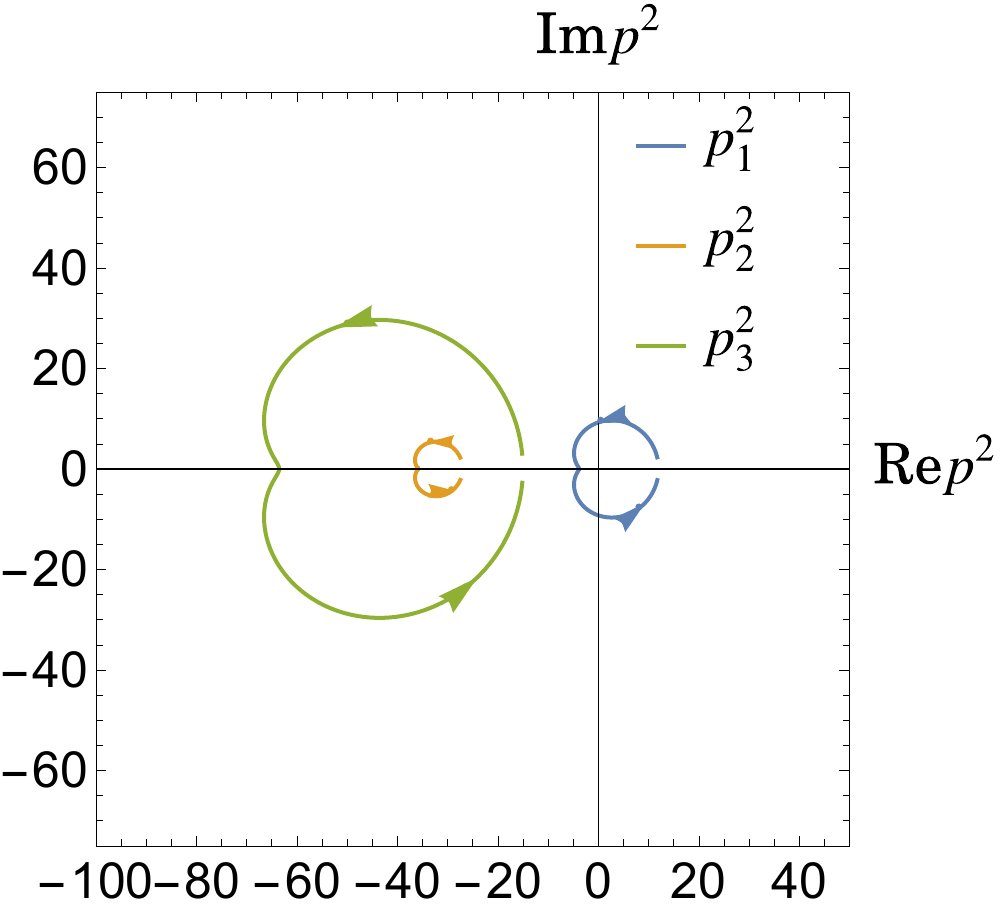}
    \caption{Example analytic continuation involving three external energies. We start at the kinematic point $p_1=(4,2,0,0), p_2=(3,6,0,0)$, and $p_3=(-7,-8,0,0)$, where we have $p_1^2 = 12,~p_2^2 = -27$, and $p_3^2=-15$. We rotate the energies by $E_j \to 0.1 + 0.9 e^{i\pi s} \cos(\pi s) E_j$ with $0 \le s< 1$. During this rotation the positive invariant $p_1^2$ circles its branch point at $p_1^2 =0$, thus taking us from $R^1 \to R^\star \to R^1$, but changing the sign of the corresponding $i \eps$. The small gaps at the beginning and end of the paths represent the $\pm i\eps$.
    }
    \label{fig:EandSRot}
\end{figure}

In general, there are many ways to rotate external energies to get from a region $R$ to the Euclidean region. For example, if the momenta in $R$ all take non-exceptional values, one can uniformly lower the energies $E_j \to \alpha E_j$ with $\alpha < 1$. Eventually, at some point $\alpha_\text{min}$ all the invariants become negative. One can then rotate the energies in the complex plane around $\alpha_\text{min} E_j$ and return to $\alpha=1$ on the opposite side of the real energy axis. This procedure respects energy-momentum conservation everywhere along the path. One only has to be careful that the invariants do not encircle their branch points twice.
A concrete example involving three momenta that follows a path homotopic to the one described in this paragraph is shown in Fig~\ref{fig:EandSRot}. We construct a number of similar paths for the examples we consider in Section~\ref{sec:examples}.

Let us now assume that an appropriate analytic continuation in the energies has been chosen, which takes us from a region $R_+$ to the corresponding region $R_-$ (where all Mandelstam invariants have the same sign, but each $+i\eps$ has been changed to $-i\eps$). Then the difference between $\cM$ before and after this analytic continuation should match the total discontinuity of a Feynman integral in the region $R$ using the covariant cutting rules:
\begin{equation}
    \left[
    \tDisc
    \cM \right]_{R_+}= \cM_{R_+} - \cM_{R_-} = \sum_{\text{cuts}} \cM_{R_{+|-}}
\label{discR}
\end{equation}
We emphasize the right side of this equation involves a sum over all cuts (in all channels), as explicitly given in~Eq.~\eqref{Cutcov}. When we cut a set of propagators, we replace each one by
\begin{equation}
    \text{cut}: \frac{1}{p^2-m^2+i\eps}  \to -2\pi i \delta(p^2-m^2) \theta(p_0)
\end{equation}
and use $+i\eps$ for all propagators before the cut and $-i\eps$ for all propagators after the cut, as implied by the subscript on $R_{+|-}$.

We would now like to derive a concrete relation between $\disc_s\cM$, and the cuts of $\cM$. The discontinuity of $\cM$ with respect to $s$ corresponds to analytically continuing $\cM$ from being evaluated at $s+i\eps$ to being evaluated at $s - i\eps$, while the other invariants remain unchanged. 
Let us denote the region in which $s>0$ and all other kinematic invariants are negative by $R^s$. As only the invariant $s$ is positive in this region, all the nonzero cuts in the sum in Eq.~\eqref{discR} are in the $s$-channel. As a result, we have
\begin{equation}
    \left[\tDisc
    \cM \right]_{R^s}= \cM_{R^s_+} - \cM_{R^s_-} = \sum_{\text{cuts}} \cM_{R_{+|-}}=  \sum_{\text{cuts in } s} \cM_{R_{+|-}}\,.
\label{discs}
\end{equation}
To further connect this sum of cut integrals to $\disc_s \cM$, we must show that the analytic continuation corresponding to $\tDisc$ in this region  encircles a  branch point in only in $s$, and in no other invariants. This turns out to be easiest to see in TOPT, which we turn to now.

\subsection{Discontinuities in TOPT \label{sec:discTOPT}}

In TOPT all internal lines are on-shell with positive energy and real masses ($q_0 >0$ and $q^2\ge 0$). External lines, however, have no such restriction; they can have $p^2 <0$ if the diagram is meant to be embedded in a larger diagram (for example, the off-shell photon in deep-inelastic scattering is spacelike), and incoming external particles can have negative energy if they correspond to outgoing particles. 

Because we are ultimately interested in the analytic properties of Feynman integrals as functions of external energies, it is helpful to separate out the contributions to TOPT propagators from internal and external lines. In particular, we can put each TOPT propagator in the form $\smash{\blue{1/(E_P - \omega_q + i\eps)}}$, where  $E_P$ corresponds to a sum over external energies, and
$\smash{\omega_q = \sum_j \omega_j}$ is a sum over particles in internal lines, where $\smash{\omega_j = \sqrt{\vphantom{A^A}\smash{ \vec{q_j}^{ 2} + m_j^2} }}$.

Consider for example the one-loop TOPT graph, with all $E_i >0$:
\begin{equation}
\begin{gathered}
\begin{tikzpicture}[baseline=-3.5]
\node at (0,0) {
\parbox{30mm} {
\resizebox{30mm}{!}{
 \fmfframe(0,00)(0,0){
 \begin{fmfgraph*}(80,50)
    \fmfstraight
	\fmfleft{L1}
	\fmfright{R1,R2}
	\fmf{plain_arrow,label=$p_1$}{L1,v1}
	\fmf{plain,tension=0.3,label=$q_1$}{v1,v2}
	\fmf{plain,tension=0.3,label=$q_3$,l.s=right}{v2,v3}
	\fmf{plain,tension=0.3,label=$q_2$,l.s=right}{v3,v1}
	\fmf{plain_arrow,tension=0.35}{v2,R1}
	\fmf{plain_arrow,tension=1.25}{v3,R2}
	\fmfv{label=$p_2$}{R2}	
	\fmfv{label=$p_3$}{R1}
\end{fmfgraph*}
}}}};
\draw[dashed, line width = 1, dashed, darkgreen] (0.3,-1) to (0.3,1);
\draw[dashed, line width = 1, dashed, darkgreen] (0.775,-1) to (0.775,1);
% \node[above,darkblue] at (0,0.4) {$+i \eps$};
% \node[below,darkblue] at (0,-0.4) {$+i \eps$};
\end{tikzpicture}
\end{gathered}
={\blue{\frac{1}{E_1 - (\omega_1  +\omega_2) + i\eps}}}
{\blue{\frac{1}{(E_1-E_3) - (\omega_2  +\omega_3) + i\eps}}}\,.
\label{tri1}
\end{equation}
In the first propagator, $E_P = E_1=E_2+E_3$ and $\omega_q = \omega_1+\omega_2$ while in the second propagator $E_P = E_1-E_3=E_2$ and $\omega_q = \omega_2 + \omega_2$. If we had drawn $p_2$ and $p_3$ as incoming lines with negative energy, the diagram would have been more awkward to draw, but we would have found an equivalent expression:
\begin{equation}
\begin{gathered}
\begin{tikzpicture}[baseline=-3.5]
\node at (0,0) {
\parbox{30mm} {
\resizebox{30mm}{!}{
 \fmfframe(0,00)(0,0){
 \begin{fmfgraph*}(80,50)
    \fmfstraight
	\fmfleft{L1,L2,L3}
	\fmfright{R1,R2}
	\fmf{plain_arrow,label=$p_1$}{L2,v1}
	\fmf{plain,tension=0.3}{v1,v2}
	\fmf{plain,tension=0.3}{v2,v3}
	\fmf{plain,tension=0.3}{v3,v1}
	\fmf{phantom,tension=0.4}{v2,R1}
	\fmf{phantom,tension=1.2}{v3,R2}
	\fmffreeze
	\fmf{plain_arrow,label=$p_3$}{L1,v2}
	\fmf{plain_arrow,label=$p_2$,l.s=left}{L3,v3}
% 	\fmfv{label=$p_3$}{L1}	
% 	\fmfv{label=$p_2$}{L3}
\end{fmfgraph*}
}}}};
\draw[dashed, line width = 1, dashed, darkgreen] (0.3,-1) to (0.3,1);
\draw[dashed, line width = 1, dashed, darkgreen] (0.8,-1) to (0.8,1);
% \node[above,darkblue] at (0,0.4) {$+i \eps$};
% \node[below,darkblue] at (0,-0.4) {$+i \eps$};
\end{tikzpicture}
\end{gathered}
={\blue{\frac{1}{E_1 - (\omega_1  +\omega_2) + i\eps}}}
{\blue{\frac{1}{(E_1+E_3) - (\omega_2  +\omega_3) + i\eps}}}\,.
\end{equation}
The value of the diagram is the same since we have flipped $E_3 \to -E_3$.

For a general TOPT graph, the energies $E_I$ appearing in the amplitude have a natural sequence. We begin with the total initial-state energy on the far left. Each time a vertex connecting to an external momentum is passed, the external energy is either added, if it is incoming, or subtracted, it if it is outgoing. If the vertex is purely internal, then $E_P$ does not change. For example, consider this graph:
\vspace{5pt}\begin{equation}
\begin{gathered}
\begin{tikzpicture}[baseline=-3.5]
\node at (0,0) {
\parbox{30mm} {
\resizebox{30mm}{!}{
 \fmfframe(0,00)(0,0){
 \begin{fmfgraph*}(80,50)
    \fmfstraight
	\fmfleft{L1,L2}
	\fmfright{R1,R2,R3}
	\fmf{phantom}{L2,v1}
	\fmf{plain_arrow}{L1,v7}
	\fmf{phantom}{v3,R3}
	\fmf{plain_arrow}{v4,R2}
	\fmf{plain_arrow}{v5,R1}
	\fmf{plain,tension=0.6}{v1,v2,v3,v4,v5,v7,v1}
	\fmffreeze
	\fmf{plain_arrow}{L2,v3}
	\fmf{plain}{v1,vx}
	\fmf{phantom,tension=1.5}{vx,vy}
	\fmf{plain_arrow,tension=0.7}{vy,R3}
	\fmf{plain,tension=0.3}{v2,v6}
	\fmf{plain,tension=0.3}{v7,v6}
	\fmf{plain,tension=0.8}{v5,v6}
 	\fmfv{label=$p_5$}{R3}	
 	\fmfv{label=$p_4$}{R2}	
 	\fmfv{label=$p_3$}{R1}
 	\fmfv{label=$p_2$}{L2}	
 	\fmfv{label=$p_1$}{L1}
\end{fmfgraph*}
}}}};
\draw[dashed, line width = 1, dashed, darkgreen] (-0.7,-1) to (-0.7,1);
\draw[dashed, line width = 1, dashed, darkgreen] (-0.3,-1) to (-0.3,1);
\draw[dashed, line width = 1, dashed, darkgreen] (0.2,-1) to (0.2,1);
\draw[dashed, line width = 1, dashed, darkgreen] (0.6,-1) to (0.6,1);
\draw[dashed, line width = 1, dashed, darkgreen] (0.85,-1) to (0.85,1);
\draw[dashed, line width = 1, dashed, darkgreen] (1.05,-1) to (1.05,1);
\end{tikzpicture}
\end{gathered}
% \quad
% \sim \frac{1}{-E_5-\omega_1}
% \frac{1}{E_1-E_5-\omega_2}
% \frac{1}{E_1-E_5-\omega_3}
% \frac{1}{E_1-E_5-\omega_3}
\vspace{5pt}\end{equation}
The initial energy is $E_1+E_2$ and the energy past the first vertex is $E_1+E_2+E_5+\omega_1+\omega_2+\omega_3$ for some internal energies $\omega_j$; this first propagator depends on the difference between these energies $E_P=-E_5$. The sequence of $E_P$ values as we move forward in time is 
\begin{equation}
    -E_5,~~  
    E_1\!-\!E_5,~~
    E_1\!-\!E_5,~~ 
    E_1\!-\!E_5,~~
    E_1\!-\!E_5\!-\!E_3,~~
    E_1\!-\!E_5\!-\!E_3\!+\!E_2\,.
    \label{Eseq}
\end{equation}
If we took all momenta to be incoming, then we would flip the sign of $E_3, E_4$, and $E_5$ so all the signs in Eq.~\eqref{Eseq} would be positive. 
The corresponding sequence is
\begin{equation}
    5 \to 1 \to 3 \to 2 \to 4 \, .
\end{equation}
If we use energy conservation to rewrite the energy sum (i.e. $E_5 = -E_4 -E_3 -E_2 -E_1)$, the sequence would be the same, in the opposite direction: 
$5 \leftarrow 1 \leftarrow 3 \leftarrow 2 \leftarrow 4$. The fact that the energies appearing in each successive propagator are a subset of the energies that appeared the preceding propagators (or vice versa) will be important to proving the Steinmann relations in Section~\ref{sec:sequential}.

Each energy $E_I$ is the energy of a four-vector $P_I^\mu =\sum_{i\in I} \pm p_i^\mu$. Thus there is a one-to-one correspondence between invariants $s_I = P_I^2$ and these energies. 
A TOPT propagator  ${\blue{1/(E_I - \omega_q + i\eps)}}$ can only become singular  when $E_I = \omega_q$, which only happens if $s_I>0$. To check this claim, note that the three-momentum $\vec P_I$ is the same as the sum of the three-momenta of all the internal particles contributing to $\omega_q$, namely $\vec P_I = \sum_j \vec{q}_j$. So we have two four-vectors, $\smash{P^\mu_I=(E_I,\vec{P}_I)}$ and $\smash{q^\mu = (\omega_q, \vec{P}_I)}$, with the same three-momentum. Recall that $\omega_q$ is the sum of the (positive) energies of the on-shell internal lines. Thus, the four-vector $q^\mu$ must be timelike, $q^2 >0$, since it corresponds to the sum of four-momenta of physical on-shell particles. Therefore, $P^\mu_I$ must be timelike when $E_I =\omega_q$. So if $s_I =P_I^2 <0$ then $E_I \ne \omega_q$. Thus the TOPT propagators can go on-shell only in the kinematical regions where there are singularities in the full Feynman integral, namely when $s_I>0$. As a corollary, we can drop the $+i\eps$ in any TOPT propagator corresponding to a negative invariant.

Now let us discuss how to take the discontinuity of a TOPT graph. A TOPT graph is a product of propagators of the form ${\blue{1/(E_I - \omega_q + i \eps)}}$. To take the discontinuity in the channel $s_I$ associated with $E_I$, we want to analytically continue $E_I$ around the pole of this propagator. More precisely, we want to continue $E_I$ around the branch point $E_I^\star$ at the end of the line of possible values of $\omega_q$ for a given external momentum. This branch point  $E_I^\star$ is at least as large as the magnitude of the momentum in the channel, $E_I^\star \ge |\vec{P}_I|$ but can be strictly larger, for example, if the internal lines are massive. The analytic continuation between $R_+$ and $R_-$ should have all the energies pass around their branch points, holding the three-momenta fixed and respecting energy conservation. 

Taking the difference between a single TOPT propagator before and after this analytic continuation gives
 \begin{equation}
\tDisc  {\blue{   \frac{1}{E_I - \omega_q + i\eps}}}
= {\blue{   \frac{1}{E_I - \omega_q + i\eps} }} - \frac{1}{E_I - \omega_q - i\eps} = {\green{ -2\pi i \delta(E_I-\omega_q)}}\,,
 \end{equation} 
as expected. %
Similarly, taking the difference between a generic TOPT graph $M$ before and after analytically continuing from $R \to R^\star \to R$ using a path that encircles the branch points in all of the energies, we get 
\begin{equation}
      \tDisc \toptM = 
       \sum_j
\blue{\frac{1}{E_{I_1} - \omega_1  + i \eps} } \cdots   \green{ (-2\pi i) \delta(E_{I_j}-\omega_j) }\cdots \red{\frac{1}{E_{I_n} - \omega_n - i \eps} } \, .\label{Maadisc} 
\end{equation}
If we sum over all TOPT diagrams with the same topology, this reproduces the covariant cutting rules for the total discontinuity of the corresponding Feynman integral $\cM$. That is, we have shown that Eq.~\eqref{Cutcov} holds with the left-hand side explicitly written as a discontinuity, and have thereby rederived Eq.~\eqref{discR} using TOPT.

Eqs.~\eqref{discR} and~\eqref{Maadisc} hold in any region $R$. Let us now focus on the region $R^s$, where only the Mandelstam invariant $s$ is positive, and all other invariants are negative. 
Since we have shown that singularities in TOPT diagrams only arise when the energy and corresponding invariant are positive ($P^0 > 0$ and $s = P^2 >0$), in the region $R^s$ there can only be singularities associated with $s$. In other words, as we continue from $R^s_+$ to $R^\star$ and back to $R^s_-$, we can only pass around branch points associated with $s$. This is what we set out to show at the end of the last subsection.  
As a result, we can now write
\begin{equation}
\left[\disc_s \cM \right]_{R^s} = \left[\tDisc \cM \right]_{R^s} =
  \sum_{\text{all cuts}} \cM_{R_{+|-}}
= \sum_{\text{cuts in } s} \cM_{R_{+|-}} \, .
\label{discRs}
\end{equation}
Stated more formally, what we have shown is that the analytic continuation used to compute $\disc_s \cM$ is homotopic to the path used to compute $\tDisc \cM$ in the region $R^s$.

We would next like to generalize this formula to the case of sequential discontinuities, in the same or different channels. Unfortunately, we cannot simply repeat the procedure that allowed us to compute the first discontinuity. The problem is that this first discontinuity takes the difference of two functions on the branch cut, and thus seems to be only defined on the branch cut itself. For example, $\disc \ln^2(s) =4\pi i \ln|s|\Theta(-s)$ is only defined for negative real $s$, where the branch cut is. In addition, when we take a cut, we turn all the propagators beyond the cut from $+i\eps$ to $-i\eps$. What is then the right way to cut a $-i\eps$ propagator?
To proceed, we will now describe a more sophisticated set of mathematical tools that will allow us to analytically continue Feynman integrals beyond the cut plane. This will allow us to take sequential discontinuities of Feynman integrals.

\section{Discontinuities as monodromies \label{sec:math}}
The $\pm i\eps$ notation in Feynman propagators is sufficient to compute single discontinuities of Feynman integrals, because this first discontinuity computes the difference between the value of the integral on different sides of a branch cut. For sequential discontinuities, we must explore a larger swath of the analytic structure of the various polylogarithmic functions that appear in a given Feynman integral.\footnote{While more general types of functions are known to appear in Feynman integrals, we leave these generalizations to future work.} The $\pm i \eps$ notation is not sufficient to describe this structure. Thus, in this section we review how polylogarithmic functions can be analytically continued beyond the principal branch, and how the resulting functions can be related back to the $\pm i \eps$ prescription. We also discuss how these types of analytic continuations can be carried out on TOPT propagators. 

\subsection{Warm-up: the natural logarithm}
Consider first the natural logarithm. It can be defined in the region $|s-1|<1$ by the sum
\begin{align} \label{eq:log_sum_def}
\ln s \equivD -\sum_{n=1}^\infty \frac{1}{n}(1-s)^n \qquad\text{for    }|s-1|<1\,.
\end{align}
To define $\ln s$ outside the region $|s-1|<1$, one can series expand Eq.~\eqref{eq:log_sum_def} around points other than $s=1$ that are within the original region of convergence to find sum representations that are valid beyond this region. Iterating this procedure, one can extend the function $\ln s$ to the entire complex plane, excluding a curve going from the origin to infinity (the branch cut). This is called the cut complex plane. Since the cut complex plane is simply connected, this analytic continuation is uniquely defined, once the location of the branch cut has been chosen.
While the shape of this branch cut is in principle arbitrary, some of this arbitrariness can be removed if we ask that the continued logarithm satisfy the reality property \(f(\bar{s}) = \overline{f(s)}\). The standard branch cut choice for the logarithm, going from 0 to $-\infty$ along the real $s$ axis, is consistent with this requirement. We call $\ln s$ with this choice of branch cut the {\it principal branch} of the logarithm.

With the standard placement of the branch cut for $\ln s$ along the negative real axis, the value of $\ln s$ for negative real $s$ is usually defined to mean the function produced by analytic continuation going counterclockwise from the positive real axis. Moreover, the discontinuity of the logarithm, which computes the difference between the value of this function just above and below the negative real axis, is given by
\begin{equation}
\disc_s \ln s =   \ln(s+i\eps) - \ln(s-i\eps) = 2\pi i\, \Theta(-s)\,. \label{logrel}
\end{equation}
The fact that this discontinuity is nonzero for negative values of $s$ illustrates the ambiguity in defining $\ln s$ on this part of the real line. The non-analytic Heaviside function  $\Theta(-s)$ should be thought of as an indication of the domain on which the discontinuity is defined: the right-hand side is {\it only} defined for real $s<0$; it is not a well-defined function on the rest of the complex plane. This is consistent with the way discontinuities were calculated in the previous section, as the only way to analytically continue a function back to the same point in the cut complex plane is if we start and end on the cut.

The $\pm i\eps$ notation is sufficient for indicating which side of a branch cut we are on when we restrict ourselves to the principal branch of a function. However, when taking additional discontinuities, the $\pm i\eps$ notation and the associated non-analytic theta function are problematic. The single logarithm is a bit too simple, but already $\ln^2 s$ demonstrates the problem. Its discontinuity is
\begin{equation}
    \disc_s \ln^2 s = \ln^2(s+i\eps) - \ln^2(s-i\eps)
    =4\pi i \ln |s|\,\Theta(-s) \, .
\end{equation}
As with $\ln s$, the discontinuity of $\ln^2 s$ is only nonzero for real $s<0$, since otherwise $\ln^2(s+i\eps)$ and $\ln^2(s-i\eps)$ agree. But if this discontinuity is only nonzero on the negative real axis, further analytic continuations are ambiguous, and correspondingly so are sequential discontinuities. 

To proceed, we note that an alternative way to define the logarithm (other than Eq.~\eqref{eq:log_sum_def}) is through the contour integral
\begin{equation}
    \ln s = \int_{1}^s \frac{dx}{x} \label{logint} \, .
\end{equation}
The integration is to be performed along any contour within the cut complex plane that goes from $1$ to $s$. This definition agrees with the series definition and analytic continuation. The discontinuity across the branch cut can then be computed as
\begin{equation} \label{logdisc}
  \operatorname{Disc_s} \ln s = \int_1^{s + i\eps} \frac {d x}{x} - \int_1^{s - i\eps} \frac {d x}{x} = \int_{\circlearrowleft_{0}} \frac {d x}{x} = 2 \pi i \,,
\end{equation}
where $\circlearrowleft_{0}$ is the infinitesimal contour that wraps around the origin once counterclockwise. For other functions, like $\ln^3 s$ or the dilogarithm $\Li_2(s)$, the discontinuity will not be constant. In such cases we can consider further discontinuities. To do so we need to consider the maximal analytic continuation of our functions, in which we do not restrict their domain to the cut complex plane.

A clue to how to proceed is given by the closed contour $\circlearrowleft_{0}$ in Eq.~\eqref{logdisc}, which
apparently passes right through the cut. Indeed, although the integral computation agrees with the discontinuity across the cut, what it is actually computing is the difference between the value of the function on two sheets of a Riemann surface; %
the location of the branch cut is immaterial. The only invariant is the location of the branch point, at $s=0$ for the logarithm. This is the unmovable singularity of the integrand.

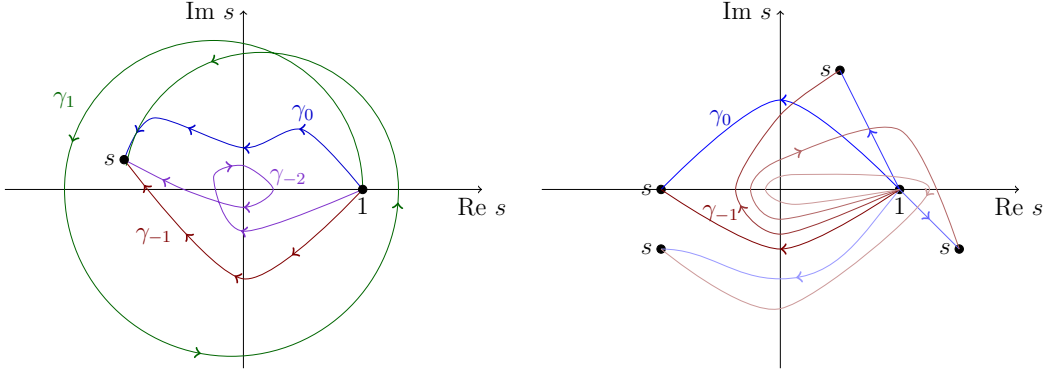
\begin{figure}[t]
    \centering
\resizebox{7cm}{!}{
\begin{tikzpicture}
[decoration={markings,
 mark=between positions 0.3 and 0.9 step 0.2 with {\arrow[line width=1pt]{>}}}]
\draw[->] (-4,0) -- (4,0) coordinate (xaxis);
\draw[->] (0,-3) -- (0,3) coordinate (yaxis);
\draw [darkblue,postaction=decorate] plot [smooth] coordinates {(2,0) (1,1) (0,0.7) (-1.5,1.2) (-2,0.5)};
\draw [darkred,postaction=decorate] plot [smooth] coordinates {(2,0)  (0,-1.5) (-2,0.5)};
\draw [darkpurple,postaction=decorate] plot [smooth] coordinates {(2,0) (0,-0.7) (-0.5,0.2) (0,0.4) (0.5,0) (0,-0.3) (-1,0) (-2,0.5)};
\draw [darkgreen,postaction=decorate] (2,0) arc (0:180:2.5) arc (180:360:2.8) arc (0:170:2.3);
\draw [fill=black] (2,0) circle[radius = 2 pt];
\node[below] at (2,0) {$1$};
\draw [fill=black] (-2,0.5) circle[radius = 2 pt];
\node[left] at (-2,0.5) {$s$};
\node[darkblue, above] at (1,1) {$\gamma_0$};
\node[darkgreen] at (-3,1.5) {$\gamma_1$};
\node[darkred, below] at (-1.5,-0.5) {$\gamma_{-1}$};
\node[darkpurple, right] at (0.3,0.2) {$\gamma_{-2}$};
\node[below] at (xaxis) {$\text{Re } s$};
\node[left] at (yaxis) {$\text{Im } s$};
\end{tikzpicture}
}
\resizebox{7cm}{!}{
\begin{tikzpicture}
[decoration={markings,
 mark=at position 0.5 with {\arrow[line width=1pt]{>}}}]
\draw[->] (-4,0) -- (4,0) coordinate (xaxis);
\draw[->] (0,-3) -- (0,3) coordinate (yaxis);
\node[below] at (xaxis) {$\text{Re } s$};
\node[left] at (yaxis) {$\text{Im } s$};
\draw [fill=black] (2,0) circle[radius = 2 pt];
\node[below] at (2,0) {$1$};
\draw [fill=black] (-2,0) circle[radius = 2 pt];
\node[left] at (-2,0) {$s$};
\draw [fill=black] (-2,-1) circle[radius = 2 pt];
\node[left] at (-2,-1) {$s$};
\draw [fill=black] (1,2) circle[radius = 2 pt];
\node[left] at (1,2) {$s$};
\draw [fill=black] (3,-1) circle[radius = 2 pt];
\node[left] at (3,-1) {$s$};
\draw [blue!100,postaction=decorate] plot [smooth] coordinates {(2,0) (0,1.5) (-2,0)};
 \draw [blue!80,postaction=decorate] plot coordinates {(2,0) (1,2)};
\draw [blue!60,postaction=decorate] plot coordinates {(2,0) (3,-1)};
\draw [blue!40,postaction=decorate] (2,0) [out=-130,in=0] to (0,-1.5) [out=180,in=0] to (-2,-1);
\draw [darkred!100,postaction=decorate] plot [smooth] coordinates {(2,0) (0,-1) (-2,0)};
\draw [darkred!80,postaction=decorate] plot [smooth] coordinates {(2,0) (0,-0.75) (-0.75,0) (0,1.25) (1,2)};
\draw [darkred!60,postaction=decorate] plot [smooth] coordinates {(2,0) (0,-0.5) (-0.5,0) (0,0.5) (2,1) (3,-1)};
\draw [darkred!40,postaction=decorate] plot [smooth] coordinates {(2,0) (0.25,-0.25) (-0.25,0) (0.25,0.25) (2.5,0) (0,-2) (-2,-1)};
\node[darkblue] at (-1,1.2) {$\gamma_0$};
\node[darkred] at (-1,-0.4) {$\gamma_{-1}$};
\end{tikzpicture}
}
    \caption{The logarithm can be defined as an integral along a path $\ln_\gamma s = \int_\gamma \frac{dx}{x}$, where the paths begin at $x=1$ and end at $x=s$. The value of $\ln_\gamma s$ depends on the number of times the integration contour wraps around the branch point at the origin. We define families of paths by $\gamma_n$ where $n$ denotes the number of times the path circles the origin. 
    The family labelled $\gamma_0$ is defined to give the principal branch of the logarithm. On the negative real $s$ axis $\ln_{\gamma_0} s = \ln(s + i \eps)$ and $\ln_{\gamma_{-1}} s =\ln(s - i \eps)$.}
    \label{fig:logpaths}
\end{figure}
We can extend the definition of the logarithm in Eq.~\eqref{logint} beyond the cut complex plane by simply writing
\begin{equation} \label{loggammadef}
    \ln_\gamma\!s = \int_{\gamma} \frac{dx}{x} \, ,
\end{equation}
 where  the integration contour $\gamma$ can be any path from $1$ to $s$ that does not pass through the origin. This is the maximal analytic continuation of $\ln s$. 
The domain of the maximal analytic continuation in this case is an infinite number of copies of the complex plane with a branch point at $s = 0$. These additional copies can be accessed by integration contours that wrap around this branch point a given number of times. By considering all such paths, we obtain an infinite number of values for $\ln s$ that differ by multiples of $2 \pi i$. This is illustrated in Fig.~\ref{fig:logpaths}, where we denote by $\gamma_n$ equivalence classes of paths that end at $s$ after wrapping around the origin $n$ times in the counterclockwise direction. The principal branch
of the logarithm corresponds to paths  that never cross the negative real $s$ axis.

The infinite tower of values associated with $\ln s$ can be thought of as being generated by the closed integration contour around the branch point at the origin. This integral is referred to as the monodromy of $\ln s$ around the origin, and constitutes the only element of the natural logarithm's monodromy group. The discontinuities of polylogarithms can be computed in terms of their monodromies; for instance, in our new notation the discontinuity across the branch cut of $\ln s$ becomes
\begin{equation}
    \ln_{\gamma_0}\!s - \ln_{\gamma_{-1}}\!s = \int_{\circlearrowleft_{0}} \frac{dx}{x} = 2\pi i \, ,
    \label{cauchy}
\end{equation}
where the integral over $\circlearrowleft_{0}$ is the monodromy. To connect the monodromy picture to the cut-plane picture, we now identify \begin{align}
    \ln(s+i\eps) = \ln_{\gamma_0}\!s \, , \qquad  \ln(s-i\eps) = \ln_{\gamma_{-1}}\! s \,  . \label{eq:log_defined_by_regions}
\end{align} 
To be clear, $\ln(s\pm i\eps)$ on the left side of these equations means we approach the real $s$ axis from above or below on the principal branch of the logarithm on the cut complex plane. The logarithms on the right hand side are defined through contours and have no branch cut --- the function $\ln_\gamma s$ is analytic on the negative real $s$ axis (and everywhere else) as long as the path $\gamma$ is deformed smoothly to change $s$. 
With this identification, Eq.~\eqref{cauchy} then agrees with Eq.~\eqref{logrel} up to the  theta function. Indeed, the discontinuity defined in terms of the monodromy is an analytic function, while the difference using the principal branch of the logarithm  comes with a non-analytic $\Theta(-s)$. 

If we adopt the relations in Eq.~\eqref{eq:log_defined_by_regions} as analytic generalizations of $\ln(s+i\eps)$ and $\ln(s-i\eps)$, we can easily compute discontinuities of powers of logarithms by simply substituting in Eq.~\eqref{cauchy}. For instance,
\begin{equation}
    \disc_s \ln^2(s+i\eps) = \ln^2(s+i\eps) - \ln^2(s-i\eps) = (2\pi i) \left[2\ln(s+i\eps) - 2\pi i\right] \label{discl2}
\end{equation}
and
\begin{align}
    \disc_s \ln^3(s+i\eps)&= \ln^3(s+i\eps) - \ln^3(s-i\eps) \nonumber \\
    &=(2\pi i)\left[3\ln^2(s+i\eps) -6 \pi i \ln (s+i \eps)- 4\pi^2\right] \label{discsub} \, .
\end{align}
We can now proceed to take additional discontinuities by subtracting from the function its value with all $+i\eps$ switched to $ -i\eps$. We then find
\begin{equation}
    \disc_s \disc_s \ln^3(s+i\eps) = (2\pi i)^2 \left[6 \ln(s+i\eps)-12 \pi i \right] \label{ddlog}
\end{equation}
and
\begin{equation}
    \disc_s \disc_s \disc_s \ln^3(s+i\eps) = 6(2\pi i)^3 \,.
\end{equation}
If we take any further discontinuities of $\ln^3(s+i\eps)$ we get zero. It is worth emphasizing here that $\disc$ does {\it not} in general satisfy the product rule
\begin{equation}
 \disc (A B)\ne A \, \disc B + B \,\disc A \, .
 \end{equation}
The discontinuity operator computes a finite difference around a branch point, which is not an infinitesimal differential in any sense.

In summary, we have seen that for powers of logarithms, we can compute sequential discontinuities by identifying the $\pm i\eps$ prescription with integration contours that end on different Riemann sheets, and the discontinuity across the cut with the monodromy around the branch point. In general, the transcendental functions that show up in scattering amplitudes are more complicated than logarithms, and depend on many Mandelstam invariants with many branch points. Understanding the monodromy group of these more complicated functions will help us untangle their analytic structure, and thereby help us compute their sequential discontinuities. Correspondingly, we now turn to  a systematic procedure for computing the generators of the monodromy group associated with a general polylogarithmic function.

\subsection{The monodromy group}
\label{sec:monodromy_group}
Given a function defined by a contour integral, we can determine the effect of an analytically continuing around one of its branch points by integrating along a closed contour that encircles this branch point. The integrals along these closed contours are referred to as the monodromies of the function, and form a group. By computing an explicit representation of this group, we can compute the value of this function anywhere in its maximally analytically continued domain. We illustrate how this group can be systematically computed, by working through some examples.

\paragraph{One branch point}~\\
Let us first return to the example of powers of logarithms $\ln^n s$, for any positive integer $n$. As the discontinuities of $\ln^n s$ involve lower powers of $\ln s$, we consider all powers up to $n$ simultaneously. The total differential of these functions is
\begin{equation}
  d \left(\frac {\ln^n s}{n!} \right) =  \left(\frac {\ln^{n-1}s}{(n-1)!} \right) \frac {d s} s\, , \label{logdrel}
\end{equation}
where we have normalized $\ln^n s$ by a factor of $n!$ for convenience. Let's take $n=3$ for concreteness and collect the functions that appear in the derivatives of $\ln^3 s$ into a vector
\begin{equation}
    \cV \equivD \begin{pmatrix}
1 & \ln s  & \frac{1}{2} \ln^2 s & \frac{1}{3!} \ln^3 s
  \end{pmatrix}.
\end{equation}
The differential relations in Eq.~\eqref{logdrel} can then be put in the matrix form
\begin{equation}
    d \cV = \cV \cdot \omega \label{Vandomega},
\end{equation}
where the \emph{connection} \(\omega\) is an \((n + 1) \times (n + 1)\) matrix defined on \(\mathbb{C}^*\equivD \mathbb{C}\backslash\{0\}\) whose entries are one-forms:
\begin{equation}
\omega
 =
  \begin{pmatrix} 0 & \frac{ds}{s}  & 0 & 0 \\
   0 & 0 & \frac{ds}{s}  & 0  \\
   0 & 0 & 0 & \frac{ds}{s} \\
   0 & 0 & 0 & 0
  \end{pmatrix}.
  \label{omegalog}
\end{equation}
As we analytically continue $\ln^n s$ around $s=0$, the vector of functions $\cV$ will mix with other functions that, like $\cV$, satisfy the differential equation in Eq.~\eqref{Vandomega}. These other functions have lower transcendental weight, and the mixing coefficients will be proportional to powers of $i \pi$. Thus, general solutions to Eq.~\eqref{Vandomega} will contain all the possible information about the monodromies of the function.

As there are $n+1$ independent solutions to Eq.~\eqref{Vandomega}, we can group these solutions into an upper-triangular matrix $\sM$ called the \emph{variation matrix}, which we normalize to have $1$'s along the diagonal. The variation matrix on the principal branch of the logarithm for $n=3$ can be written as
\begin{equation} \label{varmat3}
    \sM_{\gamma_0}(s) = \begin{pmatrix}
    1 & \ln s  & \frac{1}{2} \ln^2 s  & \frac{1}{3!} \ln^3 s\\
     0 &1 & \ln s  & \frac{1}{2} \ln^2 s  \\
        0&0&1 & \ln s \\
         0&0&0& 1
    \end{pmatrix}.
\end{equation}
Variation matrices have a close connection to the coproduct structure often utilized in Feynman integral calculations. Further discussion of this connection is given in Appendix~\ref{subsec:coaction}.

To extend the variation matrix in Eq.~\eqref{varmat3} beyond the cut complex plane, we need to determine the effect of deforming the integration contour defining its entries around their branch points. 
Although this extension changes the value of the function at $s$, the differentials of the function will still be related by the differential equation Eq.~\eqref{Vandomega}. Since the general solution to this differential equation are linear combinations of the rows of the variation matrix, we can interpret the action of the monodromy as multiplication of the variation matrix by another matrix, the monodromy matrix.

The most general solution to the differential equation in Eq.~\eqref{Vandomega} is given by 
\begin{equation}
\sM_\gamma(s) =  \mathcal{P} \exp \left( \int_\gamma \omega \right) \, , \label{Mofomega}
\end{equation}
where \(\mathcal{P} \exp(\int_\gamma \omega)\) is a path-ordered exponential
along the path $\gamma$ starting at $1$ and ending at $s$. For a given contour from $a$ to $b$, the path-ordered exponential is defined by
\begin{equation} \label{eq:path_ordered_exp}
  \mathcal{P} \exp \left( \int_a^b \omega \right) = 1 + \int_a^b \omega + \int_a^b \omega \circ \omega + \cdots\,
\end{equation}
where  \(\int_a^b \omega \circ \omega\) denotes an iterated integral.
 Since $\omega$ is a matrix, $\omega\circ\omega$ implies matrix multiplication:
\begin{equation}
  \int_a^b \omega \circ \omega = \int_{a \leq t_1 \leq t_2 \leq b} \omega_{ik}(t_1) \omega_{kj}(t_2) \, .
\end{equation}
Here we have made the matrix indices explicit for clarity, and $k$ is to be summed over. Note that the expansion in powers of \(\omega\) is finite since \(\omega\) is nilpotent. 
 
 For differential forms in several variables \(x = (x_1, \dotsc, x_n)\), these iterated integrals are defined as follows.  First, we choose a path \(\gamma\) parametrized by \(t \in [0, 1]\) and defined by \((x_1(t), \dotsc, x_n(t))\).  Then, given some differential forms \(\xi_1(x), \dotsc, \xi_k(x)\) in the variables \(x\), we can pull them back to the path \(\gamma\), whereupon they become differential forms \(\gamma^* \xi_i(t)\) in the variable \(t\) parameterizing the path.  The iterated integral of these forms along \(\gamma\) is defined as
\begin{equation}
     \int_\gamma \xi_1(x) \circ \cdots \circ \xi_k(x) =
     \int_{0 \leq t_1 \leq \cdots \leq t_k \leq 1} \gamma^* \xi_1(t_1) \cdots \gamma^* \xi_k(t_k).
     \label{eq:pathParametrization}
\end{equation}
We discuss how to evaluate integrals of this type in more detail in Appendix~\ref{app:Phi2}.

Given an integration contour $\gamma$ that ends at $x$, the value of $\mathscr{M}_\gamma(x)$ can be computed by integrating the path-ordered exponential in Eq.~\eqref{Mofomega}.
We can split up any path $\gamma$ between the basepoint (where the integration starts) and $x$
into a contour $\gamma_0$ that goes from the basepoint to $x$ without encircling any of branch points (the poles in $\omega$), and a series of contours $\{\gamma_j'\}$ that each begin and end at $x$ and encircle one of the branch points of $\omega$.
That is, we have
 $\gamma = \gamma_0 \circ \gamma'_{i_1} \circ \dots \circ \gamma'_{i_n}$, where $\gamma_a \circ \gamma_b$ denotes the composition of paths in which we first run along the path
 $\gamma_a$ and then along the path $\gamma_b$. A very useful feature of defining matrices as path-ordered exponentials is that composing two paths corresponds to matrix multiplication. So
 \begin{equation}
 \sM_\gamma = 
     \sM_{\gamma_0 \circ \gamma'_{i_1} \circ \dots \circ \gamma'_{i_n}} =  \sM_{\gamma_0} \cdot \sM_{\gamma_{i_1}'} \cdots  \sM_{\gamma'_{i_n}}
 \end{equation}
 Now, this contour can also be written 
 $\gamma = 
 (\gamma_0 \circ \gamma_{i_1}' \circ \gamma_0^{-1}) 
 \circ \cdots \circ 
 (\gamma_0 \circ \gamma_{i_n}' \circ \gamma_0^{-1}) 
 \circ \gamma_0$, where $\gamma_{i_k} = \gamma_0 \circ \gamma_{i_k}' \circ \gamma_0^{-1}$ encircles the same poles as $\gamma_{i_k}$ but starts and ends at the same basepoint as $\gamma_0$ rather than starting and ending at $x$. 
 Hence, we can also write
  \begin{equation}
      \sM_\gamma = \sM_{\gamma_{i_1} \circ \dots \circ \gamma_{i_n} \circ \gamma_0} = \sM_{\gamma_{i_1}} \cdots  \sM_{\gamma_{i_n}} \cdot  \sM_{\gamma_0} 
  \end{equation}
 where we are now {\it prepending} closed contour integrals from a common basepoint onto the integration contour before we arrive at $x$. This convention ensures that the monodromy matrices are independent of the endpoint $x$.
In summary, to compute the monodromy from $x$ along the path $\gamma_1'$ followed by $\gamma_2'$ we first multiply on the left by $\mathscr{M}_{\gamma_2}$ followed by multiplication on the left of the result by $\mathscr{M}_{\gamma_1}$ where the paths $\gamma_1$ and $\gamma_2$ start and end at the basepoint independent of $x$.

In the case of $\ln^n s$ there is only a single branch point at the origin.
The contour $\gamma_0$ can be taken to be the straight path from \(1\) to \(s\), except when $s$ lies on the negative real axis, in which case we deform the path $\gamma_0$ to go just above the branch point at zero. Then one can check that the variation matrix $\sM_{\gamma_0}$ in Eq.~\eqref{varmat3} is exactly $\mathcal{P} \exp \int_{\gamma_0} \omega$ along this path (see Eq.~\eqref{eq:principal_variation_matrix_log} below). Since there is only one branch point, 
we define paths \(\gamma_+\)  and $\gamma_-$ that encircle the origin counterclockwise or clockwise with unit radius. 
We can thus decompose a general path $\gamma$ into some number of iterations of $\gamma_+$ or $\gamma_-$, followed by $\gamma_0$, namely $\gamma = \gamma_+ \circ \dots \circ \gamma_+ \circ \gamma_0$ or $\gamma = \gamma_- \circ \dots \circ \gamma_- \circ \gamma_0$.

Given a member $\gamma_k$ of the equivalence class of contours that encircle the origin $k$ times clockwise and end at $s$, we have
\begin{equation} \label{variation_matrix_decomposition}
  \mathscr{M}_{\gamma_k}(s) = \left(\mathscr{M}_{\gamma_-} \right)^k \cdot \mathscr{M}_{\gamma_0}(s)\,.
\end{equation}
The matrix \(\mathscr{M}_{\gamma_-}\) can thus be seen to be a generator of the monodromy group, since it maps $\ln^n s$ to its value after encircling the branch point \(s = 0\) one more time. Such representations of the monodromy group generators are sometimes called monodromy matrices.

Since we have specified their integration contours, $\mathscr{M}_{\gamma_0}(s)$ and $\mathscr{M}_{\gamma_-}$ can be computed directly. To calculate \(\smash{\mathscr{M}_{\gamma_-}}\), we parametrize the path \(\gamma_-\) by \(s = \exp(-i \theta)\) for \(\theta \in [0, 2 \pi]\). This gives us \(\frac {d s} s = -i d \theta\), and thus
\begin{equation}
  \int_{\gamma_-} \underbrace{\frac {d s} s \circ \cdots \circ \frac {d s} s}_{j} = \frac {(-2 \pi i)^j}{j!}\,.
\end{equation}
The analogous set of integrals over $\gamma_0$ just return the logarithms we started with, namely
\begin{equation}
  \int_{\gamma_0} \underbrace{\frac {d s} s \circ \cdots \circ \frac {d s} s}_{j} = \frac {\ln^{j}(s)}{j!}\,.
\end{equation}
Expanding the path-ordered exponentials and evaluating the iterated integrals as described above on the connection in Eq.~\eqref{omegalog} for $\ln^3 s$, we find
\begin{equation}
 \mathscr{M}_{\gamma_-}(s)
 = \bbone + \int_{\gamma_-} \omega +  \int_{\gamma_-} \omega \circ  \omega + \int_{\gamma_-} \omega \circ  \omega \circ  \omega
 =\begin{pmatrix}
    1 & -2\pi i  & \frac{1}{2} (-2\pi i)^2  & \frac{1}{3!} (-2\pi i)^3\\
     0 &1 &  -2\pi i  & \frac{1}{2}  (-2\pi i)^2 \\
        0&0&1 & -2\pi i \\
         0&0&0& 1
    \end{pmatrix} \,
\end{equation}
and
\begin{equation}
 \mathscr{M}_{\gamma_0}(s)
 = \bbone + \int_{\gamma_0} \omega +  \int_{\gamma_0} \omega \circ  \omega + \int_{\gamma_0} \omega \circ  \omega \circ  \omega
  =\begin{pmatrix}
    1 & \ln s  & \frac{1}{2} \ln^2 s  & \frac{1}{3!} \ln^3 s\\
     0 &1 & \ln s  & \frac{1}{2} \ln^2 s  \\
        0&0&1 & \ln s \\
         0&0&0& 1
    \end{pmatrix} \, ,
  \label{eq:principal_variation_matrix_log}
\end{equation}
in agreement with Eq.~\eqref{varmat3}.

Using Eq.~\eqref{variation_matrix_decomposition}, we can then compute the effect of going around the branch point by multiplying these matrices. For example, we can calculate the first discontinuity by
\begin{equation}
    (\bbone - \sM_{\gamma_-}) \cdot \sM_{\gamma_0}(s) =
    -
    \begin{pmatrix}
    0&  -2\pi i   &  -2\pi i \ln s + \frac{(-2\pi i)^2}{2} & \frac{-2\pi i}{2} \ln^2 s +\frac{(-2\pi i)^2}{2} \ln s  + \frac{(-2\pi i)^3}{3!}
    \\
     0 & 0& -2\pi i  & -2\pi i \ln s + \frac{(-2\pi i)^2}{2} \\
        0&0&0 & -2\pi i \\
         0&0&0&0
    \end{pmatrix}\,.
\end{equation}
The discontinuity of $\ln^3\!s$ is then $3!$ times the top-right entry of this matrix, in agreement with Eq.~\eqref{discsub}.

More generally, under the action of $\mathscr{M}_{\gamma_-}$ the entry in the first row and last column of $\sM_{\gamma_0}(s)$ transforms as
\begin{align}
\mathscr{M}_{\gamma_-} \frac{\ln^n(s)}{n!}  =
\sum_{k = 0}^n \frac{\ln^{n-k}(s)}{(n-k)!} \frac {(-2 \pi i)^{k}}{k!} \, .
\end{align}
Here we are generalizing notation slightly by having $\mathscr{M}_{\gamma_-}$ act on a function rather than the variation matrix in which it is the upper-right entry. Thus, the discontinuity is
\begin{equation}
    \disc_s \ln^n\!s = (\bbone - \mathscr{M}_{\gamma_-}) \ln^n\!s = -\sum_{k = 1}^n n! \frac{\ln^{n-k}(s)}{(n-k)!} \frac {(-2 \pi i)^{k}}{k!} \, . 
\end{equation}
This agrees with what we get using the substitution $\ln(s-i\eps) = \ln(s+i\eps) - 2\pi i$, as we did for instance in Eq.~\eqref{discl2}, which gives us
\begin{align}
    \disc_s \ln^n (s + i\eps) &= \ln^n(s + i \eps) - \ln^n(s - i \eps)
    = -\sum_{k=1}^n \bin{n}{k}\ln^{n-k} (s + i \eps) (-2\pi i)^{k} \label{disclogn} \,
\end{align}
for arbitrary $n$.

Further discontinuities can be computed by acting with the same operator $\bbone - \sM_{\gamma_-}$. For later reference, we list here some general formulas that can be derived either using the substitution method or with the use of monodromy matrices:
\begin{align}
    \disc_s \disc_s \ln^n (s+i\eps) &= \ln^n(s+i \eps) -2 [\ln(s+i \eps) - 2\pi i]^n + [\ln(s+i \eps) - 4\pi i]^n \\
    & = \sum_{k=1}^n (2^k -2) \bin{n}{k}\ln^{n-k} (s+i \eps) (-2\pi i)^{k}\,. \label{twodiscs}
\end{align}
Similarly, the formula for $m$ discontinuities is
\begin{align}
    \disc_s^m \ln^n (s+i\eps) &=
    \sum_{\ell=0}^m (-1)^\ell \bin{m}{\ell} [\ln(s+i \eps) - \ell 2\pi i]^n  \\
    & = (-1)^{m} m! \sum_{k=1}^n
     \stirling{k}{m} \bin{n}{k} \ln^{n-k} (s+i \eps) (-2\pi i)^{k}\,
     \label{genlognm}
\end{align}
where
\begin{equation}
    \stirling{k}{m}   = \frac{1}{m!} \sum_{\ell=1}^m (-1)^{m-\ell} \bin{m}{\ell}  \ell^k
\end{equation}
are the Stirling numbers of second kind.  These numbers have a useful combinatorial interpretation: $\stirlings{k}{m}$ is the number of ways of partitioning a set of \(k\) elements into \(m\) non-empty sets.

\paragraph{Multiple branch points}~\\
Let us now consider an example involving two branch points, the dilogarithm
\begin{equation}
  \Li_2(s)\equivD \sum_{n = 1}^\infty \frac {s^n}{n^2}\qquad\text{for  }|s|<1 \, .
\end{equation}
Similar to the definition of the logarithm in Eq.~\eqref{eq:log_sum_def}, this power series definition is only convergent in the region \(\lvert s \rvert < 1\), but can be uniquely continued to the rest of the cut complex plane, where the branch cut is usually placed on the positive real axis running from $1$ to $\infty$. The dilogarithm can also be given by an integral definition,
\begin{equation} \label{dilog_integral_representation}
  \Li_2(s) \equivD \int_0^s \frac {d x} x \Li_1(s),\qquad\text{with}\qquad
  \Li_1(s) \equivD \int_0^s \frac{dx}{1-x} = -\ln(1-s) \,  ,
 \end{equation}
We write the integral in terms of $\Li_1(s)$ rather than $-\ln(1-s)$ to make the singularities more transparent, as $\Li_1(s)$ and $\Li_2(s)$ both have branch points at $s=1$, with a branch cut conventionally going from $1$ to $\infty$ along the positive real $s$ axis.
The standard placement of the branch cut for $\Li_n(s)$, from $1<s<\infty$ is consistent with the standard branch cut for the logarithm, $s<0$. 

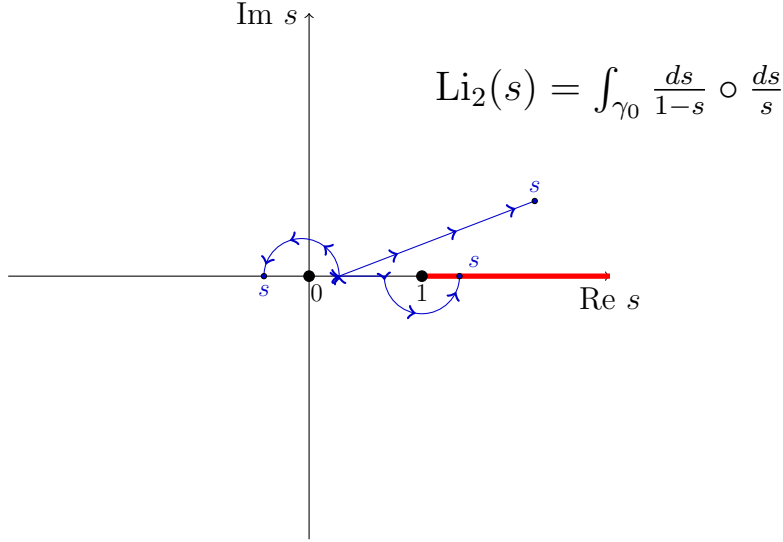
\begin{figure}[t]
    \centering
\begin{tikzpicture}
[decoration={markings,
 mark=between positions 0 and 1 step 0.3  with {\arrow[line width=1pt]{>}}}]
\draw[-> ] (-4,0) -- (4,0) coordinate (xaxis);
\draw[->] (0,-3.5) -- (0,3.5) coordinate (yaxis);
\node[below] at (xaxis) {$\text{Re } s$};
\node[left] at (yaxis) {$\text{Im } s$};
\draw [-, line width=2, red] (1.5,0) to (4,0);
\draw [fill=black] (0,0) circle[radius = 2 pt];
\node[below, scale=0.8] at (0.1,0) {$0$};
\draw [fill=black] (1.5,0) circle[radius = 2 pt];
\node[below, scale=0.8] at (1.5,0) {$1$};
\draw [-,darkblue,postaction=decorate] (0.4,0) to (1,0) arc (-180:0:0.5);
\draw [fill=darkblue] (2,0) circle[radius = 1 pt];
\node[above right, scale=0.8, darkblue] at (2,0) {$s$};
\draw [darkblue,postaction=decorate] (0.4,0) arc (0:180:0.5);
\draw [fill=darkblue] (-0.6,0) circle[radius = 1 pt];
\node[below, darkblue, scale=0.8] at (-0.6,0) {$s$};
\draw [-,darkblue,postaction=decorate] (0.4,0)  to (3,1);
\draw [fill=darkblue] (3,1) circle[radius = 1 pt];
\node[above,darkblue,scale=0.8] at (3,1) {$s$};
\node[below, scale=1.3] at (4,3) {$\operatorname{Li}_2(s) = \int_{\gamma_0} \frac{ds}{1-s} \circ \frac{ds}{s}$};
\end{tikzpicture}
    \caption{$\Li_2(s)$ has branch points at $s=0$ and $s=1$. The principal branch of $\Li_2(s)$ has a branch cut on the real line from $s=1$ to $+\infty$. The standard contour $\gamma_0$ in the analytic integral definition of $\Li_2(s)$ begins at a basepoint at $s=\eps>0$ and proceeds in a straight line to $s$, diverting in a counterclockwise path around the branch points when necessary. }
    \label{fig:li2}
\end{figure}

Using equation~\eqref{dilog_integral_representation}, we have
\begin{equation}
  d \Li_2(s) = -  \frac {d s} s \ln (1-s), \qquad
  d \ln(1-s) = - \frac {d s}{1 - s}.
\end{equation}
We can again put these relations in a matrix form
\begin{equation}
  d
  \begin{pmatrix}
    1 &
    \Li_1(s) &
   \Li_2(s)
  \end{pmatrix} =
  \begin{pmatrix}
    1 &
    \Li_1(s) &
    \Li_2(s)
  \end{pmatrix} \cdot
 \omega \end{equation}
where
\begin{align} \label{eq:dilog_connection}
\omega =  \begin{pmatrix}
    0 & \frac {d s}{1 - s} & 0 \\
    0 & 0 & \frac {d s} s \\
    0 & 0 & 0
  \end{pmatrix}
\end{align}
is defined on \(\mathbb{C}\backslash\{0, 1\}\).

For $\Li_2(s)$, we take the basepoint to be $s=0$ and the path $\gamma_0$ defining its principal branch to be the straight line from $0$ to $s$, which avoids the branch points at $0$ and $1$ with a counterclockwise detour if necessary. This is shown in Fig.~\ref{fig:li2}. Note that this contour is problematic for the differential \(\frac {d s} s\), which diverges at the lower integration bound. This can be dealt with using tangential basepoint regularization, which amounts to introducing a cutoff  $\eps$ on the lower integration limit and dropping the powers of $\ln \eps$ that result (see for instance~\cite{Panzer:2015ida}).\footnote{In more detail, this regularization implies a choice of parametrization for the path in which the tangent vector to the path at the basepoint is of length one.  The monodromy group is then defined with respect to the paths that satisfy this constraint.  In other words, we consider homotopy classes of paths which can be continuously deformed into one another with the tangent at the basepoint being kept constant.} For example,  
 \begin{align}
     \int_0^s \frac{ds}{s} \circ \frac{ds}{1-s}  =
     \int_\eps^s \frac{d s'}{1-s'} \int_\eps^{s'}\frac{d s''
     }{s''}  
     &= -\Li_2 (s) -
     \ln(1-s)\ln \frac{s}{\eps} \\ 
     &\cong  -\Li_2 (s) -
     \ln(1-s) \ln s \, ,
 \end{align}
 where $\cong$ means terms divergent in $\eps$ are dropped and then $\eps \to 0$. 
Then it is straightforward to compute the variation matrix by integrating $\omega$ along $\gamma_0$:
\begin{equation} \label{eq:dilog_variation_matric_cut_complex_plane}
  \mathscr{M}_{\gamma_0}(s) = \mathcal{P} \exp \left( \int_{\gamma_0} \omega \right) =
  \begin{pmatrix}
    1 & \Li_1(s) & \Li_2(s) \\
    0 & 1 & \ln s \\
    0 & 0 & 1
  \end{pmatrix}.
\end{equation}
Note that the this variation matrix encodes precisely the coproduct structure of $\Li_2(s)$,
\begin{equation}
  \Delta \Li_2(s) = 1 \otimes \Li_2(s) + \Li_1(s) \otimes \ln s + \Li_2(s) \otimes 1 \, ,
\end{equation}
as discussed further in Appendix~\ref{subsec:coaction}.

We would now like to extend this construction to the maximal analytic continuation of $\Li_2(s)$. As there are multiple branch points, we should in general be careful to distinguish between infinitesimal contours that encircle these branch points, and the full contours that not only wrap around these points but also start and end at our chosen basepoint of integration. For $\ln^n s$ we took the basepoint to be $1$, but for all the other functions we study in this paper we will take the basepoint to be 0 (or a small value $\epsilon$ on the positive real axis, when regularization is required). We denote the infinitesimal contour in a variable $x$ that encircles the point $p$ counterclockwise by $\circlearrowleft_p^x$. In contrast, we denote the path around $x=p$ that starts and ends at the basepoint by $\linebub{}_p^x$. When the function under study only depends on a single variable $x$, we will often drop the index indicating which variable the contour is taken in.

The contribution from moving along any contour is computed by evaluating the path-ordered exponential  \(\mathcal{P} \exp(\int_\gamma \omega)\) on the contour. For the monodromy around $0$, we find
\begin{align} \label{eq:dilog_monodromy_zero}
\sM_{\linebub{}_0}
= \sM_{\circlearrowleft_0} =
  \begin{pmatrix}
    1 & 0 & 0 \\
    0 & 1 & 2 \pi i \\
    0 & 0 & 1
  \end{pmatrix} \, .
\end{align}
To compute the monodromy matrix associated with the branch point at $1$, we first use Eq.~\eqref{eq:dilog_variation_matric_cut_complex_plane} to determine the contribution from the path between $0$ and $1$, and compute the infinitesimal  contour around $1$ as before. We find
\begin{align}
\mathscr{M}_{0 \to 1} =  \begin{pmatrix}
    1 & 0 & \frac{\pi^2}{6} \\
    0 & 1 & 0 \\
    0 & 0 & 1
  \end{pmatrix} \, , \quad \sM_{\circlearrowleft_1}&=
  \begin{pmatrix}
    1 & -2 \pi i & 0 \\
    0 & 1 & 0 \\
    0 & 0 & 1
  \end{pmatrix} \, ,
\end{align}
where we have dropped all logarithmically-divergent terms in accordance with tangential basepoint regularization. The complete path thus gives
\begin{align} \label{eq:dilog_monodromy_1}
    \sM_{\linebub{}_1}
    = \sM_{0 \to 1} \cdot \sM_{\circlearrowleft_1}  \cdot \big(\sM_{0 \to 1}\big)^{-1} = \begin{pmatrix}
    1 & -2 \pi i & 0 \\
    0 & 1 & 0 \\
    0 & 0 & 1
  \end{pmatrix} \, .
\end{align}
We highlight again that the action of the monodromy matrices  proceeds from left to right; Eq.~\eqref{eq:dilog_monodromy_1} computes the effect of moving from $0$ to $1$ along the real line, rotating counterclockwise around an infinitesimal contour centered at $1$, and then moving back to $0$.

Acting with these matrices on $\sM_{\gamma_0}$ allows us to compute any sequence of monodromies on the functions appearing in $\sM_{\gamma_0}$. For instance, prepending a monodromy around 0 to the path $\gamma_0$ gives
\begin{equation}
    \sM_{\linebub{}_0} \cdot \sM_{\gamma_0}
     =
\begin{pmatrix}
    1 & \Li_1(s) & \Li_2(s) \\
    0 & 1 & \ln s +2\pi i\\
    0 & 0 & 1
  \end{pmatrix},
\end{equation}
while prepending a contour around 1 gives
\begin{equation}
     \sM_{\linebub{}_1} \cdot \sM_{\gamma_0}
     =
\begin{pmatrix}
    1 & \Li_1(s) - 2 \pi i  & \Li_2(s) -2\pi i \ln s\\
    0 & 1 & \ln s \\
    0 & 0 & 1
  \end{pmatrix}.
\end{equation}
These matrices imply that $\Li_1(s)$ and $\Li_2(s)$ only have a monodromy around $s=1$ while $\ln s$  only has a monodromy around $s=0$, as expected. We can also now compute the sequential discontinuity of $\Li_2(s)$ by first taking the monodromy around $s=1$ and then around $s=0$. As we prepend these contours, this corresponds to
\begin{equation}
(\bbone -    \sM_{\linebub{}_1}) \cdot (\bbone -  \sM_{\linebub{}_0} )\cdot \sM_{\gamma_0}
     =
  \begin{pmatrix}
   0 & 0 & -(2\pi i )^2 \\
    0 & 0 & 0\\
    0 & 0 & 0
  \end{pmatrix}\, ,
\end{equation}
which tells us that $\disc_0 \disc_1 \Li_2 (s) =- (2\pi i)^2$. Similarly, we can compute that $(\bbone - \sM_{\linebub{}_1}) \cdot (\bbone -  \sM_{\linebub{}_0} )\cdot \sM_{\gamma_0}=0$, consistent with the fact that $\Li_2(s)$ does not have a discontinuity around $s=0$.

\paragraph{Multiple variables}~\\
Let us finally turn to an example involving multiple variables. We consider the two-variable function
\begin{equation}
  \Phi_1(z, \bar{z}) = 2 \Li_2(z) - 2\Li_2(\bar{z})-\ln(z \bar{z}) \Big[\Li_1(z) - \Li_1(\bar{z})\Big] \, .
\end{equation}
This function arises in the one-loop triangle and one-loop box integrals (see Section~\ref{sec:triangle} below). 
Here we treat $z$ and $\bar{z}$ as independent variables, so this function is analytic for $|z-\frac{1}{2}| < \frac{1}{2}$ and $|\zb-\frac{1}{2}| < \frac{1}{2}$. 
Following the same steps as in our previous examples, we first compute
\begin{equation}
  d \Phi_1 = \left(\frac {d z} z - \frac {d \bar{z}}{\bar{z}}\right) \big(\Li_1(z) + \Li_1(\bar{z}) \big) - \left(\frac {d z}{1 - z} - \frac {d \bar{z}}{1 - \bar{z}}\right) \ln(z \bar{z})\,.
\end{equation}
This can be put in the matrix form $d \mathscr{M}_{\gamma_0} = \mathscr{M}_{\gamma_0} \cdot \omega$, where
\begin{equation}
  \omega =
  \begin{pmatrix}
    0 & \frac {d z}z + \frac {d \bar{z}}{\bar{z}} & \frac {d z}{1 - z} + \frac {d \bar{z}}{1 - \bar{z}} & 0 \\
    0 & 0 & 0 & -\frac {d z}{1 - z} + \frac {d \bar{z}}{1 - \bar{z}} \\
    0 & 0 & 0 & \frac {d z}z - \frac {d \bar{z}}{\bar{z}} \\
    0 & 0 & 0 & 0
  \end{pmatrix} \, .
\end{equation}
The connection \(\omega\) is well-defined in \(\mathbb{C}^2\backslash\{z = 0, z = 1, \bar{z} = 0, \bar{z} = 1\}\), so there are now four codimension-one branching varieties. 

We can define a path $\gamma_0$ between the basepoint $(0,0)$ and $(z,\zb)$ in the same way we did for $\Li_2(s)$, namely we use straight line paths, except when $z$ or $\zb$ are on the real line outside of $(0,1)$, when we go counterclockwise around the branch points.
Integrating along this path gives the variation matrix on the principal branch. 
The result is
\begin{equation} \label{eq:principal_variation_matrix_box}
  \mathscr{M}_{\gamma_0}(z,\zb) = \mathcal{P} \exp \left( \int_{\gamma_0} \omega \right) =
  \begin{pmatrix}
    1 & \ln(z \bar{z}) & \Li_1(z) + \Li_1(\bar{z}) & \Phi_1(z, \bar{z}) \\
    0 & 1 & 0 & - \Li_1(z) + \Li_1(\bar{z}) \\
    0 & 0 & 1 & \ln(z/\bar{z}) \\
    0 & 0 & 0 & 1 
  \end{pmatrix}.
\end{equation}
Note that the antisymmetry of \(\Phi_1(z, \bar{z})\) in its arguments is encoded in the matrices \(\mathscr{M}_{\gamma_0}\) and \(\omega\) by the action of conjugation by $\operatorname{diag}(1, 1, 1, -1)$, namely
\begin{equation}
\operatorname{diag}(1, 1, 1, -1)\cdot \mathscr{M}_{\gamma_0}(z,\zb) \cdot \operatorname{diag}(1, 1, 1, -1) =\mathscr{M}_{\gamma_0}(\zb,z) \, . \label{eq:box_variation_symmetry}
\end{equation}
Further, it can be checked that the connection \(\omega\) is closed (\(d \omega = 0\)) and flat (\(d \omega - \omega \wedge \omega = 0\)). These requirements were trivially satisfied in the preceding one-variable examples, but guarantee that the functions appearing in $\smash{\mathcal{P} \exp  \int_{\gamma} \omega }$ only depend on the homotopy class of $\gamma$. 
Further discussion of this point can be found in Appendix~\ref{app:Phi2}.

We now compute the monodromy matrices associated with the branch points at $0$ and $1$ in both $z$ and $\bar{z}$ by evaluating the path-ordered exponential~\eqref{eq:path_ordered_exp} on cycles that encircle each of these four poles.
First, we compute
\begin{align}\label{eq:box_monodromies_zero}
  \sM_{\linebub{}_0^z} = \sM_{\circlearrowleft_0^z} =
  \begin{pmatrix}
    1 & 2 \pi i & 0 & 0 \\
    0 & 1 & 0 & 0 \\
    0 & 0 & 1 & 2 \pi i \\
    0 & 0 & 0 & 1
  \end{pmatrix} , \quad
\sM_{\linebub{}_0^{\bar{z}}} = \sM_{\circlearrowleft_0^{\bar{z}}} =
  \begin{pmatrix}
    1 & 2 \pi i & 0 & 0 \\
    0 & 1 & 0 & 0 \\
    0 & 0 & 1 & -2 \pi i \\
    0 & 0 & 0 & 1
  \end{pmatrix}  .
\end{align}
To compute the monodromy matrices associated with contours around $1$ we need
\begin{align}
    \sM_{0 \stackrel{z}{\to} 1} &= \begin{pmatrix}
    1 & 0 & 0 & 2 \zeta_2 \\
    0 & 1 & 0 & 0 \\
    0 & 0 & 1 & 0 \\
    0 & 0 & 0 & 1
  \end{pmatrix} \, , \quad
  \sM_{\circlearrowleft_1^z} = \begin{pmatrix}
    1 & 0 & -2 \pi i & 0 \\
    0 & 1 & 0 & 2 \pi i \\
    0 & 0 & 1 & 0 \\
    0 & 0 & 0 & 1
  \end{pmatrix} \, ,
  \\
    \sM_{0 \stackrel{\bar{z}}{\to} 1} &= \begin{pmatrix}
    1 & 0 & 0 & - 2 \zeta_2 \\
    0 & 1 & 0 & 0 \\
    0 & 0 & 1 & 0 \\
    0 & 0 & 0 & 1
  \end{pmatrix} \, , \quad
  \sM_{\circlearrowleft_1^{\bar{z}}} =
  \begin{pmatrix}
    1 & 0 & -2 \pi i & 0 \\
    0 & 1 & 0 & -2 \pi i \\
    0 & 0 & 1 & 0 \\
    0 & 0 & 0 & 1
  \end{pmatrix} \, .
\end{align}
Putting these paths together, we find
\begin{align} \label{eq:box_monodromies_1_z}
\sM_{\linebub{}_1^z} &= \sM_{0 \stackrel{z}{\to} 1} \cdot \sM_{\circlearrowleft_1^z} \cdot \big(\sM_{0 \stackrel{z}{\to} 1})^{-1}  =
 \begin{pmatrix}
    1 & 0 & -2 \pi i & 0 \\
    0 & 1 & 0 & 2 \pi i \\
    0 & 0 & 1 & 0 \\
    0 & 0 & 0 & 1
  \end{pmatrix}  , \\ \quad
  \sM_{\linebub{}_1^{\bar{z}}} &= \sM_{0 \stackrel{\bar{z}}{\to} 1} \cdot \sM_{\circlearrowleft_1^{\bar{z}}} \cdot \big(\sM_{0 \stackrel{\bar{z}}{\to} 1}\big)^{-1} =
  \begin{pmatrix}
    1 & 0 & -2 \pi i & 0 \\
    0 & 1 & 0 & -2 \pi i \\
    0 & 0 & 1 & 0 \\
    0 & 0 & 0 & 1
  \end{pmatrix} .  \label{eq:box_monodromies_1_zb}
\end{align}
Note that the matrices that encode monodromies in the variable $z$ commute with the matrices that encode monodromies in the variable $\bar{z}$. 

These matrices allow us to compute monodromies of $\Phi_1(z, \bar{z})$ and the other functions appearing in Eq.~\eqref{eq:principal_variation_matrix_box} anywhere in their domain, and therefore to compute sequential discontinuities in  $z$ or $\bar{z}$ (and correspondingly the kinematic invariants of the triangle or box diagrams). For example, to compute a sequential discontinuity in $z$ around $1$ and then $0$, we would evaluate
  \begin{align}
      \label{eq:s12_monodromy_example_2}
    (\bbone - \sM_{\linebub{}_1^z}) \cdot (\bbone- \sM_{\linebub{}_0^z}) \cdot \sM_{\gamma_0} &=
    \begin{pmatrix}
 0 & 0 & 0 & -(2\pi i)^2 \\
 0 & 0 & 0 & 0  \\
 0 & 0 & 0 & 0  \\
 0 & 0 & 0 & 0
  \end{pmatrix} .
  \end{align}
Taking these discontinuities in a different order, we get a different result
\begin{align}
\label{eq:s12_monodromy_example_1}
    (\bbone - \sM_{\linebub{}_0^z}) \cdot (\bbone- \sM_{\linebub{}_1^z}) \cdot \sM_{\gamma_0} &=
    \begin{pmatrix}
 0 & 0 & 0 & (2\pi i)^2 \\
 0 & 0 & 0 & 0  \\
 0 & 0 & 0 & 0  \\
 0 & 0 & 0 & 0
  \end{pmatrix}\,.
  \end{align}
It is also possible to take a discontinuity around both branch points by considering the monodromy matrix associated with $\infty$. We construct this monodromy matrix and discuss the full monodromy group in Appendix~\ref{sec:fundamental_group}.

As long as we analytically continue along paths which are fully contained in the Euclidean region, we never encounter branch singularities and the functions we consider are single-valued.
The variation matrix approach lends itself well to the description of single-valued functions, and in Appendix~\ref{app:singlevalue} we describe a general construction that builds a single-valued version of any generalized polylogarithm from its variation matrix.

\subsection{Monodromies of propagators}
We have seen that the $\pm i\eps$ notation is good for describing where we are on the principal branch of multivalued functions, where they describe being on opposite sides of a branch cut. We have also seen that discontinuities across the branch cut can be recast using 
monodromies around the branch point where the cut begins.

In the case of the logarithm, we recall that this amounts to identifying
\begin{equation}
{\blue {\ln(s+i\eps)}} \equivD {\blue{\ln_{\gamma_0}\!s}}\,, \qquad
{\red{\ln(s-i\eps)}} \equivD {\red{\ln_{\gamma_{-1}}\!s}} \, , \label{logascont}
\end{equation}
where $\gamma_0$ is homotopic to the straight path from 1 to $s$, and $\gamma_{-1}$ is given by a path that first crosses the real negative axis before ending at $s$, as shown in Fig.~\ref{fig:logpaths}. With these identifications, we have that $\ln(s-i\eps)=\ln(s+i\eps) - 2\pi i$ for all values of $s$. Using this identity, we can compute the discontinuity of not only $\ln(s+i\eps)$, but also $\ln(s-i\eps)$, finding
\begin{equation}
    \disc_s {\red{ \ln(s-i\eps)}} = \disc_s [{\blue{\ln(s+i\eps)}} - 2\pi i] = 2\pi i\,.
\end{equation}
This can be rewritten in a more suggestive manner:
\begin{equation}
    \disc_s {\red{\ln(s- i\eps)}}
    ={\red{\ln(s- i\eps)}} - {\purple{[ \ln(s-i \eps) - 2\pi i]}} = {\red{ \ln_{\gamma_{-1}}\!s}} - {\purple{ \ln_{\gamma_{-2}}\! s}}\,.
\end{equation}
Thus, when we take the discontinuity of $\ln(s-i\eps)$, we are not computing the difference between its value and the value of $\ln(s+i\eps)$. Rather, we are computing the difference between analytically continuing $\blue{\ln_{\gamma_0}\!s}$ around the origin of $s$ once versus twice.

For sequential discontinuities, the contour definitions are particularly helpful as they allow us to migrate away from the principal branch where $\pm i\eps$ is applicable. Recall however that all the $\pm i \eps$ displacements in Feynman integrals originate in the $\pm i \eps$ displacement of the poles in TOPT propagators. 
Thus, just as we were able to identify higher winding number versions of $\ln(s\pm i\eps)$ using different integration contours, we should be able to identify higher winding number versions of propagators.
To do so, recall that  propagator comes originally from a semi-infinite integral over time
\begin{equation}
    {\blue{\frac{1}{E + i\eps}}}  = -i\int\limits_0^\infty dt\; e^{i E t}\,, \qquad
   {\red {\frac{1}{E - i\eps}}} = -i\int\limits_0^{-\infty} dt\; e^{i E t}\,,
\end{equation}
so that
\begin{equation}
     {\blue{\frac{1}{E + i\eps}}} -{\red{ \frac{1}{E - i\eps} }}= -i\int\limits_{-\infty}^{\infty} dt\; e^{i E t} = {\green{-2\pi i \delta(E)}}\,.
\end{equation}
Thus, for the propagator the integration path goes from $t=0$ to $t=\pm \infty$ and the $\pm i \eps$ is shorthand for this integration path. We can correspondingly take sequential discontinuities of products of propagators in the same way as for logarithms. To do so, we introduce the notation
\begin{equation}
    {\green{D_j}} = {\green{-2\pi i \delta(E_j-\omega_j)}}
\end{equation}
and
\begin{equation}
\label{eq:prop_winding_n}
   {\purple{ P_j^{(n)}}} = {\blue{\frac{1}{E_j-\omega_j+i \eps}}} +n
   {\green{D_j}}
   \, ,
\end{equation}
where we call this distribution a propagator with winding number $n$. The propagators we are used to seeing correspond to ${\blue{P_j^{(0)}}} ={\blue{ \frac{1}{E_j-\omega_j+i\eps}}}$
and ${\red{P_j^{(-1)}}} ={\red{ \frac{1}{E_j-\omega_j-i\eps}}}$.

In this notation, a TOPT amplitude and its conjugate are
\begin{equation}
    \toptM   = \prod_{j=1}^n {\blue{P^{(0)}_j}} ,\quad
     \toptMb  = \prod_{j=1}^n {\red{P^{(-1)}_j}} \, .
\end{equation}
The TOPT cutting rules in Eq.~\eqref{ABDT} become
\begin{equation}
\tDisc\toptM  = 
\toptM - \toptMb 
= \sum_{j=1}^n
\Big(\prod_{k=1}^{j-1} {\blue{P^{(0)}_k}}  \Big)
{\green{D_j}}
\Big(\prod_{k=j+1}^{n}  {\red{ P^{(-1)}_k}}\Big) .\label{eq:cutting_rules_new_notation}
\end{equation}
Since $\tDisc$ is a linear operator, this can also be generalized to products of propagators with arbitrary winding numbers:
\begin{equation}
\tDisc \prod_{j=1}^n {\blue{P^{(l_j)}_j}} 
=
\sum_{j=1}^n
\Big(\prod_{k=1}^{j-1} {\purple{P^{(l_k)}_k}}  \Big) {\green{D_j}}
\Big(\prod_{k=j+1}^{n}  {\purple{ P^{(l_k-1)}_k}}\Big) \, .
\end{equation}
To take further discontinuities, we just use Eq.~\eqref{eq:prop_winding_n} to express propagators with nonzero winding number in terms of propagators with winding number 0.
Then, as in Eq.~\eqref{twodiscs},
\begin{align}
\tDisc^2 \toptM &= \tDisc \toptM - \tDisc \toptMb \label{DDPP}\\
\nonumber
&=
    ({\blue{ P_1}} \cdots {\blue{ P_n}}) 
    ~ - ~
    2({\blue{ P_1}}- {\green{D_1}}) \cdots  ({\blue{ P_n}}-{\green{D_n}})
   ~ +~  ({\blue{ P_1}}-2 {\green{D_1}}) \cdots  ({\blue{ P_n}}-2 {\green{D_n}})  \\
    &= \sum_k (-1)^k (2^k -2)\Big[ {\green{D_1}}\cdots {\green{D_k}}
    {\blue{ P_{k+1}}} \cdots {\blue{ P_n}} + \text{perms}\Big]\nonumber
\end{align}
where the sum over permutations in the last bracket corresponds to the $\bins{n}{k}$ choices for which $k$ propagators to replace with delta functions. The analog of Eq.~\eqref{genlognm} is
\begin{multline}
\label{DDPPP}
       (\tDisc)^m  {\blue{ P_1}} \cdots  {\blue{ P_n}}  =
    \sum_{\ell=0}^m (-1)^\ell \bin{m}{\ell} \Big[ ({\blue{ P_1}} - \ell  {\green{D_1}})\cdots ({\blue{ P_n}} - \ell {\green{D_n}})\Big]   \\
     = \sum_{\ell=0}^m (-1)^\ell \bin{m}{\ell}
    \sum_{k=1}^n \Big[ {\blue{ P_1}}\cdots {\blue{ P_k}} (\ell {\green{D_{k+1}}})\cdots (\ell {\green{D_n}}) + \text{perms} \Big] \,.  
\end{multline}
Although the winding numbers have been left implicit in Eq.~\eqref{DDPP} and Eq.~\eqref{DDPPP}, these equations are valid for any assignment of winding numbers.

Let us try to briefly summarize this section. We found that to take sequential discontinuities the $\pm i \eps$ language was insufficient. For a single discontinuity, one can compare a function on two sides of a branch cut on the principal branch. However, to take additional discontinuities, one needs an analytic function defined away from the cut itself. A natural way to do that is to treat the discontinuity as a monodromy around the branch point. In the monodromy language, there is no branch cut at all (the branch cut is an artifact of projecting onto a complex plane) and the discontinuity is automatically an analytic function. Moreover, monodromies can be computed in an algebraic way using a variation matrix and a connection. Finally, we saw that the monodromy picture led to a natural generalization of the $+i\eps$ propagator to a family of propagators with additional winding numbers. These propagators will be used in the derivation of the relation between multiple cuts and sequential discontinuities, to which we now return.

\section{Sequential discontinuities \label{sec:sequential}}
We saw in Section~\ref{sec:discon} that an advantage of TOPT over the covariant formalism is that one can directly identify the origin of singularities in a particular channel.
Propagators in a given TOPT diagram depend on a sequence of energies, $E_{I_1} \to \cdots \to E_{I_n}$, and each propagator will only lead to a singularity in the integration region if the corresponding energy and invariant are non-negative ($E_I \ge 0$ and $s_I = P_I^2 \ge 0$). We then saw in Section~\ref{sec:math} that, while the $\pm i \eps$ notation is sufficient to identify the two sides of a branch cut for taking a single discontinuity, for sequential discontinuities it proves useful to think in terms of branch points and monodromies. We now make use of these tools to derive formulas for the sequential discontinuities of Feynman integrals in terms of cuts. 

If we work in a region $R^s$, where only a single invariant $s = s_I=P_I^2$ with $P_I^\mu= \sum_{i \in I} P_i^\mu$ is positive, then we can drop the $i\eps$ in all TOPT propagators not involving the energy associated with this invariant. To make the equations in this section more transparent, we denote the energy and momentum associated with the $s$ channel by $E_s = E_{P_I}$ and $P_s=P_I$. In this notation, a generic TOPT diagram in $R^s$ takes the form
\begin{equation}
\toptM =     \prod_{P_i\ne P_s}\frac{1}{E_{P_i} - \omega_i}  \left[
{\blue{\frac{1}{E_s - \omega_1 + i \eps}}} \cdots {\blue{\frac{1}{E_s - \omega_n + i \eps}}}\right] \, .
\end{equation}
In this region, the discontinuity in the $s$ channel is the same as the total discontinuity:
 \begin{multline}
\left[\disc_s  \toptM \right]_{R^{s}} = 
\left[ \tDisc \toptM  \right]_{R^{s}}
\label{discMa} \\ 
=
  \prod_{P_i\ne P_s}\frac{1}{E_{P_i} - \omega_i}  
  \sum_j{ \blue{\frac{1}{E_s - \omega_1 + i \eps} }} \cdots\green{ (-2\pi i) \delta(E_s-\omega_j) }\cdots \red{\frac{1}{E_s - \omega_n - i \eps} }  \,,
 \end{multline}
The second equality comes from applying the TOPT cutting rules in Eq.~\eqref{ABDT} to all propagators, or equivalently just to the propagators involving $E_s$, as all the delta functions involving other sums of energies evaluate to zero.
Summing over all TOPT graphs with a given topology then gives the discontinuity of the corresponding Feynman integral, $[\disc_s \cM]_{R^s}$ from Eq.~\eqref{discRs}.

Before taking further discontinuities, let us pause to clarify the role being played by the region $R^s$ in Eq.~\eqref{discMa}. In principle, the discontinuity operator $\disc_s$ that appears in this equation can be applied anywhere in the maximal analytic domain of the function  $\toptM$.
On the other hand, the relation between $\disc_s \toptM$ and cut integrals in Eq.~\eqref{discMa} only holds in regions where these cuts are allowed, and only when appropriate analytic continuation paths from $R^s$ to $R^\star$ are used to take this discontinuity. This requirement, that the analytic continuation path starts in the region where the cuts are being computed and only passes through an adjacent region, will become even more important when we compute sequential discontinuities below. For instance, in the triangle and box ladder integrals we will consider in Section~\ref{sec:triangle}, we will see there are multiple ways of encircling branch points in the $z$ and $\zb$ variables used there that correspond to encircling the branch point in a given Mandelstam invariant; however, only some of these monodromies in $z$ and $\zb$ can be accessed via paths that pass through the appropriate regions. Thus, while we can compute the discontinuities of $\cM$ in arbitrary regions, these discontinuities must be evaluated in the appropriate region and using appropriate contours to be related to cuts. For instance, $\disc_s \cM$ can be computed (and will in general be nonzero) in the Euclidean region, where the cuts of $\cM$ are zero. However, it is perfectly valid for us to analytically continue the discontinuity that has been computed using the right monodromy matrices in the Euclidean region to the region $R^s$, where it must satisfy Eq.~\eqref{discMa}.

\subsection{Sequential discontinuities in the same channel} \label{sec:samechannel}
We are now ready to consider discontinuities of discontinuities.
To take a second discontinuity of Eq.~\eqref{discMa} in the $s$ channel we can simply rotate all the energies around the same
path as for the first discontinuity. This gives 
 \begin{multline}
\left[\disc_s^2 \toptM \right]_{R^{s}}
       = \prod_{P_i\ne P_s} \frac{1}{E_{P_i} - \omega_i}
       \sum_j\sum_k
       { \blue{P^{(0)}(E_s - \omega_1)}} \cdots{ \blue{P^{(0)}(E_s - \omega_{j-1})}}  \\
 \times{ {\green{ (-2\pi i) \delta(E_s-\omega_j) }}\red{P^{(-1)}(E_s-\omega_{j+1})}} \cdots {\red{P^{(-1)}(E_s-\omega_{k-1})}} \\
 \times {\green{ (-2\pi i) \delta(E_s-\omega_k) }} {\purple{P^{(-2)}(E_s-\omega_{k+1})}}\cdots
 {\purple{P^{(-2)}(E_s-\omega_{n})}}   \, .
 \end{multline}
 In words, 
 the first cut turns the $+i\eps$ propagators, denoted ${\blue{P^{(0)}}}$, to $-i\eps$ propagators, denoted ${\red{P^{(-1)}}}$. The second cut turns the ${\red{P^{(-1)}}}$ propagators into ${\purple{P^{(-2)}}}$ ones.

To make sense of the $ {\purple{P^{(-2)}(E)}}$ propagators, we rewrite them using Eq.~\eqref{eq:prop_winding_n},
\begin{equation}
    {\purple{P^{(-2)}(E)}} = {\blue{ \frac{1}{E + i\eps}}} -2 {\green{ (-2\pi i) \delta(E)}}\,.
\end{equation}
To avoid any ambiguity, we also substitute $ {\red{P^{(-1)}(E)}} = {\blue{ \frac{1}{E + i\eps}}} -{\green{ (-2\pi i) \delta(E)}}$. The result is a sum over cutting different numbers of $s$-channel propagators, in which each non-cut propagator is in the region corresponding to $+i\eps$. Explicitly, we get
\begin{align} \label{DDMsum}
      &\left[ \disc_s^2  \toptM\right]_{R^{s}}=
       \prod_{P_i\ne P} \frac{1}{E_{P_i} - \omega_j} \sum_{k=2}^n (-1)^k (2^k -2) \\
        &\quad \times \Big[
 {\green{ (-2\pi i) \delta(E_s-\omega_1) }}\cdots
 {\green{ (-2\pi i) \delta(E_s-\omega_{k}) }}
 { \blue{\frac{1}{E_s - \omega_{k+1}+i\eps}}}
 \cdots
 { \blue{ \frac{1}{E_s - \omega_n+i\eps}}}  + \text{perms} \Big]\,,  \nonumber
 \end{align}
similar to Eq.~\eqref{DDPP}.

Summing over the double discontinuities of all TOPT diagrams with the same topology, we get the double discontinuity of the associated Feynman integral. Recall that each delta function in a TOPT diagram directly corresponds to a Feynman diagram cut. As such, we can extract the combinatorial factor from Eq.~\eqref{DDMsum} and directly compute the cut Feynman diagram with all $+i\eps$ propagators. Doing so, we get
\begin{equation}
      \left[ \disc_s^2 \cM \right]_{R^{s}}=
      \left[\sum_{k=2} (-1)^k (2^k -2) \cM_{\text{$k$-cuts}}\right]_{R^s_+} \, ,
      \label{doublecut}
\end{equation}
where $\cM_{\text{$k$-cuts}}$ is the sum over all possible ways to cut $\cM$ exactly $k$ times, and $R^s_+$ indicates that all uncut propagators have $+i \eps$. Each cut should split the diagram in two, and the sum of momenta flowing across it should be $P_s$, as the cuts in all other channels vanish in $R^s$.

The formula for the triple discontinuity can be computed the same way, giving
\begin{equation}
     \left[ \disc_s^3 \cM \right]_{R^{s}}=\left[
      \sum_{k=3} (-1)^k (-3^k +3\cdot 2^k -3) \cM_{\text{$k$-cuts}}\right]_{R^s_+}\,,
      \label{triplecut}
\end{equation}
and the generalization to $m$ cuts is as in Eq.~\eqref{genlognm}:
\begin{equation}
\boxed{
     \left[ \disc_s^m \cM\right]_{R^s} = (\bbone - \sM_{\linebub{}^{s}_0})^m \cM
     =
     m!
      \sum_{k=m}^\infty
      \stirling{k}{m}
      \left(-1\right)^{m-k}
      \Big[ \cM_{\text{$k$-cuts}}\Big]_{R^s_+} }\, , \label{mcutsins}
\end{equation}
where $\stirlings{k}{m} = \frac{1}{m!} \sum_{\ell=1}^m (-1)^{m-\ell} \bins{m}{\ell}  \ell^k$ are the Stirling numbers of the second kind. We emphasize again that this relation holds when all non-cut propagators in $\cM_{\text{$k$-cuts}}$ are taken to be in the region corresponding to $+i\eps$.
We have also included the definition of the discontinuity operator in terms of $\sM_{\linebub{}^{s}_0}$, which returns the monodromy around $s=0$. More precisely, this monodromy matrix acts on the variation matrix $\sM_{\gamma_0}$, which should be computed along paths from the basepoint to $R^s$. Examples are given in Section~\ref{sec:examples}.

\subsection{Sequential discontinuities in different channels \label{sec:seqdiff}}
Next, let us consider how to take sequential discontinuities in different channels. 
Unlike the case of sequential discontinuities in the same channel, we must now analytically continue at least two different ways to isolate discontinuities in different channels. As before, we insist on using paths that rotate the external energies while leaving the external three-momenta fixed and respecting energy-momentum conservation. This gives us $n-1$ independent parameters that we can vary along each analytic continuation path, where $n$ is the number of external particles. One also must make sure that the relevant invariants only encircle their branch points once. In the examples we have explored (see Section~\ref{sec:examples}), we have not found these constraints to be overly restrictive. Nevertheless, choosing paths has to be done carefully. 
While Cauchy's residue theorem guarantees that normal contour integrals only depend on the homology class of the integration contour, iterated integrals in general depend on the homotopy class of the integration path. This means that one can in general find multiple discontinuity operators that give the same first discontinuity, but different sequential discontinuities. 
This highlights the importance of our prescription for taking discontinuities by analytically continuing through specific kinematic regions. We discuss this ambiguity in more detail in Appendix~\ref{sec:fundamental_group}.

To fix our notation, suppose we want to compute $\disc_s \disc_t \cM$, where $s= s_I = (P_I)^2$ and $t= s_J = (P_J)^2$ for sets $I$ and $J$
are different momentum invariants. We abbreviate the associated energies and momenta with $E_s = E_{P_I}$, $P_s = P_I$, $E_t = E_{P_J}$, and $P_t = P_J$. 
We also denote by $R^{ \{s,t\} }$ the region in which $s>0$, $t>0$, and all other Mandelstam invariants are real and negative.
A general TOPT amplitude with $n_s$ propagators in the $s$ channel and $n_t$ propagators in the $t$ channel in the region $R^{\{s,t\}}$ has the  form
\begin{equation}
   \left[\toptM  \right]_{R^{\{s,t\}}}=
       \prod_{P_i\notin \{P_s,P_t\}} \frac{1}{E_{P_i} - \omega_i}
       \prod_{k=1}^{n_s} {\blue{\frac{1}{E_s - \omega_k + i\eps} }}
        \prod_{\ell=1}^{n_t} {\blue{\frac{1}{E_t - \omega_\ell + i\eps} }}\,.
\end{equation}
We have dropped the $i\eps$ from all propagators in channels other than $s$ or $t$, since these will never go on shell. 

To take the discontinuity in the $t$ channel, we want to pass around the branch point at $t=0$ and no other branch points. We can do this by passing through the region $R^s$, where only $s>0$ and then back to $R^{ \{s,t\} }$ on the other side of the $t=0$ branch cut. Thus we must find a path rotating the energies, respecting energy conservation, to go from $R^{ \{s,t\} } \to R^s$ (some examples are given in Section~\ref{sec:examples}). Let us assume such a path exists. This path will encircle the branch point for $E_t$, located at the smallest value of $\omega_k$ appearing in any $E_t$ propagator, but will not encircle the branch point for $E_s$. The difference between $M$ before and after analytic continuation along this path is thus
\begin{multline}
    \left[ \disc_t \toptM  \right]_{R^{\{s,t\}}}
    =
       \prod_{P_i\notin \{P_s,P_t\}} \frac{1}{E_{P_i} - \omega_i}
\prod_{k=1}^{n_s} {\blue{\frac{1}{E_s - \omega_k + i\eps} }}
 \\
\times \sum_{\ell=1}^{n_t} { \blue{\frac{1}{E_t - \omega_1 + i \eps} }} \cdots\green{ (-2\pi i) \delta(E_t-\omega_\ell) }\cdots \red{\frac{1}{E_t - \omega_{n_t} - i \eps} } \, .
\end{multline}
Again, the propagators not in the $s$ channel will remain unaffected since
our analytic continuation path has gone from $R^{\{s,t\}}\to R^t \to R^{\{s,t\}}$. 

We can take a discontinuity in the $s$ channel in an analogous way, using an analytic continuation path in energy that encircles the branch point for $E_s$ while going from $R^{\{s,t\}}\to R^s \to R^{\{s,t\}}$. This allow us to compute
\begin{multline}
\left[\disc_s \disc_t \toptM \right]_{R^{\{s,t\}}}  =
       \prod_{P_i\ne P_s,P_t} \frac{1}{E_{P_i} - \omega_i}  \\
 \times \sum_{k=1}^{n_s}{ \blue{\frac{1}{E_s - \omega_1 + i \eps} }} \cdots\green{ (-2\pi i) \delta(E_s-\omega_k) }\cdots \red{\frac{1}{E_s - \omega_{n_s} - i \eps} }  \\
 \times \sum_{\ell=1}^{n_t} { \blue{\frac{1}{E_t - \omega_1 + i \eps} }} \cdots\green{ (-2\pi i) \delta(E_t-\omega_\ell) }\cdots \red{\frac{1}{E_t - \omega_{n_t} - i \eps} } \, . \label{eq:discST_TOPT}
\end{multline}
Like before, when we take the $s$-channel discontinuity, the $t$-channel propagators are unaffected since we have not gone around the branch point at $t=0$.

We cannot immediately sum over TOPT diagrams in Eq.~\eqref{eq:discST_TOPT} to get a Feynman integral, since it is not clear
which Feynman propagators should get $+i\eps$ and which should get $-i\eps$. 
To remedy the problem, we rewrite each diagram in terms of all $+i\eps$ propagators as we did for the sequential discontinuities in Section~\ref{sec:samechannel}. This gives
\begin{multline}
    \left[ \disc_s \disc_t  \toptM\right]_{R^{\{s,t\}}}=
       \prod_{P_i\ne P_s,P_t} \frac{1}{E_{P_i} - \omega_i}
     \\
     \times \sum_{k=1}^{n_s}(-1)^k\Big[
 {\green{ (-2\pi i)^k \delta(E_s-\omega_1) }}\cdots
 {\green{  \delta(E_s-\omega_k}) }
 { \blue{\frac{1}{E_s - \omega_{k+1}+i\eps}}}
 \cdots
 { \blue{ \frac{1}{E_s - \omega_{n_s}+i\eps}}}  + \text{perms} \Big]
 \\
        \times \sum_{\ell=1}^{n_t} (-1)^\ell\Big[
 {\green{ (-2\pi i)^\ell \delta(E_t-\omega_1) }}\cdots
 {\green{ \delta(E_t-\omega_{\ell}) }}
 { \blue{\frac{1}{E_t - \omega_{\ell+1}+i\eps}}}
 \cdots
 { \blue{ \frac{1}{E_t - \omega_{n_t}+i\eps}}}  + \text{perms} \Big] \, 
 \end{multline}
After summing over all TOPT diagrams with the same topology, we get
 \begin{equation}
 \boxed{
       \left[\disc_s\disc_t \cM\right]_{R^{\{s,t\}}} =
     \left[\sum_{k=1}^\infty \sum_{\ell=1}^\infty (-1)^{k+\ell}
     \cM_{\text{\{$k$ cuts in $s$, $\ell$ cuts in $t$\}}} \right]_{R^{\{s,t\}}_+}
     }
     \label{ddstRst1}
\end{equation}
where the sum is over all diagrams with $k\ge 1$ cuts in the $s$-channel and $\ell \ge 1$ cuts in the $t$ channel, and all propagators are assigned $+i\eps$.

One should think of Eq.~\eqref{ddstRst1} as applying at an implicit phase-space point in the physical region where the cuts are to be computed. One can analytically continue the resulting cut graphs to any region one wants, such as the Euclidean region, but the result will not be the same as evaluating the cut graphs at a phase-space point in the Euclidean region. This is because the theta functions associated with the original region determine whether the cut vanishes, rather than by the kinematics of the new region. In other words, one cannot evaluate some of the cuts at a phase space point in $R^s$ and others at a phase space point in $R^t$. Thus, our formula is derived assuming we want to relate cuts and discontinuities at a single phase space point in $R^{\{s,t\}}$. You can use a region other than $R^{\{s,t\}}$ (such as $R^{\{s,t,u\}}$), 
as long as the paths in analytic continuation between these regions exist.

In terms of monodromy matrices, this sequential discontinuity can be computed as
\begin{equation}
\label{Rstmono}
 \left[\disc_s\disc_t \cM\right]_{R^{\{s,t\}}} =(\bbone - \sM_{\linebub{}_{0}^t})(\bbone - \sM_{\linebub{}_{0}^s}) \cM \,, 
\end{equation}
where we recall that the action of these monodromy matrices should be read left to right (unlike discontinuity operators). The variation matrix $\cM$ should be evaluated along paths from the basepoint to the region $R^{\{s,t\}}$. The monodromy matrices are computed from the basepoint and the monodromies are prepended to the path $\gamma$ ending in $R^{\{s,t\}}$. Alternatively, one can apply the monodromy matrices in some other region, such as $R^\star$ and then continue to $R^{\{s,t\}}$; since we are prepending the monodromies, whether we continue before or after we prepend them gives the same answer. However, we highlight again that the same is {\it not} true of cuts---for instance, all cuts evaluate to zero in $R^\star$.

One can generalize this formula to apply to $m_i$ discontinuities in channel $i$ without additional complication:
\begin{equation}
\boxed{
\begin{aligned}
\label{masterformula}
&    \left[(\disc_{s_1})^{m_1}
\cdots (\disc_{s_n})^{m_n} \cM\right]_{R^{\{{s_1},\cdots,{s_n}\}}} 
        \\
&= 
(-1)^{N_{\text{discs}} -N_{\text{cuts}}}
    m_1! \cdots m_n!            
\sum_{k_1=m_1}^\infty
\stirling{k_1}{m_1}
\cdots 
\sum_{k_n=m_n}^\infty
\stirling{k_n}{m_n}
\Big[ 
\cM_{
\bigg\{
\substack{
k_1~\text{cuts in $s_1$} \\[.16cm]
\smash{\vdots} \\
k_n~\text{cuts in $s_n$} \\
}
\bigg\}
}
      \Big]_{R^{\{s_1,\cdots,s_n\}}_+}
      \end{aligned}
      }
  \end{equation}
where
\begin{equation}
  N_{\text{discs}} = m_1+ \cdots+  m_n 
  \quad \text{and} \quad N_{\text{cuts}} = k_1+ \cdots+ k_n \,.
\end{equation}
This is the master formula for computing any number of sequential discontinuities in any channels.

One can even go one step farther and generalize from $s_i$ being individual invariants to being sets of invariants. For example, we might have a set $\set_i = \{s,t\}$. Then the discontinuity in $\set_i$ is computed by taking the monodromy from a region $R^{\set_i}$ where the invariants in $\set_i$ are positive through the Euclidean region and back.
Then
\begin{equation}
    \left[\disc_{\set_i} \cM\right]_{R^{\set_i}}
    =(\bbone - \sM_{\linebub_{\set_i}} ) \cM = \sum_{j}
    \left[\cM_{\text{cuts in $s_j \in \set_i$}}\right]_{R^{\set_i}_+}
\end{equation}
The generalization to multiple sets and multiple discontinuities is
\begin{multline}
\frac{1}{m_1!} \cdots \frac{1}{m_n!}   \left[(\disc_{\set_1})^{m_1}
\cdots (\disc_{\set_n})^{m_n} \cM\right]_{R^{\cup{\set_i}}} 
       \\ 
=(-1)^{N_{\text{discs}} -N_{\text{cuts}}}            
\sum_{k_1=m_1}^\infty
\stirling{k_1}{m_1}
\cdots 
\sum_{k_n=m_n}^\infty
\stirling{k_n}{m_n}
\Big[ 
\cM_{k_j~\text{cuts from set}~\set_j} \Big]_{R^{\cup{\set_i}}_+} 
\end{multline}
where $\disc_{\set_j}$ is taken between the region $R^{\cup{\set_i}}$ where all invariants in any set $\set_i$ are positive to a region$R^{\cup{\set_i}/\set_j}$ where all the invariants have the same sign as in $R^{\cup{\set_i}}$ except for those in $\set_j$, which are negative. An example of this type of set discontinuity is given in Eq. \eqref{R123toR1} below.

 In~\cite{Abreu:2014cla}, a different prescription for calculating sequential discontinuities in different channels was proposed. Their proposal was that $\disc_s \disc_t \cM$ should be computed by first calculating $\disc_t \cM$ in $R^t$, and then  analytically continue to $R^{\{s,t\}}$ before computing $\disc_s$.
 They defined these discontinuities as the difference between a function on different sides of a branch cut. Using the language of monodromies around a branch point rather than discontinuities across branch cuts, this can be interpreted to mean first prepending a monodromy matrix around $t=0$ to a path going into $R^t$ and then extending the path into $R^{\{s,t\}}$. Since the monodromy matrix is independent of the endpoint of the integration, this is the same as simply computing the discontinuity in $t$ in the region $R^{\{s,t\}}$ to begin with.
 No details were given in~\cite{Abreu:2014cla} for how to choose paths for analytic continuation.

 As for the cuts, the prescription given in~\cite{Abreu:2014cla} for how to compute sequential cuts involves an algorithm with tuples of black and white dots that determines whether $+i\eps$ or $-i\eps$ should be chosen. For the examples they considered, this algorithm worked. However, in more complicated cases, it may not correctly account for the discontinuity of $-i \eps$ propagators that appear after a first discontinuity. The main difference, however, is that~\cite{Abreu:2014cla} excluded from consideration cases where sequential discontinuities were taken in the same channel. Our formulas allow for any number of discontinuities in any channels, with no restrictions.

\subsection{Steinmann relations}
Finally, let us connect to the Steinmann relations.
One of the important implications of Eq.~\eqref{ddstRst1} is that $\left[\disc_s\disc_t \cM\right]_{R^{\{s,t\}}}$ can only be nonzero when there exists at least one TOPT diagram in which both $E_s$ and $E_t$ appear. However, it is a general feature of TOPT that whenever two energies $E_t$ and $E_s$ appear in the propagators of a single diagram, one must depend on a subset of the energies that appear in the other (e.g. $E_s = E_1+E_2+E_3$ and $E_t = E_1+E_2$). It follows that $\left[\disc_s\disc_t \cM\right]_{R^{\{s,t\}}}$ will vanish whenever $s$ and $t$ involve partially overlapping sets of energies.  More precisely, recall from the beginning of this section that $s=(\sum_{i \in I} P_i)^2$ and $t=(\sum_{i \in J} P_i)^2$. Then,
 \begin{equation} \label{FeynmanStenimann}
\left[ \disc_s \disc_t \cM \right]_{R^{\{s,t\} }} = 0 \quad\text{if}\quad
I \not\subset J \quad\text{and} \quad J \not\subset I \, .
 \end{equation}
 This is a version of the Steinmann relations, which state that the double sequential discontinuity in such overlapping channels must vanish, which we have thus proven at the level of Feynman integrals.

It is worth emphasizing two conditions that are necessary for our proof of the Steinmann relations to hold. First, the region $R^{\{s,t\}}$, where all invariants other than $s$ and $t$ are negative and all momenta are real, must exist.
The existence of such regions is consistent with the assumptions of axiomatic field theory, where all particles are massive; however, when there are massless external particles, the on-shell constraint may mean the region $R^{\{s,t\}}$ is empty. In such a case, we cannot immediately apply our formulas.

Second, we go around the poles in the TOPT propagators by continuing the external energies, holding the external three-momenta fixed.
This allowed us to isolate the singularities, since the internal energies $\omega_k$ depend only on the external three-momenta, which are held fixed during the analytic continuation. If one tries to impose a constraint on some of the external momenta, such as fixing their masses to zero or some other value, then one must also rotate the external momenta to maintain the mass-shell condition. In such cases, finding the singular variety for the TOPT propagators is more complicated and our derivation also does not immediately apply.

Because of these preconditions, the Steinmann relations in Eq.~\eqref{FeynmanStenimann} do not restrict all possible double discontinuities in partially-overlapping  channels.  
In particular, they do not apply to discontinuities on sheets that are far removed from the physical sheet; they only hold at \emph{real} kinematic points, in the physical region. This subtlety appears, for instance, in the one-loop box with massless internal and external legs. This box is infrared divergent. In $d=4-2\eps$ dimensions it has the expansion~\cite{Duplancic:2000sk}
\begin{equation} \label{M0m}
    \cM^{0m} = \frac{1}{st}
    \left[\frac{4}{\eps^2} - \frac{2}{\eps}\left(\ln\frac{-s}{\mu^2} + \ln\frac{-t}{\mu^2}\right) + 2\ln\frac{-s}{\mu^2} \ln\frac{-t}{\mu^2} - \pi^2 + \cO(\eps)\right]
\end{equation}
where $s=(p_1+p_2)^2$ and $t=(p_2+p_3)^2$ partially overlap.
The $\cO(\eps^0)$ term has a $\ln(-s)\ln(-t)$ component that has a nonzero sequential discontinuity in $s$ and $t$. With massless external lines, the region $R^{\{s,t\}}$ does not exist, so there is no contradiction with our formula. This observation is consistent with results from $S$-matrix theory; since $s$ and $t$ can only simultaneously vanish outside of the physical region, the Steinmann relations do not apply~\cite{Stapp:1971hh}.

If {\it internal} particles are massless, our sequential discontinuity formulas in Eq.~\eqref{mcutsins} and Eq.~\eqref{ddstRst1}, and correspondingly the Steinmann relations in Eq.~\eqref{FeynmanStenimann}, should still apply. The key problem with massless {\it external} particles is that the massless condition constrains the surface of maximal analytic continuation; massless internal particles impose no such constraint. Nevertheless, 
with massless internal particles, certain cuts also have to be treated with care when applying the Steinmann relations (as explained, for instance, in~\cite{Cahill:1973qp}). When two overlapping momentum channels only depend on a single common momentum, cutting both channels can lead to a three-point vertex in which an external state decays into a pair of internal physical states. Some discussion of these vertices is given in Appendix~\ref{sec:massless3pt}. In \(S\)-matrix theory, external states are stable and massless three-point vertices do not appear.

Finally, let us highlight the fact that the right side of Eq.~\eqref{masterformula} does not know anything about the order of the discontinuities begin taken on the left side. This implies that the Steinmann relations force any sequence of discontinuities involving partially-overlapping channels to vanish, even if these partially-overlapping discontinuities are separated by a long sequence of unrelated discontinuities. This is related to the fact that Eq.~\eqref{masterformula} only governs discontinuities that are computed at a phase-space point in which all the relevant cuts are accessible, and holding all other variables fixed~\cite{Stapp:1971hh}. Thus, in many cases the relevant region may not correspond to real kinematics, in which case this restriction does not immediately apply.

\section{Examples}
\label{sec:examples}
In this section, we consider a number of examples in which we can check the general relations between cuts and discontinuities developed in the previous sections.

\subsection{Bubbles}
The first examples we consider are sequences of bubbles. The single bubble integral with massless internal lines in $d=4-2\epsilon$ dimensions evaluates to
\begin{align}
\cM_1^{\text{bare}} =\hspace{-4pt} \begin{gathered}
\begin{tikzpicture}[baseline=-3.5]
\node at (0,0) {
\parbox{30mm} {
\resizebox{30mm}{!}{
 \fmfframe(0,00)(0,0){
 \begin{fmfgraph*}(80,40)
	\fmfleft{L1}
	\fmfright{R1}
	\fmf{plain,label=$p$}{L1,v1}
	\fmf{plain,left=1,tension=0.5,label=$k$}{v1,v2}
	\fmf{plain,left=1,tension=0.5,label=$p-k$}{v2,v1}
	\fmf{plain}{v2,R1}
\end{fmfgraph*}
}}}};
\end{tikzpicture}
\end{gathered}
\hspace{-4pt}&= \mu^{4-d}\int \frac{d^d k}{i(2\pi)^d} \frac{1}{k^2+i\eps} \frac{1}{(p-k)^2+i\eps} \\&= - \frac{1}{16\pi^2}\left[-\frac{1}{\epsilon} + \ln\left(\frac{-s}{\widetilde{\mu}^2} - i\eps \right)-2\right] \, ,
\end{align}
where $s=p^2$ and $\widetilde{\mu}^2 = 4 \pi e^{-\gamma_E} \mu^2$.  The
counterterm graph is analytic,
so we add it to remove the UV divergence and the algebraic part of the integral (the $-2$ contribution), giving a simpler answer for the renormalized amplitude:
\begin{equation}
    \cM_1 = -\frac{1}{16\pi^2}\ln \left(\frac{-s}{\widetilde{\mu}^2}-i\eps \right)\,.
\end{equation}
The cut through the bubble is finite in four dimensions:
\begin{align}
\cM_1^\text{cut} &= \hspace{-4pt}
\begin{gathered}
\begin{tikzpicture}[baseline=-3.5]
\node at (0,0) {
\parbox{30mm} {
\resizebox{30mm}{!}{
 \fmfframe(0,00)(0,0){
 \begin{fmfgraph*}(80,40)
	\fmfleft{L1}
	\fmfright{R1}
	\fmf{plain,label=$p\rightarrow$}{L1,v1}
	\fmf{plain,left=1,tension=0.5}{v1,v2,v1}
	\fmf{plain}{v2,R1}
\end{fmfgraph*}
}}}};
\draw[dashed, line width = 1, dashed, darkgreen] (0,0.8) to (0,-0.8);
\end{tikzpicture}
\end{gathered}
\hspace{-4pt} \\
&= \int \frac{d^4 k}{i(2\pi)^4} \left(- 2\pi i\right) \delta(k^2)\Theta(k^0) \left(- 2\pi i\right) \delta[(p-k)^2] \Theta(p^0-k^0) = \frac{i}{8\pi} \Theta(s)\,.
\end{align}
Here we have assumed $p^0 >0$. If $p^0 < 0$, this cut vanishes but the cut with energy flowing in the opposite direction compensates and gives the same result.
$\cM_1$ has a branch cut on the positive real line in the $s$ plane. The discontinuity across this branch cut is
\begin{equation}
    \disc_s \cM_1  = -
   \frac{1}{16\pi^2} \left(-2\pi i\right)\Theta(s) = \cM_1^\text{cut} \, , \label{singlediscM1}
\end{equation}
in agreement with the covariant cutting rules and the optical theorem. Similarly, the monodromy computed around the branch point at $s=0$, 
\begin{equation}
    (\bbone - \sM_{\linebub{}^{s}_0})\cM_1 = \frac{i}{8\pi}\,,
\end{equation}
gives the same answer in $R^s$, where $s>0$.

\paragraph{Sequential discontinuities in the same channel}~\\
Now we consider an example that has a nonzero sequential discontinuity in a single channel. We keep the propagators in the loops massless, but give the internal lines connecting the bubbles a mass $m$ so that we can ignore their discontinuities for $m> \sqrt{s}$.
 The chain of three bubbles is given by
\begin{align}
\cM_3 &=\hspace{-5pt}
\begin{gathered}
\begin{tikzpicture}[baseline=-3.5]
\node at (0,0) {
\parbox{80mm} {
\resizebox{80mm}{!}{
 \fmfframe(0,00)(0,0){
 \begin{fmfgraph*}(200,50)
 \fmfkeep{loop3}
	\fmfleft{L1}
	\fmfright{R1}
	\fmf{plain,label=$p\rightarrow$,l.s=right}{L1,v1}
	\fmf{plain,left=1,tension=0.5}{v1,v2,v1}
	\fmf{plain,label=$m$,l.s=left}{v2,v3}
	\fmf{plain,left=1,tension=0.5}{v3,v4,v3}
	\fmf{plain,label=$m$,l.s=left}{v4,v5}
	\fmf{plain,left=1,tension=0.5}{v5,v6,v5}
	\fmf{plain}{v6,R1}
	\fmfv{label=$A$,l.angle=0,l.d=12}{v1}
	\fmfv{label=$B$,l.angle=0,l.d=12}{v3}
	\fmfv{label=$C$,l.angle=0,l.d=12}{v5}
\end{fmfgraph*}
}}}};
\end{tikzpicture}
\end{gathered}
 \hspace{-5pt} \\
 &=\frac{1}{(-16\pi^2)^3}\left( \frac{1}{s-m^2}\right)^{2} \ln^3 \left(\frac{-s}{\mu^2} - i\eps \right)\,.
\end{align}
Since this is just a product of logarithms, the discontinuities in $s$ are simple to calculate using Eq.~\eqref{genlognm}. We find
\begin{equation}
    \disc_s \cM_3 =\frac{2\pi i}{(16\pi^2)^3}\left( \frac{1}{s-m^2}\right)^2
    \left[ 3  \ln^2 \left(\frac{-s}{\mu^2} - i\eps \right) + 6 \pi i  \ln \left(\frac{-s}{\mu^2} - i\eps \right)
    -4 \pi^2 \right]\,, \label{DsM3}
\end{equation}
\begin{equation}
    \disc_s \disc_s \cM_3
=
 - \frac{ (2\pi i)^2}{(16\pi^2)^3}\left( \frac{1}{s-m^2}\right)^2
\left[ 6  \ln \left(\frac{-s}{\mu^2} - i\eps \right) +  12 \pi i \right] \, , \label{DsDsM3}
\end{equation}
and
\begin{equation}
    \disc_s \disc_s \disc_s \cM_3 =\frac{6 (2\pi i)^3}{(16\pi^2)^3}\left( \frac{1}{s-m^2}\right)^2 \,. \label{DsDsDsM3}
\end{equation}
We expect these discontinuities to be related to cuts by Eq.~\eqref{mcutsins}. 
  
Assuming $p^0>0$ and $s>0$, and using all $+i\varepsilon$ propagators, the cut through loop A is given by
\begin{align}
\cM_{3A}^\text{cut} &=\hspace{-5pt}
\begin{gathered}
\begin{tikzpicture}[baseline=-3.5]
\node at (0,0) {
\parbox{80mm} {
\resizebox{80mm}{!}{
 \fmfframe(0,00)(0,0){
 \fmfreuse{loop3}
}}}};
\draw[dashed, line width = 1, dashed, darkgreen] (-2.2,0.8) to (-2.2,-0.8);
\end{tikzpicture}
\end{gathered}\nonumber \\
&=\frac{1}{(-16\pi^2)^3}\left( \frac{1}{s-m^2}\right)^2 (-2\pi i) \ln^2\left(\frac{-s}{\mu^2} - i\eps \right) \, .
\end{align}
The cuts of the second and third loop give identical results since we always assign uncut propagators $+i \eps$. Thus, we have $\cM_{3B}^\text{cut} = \cM_{3C}^\text{cut} = \cM_{AC}^\text{cut}$.
There are also three diagrams involving two cuts. Cutting loops A and B gives
\begin{align}
\cM_{3AB}^\text{cut} &= \hspace{-5pt}
\begin{gathered}
\begin{tikzpicture}[baseline=-3.5]
\node at (0,0) {
\parbox{80mm} {
\resizebox{80mm}{!}{
 \fmfframe(0,00)(0,0){
 \fmfreuse{loop3}
}}}};
\draw[dashed, line width = 1, dashed, darkgreen] (-2.2,0.8) to (-2.2,-0.8);
\draw[dashed, line width = 1, dashed, darkgreen] (0,0.8) to (0,-0.8);
\end{tikzpicture}
\end{gathered} \nonumber \\
 &=\frac{(-2\pi i)^2}{(-16\pi^2)^3}\left( \frac{1}{s-m^2}\right)^2 \ln \left(\frac{-s}{\mu^2} - i\eps \right)\,.
\end{align}
The other diagrams involving two cuts give identical results: $\cM_{3AC}^\text{cut}=\cM_{3BC}^\text{cut}=\cM_{3AB}^\text{cut}$. The triple cut is
\begin{align}
\cM_{3ABC}^\text{cut} &= \hspace{-5pt}
\begin{gathered}
\begin{tikzpicture}[baseline=-3.5]
\node at (0,0) {
\parbox{80mm} {
\resizebox{80mm}{!}{
 \fmfframe(0,00)(0,0){
 \fmfreuse{loop3}
}}}};
\draw[dashed, line width = 1, dashed, darkgreen] (-2.2,0.8) to (-2.2,-0.8);
\draw[dashed, line width = 1, dashed, darkgreen] (0,0.8) to (0,-0.8);
\draw[dashed, line width = 1, dashed, darkgreen] (2.2,0.8) to (2.2,-0.8);
\end{tikzpicture}
\end{gathered}
\hspace{-5pt} \\
&=\frac{(-2\pi i)^3}{(-16\pi^2)^3}\left( \frac{1}{s-m^2}\right)^2\,.
\end{align}
We can now compute the right side of Eq.~\eqref{mcutsins}. For $m=1$, we get
\begin{align}
      - \sum_{k=1}^\infty (-1)^{k} \stirling{k}{1} \cM_3^{\text{($k$-cuts)}} & = \cM_3^{\text{(1-cut)}} - \cM_3^{\text{(2-cuts)}} + \cM_3^{\text{(3-cuts)}} \\ &\hspace{-2.4cm} =  (\cM_{3A}^\text{cut}+\cM_{3B}^\text{cut}+\cM_{3C}^\text{cut}) - (\cM_{3AB}^\text{cut}+\cM_{3AC}^\text{cut}+\cM_{3BC}^\text{cut}) + \cM_{3ABC}^\text{cut} 
      \\
      &\hspace{-2.4cm}  =
      \frac{\left(-2\pi i\right)}{(-16\pi^2)^3}\left( \frac{1}{s-m^2}\right)^2 
    \left[ 3  \ln^2 \left(\frac{-s}{\mu^2} - i\eps \right) + 6 \pi i \ln \left(\frac{-s}{\mu^2} - i\eps \right)
    - 4\pi^2 \right] \,. 
\end{align}
This agrees with $\disc_s {\cM_3}$, as expected. Similarly, for $m=2$ and $m=3$ we get 
\begin{equation}
      2 \sum_{k=2}^\infty (-1)^k \stirling{k}{2} \cM_3^{\text{($k$-cuts)}} = 2 (\cM_{3AB}^\text{cut}+\cM_{3AC}^\text{cut}+\cM_{3BC}^\text{cut}) - 6 \cM_{3ABC}^\text{cut}\,.
\end{equation}
and
\begin{equation}
    - 3!\sum_{k=3}^\infty (-1)^k \stirling{k}{3} \cM_3^{\text{($k$-cuts)}} = 6\cM_{3ABC}^\text{cut} \, .
\end{equation}
It can be checked that these quantities agree with the discontinuities computed in Eq.~\eqref{DsDsM3} and Eq.~\eqref{DsDsDsM3}. 

One can similarly check that the relation in Eq.~\eqref{mcutsins} holds for the $m^{\text{th}}$ discontinuity of the $n$-loop bubble chain. This is not particularly surprising, since the algebra involved is essentially the same as the algebra used to derive equations like Eq.~\eqref{triplecut}.

\begin{figure}[t]
    \centering
\begin{tikzpicture}[baseline=-3.5]
\node at (0,0) {    \includegraphics[width=6cm]{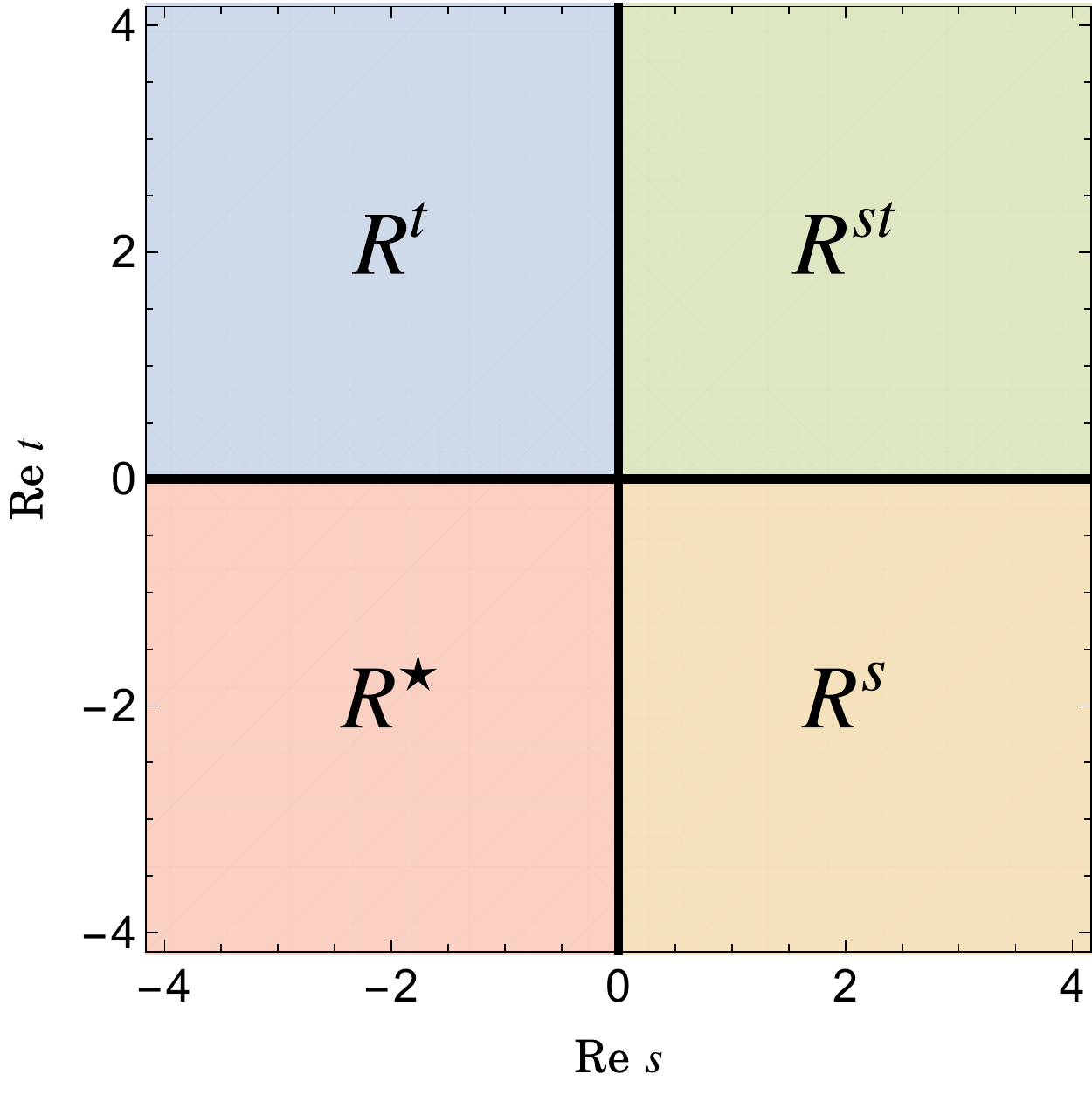}
\hspace{1cm}
    \includegraphics[width=8cm]{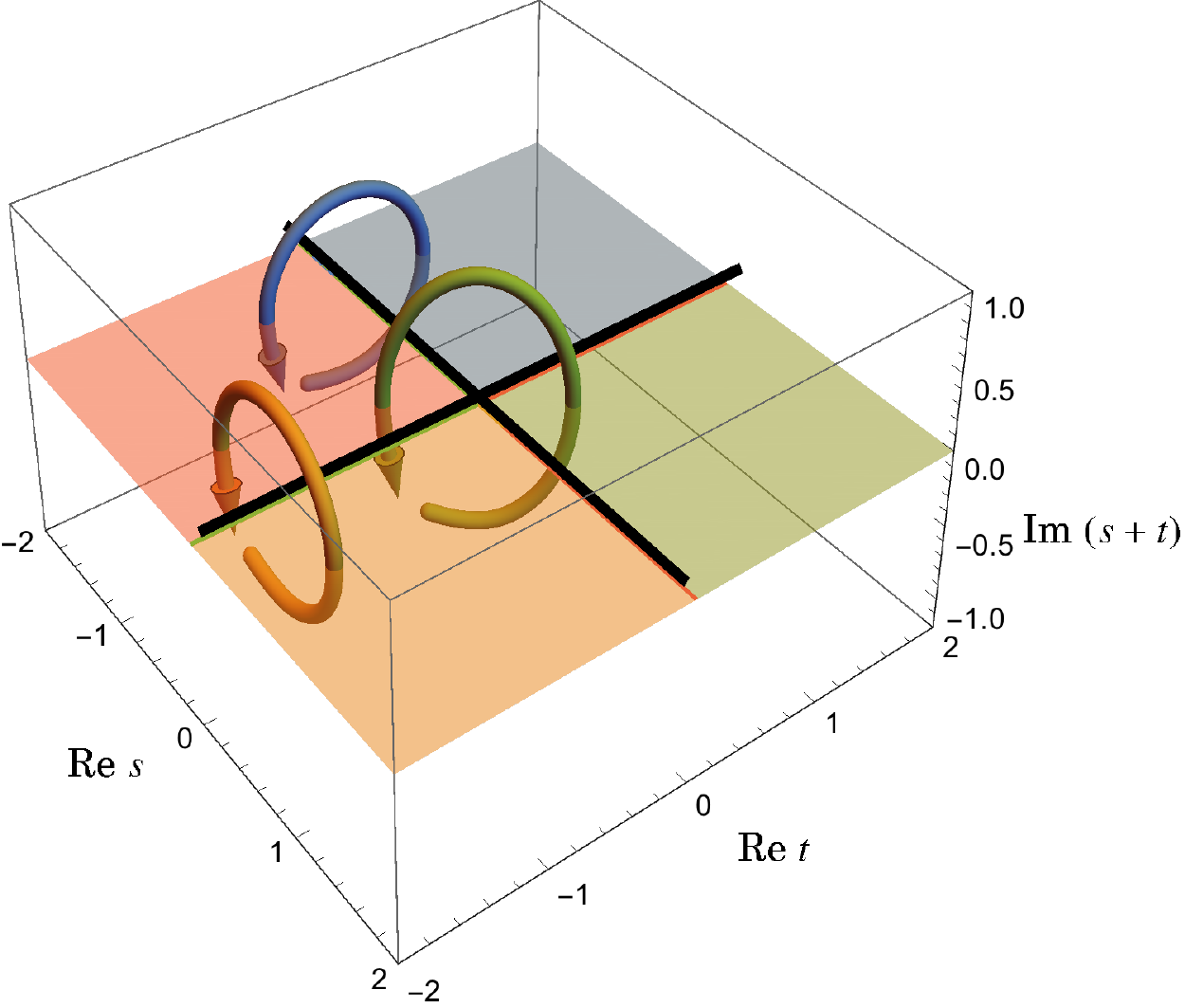}
    };
    \node[scale=1] at (2.6,-0.8) {$R^s$};
    \node[scale=1] at (3.2,2) {$R^t$};
    \node[scale=1] at (4.8,0.5) {$R^{st}$};
    \node[scale=1] at (1,1) {$R^\star$};
    \node[scale=1,darkred] at (5,1.7) {$s = 0$};
    \node[scale=1,darkred] at (4.8,-0.8) {$t = 0$};
     \end{tikzpicture}
    \caption{The function $\ln(-s)\ln(-t)$ has branch hypersurfaces at $s=0$ and $t=0$, shown in black. The Euclidean region $R^\star$ corresponds to $s<0$ and $t<0$. We can compute discontinuities in $s$ and $t$ of $\cM$ by rotating around the branch points as indicated by the curves on the right. These curves pass out of the real $s,t$ plane.}
    \label{fig:stsurf}
\end{figure}
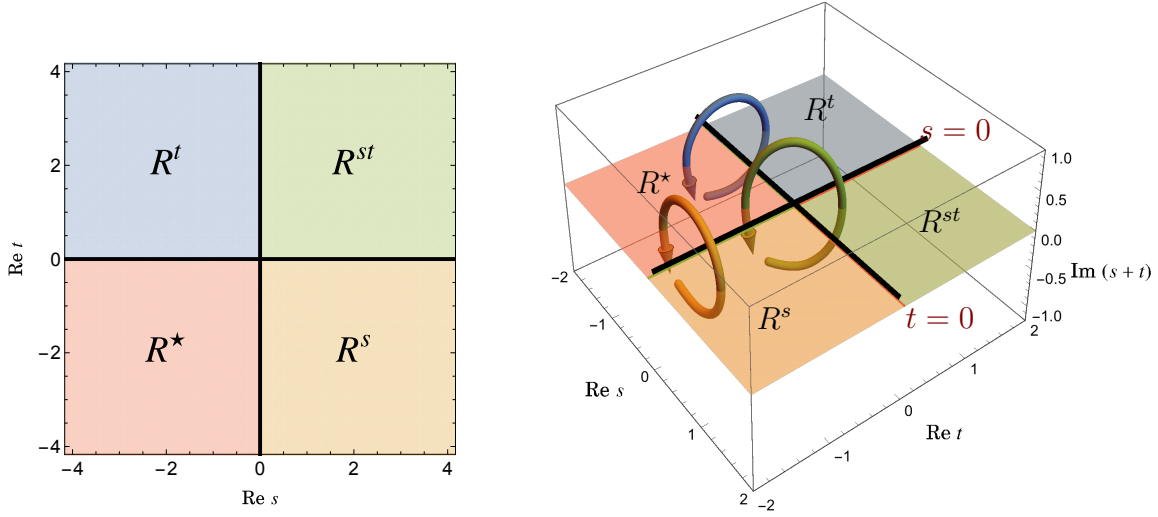

\paragraph{Sequential discontinuities in different channels}~\\
We now turn to an example involving discontinuities in different channels. We consider the diagram
\begin{equation}
\cM_{st} =
\begin{gathered}
\begin{tikzpicture}[baseline=-3.5]
\node at (0,0) {
\parbox{30mm} {
\resizebox{30mm}{!}{
 \fmfframe(0,00)(0,0){
 \begin{fmfgraph*}(80,60)
	\fmftop{s1,s4}
	\fmfbottom{s2,s3}
	\fmf{plain,label=$P_s$}{s1,v1}
	\fmf{plain,left=1,tension=0.5}{v1,v2,v1}
	\fmf{plain}{v2,c}
	\fmf{plain}{c,s2}
    \fmf{plain}{c,s3}
	\fmf{plain}{c,w1}
	\fmf{plain,left=1,tension=0.5}{w1,w2,w1}
	\fmf{plain,label=$P_t$}{w2,s4}
\end{fmfgraph*}
}}}};
\end{tikzpicture}
\end{gathered}
= \frac{1}{256 \pi^4} \ln \left(\frac{-s}{\mu^2}  - i\eps \right) \ln \left( \frac{-t}{\mu^2} - i\eps \right) \, , 
\end{equation}
where $s=P_s^2$ and $t=P_t^2$.  This function has branch points at $s=0$ and at $t=0$. In the space of complex $s$ and $t$, these branch points correspond to one-dimensional complex hypersurfaces. We have depicted this in Fig.~\ref{fig:stsurf}.

The connection and variation matrix for this function in the Euclidean region where $s<0$ and $t<0$ are
\begin{equation}
    \omega = \begin{pmatrix} 
    0 & \frac{ds}{s} & \frac{dt}{t} & 0 \\
    0 & 0 & 0 & \frac{dt}{t} \\
    0 & 0 & 0 & \frac{ds}{s} \\
    0 & 0 & 0 & 0
    \end{pmatrix},\qquad
     \sM_{\gamma_0} = \begin{pmatrix} 
    1 & \ln(-s) & \ln(-t)&  \ln (-s) \ln (-t) \\
    0 & 1 & 0 & \ln (-t) \\
    0 & 0 & 1 & \ln (-s) \\
    0 & 0 & 0 & 1
    \end{pmatrix}   ,
\end{equation}
and the monodromy matrices are
\begin{equation}
    \sM_{\linebub_0^s} = \begin{pmatrix} 
    1 & 2\pi i & 0 & 0 \\
    0 & 1 & 0 & 0 \\
    0 & 0 & 1 & 2 \pi i \\
    0 & 0 & 0 & 1
    \end{pmatrix},\qquad
     \sM_{\linebub_0^t} = \begin{pmatrix} 
    1 & 0 & 2\pi i&  0 \\
    0 & 1 & 0 & 2\pi i \\
    0 & 0 & 1 & 0 \\
    0 & 0 & 0 & 1
    \end{pmatrix}   .
\end{equation}
The variation matrix in a region with $s>0$ and/or $t>0$ is the same with $\ln(-s) \to \ln(-s- i\eps)$ and/or $\ln(-t) \to \ln(-t- i\eps)$.

We can compute $\disc_s \disc_t \cM_{st}$ by computing monodromies around the branch points at $s=0$ and $t=0$. First, the discontinuity in $s$ gives
\begin{equation}
    \disc_s \cM_{st} =
 (\bbone -\sM_{\linebub{}^{s}_{0}})\cM_{st} =
    \frac{-2\pi i}{256 \pi^4}  \ln \left( \frac{-t}{\mu^2} - i\eps \right)\,.
    \label{dsMrs}
\end{equation}
Computing the discontinuity in $t$ of this quantity gives 
\begin{equation}
\disc_t  \disc_s \cM_{st}=
(\bbone -\sM_{\linebub{}^{s}_{0}})
(\bbone -\sM_{\linebub{}^{t}_{0}})\cM_{st}
=
\frac{(-2\pi i)^2}{256 \pi^4}\,. \label{dsdtus}
\end{equation}
To compute the cuts, we must be in the region $R^{\{s,t\}}$ where neither cut vanishes. There, we find
\begin{equation}
\left[\cut_{st} \cM_{st} \right]_{R^{\{s,t\}}}=\frac{(-2\pi i)^2}{256 \pi^4} \,. \label{cutsst}
\end{equation}
We see that the cut and the sequential discontinuity agree, as they should according to Eq.~\eqref{ddstRst1}.

We can also compute the total discontinuity of this function in $R^{\{s,t\}}$, 
\begin{align}
 \left[\tDisc \cM_{st}\right]_{R^{\{s,t\}}} &= \cM_{st} - \cMb_{st} \nonumber \\
 &= \frac{1}{256 \pi^4} \left[\ln \left(\frac{-s}{\mu^2}  - i\eps \right) \ln \left( \frac{-t}{\mu^2} - i\eps \right) -\ln \left(\frac{-s}{\mu^2}
 + i\eps \right) \ln \left( \frac{-t}{\mu^2} + i\eps\right) \right] \nonumber\\
 &=  \frac{-2\pi i}{256 \pi^4}  \left[ \ln \left(\frac{-s}{\mu^2}  - i\eps \right)+\ln \left(\frac{-t}{\mu^2}  + i\eps \right)\right] 
 \end{align}
 in agreement with the standard cut prescription, where the $i\eps$ is flipped on the $\ln(-t)$ propagator because it comes after the $s$ cut. We can also write this in our standardized form, where the $i\eps$ are homogeneous:
 \begin{equation}
  \left[\tDisc \cM_{st}\right]_{R^{\{s,t\}}}=  \frac{-2\pi i}{256 \pi^4}  \left[ \ln \left(\frac{-s}{\mu^2}  - i\eps \right)+\ln \left(\frac{-t}{\mu^2}  - i\eps \right) + 2\pi i\right]%
 \label{discst}
 \,.
\end{equation}
According to Section~\ref{sec:seqdiff}, this should match the function returned by the operator $\disc_{\{s,t\}}$, which corresponds to analytically continuing around both the branch points $s=0$ and $t=0$ along a path $R^{\{s,t\}} \to R^\star \to R^{\{s,t\}}$, as depicted by the green curve in Fig.~\ref{fig:stsurf}. The result is
\begin{align}
    \left[\disc \cM_{st} \right]_{R^{\{s,t\}}} &= (\bbone -\sM_{\linebub{}^{s}_0} \sM_{\linebub{}^{t}_0}) \cM_{st}\, \\
    &= \frac{-2\pi i}{256 \pi^4}  \left[ \ln \left(\frac{-s}{\mu^2}  - i\eps \right)+\ln \left(\frac{-t}{\mu^2}  - i\eps \right) + 2 \pi i \right] \, ,
\end{align}
in agreement with Eq.~\eqref{discst}.

\subsection{Triangles and Boxes \label{sec:triangle}}
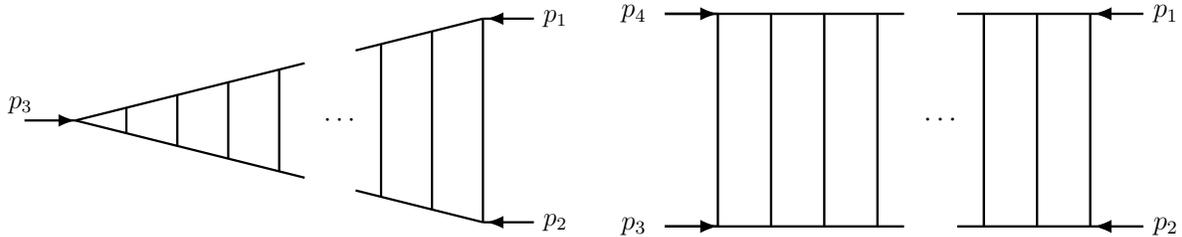
\begin{figure}[t]
    \centering
      \resizebox{80mm}{!}{
    \begin{tikzpicture}[baseline=-3.5,scale=0.8,>=latex]
        \draw[->-,line width = 1] (0,0) -- (1,0);
        \draw[line width = 1] (1,0) -- (5.5,1.125);
        \draw[line width = 1] (6.5,1.375) -- (9,2);
        \draw[line width = 1] (2,0.25) -- (2,-0.25);
        \draw[line width = 1] (3,0.5) -- (3,-0.5);
        \draw[line width = 1] (4,0.75) -- (4,-0.75);
        \draw[line width = 1] (5,1) -- (5,-1);
        \draw[line width = 1] (7,1.5) -- (7,-1.5);
        \draw[line width = 1] (8,1.75) -- (8,-1.75);
        \draw[line width = 1] (1,0) -- (5.5,-1.125);
        \draw[line width = 1] (6.5,-1.375) -- (9,-2);
        \draw[line width = 1] (9,2) -- (9,-2);
        \draw[-<-,line width = 1] (9,2) -- (10,2);
        \draw[-<-,line width = 1] (9,-2) -- (10,-2);
        \node[text width=0.5cm] at (0,0.3) {$p_3$};
        \node[text width=0.5cm] at (10.5,2.0) {$p_1$};
        \node[text width=0.5cm] at (10.5,-2) {$p_2$};
        \node[text width=1cm] at (6.5,0) {$\cdots$};
\end{tikzpicture}
}
    \resizebox{80mm}{!}{
    \begin{tikzpicture}[baseline=-3.5,scale=0.8,>=latex]
        \draw[line width = 1] (1,2) -- (5.5,2);
        \draw[->-,line width = 1] (1,2) -- (2,2);
        \draw[line width = 1] (6.5,2) -- (9,2);
        \draw[line width = 1] (2,2) -- (2,-2);
        \draw[line width = 1] (3,2) -- (3,-2);
        \draw[line width = 1] (4,2) -- (4,-2);
        \draw[line width = 1] (5,2) -- (5,-2);
        \draw[line width = 1] (7,2) -- (7,-2);
        \draw[line width = 1] (8,2) -- (8,-2);
        \draw[line width = 1] (1,-2) -- (5.5,-2);
        \draw[->-,line width = 1] (1,-2) -- (2,-2);
        \draw[line width = 1] (6.5,-2) -- (9,-2);
        \draw[line width = 1] (9,2) -- (9,-2);
        \draw[-<-,line width = 1] (9,2) -- (10,2);
        \draw[-<-,line width = 1] (9,-2) -- (10,-2);
        \node[text width=0.5cm] at (0.5,2) {$p_4$};
        \node[text width=0.5cm] at (0.5,-2) {$p_3$};
        \node[text width=0.5cm] at (10.5,2.0) {$p_1$};
        \node[text width=0.5cm] at (10.5,-2) {$p_2$};
        \node[text width=1cm] at (6.5,0) {$\cdots$};
\end{tikzpicture}
}
    \caption{The $L$-loop triangle and box ladder integrals. We take all momenta incoming with $p_3$ along the long direction of the triangle. For the box ladders,  $s=(p_1+p_2)^2$ and $t=(p_2+p_3)^2$.}
    \label{fig:tribox}
\end{figure}

Next we consider the triangle and box ladder integrals, with massless internal lines and massive external lines. These integrals are known to all loop orders~\cite{Usyukina:1993ch}, and can be treated simultaneously because they give rise to the same transcendental function at each order. For simplicity, we concentrate mostly on the triangle ladders, and comment on the box ladders at the end of the section. Our momentum labeling convention is shown in Fig.~\ref{fig:tribox}. All momenta are incoming, and we have $\sum p_i^\mu = 0$. 

\subsubsection{Triangle kinematics}
For the triangle integrals, we follow the conventions  of~\cite{Abreu:2014cla} and~\cite{Chavez:2012kn}. Since all internal lines are massless, the amplitude depends only on ratios of the invariants $p_1^2$ , $p_2^2$, and $p_3^2$. These kinematics can be parametrized using the variables $u$, $v$, $z$, and $\zb$, defined as
\begin{equation}
    u\equivD \frac{p_2^2}{p_1^2} = z \zb \quad \text{and}\quad v\equivD \frac{p_3^2}{p_1^2} = (1-z)(1-\zb)\,, \label{uvdef}
\end{equation}
where we choose
\begin{gather}
    z = \frac{1+u-v + \sqrt{1+ u^2 + v^2 - 2 u v - 2 u - 2 v}}{2},     \label{zdef}
    \\
    \zb = \frac{1+u-v - \sqrt{1+ u^2 + v^2 - 2 u v - 2 u - 2 v}}{2} \, .
    \label{zbdef}
\end{gather}
This corresponds to the convention that $\zb \le z$ for real kinematics. The triangle ladders are invariant under the $\mathbb{Z}_2$ symmetry $z \leftrightarrow \zb$.

For these integrals, it is possible to find real phase-space points with any pattern of signs for the invariants $p_1^2$, $p_2^2$, and $p_3^2$.
We denote the region where $p_1^2>0$ and $p_2^2, p_3^2 <0$ by $R^1$. In this region, $z$ and $\zb$ are real, and $\zb < 0$ while $1 <z$. 
Similarly, we denote the region in which $p_2^2 >0$ and $p_1^2, p_3^2<0$ by $R^2$, and here we have $\zb < 0 < z < 1$. Finally, we denote by $R^3$ the region where $p_3^2 >0$ and $p_1^2, p_2^2 <0$, which implies $0<\zb<1<z$. We also consider dual regions in which two invariants are positive, such as $R^{23}$, where $p_1^2<0$ and $p_2^2, p_3^2 >0$, and so on. 
Since taking 
$p_j^2 \to - p_j^2$ for all $j$ leaves $u$ and $v$ invariant (and therefore also $z$ and $\zb$), any function of $u$ and $v$ has the same form in a given region and the dual region in which all invariants have the opposite sign. For example, functions of $u$ and $v$ take the same form in $R^{23}$ and $R^1$. It is nevertheless important to distinguish a region from its dual because cuts can only be nonzero for positive invariants.

\begin{figure}[t]
    \centering
    \includegraphics[width=8.2cm]{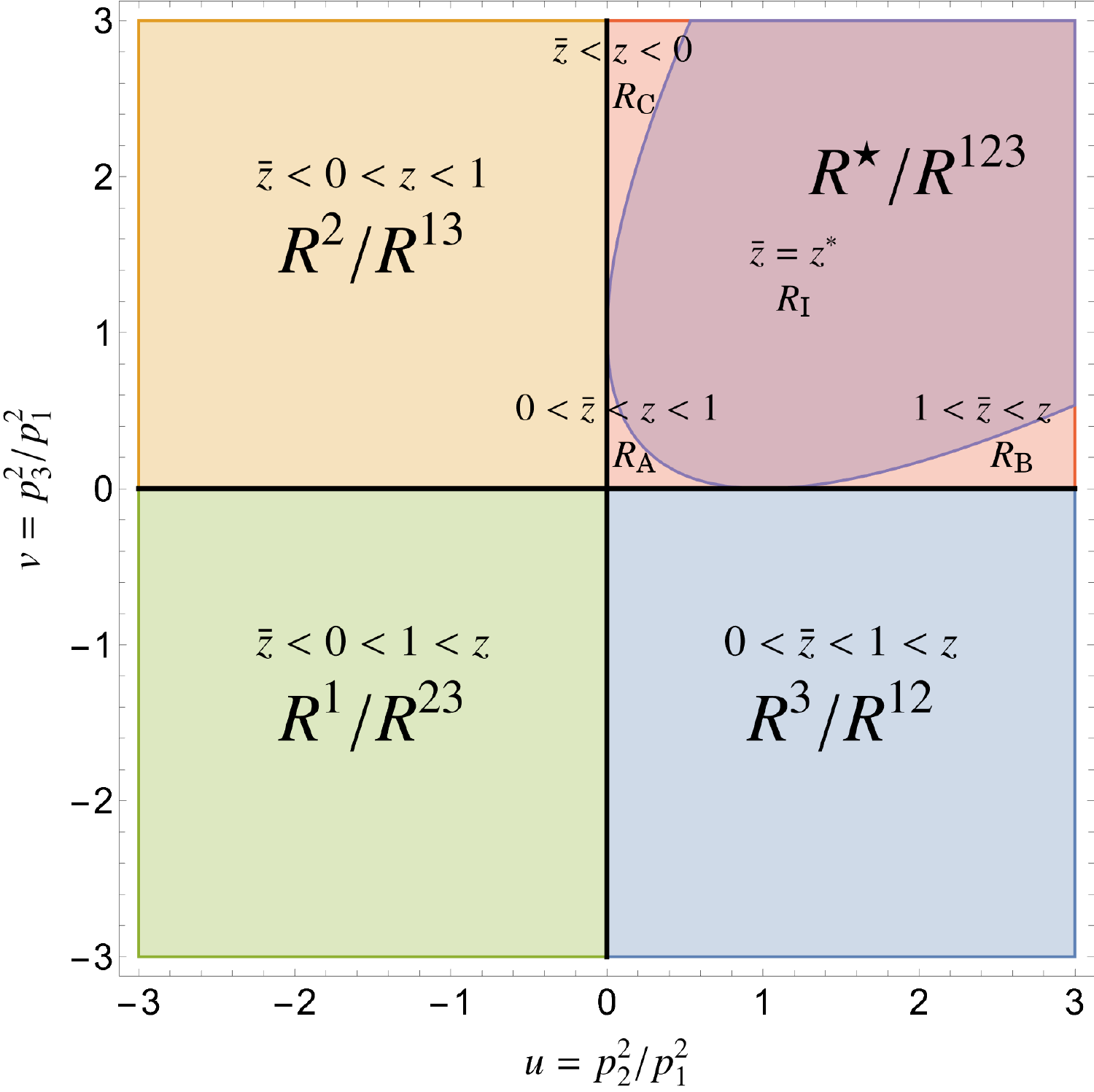}
    \includegraphics[width=8cm]{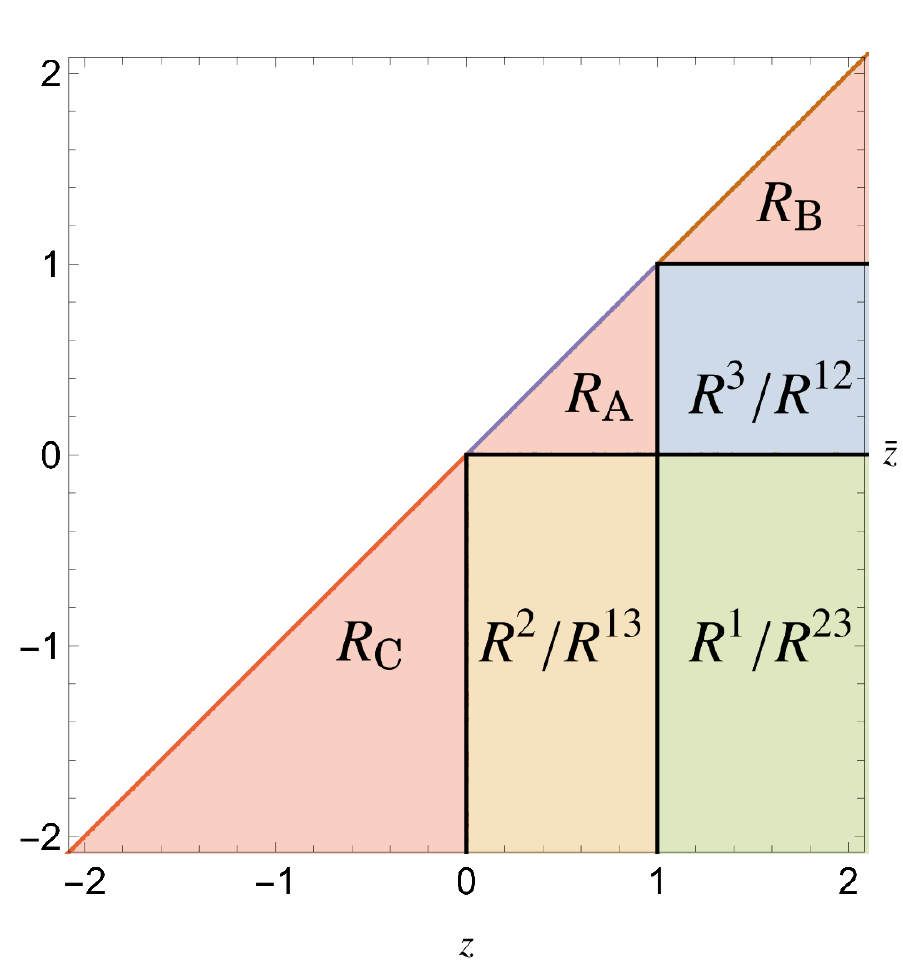}
    \caption{The triangle ladder integrals we consider depend only on $u=p_2^2/p_1^2$ and $v=p_3^2/p_1^2$, or equivalently on $z$ and $\zb$. The different regions in $u, v$ and $z,\zb$ space correspond to regions in which the Mandelstam invariants have different relative signs. For instance, in $R^1$ the invariants satisfy $p_1^2 >0$, $p_2^2<0$, and $p_3^2<0$. The Euclidean region, where $p_j^2 <0$ for all $j$, has four further subregions, described in the text.}
    \label{fig:uvzz}
\end{figure}

The Euclidean region, where all invariants are negative, is denoted $R^\star$. The Euclidean region has a number of subregions, based on the relative sizes of the $p_j^2$ invariants (or equivalently of $z$ and $\zb$). Of particular importance is the region $R_A^\star$, which corresponds to real values $0<\zb<z<1$. The functions $\ln z, \ln \zb$, $\Li_n z$, and $\Li_n \zb$ are all analytic in this region. Region $R_C^\star$ corresponds to real $\zb < z <0$, and region  $R_B^\star$ corresponds to real $1<\zb<z$. Finally, region $R_{\text{I}}^\star$
involves complex $z$ and $\zb$ that are related by complex conjugation, namely $\zb = z^*$. All of these regions correspond to two-dimensional slices of the four-dimensional space of complex $z$ and $\zb$, in which all the invariants $p_i^2$ are real. The dual of the Euclidean region, where all invariants are negative, is denoted $R^{123}$ and also has subregions corresponding to $R_A^\star$, $R_B^\star$ and $R_C^\star$.
A summary of the regions is shown in Fig.~\ref{fig:uvzz}.

To take sequential discontinuities of Feynman integrals, we analytically continue around branch points where Mandelstam invariants vanish. This analytic continuation takes us into different kinematic regions; for example, to take $\big[\disc_{p_1^2}\big]_{R^1}$ we need to analytically continue from $R^1$ to $R^\star$ and back. Our formula relating cuts and discontinuities assumes that we rotate the energies while preserving $E_1+E_2+E_3=0$ and holding all three-momenta fixed. Thus, we can set $E_3=-E_1-E_2$ and $\vec{p}_3 = -\vec{p}_1-\vec{p}_2$ and work in a frame where all momenta are aligned in the $x$ direction. Then, rescaling these momenta so that $p_1^x=1$, we can solve for $E_1$ and $E_2$ in terms of $z$, $\zb$, and the remaining unfixed momentum component $p_2^x$:
\begin{equation}
    E_1= \frac{-2 p_2^x - (z+\zb)}{\zb - z}, \qquad
    E_2 = \frac{2 z\zb + p_2^x (z+\zb)}{\zb - z} \, . \label{eq:energy_framing}
\end{equation}
One can use these equations to translate a given path in $z$ and $\zb$ to an acceptable path in energy for a given value of $p_2^x$. It turns out, however, that an analytic continuation path cannot be found between any pair of regions. For example, we cannot go from $R^1$ to $R_{\text{A}}^\star$. To see this, note that in these coordinates, the invariants are given by
\begin{equation}
    p_1^2 = \frac{4(p_2^x + z)(p_2^x + \zb)}{(z-\zb)^2},\quad
    p_2^2 = z\zb p_1^2,\quad  p_3^2 = (1-z)(1-\zb) p_1^2 \, .
\end{equation}
In $R_{\text{A}}^\star$, all the $p_j^2$ are negative. For a fixed value of $p_2^x>0$, this constraint is impossible to satisfy, as $z>\zb>0$ in $R_{\text{A}}^\star$, which implies $p_1^2>0$. In fact, we need $-1<p_2^x<0$ to get to $R_{\text{A}}^\star$. But then, in $R^1$ where $0<\zb<1<z$, we must have $p_2^x + z>0$ and $p_2^x + \zb<0$, and so $p_1^2 <0$. But this is a contradiction, since $p_1^2$ must be positive in $R^1$. Thus, we cannot go from $R^1$ to $R_{\text{A}}^\star$.

In addition to making sure the path exists, one must check that the path only encircles the desired branch points in the invariants once. For example, in particle $j$'s rest frame,  $E_j\to e^{2\pi i} E_j$ would not be an acceptable path, as it would encircle the branch point in $p_j^2$ twice.

\begin{figure}[t]
    \centering
\begin{tikzpicture}
\hspace{-10mm}
\node at (0,0) {    
\includegraphics[width=7.8cm]{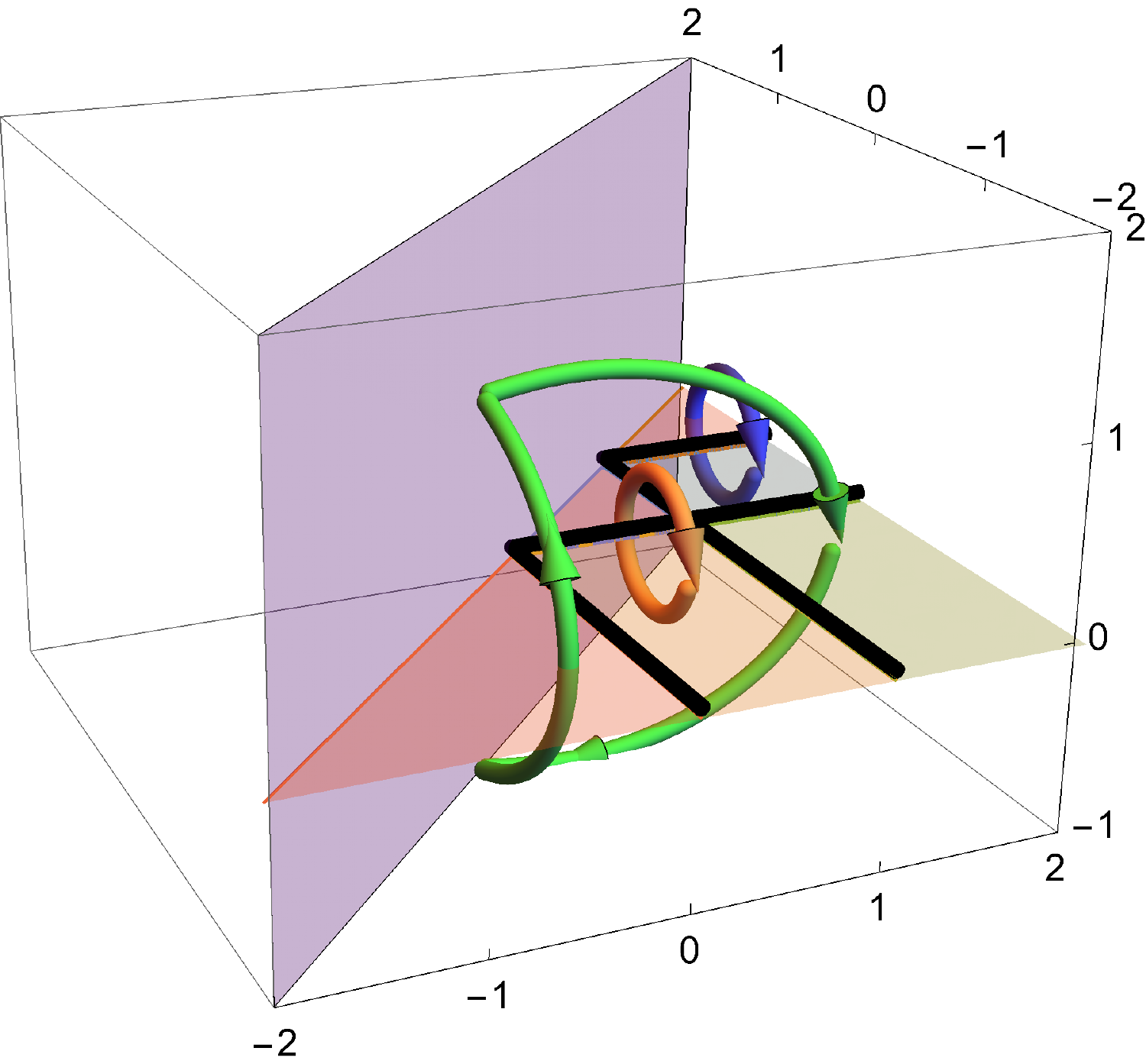}
    };
    \node[left,scale=0.8,darkred] at (-0.4,-1.2) {$R_{\text{C}}$};
    \node[left,scale=0.8,darkred] at  (1.3,-0.8) {$R^2$};
    \node[left,scale=0.8,darkred] at  (2.8,-0.4) {$R^1$};    
    \draw[->]  (1.6,1.8) to  (1.4,0.4);
    \node[above,scale=0.8,darkred] at  (1.6,1.8) {$R^3$};
    \draw[->]  (0,1.5) to  (0.2,0.2) ;
    \node[above,scale=0.8,darkred] at (0,1.5) {$R_{\text{A}}$};
    \draw[->]  (0.5,1.7) to  (0.65,0.7) ;
    \node[above,scale=0.8,darkred] at (0.5,1.7) {$R_{\text{B}}$};
    \node[scale=1,rotate=10] at (1,-3.2) {$\re z$};
    \node[scale=1,rotate=85] at (4.0,0) {$\im(z+ \zb)$};
    \node[scale=1,rotate=-20] at (2.5,3.2) {$\re \zb$};
    \node[scale=0.8] at (1.1,-1.5) {$z = 0$};
    \node[scale=0.8] at (2.4,-1.2) {$z = 1$};
    \node[right,scale=0.8] at (2.2,0.4) {$\zb = 0$};
    \node[right,scale=0.8] at (1.6, 0.8) {$\zb = 1$}; 
\end{tikzpicture}
\hspace{-15mm}
\begin{tikzpicture}
\node at (0,0.1) {
\includegraphics[width=8.4cm]{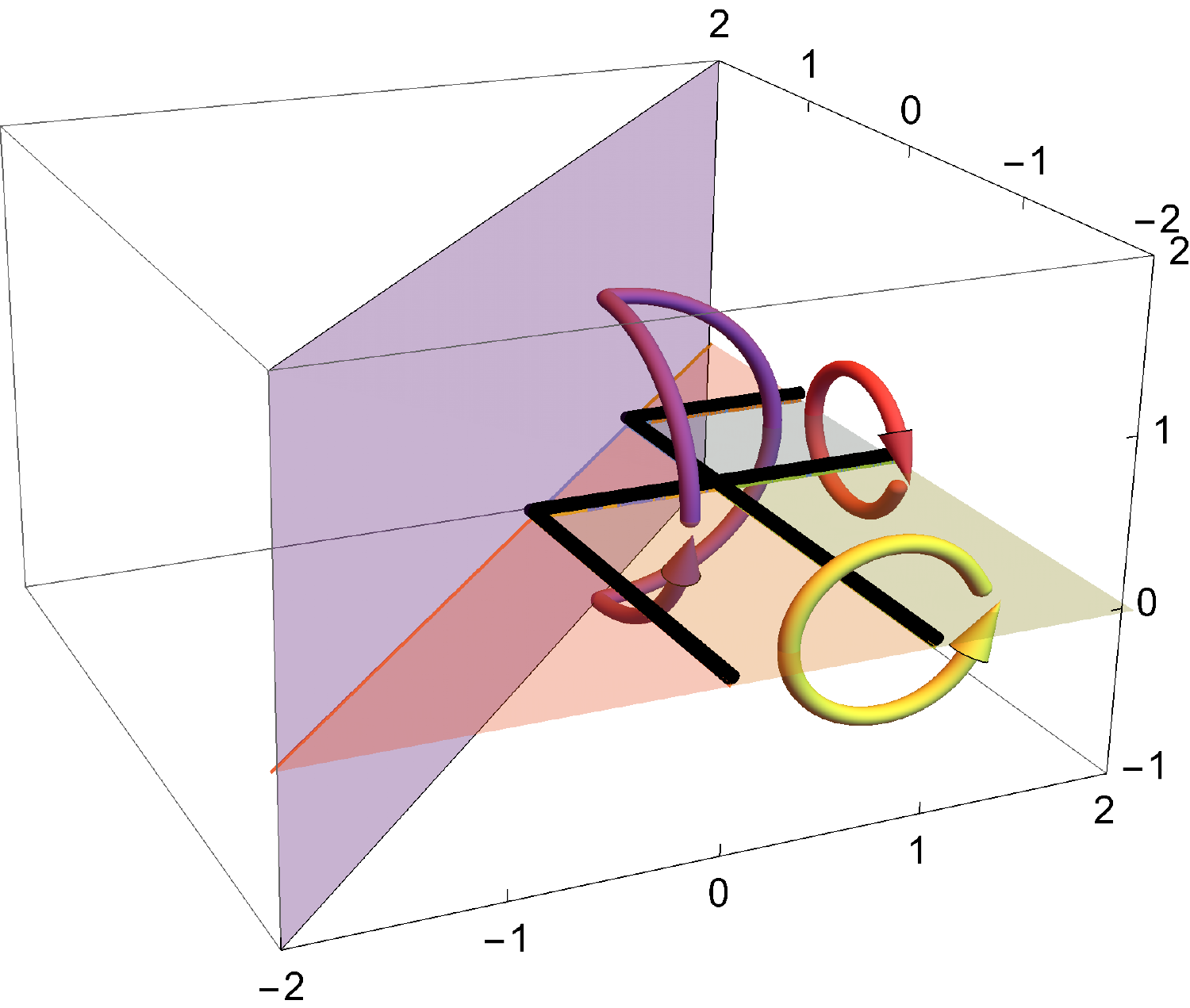}
    };
  \begin{scope}[shift={(0.3,0.2)}]
    \draw[->]  (0,-2) to  (0.7,-0.7);
    \node[below,scale=0.8,darkred] at  (0,-2) {$R^2/R^{13}$};
    \draw[->]  (2.9,0.7) to  (2,-0.2);
    \node[above,scale=0.8,darkred] at  (2.9,0.7) {$R^1/R^{23}$};    
    \draw[->]  (1.3,1.8) to  (1.1,0.4);
    \node[above,scale=0.8,darkred] at  (1.3,1.8) {$R^3/R^{12}$};
    \node[scale=1,rotate=10] at (1,-3.2) {$\re z$};
    \node[scale=1,rotate=85] at (4.0,0) {$\im(z+ \zb)$};
    \node[scale=1,rotate=-20] at (2.5,3.0) {$\re \zb$};
    \end{scope}
\end{tikzpicture}
    \caption{Analytic continuation of three-point and four-point ladder diagrams takes place in the four-dimensional space of complex $z$ and $\zb$. 
    Sample paths of analytically continuing the energies are shown. The figure on the left depicts contours that are relevant for computing sequential discontinuities in a single channel: 
    ${\green{R^1\to R_{\text{C}}^\star \to R^1}}$, ${\color{orange}{R^2 \to R_{\text{A}}^\star\to R^2}}$ and
    ${\blue{R^3 \to R_{\text{B}}^\star \to R^3}}$. The figure on the right depicts paths relevant for computing sequential discontinuities in different channels: ${\color{darkyellow}{R^{23}  \to R^{2} \to R^{23} }}$, 
    ${\color{darkpurple}{R^{13}  \to R^{3} \to R^{13} }}$
    and ${\red{R^{23}\to R^3 \to R^{23}}}$. These paths each encircle some combination of the branch hypersurfaces shown as black lines, corresponding to where $z$ or $\zb$ are equal to either 0 or 1.}
    \label{fig:zz3d}
\end{figure}
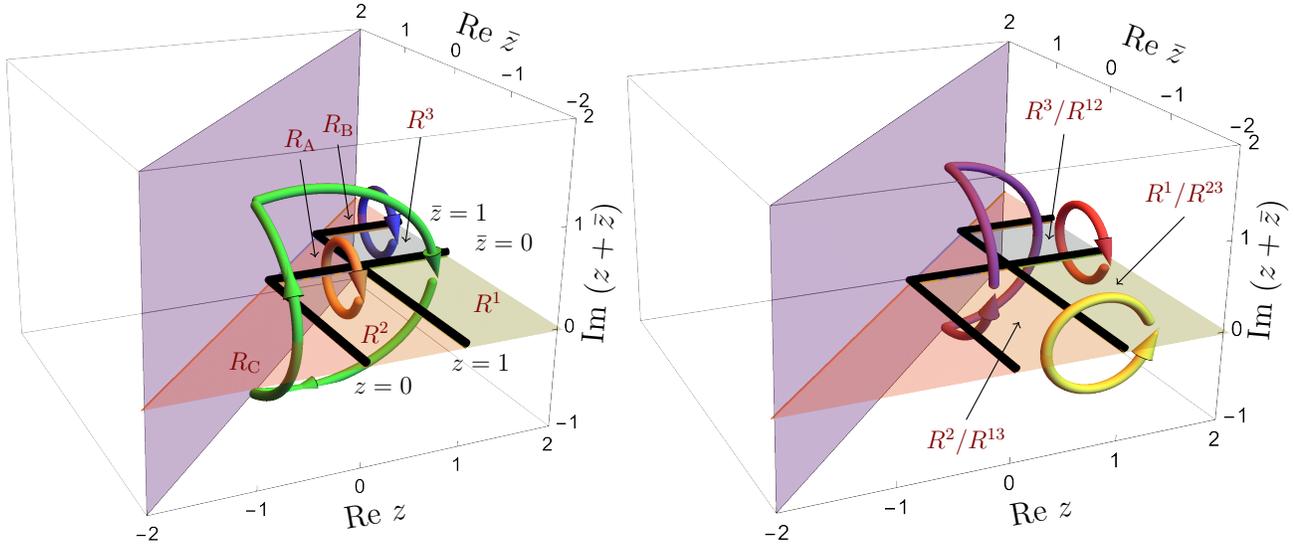

Some paths that satisfy all of these constraints are shown in Fig.~\ref{fig:zz3d}. For example, we show a path from $R^2\to R^\star_\text{A} \to R^2$. It is also possible to construct a path from  $R^2\to R^\star_\text{C} \to R^2$. Conversely, no path exists from $R^2$ to $R^*_\text{B}$, by the same type of argument that showed the impossibility of analytically continuing between $R^1$ and $R_{\text{A}}^\star$. We also show a path that starts and ends in $R^1$, after passing through $R^\star_\text{C}$. When this path intersects the $\re z = \re \zb$ plane, the branch cut in the square root that distinguishes $z$ and $\zb$ is crossed. This path can be viewed as going around $z=0$ and $z=1$, or as going around $z=\infty$.
The right side of this figure shows paths between other regions, such as $R^{23} \to R^3 \to R^{23}$. The existence of such paths is required to take sequential discontinuities in $p_2^2$ and $p_3^2$.  

Having constructed these paths, we can enumerate the monodromies corresponding to each of the discontinuities we're interested in computing. For sequential  discontinuities in a single channel, we find
\begin{subequations}
\label{discdelta}
\begin{align}
    [\disc_{p_1^2}]_{R^1} &= \bbone - 
   \mathscr{M}_{\linebubR{}^z_0}\cdot
 \mathscr{M}_{\linebubR{}^z_1}=
 \bbone - \mathscr{M}^{\linebubDown z}_\infty
 \, ,
    \\
    [\disc_{p_2^2}]_{R^2} &=\bbone -\sM_{\linebub{}_0^{\zb}} \, ,
    \\
     [\disc_{p_3^2}]_{R^3} &=\bbone -\sM_{\linebub{}_1^{\zb}}\, .
\end{align}
\end{subequations}
In each case, there are two choices of Euclidean region that we can pass through (e.g. $R^2 \to R^*_\text{A} \to R^2 $ or $R^2 \to R^*_\text{C} \to R^2$). This choice amounts to permuting $z\leftrightarrow\zb$. 
The monodromy matrix $\smash{\mathscr{M}^{\linebubDown z}_\infty}$ corresponds to going around infinity counterclockwise, where infinity is approached along some angle that goes below the real line. This implies that the contour around infinity crosses the branch cut on the negative real axis before the one on the positive real axis. This choice to go below the real axis corresponds to taking $p_1^2$ to have a small positive imaginary part, which endows $z$ with a small negative imaginary part, as per Eq.~\eqref{zdef}. This monodromy matrix is computed in Appendix~\ref{sec:fundamental_group}.

To compute sequential discontinuities in different channels, we consider analytic continuation paths from regions with multiple positive invariants to regions in which one of these invariants has the opposite sign. To construct the discontinuity operator corresponding to each of these analytic continuations, we need to determine which branch points in $z$ and $\zb$ the path encircles.
Let us illustrate how this can be done for the path from $R^{12} \to R^2 \to R^{12}$, which computes a discontinuity in $p_1^2$ in the region $R^{12}$. We first take the differential of Eq.~\eqref{uvdef}:
\begin{align} \label{eq:monodromy_constraint_1}
    d \ln z + d \ln \zb &= d \ln p_2^2 - d \ln p_1^2 \, , \\
    d \ln (1-z) + d \ln (1-\zb)  &= d \ln p_3^2 - d \ln p_1^2\, . \label{eq:monodromy_constraint_2}
\end{align}
Since we are considering a discontinuity in $p_1^2$, our path $\gamma$ must satisfy
\begin{gather}
    \oint_\gamma d \ln p_1^2 = 2 \pi i, \qquad
    \oint_\gamma d \ln p_2^2 = \oint_\gamma d \ln p_3^2 = 0\,.
\end{gather}
Eqs.~\eqref{eq:monodromy_constraint_1} and~\eqref{eq:monodromy_constraint_2} then imply that
\begin{gather}
    \oint_\gamma (d\ln z + d\ln \zb) = -2 \pi i\,, \qquad
    \oint_\gamma (d\ln (1-z) + d\ln(1-\zb)) = -2 \pi i\,. \label{eq:R12toR1toR1constraint}
\end{gather}
We furthermore have that $0 < \zb < 1 < z$ in $R^{12}$, while $\zb < 0 < z < 1$ in $R^2$. This suggests that $z$ should encircle 1 while $\zb$ should encircle 0 along this path. We see that this can be achieved in a manner consistent with Eq.~\eqref{eq:R12toR1toR1constraint} if both of these branch points are encircled clockwise. Thus, we conclude that $[\disc_{p_1^2}]_{R^{12}} =\bbone- 
       \sM_{\linebubR_{0}^{\zb}}\cdot
       \sM_{\linebubR_{1}^{z}}$.

Using similar reasoning, we compute the discontinuity operators in each of the regions involving two positive invariants to be
\begin{subequations}
\label{sdisc}
\begin{align}
      [\disc_{p_2^2}]_{R^{23}} &=  \bbone-\sM_{\linebub{}_0^{\zb}}\hspace{38pt},  \qquad\qquad
      [\disc_{p_3^2}]_{R^{23}} = \bbone-\sM_{\linebub_1^{z}},\\
      [\disc_{p_1^2}]_{R^{13}} &= \bbone-
      \sM_{\linebubR_{0}^{\zb}} \cdot \sM_{\linebubR_{1}^{z}},
      \qquad\qquad
      [\disc_{p_3^2}]_{R^{13}} = \bbone-\sM_{\linebub_1^{z}},\\
      [\disc_{p_1^2}]_{R^{12}} &=\bbone- 
       \sM_{\linebubR_{0}^{\zb}}\cdot
        \sM_{\linebubR_{1}^{z}},
        \qquad\qquad
      [\disc_{p_2^2}]_{R^{12}} =\bbone- \sM_{\linebub{}_0^{\zb}}.
\end{align}
\end{subequations}
In contrast to the first discontinuity, the region that we pass through is completely fixed, so there is only a single correct monodromy matrix in each of these cases. The paths corresponding to these discontinuity operators are depicted in Fig.~\ref{fig:zz3d}. 

One can also consider other analytic continuation paths, such as $R_C^{123} \to R^1 \to R_C^{123}$ (not shown in the figure). Such a path exists and gives us the discontinuity with respect to the pair of invariants $\set_{23} = \{p_2^2,p_3^2\}$. This path encircles $z=0$ and $z=1$, so
\begin{equation}
    [\disc_{\set_{23}}]_{R^{123}_C} = \bbone-\sM_{\linebub_1^{z}}\cdot \sM_{\linebub_0^{z}}
    \label{R123toR1}\, .
\end{equation}
Other paths that encircle the branch points of more than one invariant are also possible.

It is easiest to compute the monodromy matrices in one region and then continue the result to the other regions. The most natural region to use is  $R_A^\star$, since $0<\zb<z<1$ so all of $\ln z, \ln \zb, \Li_n(z)$ and $\Li_n(\zb)$ are analytic there. 
To evaluate the matrices for real values of $z$ and $\zb$ below 0 or above 1, we need to be careful about which side of the branch cuts we are on. In the region $R^i$, where only $p_i^2>0$ and the other squared momenta are negative, we assign $p_i^2$ a small positive imaginary part. It can be checked using Eqs.~\eqref{zdef} and~\eqref{zbdef} that this corresponds to giving $z$ and $\zb$ the following small imaginary parts in these regions:
\begin{subequations}
\label{eq:region_small_imaginary_parts}
\begin{align}
    R^1, R^{12}, R^{13}, R^{123}:&\quad z \to z- i\eps, \quad \zb \to \zb + i \eps \, ,\\
    R^2,R^3, R^{23}:&\quad z \to z+ i\eps,\quad  \zb \to \zb - i \eps \, .
\end{align} 
\end{subequations}
These assignments allow us to evaluate the variation matrix and monodromy matrices in the different regions. 

\subsubsection{One loop}
\label{sec:one_loop_triangle}

The one-loop triangle with all massless internal lines is finite in four dimensions. In the region $R^\star_\text{I}$, where all invariants are  negative and $\zb = z^*$, the Feynman integral is
\begin{align}
T_1 =
\begin{gathered}
\begin{tikzpicture}[baseline=-3.5]
\node at (0,0) {
\parbox{30mm} {
\resizebox{30mm}{!}{
 \fmfframe(0,00)(0,0){
 \begin{fmfgraph*}(80,50)
    \fmfstraight
	\fmfleft{L1}
	\fmfright{R1,R2}
	\fmf{plain_arrow,label=$p_1$}{L1,v1}
	\fmf{plain,label=$q_1$,tension=0.5}{v1,v2}
	\fmf{plain,label=$q_3$,l.s=right,tension=0.5}{v2,v3}
	\fmf{plain,label=$q_2$,l.s=right,tension=0.5}{v3,v1}
	\fmf{plain_arrow}{R1,v2}
	\fmf{plain_arrow}{R2,v3}
	\fmfv{label=$p_2$}{R2}
	\fmfv{label=$p_3$}{R1}
\end{fmfgraph*}
}}}};
\end{tikzpicture}
\end{gathered}
\hspace{12pt}&= \int\frac{d^4 k}{i(2\pi)^4}\frac{1}{k^2+i\eps} \frac{1}{(p_2-k)^2 +i\eps} \frac{1}{(p_3+k)^2+i\eps} \nonumber\\
&=\frac{1}{16 \pi^2 p_{1}^{2}} \frac{1}{z-\zb}  \Phi_1(z,\zb) \, , \label{eq:one_loop_triangle_def}
\end{align} 
where
\begin{equation}
    \Phi_1(z,\zb)=2\Li_2(z)-2\Li_2(\zb)+\ln (z \zb) \ln \left(\frac{1-z}{1-\zb}\right)\,. \label{cP2}
\end{equation}
In the regions $R^\star_\text{I}$ and $R^\star_\text{A}$, this function is analytic.

The variation matrix for $\Phi_1$ was given in Eq.~\eqref{eq:principal_variation_matrix_box}:
\begin{equation}
  \mathscr{M}_{\gamma_0}^{R_{\text{A}}^\star} =
  \begin{pmatrix}
    1 & \ln z+ \ln \zb & \operatorname{Li}_1(z) + \operatorname{Li}_1(\bar{z}) & \Phi_1(z, \bar{z}) \\
    0 & 1 & 0 & - \operatorname{Li}_1(z) + \operatorname{Li}_1(\bar{z}) \\
    0 & 0 & 1 & \ln z - \ln \zb \\
    0 & 0 & 0 & 1
  \end{pmatrix}. \label{eq:repeat_box_variation_matrix}
\end{equation}
Here $\gamma_0$ is the straight-line path from the basepoint $(0,0)$ to $(z,\zb)$. In the region $R_{\text{A}}^\star$, the variation matrix is analytic. In other regions, it has the same form with $z$ and $\zb$ on the appropriate sides of their branch cuts as determined by the displacements in Eq.~\eqref{eq:region_small_imaginary_parts}.

Using the monodromy matrices in Eqs.~\eqref{eq:box_monodromies_zero} and~\eqref{eq:box_monodromies_1_z}, we can calculate the differences of paths relevant to evaluating the discontinuities in Eq.~\eqref{discdelta}. We find
\begin{equation}
    (\bbone- \sM_{\linebub{}_0^z})\Phi_1  = 2\pi i\Big[\Li_1(z) - \Li_1(\zb)\Big] %
    \,,\qquad
     (\bbone- \sM_{\linebub{}_1^z})\Phi_1 =  2\pi i \Big[\ln z - \ln \zb\Big]
     \,,
\end{equation}
and
\begin{equation}
    (  \bbone - \mathscr{M}^{\linebubDown z}_\infty  ) \Phi_1 = -
     2\pi i\Big[\Li_1(z) - \Li_1(\zb) +\ln z - \ln \zb+ 2\pi i\Big].
\end{equation}
Rewriting these results in terms of logarithms with manifestly positive arguments in the relevant region, which in the case of $[\disc_{p_1^2} T_1]_{R^1}$ means replacing
\begin{equation}
    \Li_1(z-i\eps) - \Li_1(\zb) +\ln z - \ln (\zb+i\eps)+ 2\pi i = 
    -\ln \left[\frac{(z-1)(-\zb)}{(1-\zb) z }\right]\, ,
\end{equation} 
we have
\begin{subequations}
\begin{align}
       \left[\disc_{p_1^2} T_1\right]_{R^1} &= \frac{1}{16 \pi^2 p_{1}^{2}} \frac{2\pi i}{z-\zb}  \ln \left[\frac{(z-1)(-\zb)}{(1-\zb) z }\right], \label{eq:DiscT1R1}\\
        \left[ \disc_{p_2^2} T_1\right]_{R^2} &= \frac{1}{16 \pi^2 p_{1}^{2}} \frac{2\pi i}{z-\zb}  \ln \left[\frac{1-\zb}{1-z }\right],\label{eq:DiscT1R2}\\
        \left[ \disc_{p_3^2} T_1\right]_{R^3} &= \frac{1}{16 \pi^2 p_{1}^{2}} \frac{2\pi i}{z-\zb}  \ln \left[\frac{z}{ \zb }\right]. \label{eq:DiscT1R3}
\end{align}
\end{subequations}
As an initial cross check, we note that these discontinuities map to each other under the dihedral symmetry that permutes the legs of the one-loop triangle. Both the rational part and the transcendental part of these functions pick up a sign under odd permutations of the legs; for instance, under \(p_2 \leftrightarrow p_3\), we have \(z \to 1 - z\) and \(\bar{z} \to 1 - \bar{z}\) in the logarithms, while $(z-\zb) \to -(z-\zb)$ in the rational prefactor. The action of this symmetry is discussed in detail in Appendix~\ref{sec:symmetry}.

The corresponding cuts must be computed in the appropriate region. For example,  the cut in $p_1^2$ requires $p_1^2 > 0$, and evaluates to
\begin{align}
\label{cutp1}
    \cut_{p_1^2} T_1 &= \frac{1}{16 \pi^2 p_{1}^{2}} \frac{2 \pi i}{z-\zb}
    \Big\{\ln[-\zb(1-z)-i\eps 
    ] -\ln[- z(1-\zb)-i\eps 
    ]\Big\}  \Theta(p_1^2) \, .
\end{align}
In region $R^1$, this can be written
\begin{align}
    \left[ \cut_{p_1^2} T_1 \right]_{R^1}&= \frac{1}{16 \pi^2 p_{1}^{2}} \frac{2 \pi i}{z-\zb}
    \ln\left[\frac{(z-1)(-\zb)}{(1-\zb) z} \right]  %
    \,,
\end{align}
matching the discontinuity in Eq.~\eqref{eq:DiscT1R1} as well as the corresponding expression in \cite{Abreu:2014cla}. The cuts in $p_2^2$ and $p_3^2$ can similarly be computed, and agree with the discontinuities in Eqs.~\eqref{eq:DiscT1R2} and~\eqref{eq:DiscT1R3}, and with the results of~\cite{Abreu:2014cla}.

We can also compute the discontinuity in $p_2$ and $p_3$, using Eq.~\eqref{R123toR1}. This gives 
\begin{align}
     [\disc_{\set_{23}}T_1]_{R^{123}_C}
    &=\frac{1}{16 \pi^2 p_{1}^{2}} \frac{2\pi i}{z-\zb}  
    \left[2\pi i + \Li_1(z) - \Li_1(\zb) + \ln(z-i\eps) - \ln(\zb+i\eps)\right]\\
     &=\frac{1}{16 \pi^2 p_{1}^{2}} \frac{2\pi i}{z-\zb}  
     \ln\left[\frac{(1-\zb)(-z)}{(1-z)(-\zb)}\right]\, .
\end{align}
We should compare to the sum of the cuts in $p_2$ and $p_3^2$ which can be deduced from Eqs.\eqref{eq:DiscT1R2} and \eqref{eq:DiscT1R3}:
\begin{equation}
    \left[\cut_{p_2^2}T_1 +  \cut_{p_3^2}T_1 \right]_{R_C^{123}} =\frac{1}{16 \pi^2 p_{1}^{2}} \frac{2\pi i}{z-\zb} \left[ \ln\frac{1-\zb}{1-z} + \ln\frac{z}{\zb}\right]\,.
\end{equation}
Again, we see the discontinuities and cuts agree. 

A similar example involves going from $R^{123}_{A} \to R^\star_A \to R^{123}_{A}$. A path between these regions exists that does not go around any branch points. So 
$ [\disc_{\set_{123}}T_1]_{R^{123}_C}=0$. In $R^{123}_A$ the sum of the cuts also vanishes, although each individual cut does not. In other words, total discontinuity in the dual Euclidean region vanishes, but the discontinuities in separate channels do not. In contrast, in the Euclidean region $R^{123}_A$, all the cuts vanish individually (and the total discontinuity is still zero, using the same path).  

To take sequential discontinuities in a single channel, we iterate the monodromies in Eq.~\eqref{discdelta}. We find that these double discontinuities vanish in all channels, 
\begin{equation} \label{all2p0}
    \left[\disc_{p_j^2} \disc_{p_j^2} T_1\right]_{R^j} = 0 \qquad \forall j\,.
\end{equation}
This is consistent with our expectations, since the triangle has at most one cut in each channel. We can also consider sequential discontinuities of the triangle in different channels, such as $\disc_{p_2^2} \disc_{p_3^2} T_1$. The corresponding double cut in $p_2^2$ and $p_3^2$ can be computed in the region $R^{23}$, where $p_2^2 >0, p_3^2 >0$, and $p_1^2 <0$. Using the discontinuity operators defined in Eq.~\eqref{sdisc}, we find
\begin{equation}\label{eq:one_loop_double_disc_1}
      [\disc_{p_3^2}\disc_{p_2^2} T_1]_{R^{23}} = (\bbone-\sM_{\linebub{}_1^{z}})
       (\bbone-\sM_{\linebub{}_0^{\zb}})T_1=
        \frac{1}{16 \pi^2 p_{1}^{2}} \frac{(2\pi i)^2}{z-\zb}\,.
\end{equation}
Notice that we could have equivalently taken these discontinuities in the other order, as both sequences of discontinuities are related to the same cut integrals by Eq.~\eqref{ddstRst1}; that is, we have $[\disc_{p_3^2}\disc_{p_2^2} T_1]_{R^{23}} = [\disc_{p_2^2}\disc_{p_3^2} T_1]_{R^{23}}$.  Similarly, we find 
\begin{equation}\label{eq:one_loop_double_disc_2}
      [\disc_{p_1^2}\disc_{p_2^2} T_1]_{R^{12}} = [\disc_{p_2^2}\disc_{p_1^2} T_1]_{R^{12}} =
       - \frac{1}{16 \pi^2 p_{1}^{2}} \frac{(2\pi i)^2}{z-\zb}\,
\end{equation}
and
\begin{equation}\label{eq:one_loop_double_disc_3}
      [\disc_{p_1^2}\disc_{p_3^2} T_1]_{R^{13}} = [\disc_{p_3^2}\disc_{p_1^2} T_1]_{R^{13}} =
       - \frac{1}{16 \pi^2 p_{1}^{2}} \frac{(2\pi i)^2}{z-\zb}\,.
\end{equation}
Notice the additional minus sign in both of these expressions compared to Eq.~\eqref{eq:one_loop_double_disc_1}. As discussed in Appendix~\ref{sec:symmetry}, 
these relative signs are expected from the invariance of the triangle integral under permutations of its external legs.

To illustrate the importance of using the specific operators in Eq.~\eqref{sdisc} for computing sequential discontinuities in different channels, we can see what happens if we instead use the discontinuity operators from Eq.\eqref{discdelta}. In the case of $\disc_{p_3^2}\disc_{p_2^2}T_1$ we would have found
\begin{equation}\begin{split}
 [\disc_{p_3^2}[\disc_{p_2^2} T_1]_{R^2}]_{R^3}  &=
       (\bbone-\sM_{\linebub{}_1^{\zb}})(\bbone-\sM_{\linebub{}_0^{\zb}})T_1 =
       \frac{1}{16\pi^2 p_{1}^{2}} \frac{(2\pi i)^2}{z-\zb}\,,\\
    [\disc_{p_2^2}[\disc_{p_3^2} T_1]_{R^3}]_{R^2}   &=
       (\bbone-\sM_{\linebub{}_0^{\zb}})(\bbone-\sM_{\linebub{}_1^{\zb}})T_1 =- 
        \frac{1}{16 \pi^2 p_{1}^{2}} \frac{(2\pi i)^2}{z-\zb}\,.
\end{split}\end{equation}
The results differ by a sign. This highlights the importance of computing the discontinuities by analytically continuing from the region in which the cuts are being computed into adjacent regions.

Let us also reiterate that all the discontinuities we consider are computed along paths in external  energies such that energy is conserved. If one tries instead to do what may seem more natural, by continuing the Lorentz invariants directly, one can run into trouble. For example, by continuing $z$ and $\zb$ one can easily go from $R^{123}_A \to R^{23} \to R^{123}_A$ by passing around $\zb=0$ and $z=1$. The discontinuity along this path is
\begin{equation}
 (\bbone-\sM_{\linebub_0^{\zb}}\cdot\sM_{\linebub_1^{z}}) T_1 =\frac{1}{16 \pi^2 p_{1}^{2}} \frac{2\pi i}{z-\zb}  
    \left[\Li_1(z) - \Li_1(\zb) + \ln(z) - \ln(\zb)+ 2\pi i \right]\,.
\end{equation}
This is analytic in $R^{123}_A$, but differs from $\cut_{p_1^2} T_1$ in $R^{123}_A$ in  Eq.~\eqref{cutp1} by the extra $2\pi i$. Thus, specifying the regions of interest is not in general enough: one must also know how to connect them. 

\subsubsection{Two loops}
Next, we consider the two-loop triangle. As before, all internal lines are taken to be massless. The Feynman integral evaluates to
\begin{equation}\begin{split}
T_2 &= \hspace{-6pt}
\begin{gathered}
\begin{tikzpicture}[baseline=-3.5]
\node at (0,0) {
\parbox{50mm} {
\resizebox{50mm}{!}{
 \fmfframe(0,00)(0,0){
 \begin{fmfgraph*}(200,100)
     \fmfstraight
	\fmfleft{L1}
	\fmfright{R1,R2}
	\fmf{plain_arrow,label=$p_3$}{L1,v1}
	\fmf{plain,label=$1$,tension=0.5,l.s=right}{v1,v2}
	\fmf{plain,label=$3$,l.s=right,tension=0}{v2,v3}
	\fmf{plain,label=$2$,l.s=right,tension=0.5}{v3,v1}
	\fmf{plain,label=$4$,tension=0.5}{v2,v4}
	\fmf{plain,label=$6$,tension=0.5}{v4,v5}
	\fmf{plain,label=$5$,tension=0.5,l.s=right}{v5,v3}
	\fmf{plain_arrow,tension=1.3}{R1,v4}
	\fmf{plain_arrow,tension=1.3}{R2,v5}
	\fmfv{label=$p_1$}{R2}	
	\fmfv{label=$p_2$}{R1}
\end{fmfgraph*}
}}}};
\end{tikzpicture}
\end{gathered}
\\
&=\int \frac{d^{4} k_1}{i(2\pi)^4} \int\frac{d^{4} k_2}{i(2\pi)^4}  \frac{1}{k_{1}^{2}\left(p_{3}-k_{1}\right)^{2}\left(k_{1}+p_{1}\right)^{2} k_{2}^{2}\left(p_{3}-k_{2}\right)^{2}\left(k_{1}-k_{2}\right)^{2}} \\
&=\frac{1}{\left(4\pi\right)^4 p_1^2 p_3^2} \frac{1}{(z-\zb)} \Phi_2(z,\zb)
\end{split}
\label{eq:T2}
\end{equation}
where in the region $R^\star_\text{A}$ the function $\Phi_2(z,\zb)$ takes the form 
\begin{equation}
   \Phi_2(z, \zb)=6 [\Li_4(z)-\Li_4(\zb)]
    -3 \ln (z \zb)    [\Li_3(z) - \Li_3(\zb)]
    +\frac{1}{2} \ln^{2}(z \zb)    [\Li_2(z) - \Li_2(\zb)]\,, \label{FRsI}
\end{equation}
and as before $z$ and $\zb$ satisfy the relations in Eqs.~\eqref{uvdef}, \eqref{zdef}, and \eqref{zbdef}. The variation matrix for this integral is described in Appendix~\ref{app:Phi2}, where the relevant monodromy matrices are also presented.

We first compute the single discontinuities, using the operators in Eq.~\eqref{discdelta}:
\begin{subequations}
\begin{align}
       \left[\disc_{p_1^2} \Phi_2\right]_{R^1} &=
        (2\pi i) \times \Big\{ 3 \Li_3(\zb) - 3\Li_3(z)  + \big(\ln z  + \ln \zb - i \pi \big) \big(\Li_2(z) - \Li_2(\zb) \big) \nonumber \\
        &\hspace{2.8cm} + \frac 1 2 \ln z \, \big( \ln z - \ln \zb + 2 \pi i \big) \big( \ln \zb - 2 \pi i \big)\Big\} \,,
      \\
         \left[\disc_{p_2^2} \Phi_2\right]_{R^2} &= (2\pi i) \times \Big\{ 3\Li_3(z) - 3\Li_3(\zb)
      -  \big(\ln z + \ln \zb + i \pi \big) [\Li_2(z) - \Li_2(\zb)] \Big\},\label{discR2}\\
      \left[\disc_{p_3^2} \Phi_2\right]_{R^3} &=  (2\pi i)\times \Big\{-\frac{1}{2} \ln z \ln \zb \, \big( \ln z- \ln \zb \big)\Big\}.
       \label{discR3}
\end{align}
\end{subequations} 
 All the explicit factors of $i \pi$ in these expressions can be absorbed into polylogarithms that are manifestly real in the appropriate region (taking into account Eq.~\eqref{eq:region_small_imaginary_parts}). The resulting expressions agree with the cuts computed in Eqs. (5.26),  (5.37) and (5.41) of~\cite{Abreu:2014cla}.

The sequential discontinuities in these channels can be computed using the same monodromy matrices. We find
\begin{subequations}
\begin{align}
\left[\disc_{p_1^2} \disc_{p_1^2} \Phi_2\right]_{R^1} &= 
    (2\pi i)^2    \Big\{\Li_2(z) - \Li_2(\zb) + \frac 1 2 \big(\ln z - \ln \zb + 2 \pi i \big) \big(\ln z + \ln \zb \big) \Big\}\,, \label{disc2R1} \\
     \left[\disc_{p_2^2} \disc_{p_2^2} \Phi_2\right]_{R^2} &= (2\pi i)^2   \Big\{\Li_2(z) - \Li_2(\zb)\Big\}\,, \label{disc2R2} \\
     \left[\disc_{p_3^2} \disc_{p_3^2} \Phi_2\right]_{R^3} &=  0\,. \label{disc2R3}
\end{align}
\end{subequations}
Note that the right side of Eq.~\eqref{disc2R1} can be rewritten as $\Li_2(1/\zb) - \Li_2( 1/z)$ in $R^1$, and thus $\big[\disc_{p_1^2} \disc_{p_1^2} \Phi_2\big]_{R^1}$ and $\big[\disc_{p_2^2} \disc_{p_2^2} \Phi_2\big]_{R^1}$ get mapped to minus each other under the permutation $p_1 \leftrightarrow p_2$, which corresponds to $z \to 1/z, \zb \to 1/\zb$. This is consistent with what we expect from Appendix~\ref{sec:symmetry}. The triple discontinuities all vanish,
 \begin{equation}
      \left[\disc_{p_j^2} \disc_{p_j^2} \disc_{p_j^2} \Phi_2 \right]_{R^j} = 0\,, \label{ddR3zero}
\end{equation}
in accordance with the fact that there aren't three cuts in any of the channels. 

These sequential cuts in the same channel have not been computed before to our knowledge. To do so, we regulate the IR divergence of the cuts by giving the lines labeled $4$ and $5$ in the figure below with a small mass $m_\text{reg}$, and work to leading power in $m_\text{reg}$. In region $R^3$, we find
\begin{equation}
\left[T_{(12),(45)}^\text{cut} \right]_{R^3}
= \hspace{-6pt}
\begin{gathered}
\begin{tikzpicture}[baseline=-3.5]
\node at (0,0) {\parbox{40mm} {\resizebox{40mm}{!}{
 \fmfframe(0,00)(0,0){ \begin{fmfgraph*}(140,70)
     \fmfstraight
	\fmfleft{L1}
	\fmfright{R1,R2}
	\fmf{plain_arrow,label=$p_3$}{L1,v1}
	\fmf{plain,label=$1$,tension=0.5,l.s=right}{v1,v2}
	\fmf{plain,label=$3$,l.s=right,tension=0}{v2,v3}
	\fmf{plain,label=$2$,l.s=right,tension=0.5}{v3,v1}
	\fmf{plain,label=$4$,tension=0.5}{v2,v4}
	\fmf{plain,label=$6$,tension=0.5}{v4,v5}
	\fmf{plain,label=$5$,tension=0.5,l.s=right}{v5,v3}
	\fmf{plain_arrow,tension=1.3}{R1,v4}
	\fmf{plain_arrow,tension=1.3}{R2,v5}
	\fmfv{label=$p_1$}{R2}	
	\fmfv{label=$p_2$}{R1}  \end{fmfgraph*} }}}};
\draw[dashed, line width = 1, dashed, darkgreen] (0,-0.8) to (0,0.8);
\draw[dashed, line width = 1, dashed, darkgreen] (1.0,-0.8) to (1.0,0.8);
\end{tikzpicture}
\end{gathered}
\hspace{6pt}= -2T^\text{cut}_{2}
\end{equation}
where
\begin{equation}
   T_{2}^\text{cut} =\frac{1}{64 p_1^2 p_{3}^{2} \pi^{2}(z-\zb)}
   \ln \frac{m_\text{reg}^2}{p_3^2} \ln \frac{z}{\zb} \, .
   \label{eq:T2cutmassreg}
\end{equation}
The other cuts give multiples of this expression. In particular, we find
\begin{align}
\left[T_{(12),(135)}^\text{cut} \right]_{R^3} &=\hspace{-6pt}
\begin{gathered}
\begin{tikzpicture}[baseline=-3.5]
\node at (0,0) {\parbox{40mm} {\resizebox{40mm}{!}{
 \fmfframe(0,00)(0,0){ \begin{fmfgraph*}(140,70) \end{fmfgraph*}}}}};
\draw[line width = 1, dashed, darkgreen] (0,-0.8) to (0,0.8);
\draw[line width = 1, dashed, darkgreen,smooth] plot [smooth] coordinates {(0.2,-0.8) (0.3,-0.1) (0.6,0.2) (0.7,0.8)};
\end{tikzpicture}
\end{gathered}
\hspace{6pt}= 0 \, , \label{cutp32triple21} \\[10pt]
\left[ T_{(12),(234)}^\text{cut}\right]_{R^3} &=\hspace{-6pt} \begin{gathered}
\begin{tikzpicture}[baseline=-3.5]
\node at (0,0) {\parbox{40mm} {\resizebox{40mm}{!}{
 \fmfframe(0,00)(0,0){ \begin{fmfgraph*}(140,70)
     \fmfstraight
	\fmfleft{L1}
	\fmfright{R1,R2}
	\fmf{plain_arrow,label=$p_3$}{L1,v1}
	\fmf{plain,label=$1$,tension=0.5,l.s=right}{v1,v2}
	\fmf{plain,label=$3$,l.s=right,tension=0}{v2,v3}
	\fmf{plain,label=$2$,l.s=right,tension=0.5}{v3,v1}
	\fmf{plain,label=$4$,tension=0.5}{v2,v4}
	\fmf{plain,label=$6$,tension=0.5}{v4,v5}
	\fmf{plain,label=$5$,tension=0.5,l.s=right}{v5,v3}
	\fmf{plain_arrow,tension=1.3}{R1,v4}
	\fmf{plain_arrow,tension=1.3}{R2,v5}
	\fmfv{label=$p_1$}{R2}	
	\fmfv{label=$p_2$}{R1}  \end{fmfgraph*} }}}};
\draw[line width = 1, dashed, darkgreen] (0,-0.8) to (0,0.8);
\draw[line width = 1, dashed, darkgreen,smooth] plot [smooth] coordinates {(0.5,-0.8) (0.5,-0.1) (0.2,0.1) (0.2,0.8)};
\end{tikzpicture}
\end{gathered}
\hspace{6pt}= 0 \, ,
\label{cutp32triple22}
\end{align}
and
\begin{align}
\left[T_{(45),(135)}^\text{cut}\right]_{R^3} &= \hspace{-6pt}
\begin{gathered}
\begin{tikzpicture}[baseline=-3.5]
\node at (0,0) {\parbox{40mm} {\resizebox{40mm}{!}{
 \fmfframe(0,00)(0,0){ \begin{fmfgraph*}(140,70) \end{fmfgraph*}}}}};
\draw[dashed, line width = 1, dashed, darkgreen] (1.2,-0.8) to (1.2,0.8);
\draw[line width = 1, dashed, darkgreen,smooth] plot [smooth] coordinates {(0.2,-0.8) (0.3,-0.1) (0.6,0.2) (0.7,0.8)};
\end{tikzpicture}
\end{gathered}
\hspace{6pt}= T^\text{cut}_{2} \, , \label{cutp32triple11}
 \\[10pt]
\left[ T_{(45),(234)}^\text{cut}\right]_{R^3} &= \hspace{-6pt}\begin{gathered}
\begin{tikzpicture}[baseline=-3.5]
\node at (0,0) {\parbox{40mm} {\resizebox{40mm}{!}{
 \fmfframe(0,00)(0,0){ \begin{fmfgraph*}(140,70)
     \fmfstraight
	\fmfleft{L1}
	\fmfright{R1,R2}
	\fmf{plain_arrow,label=$p_3$}{L1,v1}
	\fmf{plain,label=$1$,tension=0.5,l.s=right}{v1,v2}
	\fmf{plain,label=$3$,l.s=right,tension=0}{v2,v3}
	\fmf{plain,label=$2$,l.s=right,tension=0.5}{v3,v1}
	\fmf{plain,label=$4$,tension=0.5}{v2,v4}
	\fmf{plain,label=$6$,tension=0.5}{v4,v5}
	\fmf{plain,label=$5$,tension=0.5,l.s=right}{v5,v3}
	\fmf{plain_arrow,tension=1.3}{R1,v4}
	\fmf{plain_arrow,tension=1.3}{R2,v5}
	\fmfv{label=$p_1$}{R2}	
	\fmfv{label=$p_2$}{R1}  \end{fmfgraph*} }}}};
\draw[dashed, line width = 1, dashed, darkgreen] (1.2,-0.8) to (1.2,0.8);
\draw[line width = 1, dashed, darkgreen,smooth] plot [smooth] coordinates {(0.5,-0.8) (0.5,-0.1) (0.2,0.1) (0.2,0.8)};
\end{tikzpicture}
\end{gathered}
\hspace{6pt}= T^\text{cut}_{2} \, .
\label{cutp32triple12}
\end{align}
It follows that the sum of all double cuts in $R^3$ is exactly zero,
\begin{equation}
\label{sumiszero}
  \left[  
  \sum_\text{double cuts}
  T_2\right]_{R^3}=0 \, ,
\end{equation}
which agrees with Eq.~\eqref{disc2R3}. Note that the diagrams in Eq.~\eqref{cutp32triple11} and Eq~\eqref{cutp32triple12} both involve an isolated three-point vertex with only massless lines. 
For $d>4$ such cut graphs may be zero, while they are nonzero in $d=4$ 
(they contain integrals of the form $\int dx \delta(x) x^{\frac{d-4}{2}}$). 
If we were to set them to zero, we would get the wrong answer. This can easily be seen in the example above, as Eq.~\eqref{eq:T2cutmassreg} would give a non-vanishing result in dimensional regularization, while the graphs in Eq.~\eqref{cutp32triple21}, Eq.~\eqref{cutp32triple22}, Eq.~\eqref{cutp32triple11}, and Eq.~\eqref{cutp32triple12} would vanish. See Appendix~\ref{sec:massless3pt} for more details. 

In $R^2$, there is only one diagram. We find
\begin{equation}
    \left[ T_{(46),(136)}^\text{cut}\right]_{R^2} =\hspace{-6pt} \begin{gathered}
\begin{tikzpicture}[baseline=-3.5]
\node at (0,0) {\parbox{40mm} {\resizebox{40mm}{!}{
 \fmfframe(0,00)(0,0){ \begin{fmfgraph*}(140,70)
     \fmfstraight
	\fmfleft{L1}
	\fmfright{R1,R2}
	\fmf{plain_arrow,label=$p_3$}{L1,v1}
	\fmf{plain,label=$1$,tension=0.5,l.s=right}{v1,v2}
	\fmf{plain,label=$3$,l.s=right,tension=0}{v2,v3}
	\fmf{plain,label=$2$,l.s=right,tension=0.5}{v3,v1}
	\fmf{plain,label=$4$,tension=0.5}{v2,v4}
	\fmf{plain,label=$6$,tension=0.5}{v4,v5}
	\fmf{plain,label=$5$,tension=0.5,l.s=right}{v5,v3}
	\fmf{plain_arrow,tension=1.3}{R1,v4}
	\fmf{plain_arrow,tension=1.3}{R2,v5}
	\fmfv{label=$p_1$}{R2}	
	\fmfv{label=$p_2$}{R1}  \end{fmfgraph*} }}}};
\draw[line width = 1, dashed, darkgreen,smooth] plot [smooth] coordinates {(-0.5,-0.8) (0.5,0.1) (2.2,0.5)};
\draw[line width = 1, dashed, darkgreen,smooth] plot [smooth] coordinates {(0.5,-0.8) (1,-0.2) (2.5,0.3)};
\end{tikzpicture}
\end{gathered}
\hspace{6pt}=
- \frac{1}{256 \pi^{2} p_{1}^{2} p_3^2} \frac{2 \Li_{2}(z)-2 \Li_2(\zb)}{z-\zb}\,. \label{eq:p2_channel_cut}
\end{equation}
Comparing to Eq.~\eqref{discR2}, we then find
\begin{equation}
    \left[  
  \sum_\text{double cuts}
  T_2\right]_{R^2} =
  \left[ \disc_{p_2^2} \disc_{p_2^2} T_2(z,\zb)\right]_{R^2} = 2\left[ T_{(46),(136)}^\text{cut}\right]_{R^2}\, ,
\end{equation}
in agreement with Eq.~\eqref{doublecut}. The sum of double cuts in the $p_1^2$ channel are related by $z \leftrightarrow 1/z$, $\zb \leftrightarrow 1/\zb$ to the sum of double cuts in the $p_2^2$ channel, and thus the sum of double cuts in $R^1$ is related to the sequential discontinuity computed in Eq.~\eqref{disc2R1} by the same combinatorial factor.
These provide highly nontrivial checks of Eq.~\eqref{doublecut}.

Finally, we compute the sequential discontinuities in different channels.
We find
\begin{subequations}
\begin{align}
\left[\disc_{p_3^2} \disc_{p_1^2} \Phi_2\right]_{R^{13}} &= (2\pi i)^2    \Big\{- \frac{1}{2} \ln^2 z +\ln z \ln (\zb+i\eps)  - i \pi \ln z \Big\}, \label{eq:disc3disc1_two_loops} \\
     \left[\disc_{p_2^2} \disc_{p_1^2} \Phi_2\right]_{R^{12}} &= (2\pi i)^2   \Big\{ \Li_2(\zb) - \Li_2(z-i\eps) -\frac{1}{2} \ln^2 z + \ln z \ln \zb - i \pi  \ln z \Big\} \, ,
     \label{eq:disc2disc1_two_loops}
     \\
         \left[\disc_{p_2^2} \disc_{p_3^2} \Phi_2\right]_{R^{23}} &= (2\pi i)^2   \Big\{\frac{1}{2} \ln^2 z - \ln z \ln (\zb - i\eps)- i \pi  \ln z \Big\} \, .
\label{eq:disc2disc3_two_loops} 
\end{align}
\end{subequations}
We believe these agree with the results in~\cite{Abreu:2014cla}.\footnote{These equations differ slightly from Eqs.~(6.4) and (6.5) in~\cite{Abreu:2014cla}. However, summing the results from their Appendix D, we believe their (6.4) should agree with our Eq.~\eqref{eq:disc3disc1_two_loops}. For $\disc_{p_2^2} \disc_{p_1^2} \Phi_2$, we find that summing their expressions with some typos corrected gives $\smash{(2\pi i)^2   \left\{
    \Li_2(\zb) + \Li_2(1-z) + \ln(z-1) \ln z - \frac{1}{2} \ln^2 z+\ln z \ln \zb - \frac{\pi^2}{6}\right\}}$, which agrees with Eq.~\eqref{eq:disc2disc1_two_loops}.}
Recall that~\cite{Abreu:2014cla} uses a different cut prescription, which involves both $-i\eps$ and $+i\eps$ propagators, and that they use dimensional regularization and so massless three-point vertices vanish. For reasons discussed in Appendix~\ref{sec:massless3pt}, we believe it is safer to use a mass regulator. With our $+i\eps$ convention, the double-cut graphs in $R^{12}$ give
\begin{align}
    \left[\Phi_{2}^{\text{(2-cuts)}}\right]_{R^{12}}
    & = 
    \hspace{-6pt} \begin{gathered}
\begin{tikzpicture}[baseline=-3.5]
\node at (0,0) {\parbox{40mm} {\resizebox{40mm}{!}{
 \fmfframe(0,00)(0,0){ \begin{fmfgraph*}(140,70)
     \fmfstraight
	\fmfleft{L1}
	\fmfright{R1,R2}
	\fmf{plain_arrow,label=$p_3$}{L1,v1}
	\fmf{plain,label=$1$,tension=0.5,l.s=right}{v1,v2}
	\fmf{plain,label=$3$,l.s=right,tension=0}{v2,v3}
	\fmf{plain,label=$2$,l.s=right,tension=0.5}{v3,v1}
	\fmf{plain,label=$4$,tension=0.5}{v2,v4}
	\fmf{plain,label=$6$,tension=0.5}{v4,v5}
	\fmf{plain,label=$5$,tension=0.5,l.s=right}{v5,v3}
	\fmf{plain_arrow,tension=1.3}{R1,v4}
	\fmf{plain_arrow,tension=1.3}{R2,v5}
	\fmfv{label=$p_1$}{R2}	
	\fmfv{label=$p_2$}{R1}  \end{fmfgraph*} }}}};
 \draw[line width = 1, dashed, darkgreen,smooth] plot [smooth] coordinates {(-0.5,-0.8) (0.3,-0.15) (2.2,-0.2)};
\draw[line width = 1, dashed, darkgreen,smooth] plot [smooth] coordinates {(-0.5,0.8) (0.3,0.15) (2.2,0.2)};
\end{tikzpicture}
\end{gathered}
\hspace{6pt} 
+
\hspace{-6pt} \begin{gathered}
\begin{tikzpicture}[baseline=-3.5]
\node at (0,0) {\parbox{40mm} {\resizebox{40mm}{!}{
 \fmfframe(0,00)(0,0){ \begin{fmfgraph*}(140,70)
     \fmfstraight
	\fmfleft{L1}
	\fmfright{R1,R2}
	\fmf{plain_arrow,label=$p_3$}{L1,v1}
	\fmf{plain,label=$1$,tension=0.5,l.s=right}{v1,v2}
	\fmf{plain,label=$3$,l.s=right,tension=0}{v2,v3}
	\fmf{plain,label=$2$,l.s=right,tension=0.5}{v3,v1}
	\fmf{plain,label=$4$,tension=0.5}{v2,v4}
	\fmf{plain,label=$6$,tension=0.5}{v4,v5}
	\fmf{plain,label=$5$,tension=0.5,l.s=right}{v5,v3}
	\fmf{plain_arrow,tension=1.3}{R1,v4}
	\fmf{plain_arrow,tension=1.3}{R2,v5}
	\fmfv{label=$p_1$}{R2}	
	\fmfv{label=$p_2$}{R1}  \end{fmfgraph*} }}}};
\draw[line width = 1, dashed, darkgreen,smooth] plot [smooth] coordinates {(0.7,-0.8) (1,-0.4) (2.2,-0.4)};
\draw[line width = 1, dashed, darkgreen,smooth] plot [smooth] coordinates {(0.7,0.8) (1,0.4) (2.2,0.4)};
\end{tikzpicture}
\end{gathered}
\hspace{40pt} \nonumber
\\[10pt]
& \hspace{40pt}+
\hspace{-6pt} \begin{gathered}
\begin{tikzpicture}[baseline=-3.5]
\node at (0,0) {\parbox{40mm} {\resizebox{40mm}{!}{
 \fmfframe(0,00)(0,0){ \begin{fmfgraph*}(140,70)
     \fmfstraight
	\fmfleft{L1}
	\fmfright{R1,R2}
	\fmf{plain_arrow,label=$p_3$}{L1,v1}
	\fmf{plain,label=$1$,tension=0.5,l.s=right}{v1,v2}
	\fmf{plain,label=$3$,l.s=right,tension=0}{v2,v3}
	\fmf{plain,label=$2$,l.s=right,tension=0.5}{v3,v1}
	\fmf{plain,label=$4$,tension=0.5}{v2,v4}
	\fmf{plain,label=$6$,tension=0.5}{v4,v5}
	\fmf{plain,label=$5$,tension=0.5,l.s=right}{v5,v3}
	\fmf{plain_arrow,tension=1.3}{R1,v4}
	\fmf{plain_arrow,tension=1.3}{R2,v5}
	\fmfv{label=$p_1$}{R2}	
	\fmfv{label=$p_2$}{R1}  \end{fmfgraph*} }}}};
\draw[line width = 1, dashed, darkgreen,smooth] plot [smooth] coordinates {(-0.5,-0.8) (0.3,-0.15) (2.2,-0.2)};
\draw[line width = 1, dashed, darkgreen,smooth] plot [smooth] coordinates {(0.7,0.8) (1,0.4) (2.2,0.4)};
\end{tikzpicture}
\end{gathered}
\hspace{6pt} 
+
\hspace{-6pt} \begin{gathered}
\begin{tikzpicture}[baseline=-3.5]
\node at (0,0) {\parbox{40mm} {\resizebox{40mm}{!}{
 \fmfframe(0,00)(0,0){ \begin{fmfgraph*}(140,70)
     \fmfstraight
	\fmfleft{L1}
	\fmfright{R1,R2}
	\fmf{plain_arrow,label=$p_3$}{L1,v1}
	\fmf{plain,label=$1$,tension=0.5,l.s=right}{v1,v2}
	\fmf{plain,label=$3$,l.s=right,tension=0}{v2,v3}
	\fmf{plain,label=$2$,l.s=right,tension=0.5}{v3,v1}
	\fmf{plain,label=$4$,tension=0.5}{v2,v4}
	\fmf{plain,label=$6$,tension=0.5}{v4,v5}
	\fmf{plain,label=$5$,tension=0.5,l.s=right}{v5,v3}
	\fmf{plain_arrow,tension=1.3}{R1,v4}
	\fmf{plain_arrow,tension=1.3}{R2,v5}
	\fmfv{label=$p_1$}{R2}	
	\fmfv{label=$p_2$}{R1}  \end{fmfgraph*} }}}};
\draw[line width = 1, dashed, darkgreen,smooth] plot [smooth] coordinates {(-0.5,0.8) (0.3,0.15) (2.2,0.2)};
\draw[line width = 1, dashed, darkgreen,smooth] plot [smooth] coordinates {(0.7,-0.8) (1,-0.4) (2.2,-0.4)};
\end{tikzpicture}
\end{gathered}
\hspace{6pt} 
\nonumber
        \\
    & = (2\pi i)^2   \Big\{ \Li_2(\zb) - \Li_2(z-i\eps) -\frac{1}{2} \ln^2 z + \ln z \ln \zb + i \pi \ln z -2 \pi i \ln \zb \Big\} \, .
\end{align}
To match onto the discontinuity in Eq.~\eqref{eq:disc2disc1_two_loops}, we must in our analysis add the three-cut graphs according to Eq.~\eqref{masterformula}.
We find
\begin{multline}
\left[\Phi_{2}^{\text{(3-cuts)}}\right]_{R^{12}}=
    \hspace{-6pt} \begin{gathered}
\begin{tikzpicture}[baseline=-3.5]
\node at (0,0) {\parbox{40mm} {\resizebox{40mm}{!}{
 \fmfframe(0,00)(0,0){ \begin{fmfgraph*}(140,70)
     \fmfstraight
	\fmfleft{L1}
	\fmfright{R1,R2}
	\fmf{plain_arrow,label=$p_3$}{L1,v1}
	\fmf{plain,label=$1$,tension=0.5,l.s=right}{v1,v2}
	\fmf{plain,label=$3$,l.s=right,tension=0}{v2,v3}
	\fmf{plain,label=$2$,l.s=right,tension=0.5}{v3,v1}
	\fmf{plain,label=$4$,tension=0.5}{v2,v4}
	\fmf{plain,label=$6$,tension=0.5}{v4,v5}
	\fmf{plain,label=$5$,tension=0.5,l.s=right}{v5,v3}
	\fmf{plain_arrow,tension=1.3}{R1,v4}
	\fmf{plain_arrow,tension=1.3}{R2,v5}
	\fmfv{label=$p_1$}{R2}	
	\fmfv{label=$p_2$}{R1}  \end{fmfgraph*} }}}};
\draw[line width = 1, dashed, darkgreen,smooth] plot [smooth] coordinates {(-0.5,-0.8) (0.3,-0.15) (2.2,-0.2)};
\draw[line width = 1, dashed, darkgreen,smooth] plot [smooth] coordinates {(0.7,-0.8) (1,-0.4) (2.2,-0.4)};
\draw[line width = 1, dashed, darkgreen,smooth] plot [smooth] coordinates {(-0.5,0.8) (0.3,0.15) (2.2,0.2)};
\end{tikzpicture}
\end{gathered}
\hspace{6pt} 
+
\hspace{-6pt} 
\begin{gathered}
\begin{tikzpicture}[baseline=-3.5]
\node at (0,0) {\parbox{40mm} {\resizebox{40mm}{!}{
 \fmfframe(0,00)(0,0){ \begin{fmfgraph*}(140,70)
     \fmfstraight
	\fmfleft{L1}
	\fmfright{R1,R2}
	\fmf{plain_arrow,label=$p_3$}{L1,v1}
	\fmf{plain,label=$1$,tension=0.5,l.s=right}{v1,v2}
	\fmf{plain,label=$3$,l.s=right,tension=0}{v2,v3}
	\fmf{plain,label=$2$,l.s=right,tension=0.5}{v3,v1}
	\fmf{plain,label=$4$,tension=0.5}{v2,v4}
	\fmf{plain,label=$6$,tension=0.5}{v4,v5}
	\fmf{plain,label=$5$,tension=0.5,l.s=right}{v5,v3}
	\fmf{plain_arrow,tension=1.3}{R1,v4}
	\fmf{plain_arrow,tension=1.3}{R2,v5}
	\fmfv{label=$p_1$}{R2}	
	\fmfv{label=$p_2$}{R1}  \end{fmfgraph*} }}}};
\draw[line width = 1, dashed, darkgreen,smooth] plot [smooth] coordinates {(-0.5,-0.8) (0.3,-0.15) (2.2,-0.2)};
\draw[line width = 1, dashed, darkgreen,smooth] plot [smooth] coordinates {(0.7,0.8) (1,0.4) (2.2,0.4)};
\draw[line width = 1, dashed, darkgreen,smooth] plot [smooth] coordinates {(-0.5,0.8) (0.3,0.15) (2.2,0.2)};
\end{tikzpicture}
\end{gathered}
\\ 
\begin{gathered}
\begin{tikzpicture}[baseline=-3.5]
\node at (0,0) {\parbox{40mm} {\resizebox{40mm}{!}{
 \fmfframe(0,00)(0,0){ \begin{fmfgraph*}(140,70)
     \fmfstraight
	\fmfleft{L1}
	\fmfright{R1,R2}
	\fmf{plain_arrow,label=$p_3$}{L1,v1}
	\fmf{plain,label=$1$,tension=0.5,l.s=right}{v1,v2}
	\fmf{plain,label=$3$,l.s=right,tension=0}{v2,v3}
	\fmf{plain,label=$2$,l.s=right,tension=0.5}{v3,v1}
	\fmf{plain,label=$4$,tension=0.5}{v2,v4}
	\fmf{plain,label=$6$,tension=0.5}{v4,v5}
	\fmf{plain,label=$5$,tension=0.5,l.s=right}{v5,v3}
	\fmf{plain_arrow,tension=1.3}{R1,v4}
	\fmf{plain_arrow,tension=1.3}{R2,v5}
	\fmfv{label=$p_1$}{R2}	
	\fmfv{label=$p_2$}{R1}  \end{fmfgraph*} }}}};
\draw[line width = 1, dashed, darkgreen,smooth] plot [smooth] coordinates {(-0.5,0.8) (0.3,0.15) (2.2,0.2)};
\draw[line width = 1, dashed, darkgreen,smooth] plot [smooth] coordinates {(0.7,-0.8) (1,-0.4) (2.2,-0.4)};
\draw[line width = 1, dashed, darkgreen,smooth] plot [smooth] coordinates {(0.7,0.8) (1,0.4) (2.2,0.4)};
\end{tikzpicture}
\end{gathered}
\hspace{6pt} 
+
\hspace{-6pt} \begin{gathered}
\begin{tikzpicture}[baseline=-3.5]
\node at (0,0) {\parbox{40mm} {\resizebox{40mm}{!}{
 \fmfframe(0,00)(0,0){ \begin{fmfgraph*}(140,70)
     \fmfstraight
	\fmfleft{L1}
	\fmfright{R1,R2}
	\fmf{plain_arrow,label=$p_3$}{L1,v1}
	\fmf{plain,label=$1$,tension=0.5,l.s=right}{v1,v2}
	\fmf{plain,label=$3$,l.s=right,tension=0}{v2,v3}
	\fmf{plain,label=$2$,l.s=right,tension=0.5}{v3,v1}
	\fmf{plain,label=$4$,tension=0.5}{v2,v4}
	\fmf{plain,label=$6$,tension=0.5}{v4,v5}
	\fmf{plain,label=$5$,tension=0.5,l.s=right}{v5,v3}
	\fmf{plain_arrow,tension=1.3}{R1,v4}
	\fmf{plain_arrow,tension=1.3}{R2,v5}
	\fmfv{label=$p_1$}{R2}	
	\fmfv{label=$p_2$}{R1}  \end{fmfgraph*} }}}};
\draw[line width = 1, dashed, darkgreen,smooth] plot [smooth] coordinates {(-0.5,-0.8) (0.3,-0.15) (2.2,-0.2)};
\draw[line width = 1, dashed, darkgreen,smooth] plot [smooth] coordinates {(0.7,-0.8) (1,-0.4) (2.2,-0.4)};
\draw[line width = 1, dashed, darkgreen,smooth] plot [smooth] coordinates {(0.7,0.8) (1,0.4) (2.2,0.4)};
\end{tikzpicture}
 \end{gathered}
=
\left(2\pi i\right)^3 \left\{\ln z - \ln \zb \right\}
\label{eq:phi23cuts}
\end{multline}
Inserting into  Eq.~\eqref{masterformula} the sum of all cuts gives
\eq{
\label{twoandthree}
    \left[\Phi_{2}^{\text{(2-cuts)}} - \Phi_{2}^{\text{(3-cuts)}}\right]_{R^{12}}
    =
    (2\pi i)^2   \Big\{ \Li_2(\zb) - \Li_2(z-i\eps) -\frac{1}{2} \ln^2 z + \ln z \ln \zb - i \pi \ln z \Big\}
}
in agreement
with the discontinuity in Eq.~\eqref{eq:disc2disc1_two_loops}.
In particular, the three-cut diagrams $\Phi_{2}^{\text{(3-cuts)}}$ containing massless three-point vertices must be added to get the correct result. We have verified this result using a mass regulator, and the technique discussed in Appendix~\ref{sec:massless3pt}.
Note that while these diagrams add up to a finite result in this case, each diagram would na\"ively be set to zero in dimensional regularization as discussed earlier, which would lead to a wrong result. Further discussion on how to calculate massless three-point vertices can be found in Appendix~\ref{sec:massless3pt}.

\subsubsection{Three loops}
\label{sec:3loopT2cuts}
It is instructive to continue to three loops.
The most interesting case is the one in which two cuts are taken in the $p_2^2$ channel, where Eq.~\eqref{doublecut} tells us we should find
\eq{
    \left[\text{Disc}^2_{p_2^2} T_3 \right]_{R^2} = \left[2 T_3^{\text{(2-cuts)}} - 6 T_3^{\text{(3-cuts)}} \right]_{R^2} \, 
}
when we assign all propagators $+i\varepsilon$. 

The three-loop triangle 
\begin{equation}
\begin{gathered}
\begin{tikzpicture}[baseline=-3.5,scale=0.8]
        \draw[line width = 1] (0,0) -- (1,0);
        \draw[line width = 1] (1,0) -- (5,1);
        \draw[line width = 1] (1,0) -- (3,0.5);
        \draw[line width = 1] (1,0) -- (5,-1);
        \draw[line width = 1] (7,1.5) -- (7,-1.5);
        \draw[line width = 1] (7,-1.5) -- (8,-1.5);
        \draw[line width = 1] (5,1) -- (7,1.5);
        \draw[line width = 1] (5,-1) -- (7,-1.5);
        \draw[line width = 1] (7,1.5) -- (8,1.5);
        \draw[line width = 1] (3,0.5) -- (7,-0.5);
        \draw[line width = 1] (5,1) -- (7,0.5);
        \draw[dashed,line width = 1,color=darkgreen] (2,1.5) -- (2,-1.5);
        \draw[dashed,line width = 1,color=darkgreen] (4,1.5) -- (4,-1.5);
        \draw[dashed,line width = 1,color=darkgreen] (6,1.5) -- (6,-1.5);
        \node[text width=0.5cm] at (0,0.3) {$p_2$};
        \node[text width=0.5cm] at (8.3,1.5) {$p_3$};
        \node[text width=0.5cm] at (8.3,-1.5) {$p_1$};
        \node[text width=0.5cm,color=darkgreen] at (2.0,1.8) {$\mc{C}_1$};
        \node[text width=0.5cm,color=darkgreen] at (4.0,1.8) {$\mc{C}_2$};
        \node[text width=0.5cm,color=darkgreen] at (6.0,1.8) {$\mc{C}_3$};
        \node[text width=0.5cm] at (1.5,0.5) {$k_1$};
        \node[text width=0.5cm] at (3.5,1) {$k_2$};
        \node[text width=1.5cm] at (3.3,-1) {$p_2-k_1$};
        \node[text width=1.5cm] at (8,-1) {$p_3+k_1$};
        \node[text width=1.5cm] at (8,0) {$p_3+k_2$};
\end{tikzpicture}
\end{gathered}
\label{fig:3loops2cuts}
\end{equation}
is given by
\begin{multline}
    T_3 = - \frac{1}{6 \left(4 \pi\right)^6 p_1^2 p_3^4 \left(z-\zb\right)} \Big\{ \left[ \text{Li}_3(z) -\text{Li}_3(\zb) \right] \ln ^3(z \zb) -12 \left[\text{Li}_4(z)-\text{Li}_4(\zb) \right] \ln^2 (z \zb)  \\ + 60 \left[ \text{Li}_5(z)- \text{Li}_5(\zb)\right]\ln (z \zb) - 120 \left[ \text{Li}_6(z)- \text{Li}_6(\zb)\right]\Big\}\,.
\end{multline}
Taking two discontinuities in the $p_2^2$ channel using Eq.~\eqref{discdelta} gives
\eq{
    \text{Disc}^2_{p_2^2} T_3 =
    \frac{1}{1024 \pi^4 p_1^2 p_3^4 \left(z-\zb\right)} \Big\{ \left[ \text{Li}_3(z) -\text{Li}_3(\zb) \right] \left[\ln (z \zb) + 2 \pi i \right] - 4 \left[\text{Li}_4(z)-\text{Li}_4(\zb) \right] \Big\}\ \, , \label{eq:three_loop_double_disc}
}
while taking three discontinuities results in
\eq{
    \text{Disc}^3_{p_2^2} T_3 =
    -
    \frac{i}{512 \pi^3 p_1^2 p_3^4 \left(z-\zb\right)} \left[ \text{Li}_3(z) -\text{Li}_3(\zb) \right] \,.
}
To facilitate the cut computation, we rewrite Eq.~\eqref{doublecut} in a way that allows us to recycle results for the single cuts of the two-loop triangle. The sum of all single cuts in the $p_2^2$ channel of the two-loop triangle $T_2\left(p_1^2,p_2^2,p_3^2\right)$, with the traditional $i\varepsilon$ prescription involving $-i\varepsilon$'s to the right of the cut, was shown in~\cite{Abreu:2014cla} to agree with the discontinuity in $p_2^2$. We can use these results if we rewrite the term corresponding to the double cut $\mc{C}_1\mc{C}_2$ in Eq.~\eqref{fig:3loops2cuts} to have $-i\varepsilon$'s to the right of the cut, adding a triple cut term to compensate for it. When doing so, we must be careful with the combinatorial factors that come along with massless three-point vertices, as these cut integrals involve delta functions with support only at integration endpoints. In Appendix~\ref{sec:massless3pt}, we show that one gets an additional factor of $\frac{1}{m!}$ compared to na\"ively evaluating these delta functions to 1, where $m$ is the number of cuts being taken. Thus, we must absorb a term $-6 T_3^{\text{(3-cuts)}}$ to correct the $+i\varepsilon$'s to $-i\varepsilon$'s in the term $2 T_3^{\mc{C}_1 \mc{C}_2}$.  
The result we want to verify is therefore
\eq{
    \left[\left(\disc_{p_2^2}\right)^2 T_3 \right]_{R^2} = 2 \left. T_3^{\mc{C}_1,\mc{C}_2} \right\vert_{-i \varepsilon\text{ on r.h.s.}} + 2 \, T_3^{\mc{C}_1,\mc{C}_3} + 2\, T_3^{\mc{C}_2,\mc{C}_3}\,.
    \label{eq:result}
}
The first two terms in this expression correspond to cutting in $\mc{C}_1$ and summing over the one-cuts of the two-loop triangle. The details of the calculation are worked out in Appendix~\ref{sec:A_3looptri}, and the result is
\begin{multline}
    \left. T_3^{\mc{C}_1,\mc{C}_2} \right\vert_{-i \varepsilon\text{ on r.h.s. of cut}} + T_3^{\mc{C}_1,\mc{C}_3}
    =
    \frac{1}{2048 \pi^4} \frac{1}{p_1^2 p_3^4 \left(z-\zb\right) }
    \Big\{-3 \left[\text{Li}_4\left(z\right) - \text{Li}_4\left(\zb\right) \right] \\
    \hspace{2cm}
    +
    \ln\left(-\frac{m^2}{p_3^2}\right)
    \left[\text{Li}_3\left(z\right) - \text{Li}_3\left(\zb\right) \right]
    -\frac{1}{2} \left[\text{Li}_2^2\left(z\right) - \text{Li}_2^2\left(\zb\right) \right]
    \Big\} \, ,
\end{multline}
where $m$ is a small mass of the line labelled as $k$ used to regulate the IR divergence of the cut graphs.
The cut $T_3^{\mc{C}_2 \mc{C}_3}$ is given by
\begin{multline}
    T_3^{\mc{C}_2,\mc{C}_3} =
    \frac{1}{2048 \pi^4 p_1^2 p_3^4 \left(z-\zb\right)}
    \Big\{ \left[- \ln \left( - \frac{m^2}{p_3^2} \right) + \ln \left(z \zb\right) + 2 \pi i \right] \left[ \text{Li}_3\left(z\right) - \text{Li}_3\left(\zb\right) \right]
    \\
    + \left[\frac{1}{2} \text{Li}_2 \left(z\right) - \frac{1}{2}  \text{Li}_2 \left(\zb\right) \right]
    - \left[ \text{Li}_4 \left(z\right) - \text{Li}_4 \left(\zb\right) \right]
    \Big\} \,.
\end{multline}
The sum of all cuts is therefore
\begin{multline}
    2 \left. T_3^{\mc{C}_1,\mc{C}_2} \right\vert_{-i \varepsilon\text{ on r.h.s.}} + 2 \, T_3^{\mc{C}_1,\mc{C}_3} + 2 \, T_3^{\mc{C}_2,\mc{C}_3}
    \\ =
    \frac{1}{1024 \pi^4 p_1^2 p_3^4 \left(z-\zb\right)} \Big\{ \left[ \text{Li}_3(z) -\text{Li}_3(\zb) \right] \left[\ln (z \zb) + 2 \pi i \right] - 4 \left[\text{Li}_4(z)-\text{Li}_4(\zb) \right] \Big\}\
\end{multline}
in agreement with Eqs.~\eqref{eq:result} and~\eqref{eq:three_loop_double_disc}.

\subsubsection{\texorpdfstring{$L$}{L} loops}
\label{sec:Lloops}
Let us now consider the $L$-loop triangle integral,
\begin{align}
T_L\left(p_1^2,p_2^2,p_3^2\right)
    & =
    \begin{gathered}
    \resizebox{80mm}{!}{
    \begin{tikzpicture}[baseline=-3.5,scale=0.8,>=latex]
        \draw[line width = 1] (0,0) -- (1,0);
        \draw[line width = 1] (1,0) -- (5.5,1.125);
        \draw[line width = 1] (6.5,1.375) -- (9,2);
        \draw[line width = 1] (2,0.25) -- (2,-0.25);
        \draw[line width = 1] (3,0.5) -- (3,-0.5);
        \draw[line width = 1] (4,0.75) -- (4,-0.75);
        \draw[line width = 1] (5,1) -- (5,-1);
        \draw[line width = 1] (7,1.5) -- (7,-1.5);
        \draw[line width = 1] (8,1.75) -- (8,-1.75);
        \draw[line width = 1] (1,0) -- (5.5,-1.125);
        \draw[line width = 1] (6.5,-1.375) -- (9,-2);
        \draw[line width = 1] (9,2) -- (9,-2);
        \draw[line width = 1] (9,2) -- (10,2);
        \draw[line width = 1] (9,-2) -- (10,-2);
        \node[text width=0.5cm] at (0,0.3) {$p_3$};
        \node[text width=0.5cm] at (10.5,2.0) {$p_1$};
        \node[text width=0.5cm] at (10.5,-2) {$p_2$};
        \node[text width=1.5cm] at (10.2,0) {$p_2-k_1$};
        \node[text width=0.5cm] at (8.5,-2.25) {$k_1$};
        \node[text width=0.5cm] at (7.5,-2) {$k_2$};
        \node[text width=0.5cm] at (1.5,-0.5) {$k_L$};
        \node[text width=1.5cm] at (9,2.5) {$p_3-k_1$};
        \node[text width=1.5cm] at (7.25,2.15) {$p_3-k_2$};
        \node[text width=1cm] at (6.5,0) {$\cdots$};
\end{tikzpicture}
}
\end{gathered}
\\ & =
    \int
    \frac{d^4 k_1}{i \left(2\pi\right)^4}
    \cdots
    \frac{d^4 k_L}{i \left(2\pi\right)^4}
    \frac{1}{k_1^2+i \varepsilon}
    \frac{1}{k_2^2+i \varepsilon}
    \cdots
    \frac{1}{k_L^2+i\varepsilon}
    \frac{1}{\left(p_2-k_1\right)^2+i\varepsilon}
    \\
    & \hspace{-1cm} \times
    \frac{1}{\left(k_1-k_2\right)^2+i \varepsilon}
    \cdots
    \frac{1}{\left(k_{L-1}-k_L\right)^2+i\varepsilon}
    \frac{1}{\left(p_3-k_1\right)^2+i\varepsilon}
    \cdots
    \frac{1}{\left(p_3-k_L\right)^2+i\varepsilon} \,. \nonumber
\end{align}
The result after performing the loop integration is~\cite{Usyukina:1993ch}
\eq{
    T_L\left(p_1^2,p_2^2,p_3^2\right)
    = -
    \frac{1}{z-\zb} \frac{1}{L! \left(4\pi\right)^{2L} p_1^2 \left(p_3^2\right)^{L-1}}
    \sum_{j=L}^{2L}
    \frac{\left(-1\right)^j j! \ln^{2L-j} \left(z \zb \right)}{\left(j-L\right)!\left(2L-j\right)!} \left[ \text{Li}_{j}\left(z\right) - \text{Li}_{j}\left(\zb \right) \right]\,,
    \label{eq:Llooptriresult}
}
with $z$ and $\zb$ defined as before. 

One thing we can immediately observe about this expression is that taking two or more discontinuities along the long axis (in the $p_3^2$ channel) gives zero. To see this, we note that taking a discontinuity in $p_3^2$ corresponds to taking a monodromy around $z=1$, which is only nonvanishing for the $\Li_j(z)$ factor in Eq.~\eqref{eq:Llooptriresult}. Using the fact that the discontinuity of $\Li_n(z)$ corresponding to encircling the branch point at $z=1$ gives $\smash{2 \pi i \frac{\ln^{n-1} z}{(n-1)!}}$, we get
\eq{
  \hspace{-.45cm} \left[ \disc_{p_3^2} T_L \left(p_1^2,p_2^2,p_3^2\right) \right]_{R^3}
    =
    \frac{-2\pi i}{z-\zb} \frac{1}{L! \left(4\pi\right)^{2L} p_1^2 \left(p_3^2\right)^{L-1}}
    \sum_{j=L}^{2L}
    \frac{\left(-1\right)^j j! \ln^{2L-j} \left(z \zb \right)}{\left(j-L\right)!\left(2L-j\right)!} \frac{\ln^{j-1}(z)}{\left(j-1\right)!} \, . \label{eq:L_loop_p3_disc}
}
In this expression, there are no longer branch points at 1 in $z$ or $\zb$. Thus, further discontinuities in $p_3^2$ vanish, 
\begin{equation}
   \left[ (\disc_{p_3^2})^2 T_L \right]_{R^3} = 0\,. \label{eq:disc_p3_twice}
\end{equation}
The sum of taking two and more cuts of the $L$-loop triangle along the long axis must correspondingly also vanish. 

We now show that taking $L$ discontinuities in the $p_2^2$ channel amounts to taking $L$ cuts in the same channel, i.e.\
\begin{equation}
   \left[\big(\disc_{p_2^2}\big)^L T_L \left(p_1^2,p_2^2,p_3^2\right) \right]_{R^2} = \left[ L! \, \text{Cut}_{\mc{C}_1, \cdots, \mc{C}_L} T_L \left(p_1^2,p_2^2,p_3^2\right) \right]_{R^2} \,. \label{eq:L_disc_cuts}
\end{equation}
We start by computing the sequential discontinuity, which amounts to taking $L$ discontinuities of the factor $\ln^{2L-j} \left(z \zb \right)$ in the expression above. Only the first term in the sum over $j$, where $j=L$, contributes to this discontinuity. The result is
\begin{align}
    \left[\text{Disc}^L_{p_2^2} T_L \left(p_1^2,p_2^2,p_3^2\right) \right]_{R^2} =
    -
    \frac{i^L}{z-\zb} \frac{1}{\left(8\pi\right)^{L} p_1^2 \left(p_3^2\right)^{L-1}}
    \left[\text{Li}_{L}\left(z\right) - \text{Li}_{L}\left(\zb \right) \right]\,.
    \label{eq:DiscTL}
\end{align}
Next, we calculate the cuts. Putting the lines corresponding to the cuts $\mc{C}_1 \cdots \mc{C}_L$ on-shell in the region $R^2$ gives the following expression:
\begin{multline}
    \text{Cut}_{\mc{C}_1, \cdots, \mc{C}_L} T_L \left(p_1^2,p_2^2,p_3^2\right)
    =
    \int
    \frac{d^4 k_1}{i \left(2\pi\right)^4}
    \cdots
    \frac{d^4 k_L}{i \left(2\pi\right)^4}
    \left(-2\pi i\right)^{2L}
    \delta \left( k_1^2 \right) \Theta\left( k_1^0\right)
    \cdots
    \delta \left( k_L^2 \right)
    \Theta\left( k_L^0\right)
    \\ \times
    \delta \left[ \left(p_2-k_1\right)^2 \right]
    \Theta\left(p_2^0-k_1^0\right)
    \delta \left[ \left(k_1-k_2\right)^2 \right]
    \Theta\left(k_1^0-k_2^0\right)
    \cdots
    \delta \left[ \left(k_{L-1}-k_L\right)^2 \right]
    \Theta\left(k_{L-1}^0-k_L^0\right)
    \\ \times
    \frac{1}{\left(p_3+k_1\right)^2}
    \frac{1}{\left(p_3+k_2\right)^2}
    \cdots
    \frac{1}{\left(p_3+k_L\right)^2}\,.
\end{multline}
We perform the energy integrals using the delta functions $\delta \left( k_1^2 \right) \cdots \delta \left( k_L^2 \right)$, and get
\begin{multline}
    \text{Cut}_{\mc{C}_1, \cdots, \mc{C}_L} T_L \left(p_1^2,p_2^2,p_3^2\right)
    =
    \int
    \frac{d^3 k_1}{\left(2\pi\right)^3 2 \omega_{k_1}}
    \frac{d^3 k_2}{\left(2\pi\right)^3 2 \omega_{k_2}}
    \cdots
    \frac{d^3 k_L}{\left(2\pi\right)^3 2 \omega_{k_L}}
    \left(2\pi i\right)^{L}
    \delta \left( p_2^2 - 2 \, p_2 \cdot k_1 \right)
    \\ \times
    \delta \left(-2 \, k_1 \cdot k_2 \right)
    \cdots
    \delta \left(-2 \, k_{L-1} \cdot k_L \right)
    \frac{1}{\left(p_3+k_1\right)^2}
    \cdots
    \frac{1}{\left(p_3+k_L\right)^2}\,.
    \label{eq:cutLloopinitial}
\end{multline}
The remaining delta functions show that this cut only has support when the momenta $k_1, \cdots k_L$ and $k_1-k_2, \cdots k_{L-1}-k_L$ are all collinear. We therefore get a product of $L-1$ massless vertices. This configuration is singular and must be treated with care, using TOPT. As explained in Appendix~\ref{sec:massless3pt}, evaluating the integrals over these remaining delta functions gives rise to a combinatorial factor of $\frac{1}{L!}$. The result of the integral, worked out in detail in Appendix~\ref{sec:massless3pt}, is
\eq{
    \text{Cut}_{\mc{C}_1 \cdots \mc{C}_{L}} T_L \left(p_1^2,p_2^2,p_3^2\right)
    =
    -
    \frac{i^L}{z-\zb} \frac{1}{L! \left(8\pi\right)^{L} p_1^2 \left(p_3^2\right)^{L-1}}
    \left[\text{Li}_{L}\left(z\right) - \text{Li}_{L}\left(\zb \right) \right]\,.
}
Comparing this result to Eq.~\eqref{eq:DiscTL}, we see that Eq.~\eqref{eq:L_disc_cuts} is indeed satisfied.

\paragraph{Sequential discontinuities of the $L$-loop box ladders}~\\
We finally comment on the sequential discontinuities of the $L$-loop box ladder,
\eq{
B_L \left(p_1^2,p_2^2,p_3^2,p_4^2,(p_1+p_2)^2,(p_2+p_3)^2\right)
    =  \hspace{-.2cm}
    \begin{gathered}
    \resizebox{80mm}{!}{
    \begin{tikzpicture}[baseline=-3.5,scale=0.8,>=latex]
        \draw[line width = 1] (1,2) -- (5.5,2);
        \draw[line width = 1] (6.5,2) -- (9,2);
        \node[text width=0.5cm] at (2.5,0.3) {$x_{\ell_L}$};
        \draw[fill=black] (2.5,0.0) circle (0.05);
        \node[text width=0.5cm] at (3.5,0.3) {$x_{\ell_{L - 1}}$};
        \draw[fill=black] (3.5,0.0) circle (0.05);
        \draw[line width = 1] (2,2) -- (2,-2);
        \draw[line width = 1] (3,2) -- (3,-2);
        \draw[line width = 1] (4,2) -- (4,-2);
        \draw[line width = 1] (5,2) -- (5,-2);
        \draw[line width = 1] (7,2) -- (7,-2);
        \draw[line width = 1] (8,2) -- (8,-2);
        \draw[line width = 1] (1,-2) -- (5.5,-2);
        \draw[line width = 1] (6.5,-2) -- (9,-2);
        \draw[line width = 1] (9,2) -- (9,-2);
        \draw[line width = 1] (9,2) -- (10,2);
        \draw[line width = 1] (9,-2) -- (10,-2);
        \node[text width=0.5cm] at (8.5,0.3) {$x_{\ell_1}$};
        \draw[fill=black] (8.5,0.0) circle (0.05);
        \node[text width=0.5cm] at (0.5,2) {$p_4$};
        \node[text width=0.5cm] at (0.5,-2) {$p_3$};
        \node[text width=0.5cm] at (10.5,2.0) {$p_1$};
        \node[text width=0.5cm] at (10.5,-2) {$p_2$};
        \node[text width=0.5cm] at (9.5,0.3) {$x_2$};
        \draw[fill=black] (9.5,0.0) circle (0.05);
        \node[text width=0.5cm] at (5.5,-3) {$x_3$};
        \draw[fill=black] (5.5,-2.7) circle (0.05);
        \node[text width=0.5cm] at (1.5,0.3) {$x_4$};
        \draw[fill=black] (1.5,0.0) circle (0.05);
        \node[text width=0.5cm] at (5.5,3) {$x_1$};
        \node[text width=1cm] at (6.5,0) {$\cdots$};
\end{tikzpicture}
} \hspace{-.3cm}
\end{gathered}
}
These ladder integrals yield the same transcendental functions as the triangle integrals.  This is easiest to see in  dual space, as first considered in~\cite{Broadhurst:1993ib}.\footnote{Dual space can be defined as follows: for any planar diagram, we associate a variable \(x_{\ell_i}\) for each loop and a variable \(x_i\) for each region between two external lines.  Then, once we pick an orientation on each of the edges, we take the momentum flowing through that edge to be the difference between the dual variable on the right and the dual variable on the left.  This ensures momentum conservation at each vertex.  In some cases, the dual variables make manifest hidden symmetries, such as the dual conformal symmetry (see~\cite{Drummond:2007aua}).}
Translating to dual space, we label the dual points corresponding to loops by $x_{\ell_i}$, and by $x_j$ for external points, with $j \in \{1,2,3,4\}$. The ladder integral is then given by
\begin{align}
    B_L &\propto
  ((x_1 - x_3)^2)^L (x_2 - x_4)^2 \\
  &\qquad \times \int \prod_{i = 1}^L \frac {d^4 x_{\ell_i}}{(x_{\ell_i} - x_1)^2 (x_{\ell_i} - x_3)^2} \frac 1 {(x_2 - x_{\ell_1})^2} \prod_{i = 1}^{L - 1} \frac 1 {(x_{\ell_i} - x_{\ell_{i + 1}})^2} \frac 1 {(x_{\ell_L} - x_4)^2}\,. \nonumber
\end{align}
This integral is invariant under conformal transformations of the dual variables \(x\), which can be shown using Lorentz invariance and the (less obvious) invariance under inversion \(x^\mu \to \frac {x^\mu}{x^2}\). By a combination of translation and inversion we can send \(x_4\) to infinity.  In this limit we have \(\smash{\frac {(x_2 - x_4)^2}{(x_{\ell_L} - x_4)^2} \to 1}\).  This is precisely the triangle ladder in dual space. The box and triangle integrals therefore give the same analytic expression, and working out the exact transformation between the two, one can show that $z$ and $\zb$ variables for the box are given in terms of the Mandelstams as
\eq{
    \label{uvdefbox}
     z \zb = \frac{p_2^2 p_4^2}{p_1^2 p_3^2}\,, \hspace{1cm} \left(1-z\right)\left(1-\zb\right) = \frac{s t }{p_1^2 p_3^2} \, ,
}
with $s=\left(p_1+p_2\right)^2$ and $t=\left(p_2+p_3\right)^2$. All of the analysis for the triangle integrals therefore extends to $L$-loop box ladders. 

We can also compute the sequential discontinuity of the box ladder integrals in the $s$ and then $t$ channels, which is expected to vanish due to the Steinmann relations. To compute this quantity, we go to the region $R^{\{s,t\}}$, where $s,t>0$ while all other invariants $p_i^2<0$. For concreteness, we consider the phase-space point 
\begin{align}
    p_1 &= (1, 5,-6,0)\, ,\quad &p_2 = (1,-6,5,0)\, ,  \qquad \\
    p_3 &= (1,7,-6,0)\, , \quad &p_4 = (-3, -4, 7, 0) \, . \qquad 
\end{align}
We can analytically continue into $R^t$ by rescaling $E_1 \to \alpha E_1$ and $E_2 \to \alpha E_2$ by $1>\alpha >0$, while keeping $E_3$ fixed and varying $E_4 = -E_1-E_2-E_3$ along with $E_1$ and $E_2$. We then return to $R^{\{s,t\}}$ by the reverse path, after encircling the branch point at $s=0$. In the $z$ and $\zb$ variables, this corresponds to analytically continuing around $z=1$. A similar path around the branch point at $t=0$ can be constructed by instead rescaling $E_2$ and $E_3$, and also corresponds to computing a monodromy around $z=1$. Since this sequence of discontinuity operators is identical to the sequence of operators used to compute sequential discontinuities in the $p_3^2$ channel of the triangle, Eq.~\eqref{eq:disc_p3_twice} confirms that the Steinmann relations are satisfied by the box ladder integral at all loop orders. This matches the Steinmann analysis carried out in~\cite{Basso:2017jwq}, where the expression that appears in Eq.~\eqref{eq:L_loop_p3_disc} was also shown to reduce to a simpler functional form (given as Eq. (19) of that paper, which has slightly different rational normalization).

\section{Conclusions}
\label{sec:conclusions}

In this paper we have analyzed the discontinuities and cuts of Feynman integrals from several points of view. We first described how to compute the imaginary part of Feynman integrals in terms of cuts, reviewing the work of Cutkosky and 't Hooft and Veltman, and also described the analogous relations in non-covariant time-ordered perturbation theory.  
These traditional approaches are based on the idea that Feynman integrals have branch cuts in physical regions, and that integrals over propagators with $+i\eps$ and $-i\eps$ displacements are on opposite sides of these branch cuts. The main focus of this paper has been to extend these methods to sequential discontinuities. The $\pm i \eps$ prescription is in general insufficient for computing more than one discontinuity, but the relevant computations can be carried out by considering monodromies around the branch points of Feynman integrals. In particular, by understanding discontinuities in terms of monodromies, we are able to homogenize the $+i\eps$ and $-i\eps$ propagators that appear after the first discontinuity by analytically continuing them into same cut complex plane. This allows subsequent discontinuities to be taken. For integrals that are expressible in terms of generalized polylogarithms, we have also described how discontinuities can be computed using variation matrices and the monodromy group.\footnote{It should be possible to extend the variation and monodromy matrix construction to elliptic polylogarithms~\cite{brown2011multiple,Broedel:2017kkb,Broedel:2018iwv}, which also appear in Feynman integral calculations. It would be interesting to see if it could also be used in conjunction with the diagrammatic coaction~\cite{Abreu:2017enx,Abreu:2017mtm,Abreu:2019eyg}.}

The main result of this paper is a formula relating the sequential discontinuities in the same or different channels around branch points associated with invariants $s_j$ to cuts:
\begin{multline}
    \frac{1}{m_1!} \cdots \frac{1}{m_n!} \left[(\disc_{s_1})^{m_1}\cdots (\disc_{s_n})^{m_n} \cM\right]_{R^{\{s_1,\cdots,s_n\}}} \\
=
\sum_{k_1=m_1}^\infty
\stirling{k_1}{m_1}
      \cdots      
      \sum_{k_n=m_n}^\infty
      \stirling{k_n}{m_n}
      (-1)^{\sum m_i - \sum k_i} 
\Big[ \cM_{\{
      k_i~\text{cuts in channel $s_i$} \}}
      \Big]_{R^{\{s_1,\cdots,s_n\}}_+} \, .
\end{multline}
It is crucial that these relations are understood to apply only in regions where all the cuts of interest are nonvanishing.  In particular, we emphasize that these discontinuities are always taken as the difference between $\cM$ evaluated at the same physical value of real external momenta on different Riemann sheets.

An important consideration that we have spent considerable time exploring is that the analytic continuations by which these discontinuities are computed must be chosen with care. Paths that are homologous but not in the same homotopy class may give different answers
(as discussed in Appendix~\ref{sec:fundamental_group}).
In addition, the derivation of our formulas is made assuming a path exists which continues the external energies, holding the three-momenta fixed and respecting energy conservation. 
We have presented many nontrivial examples of cut and discontinuity computations, and have checked that Eqs.~\eqref{mcutsins} and~\eqref{ddstRst1} hold in these examples. For each example, we have been sure to find an explicit path in energies connecting the relevant regions, and used the path to determine which branch points are encircled. If one just picks an arbitrary path between regions, the discontinuity can still be computed, but there is no guarantee of agreement with cuts (and in fact, the agreement sometimes fails). While there is undoubtedly a more covariant way to understand the constraints on the paths, in every case where we have found an explicit path in energy we have found agreement between discontinuities and cuts according to our formulas, and conversely, in cases where our formulas seem to fail, we have not been able to find an explicit path in energy between regions (so that our formulas do not apply).

An important class of sequential discontinuities described by Eq.~\eqref{ddstRst1} are those in which the discontinuity channels are partially overlapping. In these cases, this equation encodes the Steinmann relations, originally derived using axiomatic quantum field theory, which state that sequential discontinuities in partially overlapping channels must vanish. In the original $S$-matrix program, this was shown to hold for full non-perturbative $S$-matrix elements in a mass-gapped scalar quantum field theory. Our analysis implies that the Steinmann  relations in fact hold for individual Feynman integrals.\footnote{It had previously been observed that the Steinmann relations were obeyed by many of the Feynman integrals that appear in planar $\mathcal{N}=4$, insofar as these integrals appear in the space of Steinmann-satisfying hexagon functions~\cite{Caron-Huot:2016owq,Caron-Huot:2018dsv,Caron-Huot:2019bsq}.} This amounts to a proof of the Steinmann relations in perturbation theory, diagram by diagram.  Our proof requires only that the region where both channels can be simultaneously cut must exist, and that the external momenta are not constrained (for instance by being massless).

Of course, the constraint that all external lines be massive is a strong one, and excludes many theories of physical interest. As such, it would be good to understand the massless case in more depth. The tools we have developed should in principle apply to any Feynman integral, but a full analysis of the massless case involves an additional profusion of subtleties. For example, if we regulate the IR divergences of the massless box by going to $d>4$ dimensions, we get a $\ln s \ln t$ contribution (see Eq.~\eqref{M0m}), and a nonzero (and IR-finite) sequential monodromy in $s$ and $t$. However, regulating the external lines with masses, as done in the four-mass box, the sequential monodromy in $s$ and $t$ vanishes (this follows from Eq.~\eqref{all2p0}, if we use Eq.~\eqref{uvdefbox} to map the triangle to the box integral). Thus, this sequential discontinuity, despite being IR finite, is regulator-dependent. We leave further study of these subtleties to future work.

Time-ordered perturbation theory played an essential role in our derivation. There is a sense in which time-ordered perturbation theory is more physical than covariant perturbation theory, since particles are always on-shell. Indeed, the benefits of a non-covariant formulation in some other contexts are well-known, such as how light-cone perturbation theory is used to show factorization, and new uses are constantly being developed, such as for cosmological polytopes~\cite{Arkani-Hamed:2017fdk,Arkani-Hamed:2018bjr}. It would be interesting to see if Steinmann-type constraints and the monodromy group could be useful as a bootstrapping technique in cosmological contexts.

The existence of IR divergences in amplitudes involving massless particles actually facilitates the study of certain aspects of these amplitudes. The IR structure of gauge theories is particularly well understood: a scattering amplitude can be factorized into a hard part, a jet (collinear) part, and a soft part~\cite{Collins:1988ig,Collins:1989gx,
CATANI1989323,catani1998singular,Bauer:2000yr,Bauer:2002nz,Beneke:2002ni,
Feige:2013zla,Feige:2014wja}.
The hard part is IR-finite and can be interpreted as the $S$-matrix (the `hard' $S$-matrix) in a computational scheme where the soft and collinear parts are included in the asymptotic Hamiltonian~\cite{Hannesdottir:2019opa,Hannesdottir:2019rqq}. This suggests that analytic properties of the hard part alone might be amenable to the same techniques used to study massive, IR-finite theories like we have done in this paper. Indeed, the analytic properties of scattering amplitudes in planar $\cN=4$ super-Yang-Mills theory are usually studied at the level of IR-finite remainder functions, which can also be interpreted as hard $S$-matrix elements. In fact, this connection was part of the motivation for the current work.

The soft part of the scattering amplitude in theories with massless particles can also reproduce the IR-dominated non-analytic behavior of the full $S$-matrix elements. 
The soft function, which can be represented as a matrix element of Wilson lines, satisfies a renormalization group equation and can be written as
the exponential of the integral of the soft anomalous dimension~\cite{Polyakov:1980ca,Arefeva:1980zd,Dotsenko:1979wb,Brandt:1981kf,Korchemsky:1985xj,Korchemsky:1985xu,Korchemsky:1987wg}.  The soft anomalous dimension depends on kinematics and is a matrix in color space; it contains a dipole part, which is diagonal in color space, and a correction term with restricted kinematic dependence~\cite{Gardi:2009qi,Becher:2009cu}. The dipole part is determined by the cusp anomalous dimension, and is proportional to \(\sum_{i < j} \ln(\frac{-p_i \cdot p_j}{\mu^2})\), where \(\mu\) is the renormalization-group scale. The correction to the dipole formula depends only on the directions of the external momenta and not on their magnitudes; this implies that it can only
depend on rescaling-invariant cross-ratios of the form \(\rho_{i j k l} = \frac {(p_i \cdot p_j) (p_k \cdot p_l)}{(p_i \cdot p_k) (p_j \cdot p_l)}\).  This constitutes a strong constraint, and in particular implies that a soft function can never have cuts in channels with more than two particles.
Since simultaneously cutting a pair of partially-overlapping two-particle channels isolates a one-particle channel, i.e.~a decay; such partially-overlapping cuts are forbidden in theories with only stable particles. This is one way to understand the Steinmann relation in $S$-matrix theory in the soft limit. 
In contrast, in theories with massless particles, $1 \to n$ amplitudes do not have to identically vanish. Correspondingly, the articulation of the Steinmann relations for these theories proves to be more challenging. Nevertheless, the restriction to two-particle cuts in the soft limit gives a clue to how we might understand the analytic properties of the massless case. Also, since the soft function is an expectation value of a product of Wilson lines, one could ask what restrictions causality imposes on this expectation value.\footnote{While the Steinmann relations were initially studied for correlation functions of local operators, the implications of causality on non-local operators do not seem to have been studied.}

To facilitate our analysis, we have presented an introduction to the monodromies of polylogarithmic functions, drawing inspiration from~\cite{MR1265552,MR3469645}.  A central role in this analysis is played by the connection \(\omega\) and an integration contour \(\gamma\).  These ingredients are sufficient to determine a variation matrix via \(\smash{\mathscr{M}_\gamma = \mathcal{P}\exp \int_\gamma \omega}\).  The variation matrix is a homotopy functional, i.e.\ its value depends only on the homotopy class of the integration contour \(\gamma\).   In typical cases, the number of homotopy classes is infinite.  Nevertheless, in physical applications one rarely considers analytic continuations in the full domain of analyticity; in the examples we studied, it was sufficient to consider rotations in the phases of energies.
The allowed sequences of cuts correspond to non-vanishing elements in the variation matrix, while forbidden sequences of cuts correspond to vanishing elements.  

This type of reasoning, in which the vanishing of certain cuts (or sequences of cuts) is used to constrain the analytic structure of polylogarithmic scattering amplitudes and Feynman integrals, has appeared in a number of contexts (see for instance~\cite{DelDuca:2011wh,Dixon:2011pw,Duhr:2012fh,Chavez:2012kn,Dixon:2013eka,Abreu:2014cla}). These analyses are often carried out at the level of the symbol, with the resulting objects only being later upgraded to full polylogarithmic functions using the methods of~\cite{Brown:2011ik,Duhr:2011zq,Duhr:2012fh} (or more implicitly, using the methods reviewed in~\cite{Caron-Huot:2020bkp}).
It is important to note, however, that when such constraints are imposed directly at the level of the symbol, it is not always clear whether the corresponding cuts can arise in the physical region, or only outside of it. This could prove salient, as the Steinmann relations do not necessarily apply when the relevant cuts are not accessible within the physical region. 

It would be particularly interesting to understand whether the Steinmann-type constraints that prove useful in planar $\mathcal{N}=4$ ~\cite{Caron-Huot:2016owq} all correspond to cuts that are accessible within physical regions, or point to some further special property of these amplitudes. In particular, it has been observed that these constraints can be generalized to the \emph{extended} Steinmann relations, which apply to sequential discontinuities at all depths in the symbol~\cite{Caron-Huot:2018dsv,Caron-Huot:2019bsq}, and that these extended constraints exhibit intriguing connections to cluster algebras~\cite{Drummond:2017ssj}. The extended Steinmann relations have been used in conjunction with additional formal constraints, such integrability (which ensures that symbols can be upgraded to genuine functions), first entry conditions (which constrain the branch cuts that are accessible on the boundary of the Euclidean region), and last entry conditions (which constrains the derivative of these amplitudes) to formulate ans\"atze for six- and seven-particle amplitudes in this theory, which can be further constrained in special kinematic limits to determine the amplitude at a given loop order~\cite{Dixon:2013eka,Dixon:2014voa,Dixon:2014iba,Drummond:2014ffa,Dixon:2015iva,Caron-Huot:2016owq,Dixon:2016apl,Dixon:2016nkn,Drummond:2018caf,Caron-Huot:2019vjl}. These types of constraints can all be conveniently formulated in terms of the connection \(\omega\).  The integrability condition is just the requirement that \(\omega \wedge \omega = 0\), the first entry condition constrains the differentials that appear in the first row of \(\omega\), and the last entry condition constrains differentials that appear in the last column of \(\omega\).

In fact, one can consider bootstrapping Feynman integrals directly in terms of the elements of their variation matrices \(\mathscr{M}_\gamma\).\footnote{A similar idea, of using dispersion relations to complete the coproduct of a Feynman integral, was developed in~\cite{Abreu:2014cla}.} Many of the entries in the right column of \(\mathscr{M}_\gamma\) correspond to different (sequential) cut channels, and should therefore be expressible as integrals over the phase space of on-shell amplitudes.\footnote{This will not be true in channels that are only accessibly outside of physical regions.} The integrability condition \(\omega \wedge \omega = 0\) imposes linear constraints that relate these cut integrals to the other entries of \(\mathscr{M}\). Moreover, when working in terms of the connection \(\omega\), one can impose additional constraints having to do with the unipotence of its monodromy matrices, namely that property that $(\bbone- \mathscr{M}_{\linebub^x_p})^k = 0$ for some integer $k$, where this integer $k$ is related to the number of cuts one can take in channel corresponding to this monodromy. More generally, this unipotence property provides strong constraints on the underlying mixed Hodge structure of the polylogarithmic functions that arise from Feynman integrals, and it would be interesting to understand these constraints in more detail.

\section*{Acknowledgments}
\enlargethispage{20pt}
The authors would like to thank Lance Dixon, Claude Duhr, and Einan Gardi for enlightening conversations.
This project emerged from discussions at the Aspen Center for Physics, which is supported by National Science Foundation grant PHY-1607611. %
This work has been supported in part by the U.S.\ Department of Energy under contract DE-SC0013607 (MS,HH), an ERC Starting Grant (No.\ 757978) and a grant from the Villum Fonden (No.\ 15369) (JLB,AJM,CV), and a Carlsberg Postdoctoral Fellowship (CF18-0641) (AJM). JLB, AJM, and CV are grateful for the hospitality of the Harvard University Physics Department and the Harvard Center for Mathematical Sciences and Applications.

\newpage
\appendix

\section{The coproduct from variation matrices} \label{subsec:coaction}

Polylogarithms come equipped with a motivic coproduct~\cite{Goncharov:2001iea,Goncharov:2005sla} which is sometimes usefully upgraded to a coaction~\cite{Brown1102.1312,2011arXiv1101.4497D}.  The coproduct or coaction can be used to systematically decompose the analytic structure of complicated functions into simpler building blocks.
These mathematical notions have been used in a wide variety of Feynman integral calculations to constrain the functional form of the answer based on knowledge of the locations of its discontinuities (see for example~\cite{Goncharov:2010jf,Duhr:2012fh,CaronHuot:2011ky,Duhr:2012fh,Abreu:2014cla,Caron-Huot:2018dsv,Caron-Huot:2019vjl}).
In this appendix, we show how the coproduct arises naturally in the language of the variation matrices \(\mathscr{M}\).

Let us consider again the example of the dilogarithm, which has the variation matrix
\begin{equation}
  \mathscr{M} =
  \begin{pmatrix}
    1 & \operatorname{Li}_1(z) & \operatorname{Li}_2(z) \\
    0 & 1 & \ln(z) \\
    0 & 0 & 1
  \end{pmatrix}. \label{eq:appendix_dilog}
\end{equation}
A couple of observations can be made about the entries in the top row and the last column of this matrix. The first is that the product $\mathscr{M}_{1i} \mathscr{M}_{i3}$ has the same transcendental weight as the original function $\mathscr{M}_{13}$, for all $i$. Second, because of the differential equation this matrix satisfies, the entries $\mathscr{M}_{1i}$ involve the iterated integral corresponding to carrying out the first $i-1$ integrations in the definition of $\operatorname{Li}_2(z)$ (as given in Eq.~\eqref{dilog_integral_representation}), while the entries $\mathscr{M}_{i3}$ involve the iterated integral that results from dropping the first $i-1$ integrations. Following these observations, we can consider defining an operator $\Delta$ that maps $\operatorname{Li}_2(z)$ to a sum over a tensor product of these matrix entries, which we might think of as summing over the possible ways to partition the integrations in $\operatorname{Li}_2(z)$ into an initial and a final set:
\begin{equation}
  \Delta \mathscr{M}_{1 3} = \sum_{j = 1}^3 \mathscr{M}_{1 j} \otimes \mathscr{M}_{j 3}.
\end{equation}
Plugging in the functions that appear in $\mathscr{M}$, this equation becomes
\begin{equation}
  \Delta \operatorname{Li}_2(z) = 1 \otimes \operatorname{Li}_2(z) + \operatorname{Li}_1(z) \otimes \ln(z) + \operatorname{Li}_2(z) \otimes 1 \, ,
\end{equation}
which can be recognized to be precisely the coproduct of the dilogarithm, as defined in~\cite{Goncharov:2005sla}.

These observations, and the corresponding construction of the coproduct, can be extended to the general case. Namely, due to the fact that each row of $\mathscr{M}$ satisfies the same differential equation, the product $\mathscr{M}_{ij}\mathscr{M}_{jk}$ has the same transcendental weight as $\mathscr{M}_{ik}$ for all $i \leq j \leq k$. And while generic variation matrix entries $\mathscr{M}_{ik}$ involve sums of iterated integrals, the functions $\mathscr{M}_{ij}$ still correspond to carrying out (some linear combination of) the initial integrations entering $\mathscr{M}_{ik}$, while the functions $\mathscr{M}_{jk}$ still correspond to carrying out (some linear combination of) the final integrations in $\mathscr{M}_{ik}$. Correspondingly, the coproduct can be defined in terms of entries of the variation matrix by
\begin{equation} \label{appendix:coproduct_def}
  \Delta \mathscr{M}_{ik} = \sum_{j = i}^k \mathscr{M}_{i j} \otimes \mathscr{M}_{j k}\,.
\end{equation}
As indicated by the use of general indices $i$ and $k$, the coproduct can be applied to any entry of a variation matrix; however, as in~\cite{Goncharov:2005sla}, the second factor in this tensor product must be interpreted modulo factors of $i \pi$. Instances of $i\pi$ that appear in the first factor can be retained using the methods of~\cite{Brown:2011ik}.

It is worth emphasizing that the coproduct~\eqref{appendix:coproduct_def} can be applied to entries of the variation matrix in any region, and that it commutes with the action of the monodromy matrices. For instance, recall the variation matrix for the triangle and box integral from Eq.~\eqref{eq:principal_variation_matrix_box},
\begin{equation} \label{appendix:principal_variation_matrix_box}
  \mathscr{M}_{\gamma_0} =
  \begin{pmatrix}
    1 & \ln(z \bar{z}) & \operatorname{Li}_1(z) + \operatorname{Li}_1(\bar{z}) & \Phi_1(z, \bar{z}) \\
    0 & 1 & 0 & - \operatorname{Li}_1(z) + \operatorname{Li}_1(\bar{z}) \\
    0 & 0 & 1 & \ln(z/\bar{z}) \\
    0 & 0 & 0 & 1
 \end{pmatrix}\, ,
\end{equation}
where we recall that
\begin{equation}
  \Phi_1(z, \bar{z}) = -\ln(z \bar{z}) (\operatorname{Li}_1(z) - \operatorname{Li}_1(\bar{z})) + 2 (\operatorname{Li}_2(z) - \operatorname{Li}_2(\bar{z})) \, .
\end{equation}
Using Eq.~\eqref{appendix:coproduct_def}, we can easily read off the coproduct of $\Phi_1(z, \bar{z})$ from Eq.~\eqref{appendix:principal_variation_matrix_box}:
\begin{align} \label{eq:appendix_coproduct_box}
    \Delta \Phi_1(z, \bar{z}) &= 1 \otimes \Phi_1(z, \bar{z}) - \ln(z \bar{z}) \otimes \operatorname{Li}_1(z) + \ln(z \bar{z}) \otimes \operatorname{Li}_1(\bar{z}) \\
    &\quad + \operatorname{Li}_1(z) \otimes \ln(z/\bar{z}) + \operatorname{Li}_1(\bar{z}) \otimes \ln(z/\bar{z}) + \Phi_1(z, \bar{z}) \otimes 1\, . \nonumber
\end{align}
To analytically continue Eq.~\eqref{eq:appendix_coproduct_box} around one of its branch points, we can replace all of the functions in the left factor of the coproduct with the value they take after being acted on by one of the monodromy matrices. It should be clear that this results in the same coproduct that one would get from applying Eq.~\eqref{appendix:coproduct_def} directly to the variation matrix that results from the action of the monodromy matrix. Further details on the properties of the coproduct can be found in~\cite{Duhr:2014woa}.

\section{The monodromy and fundamental groups}
\label{sec:fundamental_group}

As seen in Section~\ref{sec:math}, the complete analytic structure of a collection of polylogarithms can be encoded in a set of monodromy matrices. These matrices occur in one-to-one correspondence with the location of simple poles in the integral definition of these polylogarithms, reflecting the fact that the corresponding integration contours are always homotopic to a composition of (some sequence of) closed contours that encircle individual poles, and a contour that does not cross any branch cuts. This indicates that there should be some relation between the monodromy group and the fundamental group describing the manifold on which these polylogarithms are defined, which has punctures at precisely the loci of these simple poles.

To make this connection between the monodromy and fundamental groups more explicit, we first observe that monodromy matrices can be written as the conjugation of a matrix with rational entries by a diagonal matrix whose entries are integer powers of \(2 \pi i\).  For instance, the monodromy matrices of the dilogarithm from Eq.~\eqref{eq:dilog_monodromy_zero} and Eq.~\eqref{eq:dilog_monodromy_1} can be written as
\begin{equation}
{\blue{ \mathscr{M}_{\linebub{}_0}}} =
 \begin{pmatrix}
    1 & 0 & 0 \\
    0 & 1 & 2 \pi i \\
    0 & 0 & 1
  \end{pmatrix} \\
  =
  \begin{pmatrix}
    1 & 0 & 0 \\
    0 & 2\pi i & 0 \\
    0 & 0 & (2 \pi i)^2
  \end{pmatrix}^{-1}
\cdot
  \begin{pmatrix}
    1 & 0 & 0 \\
    0 & 1 & 1 \\
    0 & 0 & 1
  \end{pmatrix} \cdot
    \begin{pmatrix}
    1 & 0 & 0 \\
    0 & 2\pi i & 0 \\
    0 & 0 & (2 \pi i)^2
  \end{pmatrix}
\label{eq:dilog_homotopy_0b}
\end{equation}
and
\begin{align}
\! {\red{ \mathscr{M}_{\linebub{}_1}}} =
\begin{pmatrix}
    1 & -2 \pi i & 0 \\
    0 & 1 & 0 \\
    0 & 0 & 1
  \end{pmatrix}
  =
  \begin{pmatrix}
    1 & 0 & 0 \\
    0 & 2\pi i & 0 \\
    0 & 0 & (2 \pi i)^2
  \end{pmatrix}^{-1} \cdot
  \begin{pmatrix}
    1 & -1 & 0 \\
    0 & 1 & 0 \\
    0 & 0 & 1
  \end{pmatrix} \cdot
   \begin{pmatrix}
    1 & 0 & 0 \\
    0 & 2\pi i & 0 \\
    0 & 0 & (2 \pi i)^2
  \end{pmatrix} .
\label{eq:dilog_homotopy_1b}
\end{align}
These conjugated matrices can be understood as furnishing a representation of the homotopy group of \(\mathbb{C} - \{0, 1\}\) by matrices in \(\mathrm{GL}(3, \mathbb{Z})\).  More explicitly, the homotopy group of \(\mathbb{C} - \{0, 1\}\) is the free group with two generators, which are associated with the homotopy classes of paths around \(z = 0\) and \(z = 1\).  Up to conjugation by $\operatorname{diag}\big(1, 2 \pi i, (2 \pi i)^2\big)$, the monodromy matrices give us an explicit representation of this group.

\begin{figure}[t]
    \centering
\newsavebox{\lbubinfp} \sbox{\lbubinfp}{$\green{\gamma_{\linebubUp}^\infty}$}
\newsavebox{\lbubinfm} \sbox{\lbubinfm}{$\purple{\gamma^{\linebubDown}_\infty}$}
\newsavebox{\lbubzero} \sbox{\lbubzero}{$\blue{\gamma_{\linebub_0}}$}
\newsavebox{\lbubone} \sbox{\lbubone}{$\red{\gamma_{\linebub_1}}$}
\newsavebox{\lbubzeroone} \sbox{\lbubzeroone}{$\red{\gamma_{\linebub_{0,1}}}$}
\resizebox{7cm}{!}{
\begin{tikzpicture}
[decoration={markings,
 mark=between positions 0.1 and 1 step 0.2  with {\arrow[line width=1pt]{>}}}]
\tikzmath{\edx=0.4;
\rad = 0.4; \ang =170; \dy = \rad*sin(\ang);
\th=70;\rr = 2.8;\yy = \rr*sin(\th); \xx = \rr*cos(\th);
\thb=62;\rrb = 3.2;\yyb = \rrb*sin(\thb); \xxb = \rr*cos(\thb);
\thR=-70;\rrR = 3.2;\yyR = \rrR*sin(\thR); \xxR = \rrR*cos(\thR);}
\draw[->] (-4,0) -- (4,0) coordinate (xaxis);
\draw[->] (0,-3.5) -- (0,3.5) coordinate (yaxis);
\draw [darkblue,postaction=decorate, thick] (\edx,0) arc (0:350:\rad);
\draw [darkred,postaction=decorate, thick] (\edx,0) to (1.8,-\dy) arc (-\ang:\ang:\rad) to (\edx,0);
\draw [darkgreen,dashed,postaction=decorate, thick] (\edx,0) to (\xx,\yy) arc (\th:360:\rr) arc (0:\th-15:\rr) to (\edx,0);
\draw [darkgreen,postaction=decorate, thick] (\edx,0) to (\xxb,\yyb) arc (220:0:0.4) arc (0:-100:0.4) to (\edx,0);
\draw [fill=black] (0,0) circle[radius = 2 pt];
\node[below, scale=0.8] at (0.1,0) {$0$};
\draw [fill=black] (2.2,0) circle[radius = 2 pt];
\node[below, scale=0.8] at (2.3,0) {$1$};
\draw [fill=black] (\xxb+0.3,\yyb+0.25) circle[radius = 2 pt];
\node[below, scale=0.8] at (\xxb+0.3,\yyb+0.25) {$\infty$};
\node[darkblue] at (-0.8,0.5) {\usebox{\lbubzero}};
\node[darkred] at (2.3,0.6) {\usebox{\lbubone}};
\node[darkgreen] at (2.5, 3) {\usebox{\lbubinfp}};
\node[darkgreen] at (-2.5, 2.2) {\usebox{\lbubinfp}};
\node[below] at (xaxis) {$\text{Re } s$};
\node[left] at (yaxis) {$\text{Im } s$};
\end{tikzpicture}}
\resizebox{7cm}{!}
{\begin{tikzpicture}
[decoration={markings,
 mark=between positions 0.1 and 1 step 0.3  with {\arrow[line width=1pt]{>}}}]
\tikzmath{\edx=0.4;
\rad = 0.4; \ang =170; \dy = \rad*sin(\ang);
\thR=-50;\rrR = 3.2;\yyR = \rrR*sin(\thR); \xxR = \rrR*cos(\thR);}
\draw[->] (-4,0) -- (4,0) coordinate (xaxis);
\draw[->] (0,-3.5) -- (0,3.5) coordinate (yaxis);
\draw [darkred,postaction=decorate, thick] (\edx,0) arc (0:-270:\rad) to (1.8,-\dy) arc (-\ang:\ang:\rad) to (-0.1,1.3*\rad) arc (-270:0:1.2*\rad);
\draw [darkpurple,postaction=decorate, thick] (\edx,0) to (\xxR,\yyR) 
arc (150:360:0.4)  arc (0:120:0.4) to (\edx,0);
\draw [fill=black] (0,0) circle[radius = 2 pt];
\node[below, scale=0.8] at (0.1,0) {$0$};
\draw [fill=black] (2.2,0) circle[radius = 2 pt];
\node[below, scale=0.8] at (2.3,0) {$1$};
\node[darkred,thick] at (2.1,0.6) {\usebox{\lbubzeroone}};
\node[darkpurple] at (2.5, -1.8) {\usebox{\lbubinfm}};
\draw [fill=black] (\xxR+0.38,\yyR-0.2) circle[radius = 2 pt];
\node[below, scale=0.8] at (\xxR+0.38,\yyR-0.2) {$\infty$};
\node[below] at (xaxis) {$\text{Re } s$};
\node[left] at (yaxis) {$\text{Im } s$};
\end{tikzpicture}
}
    \caption{Paths around \(0\), \(1\) and \(\infty\).  We depict two possible contours that go around infinity, starting at points in the upper or lower half-plane. These are each homotopic to paths around 0 and 1, but in different orders. The two contours around infinity are not homotopically equivalent. The right panel shows that the path ambiguity is present also for paths around $s=1$.}
    \label{fig:infcont}
\end{figure}
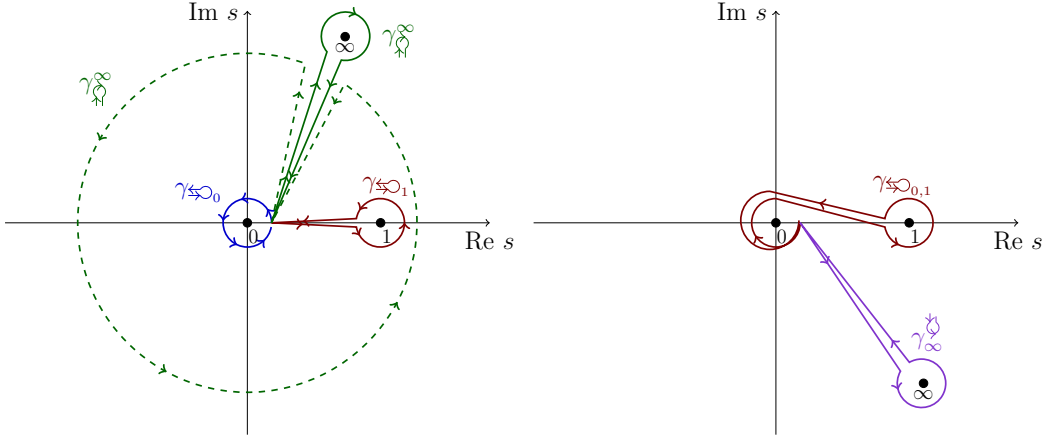

Note that this connection to the fundamental group remains valid if we compactify the complex plane by considering the monodromy matrix associated with infinity. Using the connection in Eq.~\eqref{eq:dilog_connection},
we can compute the monodromy matrix from an infinitesimal contour encircling infinity. For instance, if we integrate the dilogarithm integrand around a circular path that starts and ends at a complex point $|R|>1$, we have
\begin{equation}
    \int_{\gamma_R} \frac{d s}{1 - s} \circ \frac{d s} s = (2\pi i)^2\int_0^1 d t \int_0^t du  \frac{R e^{2\pi i u} }{1-R e^{2\pi i u}} = 2\pi^2 +2\pi i \ln\frac{R-1}{R} 
    \, .
\end{equation}
Since $\ln\frac{R-1}{R}$ is a continuous function for large $|R|$,  $\lim_{R\to\infty}\ln\frac{R-1}{R} =0$ and we get $2\pi^2$. The full matrix can be computed to be
\begin{equation}
    \mathscr{M}_{\circlearrowright_\infty}  =
  \begin{pmatrix}
    1 & -2 \pi i &  2\pi^2 \\
    0 & 1 & 2 \pi i  \\
    0 & 0 & 1  \\
  \end{pmatrix}\, .
\end{equation} 
Note that going around infinity clockwise corresponds to a counterclockwise contour around $0$ and $1$.
If we compute the matrix along a straight line path between $0$ and $R$, we get the variation matrix in Eq.~\eqref{eq:dilog_variation_matric_cut_complex_plane}:
\begin{equation}
    \mathscr{M}_{0\to R}  =
  \begin{pmatrix}
    1 & -\ln(1-R) & \Li_2(R) \\
    0 & 1 & \ln R  \\
    0 & 0 & 1  \\
  \end{pmatrix}\, .
\end{equation} 
Then, if we take $R\to \infty$ with $\im R > 0$,  we get
\begin{equation}
 {\green{ \mathscr{M}_{\linebubUp}^\infty}}=
     \mathscr{M}_{0\to R} \cdot
     \mathscr{M}_{\circlearrowright_\infty} \cdot  \mathscr{M}_{0\to R}^{-1} =
      \begin{pmatrix}
    1 & -2 \pi i & 0 \\
    0 & 1 & 2 \pi i  \\
    0 & 0 & 1  \\
  \end{pmatrix}
  = {\blue{\mathscr{M}_{\linebub{}_0}}}\cdot {\red{
 \mathscr{M}_{\linebub{}_1} }} \, . \label{eq:dilog_monodromy_infinty_plus}
\end{equation}
This monodromy around infinity can be written as the product of a monodromy around 0 and 1, since the path around infinity is homotopic to a path around 0 then around 1, as illustrated in the left part of Fig.~\ref{fig:infcont}. There, we see that the choice to take $\im R > 0$ was what determined that we encircled the branch point at 0 first, and then the branch point at 1. If we take $R\to \infty$ with $\im R < 0$ (so that the contour circles the branch point at 1 first), the monodromy matrix differs in the top-right entry
\begin{equation}
{\purple{  \sM^{\linebubDownR}_\infty}}=
     \mathscr{M}_{0\to R} \cdot
     \mathscr{M}_{\circlearrowleft_\infty} \cdot  \mathscr{M}_{0\to R}^{-1} =
      \begin{pmatrix}
    1 & 2 \pi i &  4\pi^2\\
    0 & 1 & -2 \pi i  \\
    0 & 0 & 1  \\
  \end{pmatrix}
 =
 {\red{
 \mathscr{M}_{\linebub{}_1} }}\cdot {\blue{\mathscr{M}_{\linebub{}_0}}}
\end{equation}
The result is the product of the 0 and 1 monodromies in the opposite order. This path around infinity is illustrated on the right in Fig.~\ref{fig:infcont}.

This ambiguity at $\mathcal{O}(\pi^2)$ in the monodromy matrix associated with infinity is also present for the other monodromy matrices. For example, we could have computed the monodromy around $1$ using a contour that first crosses the negative real axis before going around $1$, as illustrated on the right in Fig.~\ref{fig:infcont}. The result would have been
\begin{equation}
   {\red{\sM_{\linebub{}_{0,1}}}} = 
   {\blue{\mathscr{M}_{\linebub_0}}}^{-1} \cdot 
   {\red{ \mathscr{M}_{\linebub_1}}}\cdot {\blue{\mathscr{M}_{\linebub_0}}} =  
   \begin{pmatrix}
    1 & -2 \pi i &  4\pi^2\\
    0 & 1 & 0  \\
    0 & 0 & 1  \\
  \end{pmatrix}.
\end{equation}
The $\cO(\pi)$ terms in this monodromy matrix are the same as for $  {\red{\sM_{\linebub{}_{1}}}}$ in Eq.~\eqref{eq:dilog_homotopy_1b}, as expected from Cauchy's residue theorem, but the $\cO(\pi^2)$ terms are different.

To describe these $\cO(\pi^2)$ ambiguities more formally, consider a codimension-one branch variety defined by an equation \(f(\{s_j\}) = 0\), for some set of variables \(\{s_j\}\) which we can take to be Mandelstam invariants. To compute the monodromy around this branch variety, we find a closed path \(\gamma\) such that
\(\oint_\gamma d \ln f(\{s_j\}) = 2 \pi i\). However, as there are many paths \(\gamma\) that satisfy this requirement, there is some ambiguity in this choice.  In particular,
all the paths in the same homology class of \(\gamma\) satisfy the
same relation; however, the paths in this homology class may still be in different homotopy classes.  While the integral \(\oint_\gamma d \ln f(\{s\})\) depends only on the
homology class of \(\gamma\), the elements of the monodromy group depend on the homotopy class of
\(\gamma\).

The fundamental group and first homology group are related by
Hurewicz theorem, which states that the first homology group is the abelianization of the fundamental group. That is, given any two elements \(a\) and \(b\) of the fundamental group, we can quotient the fundamental group by the commutator subgroup generated by elements \(a b a^{-1} b^{-1}\) to obtain the homology group. The contour corresponding to the commutator \(a b a^{-1} b^{-1}\) is called a Pochhammer contour, and corresponds to a trivial element in homology. Thus, for every path \(\gamma\) which satisfies the condition \(\oint_\gamma d \ln f(\{s_j\}) = 2 \pi i\), we can find another path \(\gamma a b a^{-1} b^{-1}\) that also satisfies this relation. Moreover, as this new path belongs to a different homotopy class, it yields a different monodromy beyond $\cO(\pi)$.

Despite these ambiguities, any choice of closed contours around 0, 1, and infinity will furnish us with a representation of the fundamental group on the Riemann sphere with three marked points. For instance, we can choose the rational matrices appearing in Eqs.~\eqref{eq:dilog_homotopy_0b},~\eqref{eq:dilog_homotopy_1b}, and~\eqref{eq:dilog_monodromy_infinty_plus}, which satisfy a single multiplicative identity.
Note, however, that the contours used must all start at the same basepoint, so we cannot use the rational matrices corresponding to $\sM_{\circlearrowleft_0}$, $\sM_{\circlearrowleft_1}$, and $\sM_{\circlearrowleft_\infty}$.

For a multivariable function, like the function $\Phi_1(z,\bar{z})$ that appears in the one-loop triangle and box, we can carry out the same analysis for the contours in $z$ while holding $\bar{z}$ fixed. The contours around $z=0$ and $z=1$ were computed in Eqs.~\eqref{eq:box_monodromies_zero} and \eqref{eq:box_monodromies_1_z} to be
\begin{equation}
{\blue{\sM_{\linebub{}_0^z}}} = 
  \begin{pmatrix}
    1 & 2 \pi i & 0 & 0 \\
    0 & 1 & 0 & 0 \\
    0 & 0 & 1 & 2 \pi i \\
    0 & 0 & 0 & 1
  \end{pmatrix}, \quad
{\red {\sM_{\linebub{}_1^z} }}
=
 \begin{pmatrix}
    1 & 0 & -2 \pi i & 0 \\
    0 & 1 & 0 & 2 \pi i \\
    0 & 0 & 1 & 0 \\
    0 & 0 & 0 & 1
  \end{pmatrix} .
  \end{equation}
For the contours around infinity, a calculation analogous to the dilog case gives
  \begin{align}
   {\green{ \mathscr{M}_{\linebubUp z}^\infty}} &=
 \begin{pmatrix}
    1 & 2\pi i & -2 \pi i & -4 \pi^2\\
    0 & 1 & 0 & 2 \pi i \\
    0 & 0 & 1 & 2 \pi i \\
    0 & 0 & 0 & 1
  \end{pmatrix} 
 ={\blue{\mathscr{M}_{\linebub{}^z_0}}}
 \cdot
 {\red{
 \mathscr{M}_{\linebub{}^z_1} }},\\
 {\purple{  \sM^{\linebubDown z}_\infty}} &=
 \begin{pmatrix}
    1 & 2\pi i & -2 \pi i & 4\pi^2 \\
    0 & 1 & 0 & 2 \pi i \\
    0 & 0 & 1 & 2\pi i \\
    0 & 0 & 0 & 1
  \end{pmatrix} 
  = {\blue{\mathscr{M}_{\linebubR{}^z_0}}}\cdot{\red{ \mathscr{M}_{\linebubR{}^z_1} }}
 .
\end{align}
The monodromy matrices for contours in $\zb$ can be computed in a similar fashion, and commute with the monodromy matrices in $z$. Like for the case of the dilogarithm, each monodromy matrix gives rise to an associated rational matrix that corresponds to a generator of the fundamental group, which in this case describes the manifold corresponding to the space of complex $z$ and $\zb$ with the points 0, 1, and infinity in each variable removed.

More generally, the monodromy group describing the discontinuity of a set of polylogarithms also furnishes us with a representation of the fundamental group describing the manifold on which these polylogarithms are defined. When we consider polylogarithms that only depend on a single variable, the relevant manifold is the Riemann sphere with $n$ marked points and the fundamental group corresponds to the free group with \(n - 1\) generators. However, the fundamental group of higher-dimensional manifolds will in general be more complicated.

\section{Single-valued polylogarithms} \label{app:singlevalue}
Using the Knizhnik-Zamolodchikov equation, polylogarithms can be mapped to single-valued avatars of themselves~\cite{BROWN2004527}. In these new single-valued functions, all contributions generated by analytically continuing around branch points are systematically cancelled out by new functional dependence on variables conjugate to the variables of the original function. This type of single-valued map has proven useful in a variety of physics contexts, such as multi-Regge limits~\cite{Dixon:2012yy,DelDuca:2016lad}, the infrared structure of gauge theory~\cite{Almelid:2015jia,Almelid:2017qju}, string amplitudes~\cite{Schlotterer:2012ny}, and massless $\phi^4$ theory~\cite{Schnetz:2013hqa}. Motivated by~\cite{MR1265552,MR3469645} we show here that the same map can be constructed in terms of variation matrices.

We begin by considering a variation matrix $\mathscr{M}$ that depends on any number of variables, whose discontinuities are described by a set of monodromy matrices $\{\mathscr{M}_{\linebub,k}\}$ indexed by $k$. %
In order to construct a single-valued version of the matrix $\mathscr{M}$, we want to find a matrix that transforms in the opposite way as $\mathscr{M}$ when analytically continued around branch points. A natural object to consider is the inverse conjugate matrix $\smash{\overline{\mathscr{M}}^{\, -1}}$, namely the inverse matrix of $\mathscr{M}$ in which all variables $z_j$ have additionally been replaced by their complex conjugates $\overline{z}_j$. Under the action of the monodromy group, this pair of matrices transform as
\begin{align}
\mathscr{M} &\quad \to \quad \mathscr{M}_{\linebub,k} \cdot \mathscr{M} \, , \\
\overline{\mathscr{M}}^{\, -1} & \quad \to \quad \overline{\mathscr{M}}^{\, -1} \cdot \overline{\mathscr{M}}^{\, -1}_{\linebub,k} \, .
\end{align}
Thus, the product of these two matrices is not quite invariant under arbitrary analytic continuations, because $\smash{\overline{\mathscr{M}}^{\, -1}_{\linebub,k}} \cdot  \mathscr{M}_{\linebub,k} \neq \bbone$.

This mismatch can be fixed by decomposing the monodromy matrices as discussed in section~\ref{sec:fundamental_group}. In particular, we have
\
\begin{align}
    \mathscr{M}_{\linebub,k} = D^{-1} \cdot M_k \cdot D \, ,
\end{align}
where $D$ is a diagonal matrix whose entries are integer powers of $2 \pi i$, and $M_k$ is an element of the general linear group with rational entries. Since the action of the monodromy matrices preserves transcendental weight, the matrix $D$ (which encodes the relative weight the rows of $\mathscr{M}$) does not depend on $k$. Having made this observation, we define the single-valued matrix
\begin{align} \label{eq:single_valued}
\mathscr{M}_{\text{sv}} \equivD \overline{\mathscr{M}}^{\, -1} \cdot \overline{D}^{\, -1} \cdot D \cdot \mathscr{M} \, .
\end{align}
This matrix invariant under the action of the monodromy group, since
\begin{align}
\mathscr{M}_{\text{sv}} \quad \to \quad  \Big( \overline{\mathscr{M}}^{\, -1} \cdot \overline{D}^{\, -1} \cdot M_k^{-1} \cdot \overline{D} \Big) \cdot \overline{D}^{\, -1} \cdot D \cdot \Big( D^{-1} \cdot M_k \cdot D \cdot \mathscr{M} \Big) =  \mathscr{M}_{\text{sv}}
\end{align}
whenever $\overline{z}_j = z_j^*$.  We note that the definition~\eqref{eq:single_valued} is equivalent to the map defined in Eq.~(3.82) of~\cite{DelDuca:2016lad} using the coproduct formalism.

Let us see how this works in the case of the dilogarithm. Referring to its variation matrix $\mathscr{M}$ in Eq.~\eqref{eq:appendix_dilog}, we see that $\overline{D}^{\, -1} \cdot D = \operatorname{diag}(1,-1,1)$ and
\begin{equation}
  \overline{\mathscr{M}}^{\, -1} =
  \begin{pmatrix}
    1 & - \operatorname{Li}_1(\overline{z}) & - \operatorname{Li}_2(\overline{z})+ \operatorname{Li}_1(\overline{z}) \ln \overline{z} \\
    0 & 1 & - \ln \overline{z} \\
    0 & 0 & 1
  \end{pmatrix}.
\end{equation}
The single-valued matrix is thus given by
\begin{align}
  \mathscr{M}_{\text{sv}} =
  \begin{pmatrix}
    1 & \operatorname{Li}_1(z) + \operatorname{Li}_1(\bar{z}) & \operatorname{Li}_2(z) - \operatorname{Li}_2(\bar{z}) + \ln(z \bar{z}) \operatorname{Li}_1(\bar{z}) \\
    0 & -1 & -\ln(z \bar{z}) \\
    0 & 0 & 1
  \end{pmatrix} \, .
\end{align}
It is not hard to check that all effects of analytically continuing $z$ and $\bar{z}$ in opposite directions around any branch point cancel out in the entries of this matrix, as expected.

\section{Permutation symmetry of the triangle integral}
\label{sec:symmetry}

The one-loop triangle integral considered in Section~\ref{sec:one_loop_triangle}, given by 
\begin{equation}
    \frac{1}{16 \pi^2 p_{1}^{2}} \frac{1}{z-\zb}  \Phi_1(z,\zb)\, 
\end{equation}
where $\Phi_1(z,\zb)$ was defined in Eq.~\eqref{cP2}, respects an \(S_3\) symmetry under the permutation of its external legs.  This symmetry turns out to be realized in an interesting way, by the collusion of this integral's rational and transcendental parts.

We first discuss the rational prefactor. To determine how the quantity $p_1^2 (z-\zb)$ transforms under the permutation of external momenta, we consider the wedge product \(p_1 \wedge p_2\).  We work in the coordinate system described above Eq.~\eqref{eq:energy_framing}, where \(p_1 = (E_1, 1)\) and \(p_2 = (E_2, p_2^x)\). In terms of a pair of basis vectors $e_t$ and $e_x$, these momenta become \(p_1 = E_1 e_t + e_x\) and \(p_2 = E_2 e_t + p_2^x e_x\), and we have
\begin{equation}
    \label{eq:zzbar_wedge}
    p_1 \wedge p_2 = -\frac 1 2 p_1^2 (z - \bar{z}) e_t \wedge e_x \, .
\end{equation}
We can correspondingly use this quantity to study the transformation properties of \(p_1^2 (z - \bar{z})\). Clearly, under \(p_1 \leftrightarrow p_2\) the left-hand side of Eq.~\eqref{eq:zzbar_wedge} changes sign.  Similarly, under \(p_1 \leftrightarrow p_3\) we have \(p_1 \wedge p_2 \leftrightarrow p_3 \wedge p_2 = -p_1 \wedge p_2\).  We conclude that the representation of the symmetric group \(S_3\) when acting on \(p_1^2 (z - \bar{z})\) is the sign representation.

Before moving on to discuss the symmetries of $\Phi_1(z,\zb)$, we need to find the action of the \(S_3\) symmetry on \(z\) and \(\zb\).  From the above, we know that
\begin{equation}
    \sigma (p_1^2 (z - \bar{z})) =
    p_{\sigma(1)}^2 (\sigma(z) - \sigma(\bar{z})) =
    (-1)^{\lvert\sigma\rvert} p_1^2 (z - \bar{z}).
\end{equation}
We also know, from Eq.~\eqref{uvdef}, that under the \(p_2 \leftrightarrow p_3\) permutation we have \(z \bar{z} \leftrightarrow (1 - z)(1 - \bar{z})\). These constraints can be solved with the unique solution that \(p_2 \leftrightarrow p_3\) corresponds to \(z \leftrightarrow 1 - z\) and \(\bar{z} \leftrightarrow 1 - \bar{z}\). Similarly, one can show that \(p_1 \leftrightarrow p_2\) 
must correspond to \(z \leftrightarrow \frac 1 z\) and \(\bar{z} \leftrightarrow \frac 1 {\bar{z}}\).  The action of the remaining permutations can be determined from these two transformations.

We are now ready to study the symmetry of the transcendental part of the triangle function.  It turns out that this is related to the Bloch-Wigner function
\begin{equation}
    D(z) = \Im \operatorname{Li}_2(z) + \operatorname{arg}(1 - z) \ln(\lvert z\rvert).
\end{equation}
In particular, using
\begin{gather}
    \Im \operatorname{Li}_2(z) = \frac 1 {2 i} (\operatorname{Li}_2(z) - \operatorname{Li}_2(z^*)), \\
    \operatorname{arg}(1 - z) = \frac 1 {2 i} \ln \frac {1 - z}{1 - z^*}, \\
    \ln \lvert z\rvert = \frac 1 2 \ln(z z^*),
\end{gather}
we have
\begin{equation}
    4 i D(z) = 2 (\operatorname{Li}_2(z) - \operatorname{Li}_2(z^*)) + \ln\left(\frac{1 - z}{1 - z^*} \right) \ln(z z^*).
\end{equation}
In the region \(R_{\text{I}}^*\), where \(\bar{z} = z^*\), this gives precisely the transcendental part of the one-loop triangle, $\Phi_1(z,\zb)$.

The Bloch-Wigner function satisfies
\begin{equation}
    D(z) = -D(1 - z) = -D \left(\frac 1 z\right).
\end{equation}
These signs precisely compensate the signs arising from the action of the permutation group on the rational prefactor.  In the other regions, where \(\bar{z} \neq z^*\), the transcendental part should be thought as a function of two independent variables.  Still, the same relations hold under the transformation of both \(z\) and \(\bar{z}\).

How is the symmetry realized on the cuts?  It is instructive to consider the example of a leading singularity, where the only dependence on the kinematics is in the rational prefactor, while the transcendental part is a power of \(2 \pi i\).  By the argument above, under the action of the permutation of external legs, the rational prefactor may pick up a sign.  Hence, the residue on a given leading singularity is \emph{not} invariant under the permutation group.  However, each leading singularity locus is paired with another one with opposite residue, as required by global residue theorems.  It follows that the set of values the residue takes on all the leading singularities is invariant under the action of the permutation group.  A similar statement holds for the rest of the cuts.

\section{Variation matrix of the two-loop box} \label{app:Phi2}
In this appendix we present the connection and variation matrix for the two-loop ladder triangle/box function
\begin{equation}
     \Phi_2(z, \zb)=6 [\Li_4(z)-\Li_4(\zb)]
    -3 \ln (z \zb)    [\Li_3(z) - \Li_3(\zb)]
    +\frac{1}{2} \ln^{2}(z \zb)    [\Li_2(z) - \Li_2(\zb)] \, .
\end{equation}
The two-loop connection is
\begin{equation}
    \omega = \left(\begin{array}{c|cc|ccc|cc|c}
        0 & -\omega_1 - \overline{\omega}_1 & \omega_0 + \overline{\omega}_0 & 0 & 0 & 0 & 0 & 0 & 0 \\
        \hline
        0 & 0 & 0 & -\overline{\omega}_0 & \omega_0 & 0 & 0 & 0 & 0 \\
        0 & 0 & 0 & \omega_1 & -\overline{\omega}_1 & \omega_0 + \overline{\omega}_0 & 0 & 0 & 0 \\
        \hline
        0 & 0 & 0 & 0 & 0 & 0 & \omega_0 - \overline{\omega}_0 & -\omega_0 & 0 \\
        0 & 0 & 0 & 0 & 0 & 0 & -\overline{\omega}_0 & -\omega_0 + \overline{\omega}_0 & 0 \\
        0 & 0 & 0 & 0 & 0 & 0 & -\omega_1 & \overline{\omega}_1 & 0 \\
        \hline
        0 & 0 & 0 & 0 & 0 & 0 & 0 & 0 & \omega_0 \\
        0 & 0 & 0 & 0 & 0 & 0 & 0 & 0 & \overline{\omega}_0 \\
        \hline
        0 & 0 & 0 & 0 & 0 & 0 & 0 & 0 & 0
    \end{array}\right), \label{eq:two_loop_connection}
\end{equation}
where
\begin{gather}
    \omega_0 = \frac {d z} z, \qquad
    \omega_1 = \frac {d z}{z - 1}, \\
    \bar{\omega}_0 = \frac {d \bar{z}}{\bar{z}}, \qquad
    \bar{\omega}_1 = \frac {d \bar{z}}{\bar{z} - 1}.
\end{gather}
The connection trivially satisfies \(d \omega = 0\), and using the fact that \(\omega_0 \wedge \omega_1 = 0\), we also have that \(\omega \wedge \omega = 0\). Thus, the connection has zero curvature ($d\omega - \omega \wedge \omega = 0$).

Using this connection, we can compute the variation matrix \(\mathscr{M}_{\gamma_0}\). We encounter integrals of one-forms, which are familiar, but also iterated integrals of higher weight. As an example, consider
\eq{
    \mathscr{M}_{1,6} =
    \int^{z,\zb} \left(\omega_0 + \overline{\omega}_0 \right) \circ \left(\omega_0 + \overline{\omega}_0 \right) \, . \label{eq:iterated_integral_M16}
}
While this integral can be computed using Eq.~\eqref{eq:pathParametrization} along a concretely chosen contour, it is easier to use the fact that
any pair of one-forms \(\sigma_1\) and \(\sigma_2\) satisfies
\begin{equation}
  \int_\gamma (\sigma_1 \circ \sigma_2 + \sigma_2 \circ \sigma_1) = \left(\int_\gamma \sigma_1 \right) \left(\int_\gamma \sigma_2 \right)\,.
  \label{eq:oneforms}
\end{equation}
This allows us to rewrite Eq.~\eqref{eq:iterated_integral_M16} as
\begin{equation}
    \mathscr{M}_{1,6} =
    \frac{1}{2} \int^{z} \omega_0 \int^{z} \omega_0 +
    \int^z \omega_0 \int^{\zb} \overline{\omega}_0
    +
    \frac{1}{2} \int^{\zb} \overline{\omega}_0 \int^{\zb} \overline{\omega}_0 \, .
\end{equation}
These integrals are much simpler to evaluate, and we get
\eq{
    \mathscr{M}_{1,6} =
    \frac{1}{2} \ln^2 z + \ln z \ln \zb + \frac{1}{2} \ln^2 \zb  \, .
}
The relation in Eq.~\eqref{eq:oneforms} can be iterated, to give us
\begin{equation}
  \int_\gamma \sum_{\substack{\{j_1,\cdots,j_n\} \\ \in \text{perms of }\{1,\cdots n\}}} \sigma_{j_1} \circ \sigma_{j_2} \circ \cdots \sigma_{j_n} = \int_\gamma \sigma_1 \int_\gamma \sigma_2 \cdots \int_\gamma \sigma_n \,
\end{equation}
for $n$ one-forms $\sigma_1, \cdots, \sigma_n$, along with relations such as
\begin{equation}
  \int_\gamma (\sigma_1 \circ \sigma_2 \circ \sigma_3 + \sigma_2 \circ \sigma_1 \circ \sigma_3 + \sigma_2 \circ \sigma_3 \circ \sigma_1) = \int_\gamma \sigma_1 \int_\gamma \sigma_2 \circ \sigma_3 \,.
\end{equation}
Using these kinds of formulas, we can reduce the expressions in the calculation of the variation matrix to familiar integrals, such as
\begin{align}
    \Li_n \left(z\right) = - \int_0^z \omega_1 \circ \underbrace{\omega_0 \circ \omega_0 \cdots \circ \omega_0}_{n-1} \ \, , 
\end{align}
along with integrals that can easily be performed, such as
\begin{align}
    \int_0^z \omega_0 \circ \omega_1 \circ \omega_0 & = 2 \Li_3 \left(z\right) - \ln z \Li_2 \left(z\right) \, .
\end{align}
The iterated integrals we study have the special property that they are independent of small deformations of the integration contour which preserve its endpoints.  This is a consequence of the flatness of the connection \(\omega\) and is sometimes called integrability condition.  In our example, the integrability condition reads
\begin{equation}
  (\omega_0 + \overline{\omega}_0) \wedge (\omega_0 + \overline{\omega}_0) = 0.
\end{equation}
This condition is trivial when both forms only depend on a single variable, but imposes non-trivial restrictions when two or more variables are involved.

The result of performing the integrations is
\begin{equation}
  \mathscr{M}_{\gamma_0}(z, \bar{z}) = \left(\begin{array}{c|cc|ccc|cc|c}
    1 & \mathscr{M}_{1, 2} & \mathscr{M}_{1, 3} & \mathscr{M}_{1, 4} & \mathscr{M}_{1, 5} & \mathscr{M}_{1, 6} & \mathscr{M}_{1, 7} & \mathscr{M}_{1, 8} & \mathscr{M}_{1, 9} \\
    \hline
    0 & 1 & 0 & \mathscr{M}_{2, 4} & \mathscr{M}_{2, 5} & 0 & \mathscr{M}_{2, 7} & \mathscr{M}_{2, 8} & \mathscr{M}_{2, 9} \\
    0 & 0 & 1 & \mathscr{M}_{3, 4} & \mathscr{M}_{3, 5} & \mathscr{M}_{3, 6} & \mathscr{M}_{3, 7} & \mathscr{M}_{3, 8} & \mathscr{M}_{3, 9} \\
    \hline
    0 & 0 & 0 & 1 & 0 & 0 & \mathscr{M}_{4, 7} & \mathscr{M}_{4, 8} & \mathscr{M}_{4, 9} \\
    0 & 0 & 0 & 0 & 1 & 0 & \mathscr{M}_{5, 7} & \mathscr{M}_{5, 8} & \mathscr{M}_{5, 9} \\
    0 & 0 & 0 & 0 & 0 & 1 & \mathscr{M}_{6, 7} & \mathscr{M}_{6, 8} & \mathscr{M}_{6, 9} \\
    \hline
    0 & 0 & 0 & 0 & 0 & 0 & 1 & 0 & \mathscr{M}_{7, 9} \\
    0 & 0 & 0 & 0 & 0 & 0 & 0 & 1 & \mathscr{M}_{8, 9} \\
    \hline
    0 & 0 & 0 & 0 & 0 & 0 & 0 & 0 & 1
  \end{array}\right),
\end{equation}
where
\begingroup
\allowdisplaybreaks
\begin{align}
 \sM_{1, 2} &= -\ln (1-\zb)-\ln (1-z), \\
\sM_{1, 3} &=  \ln z + \ln \zb, \\
\sM_{1, 4} &= -\Li_2(\zb)+\ln (1-z)  \left( \ln z + \ln \zb \right)+\Li_2(z), \\
\sM_{1, 5} &= -\Li_2(\zb)-\ln (1-\zb) 
  \left( \ln z + \ln \zb \right) + \Li_2(z), \\
\sM_{1, 6} &= \frac{1}{2} \ln^2 z + \ln z \ln \zb + \frac{1}{2} \ln^2 \zb, \\
\sM_{1, 7} &= \frac{1}{2} \Big[
  6 \Li_3(\zb)
  -4 \Li_2(z) 
   \left( \ln z + \ln \zb \right)
  -2 \Li_2(\zb) \left( \ln z + \ln \zb \right)
  -\ln (1-z) \ln^2\!\zb \nonumber \\
&\hspace{2cm}  -2 \ln (1-z) \ln z \ln\zb+6
  \Li_3(z)-\ln (1-z) \ln^2\!z\Big], \\
\sM_{1, 8} &= \frac{1}{2} \Big[-6 \Li_3(z)
+4 \Li_2(\zb) \left( \ln z + \ln \zb \right)
+2 \Li_2(z) \left( \ln z + \ln \zb \right)
  +\ln(1-\zb)\ln^2\!z \nonumber\\
&\hspace{2cm}  +2 \ln (1-\zb) \ln z \ln\zb-6 \Li_3(\zb) +\ln (1-\zb) \ln^2\!\zb  \Big], \\
\sM_{1, 9} &= \Phi_2, \\
\sM_{2, 4} &= -\ln\zb, 
  \qquad\qquad~~
\sM_{2, 5} = \ln z, \\
\sM_{2, 7} &= \frac{1}{2} \ln^2\zb - \ln \zb \ln z ,
\qquad~
\sM_{2, 8} = -\frac{1}{2} \ln^2 z + \ln z \ln \zb,
\qquad \\
\sM_{2, 9} & =  \frac{1}{2} \ln z \ln^2 \zb - \frac{1}{2} \ln^2 z \ln \zb,\\
\sM_{3, 4} &= \ln(1 - z), 
 \qquad~~~
\sM_{3, 5} = -\ln (1-\zb), 
  \qquad~~
\sM_{3, 6} = \ln z + \ln \zb, \\
\sM_{3, 7} &= -\Li_2(\zb)-\ln (1-z)  \left( \ln z + \ln \zb \right)-2 \Li_2(z), \\
\sM_{3, 8} &= 2 \Li_2(\zb)+\ln (1-\zb) \left( \ln z + \ln \zb \right) + \Li_2(z), \\
\sM_{3, 9} &= 3 \Li_3(\zb)+\Li_2(z)  \left( \ln z + \ln \zb \right)
  -\Li_2(\zb)  \left( \ln z + \ln \zb \right)-3 \Li_3(z), \\
\sM_{4, 7} &= \ln z - \ln \zb, \qquad~~
\sM_{4, 8} = -\ln z, 
  \qquad\qquad
\sM_{4, 9} = \frac{1}{2} \ln^2 z - \ln z \ln \zb,\\
\sM_{5, 7} &= -\ln\zb, 
\qquad \qquad~
\sM_{5, 8} =  \ln \zb - \ln z ,
  \qquad~
\sM_{5, 9} = \frac{1}{2} \ln^2 \zb - \ln \zb \ln z ,\\
\sM_{6, 7} &= -\ln(1 - z), 
  \qquad
\sM_{6, 8} = \ln (1-\zb), 
  \qquad~~
\sM_{6, 9} = \Li_2(z)-\Li_2(\zb), \\
\sM_{7, 9} &= \ln z, \qquad\qquad\quad~
\sM_{8, 9} = \ln\zb.
\end{align}
\endgroup
The monodromy around \(z = 0\) is
\begin{equation}
    \sM_{\linebub{}_0^{z}} = \left(\begin{array}{c|cc|ccc|cc|c}
        1 & 0 & 2 \pi i & 0 & 0 & \frac 1 2 (2 \pi i)^2 & 0 & 0 & 0\\
        \hline
        0 & 1 & 0 & 0 & 2 \pi i & 0 & 0 & -\frac 1 2 (2 \pi i)^2 & 0 \\
        0 & 0 & 1 & 0 & 0 & 2 \pi i & 0 & 0 & 0 \\
        \hline
        0 & 0 & 0 & 1 & 0 & 0 & 2 \pi i & -2 \pi i & \frac 1 2 (2 \pi i)^2 \\
        0 & 0 & 0 & 0 & 1 & 0 & 0 & -2 \pi i & 0 \\
        0 & 0 & 0 & 0 & 0 & 1 & 0 & 0 & 0 \\
        \hline
        0 & 0 & 0 & 0 & 0 & 0 & 1 & 0 & 2 \pi i \\
        0 & 0 & 0 & 0 & 0 & 0 & 0 & 1 & 0 \\
        \hline
        0 & 0 & 0 & 0 & 0 & 0 & 0 & 0 & 1
    \end{array}\right) .
\end{equation}
We note that \((\bbone - \sM_{\linebub{}_0^{z}})^3 = 0\). This is consistent with \textit{three} (but not two) sequential cuts in the $p_2^2$ channel of the 2-loop triangle vanishing. 
The monodromy around \(z = 1\) is
\begin{equation}
    \sM_{\linebub{}_1^{z}} = \left(\begin{array}{c|cc|ccc|cc|c}
        1 & -2 \pi i & 0 & 0 & 0 & 0 & 0 & 0 & 0 \\
        \hline
        0 & 1 & 0 & 0 & 0 & 0 & 0 & 0 & 0 \\
        0 & 0 & 1 & 2 \pi i & 0 & 0 & 0 & 0 & 0 \\
        \hline
        0 & 0 & 0 & 1 & 0 & 0 & 0 & 0 & 0 \\
        0 & 0 & 0 & 0 & 1 & 0 & 0 & 0 & 0 \\
        0 & 0 & 0 & 0 & 0 & 1 & -2 \pi i & 0 & 0 \\
        \hline
        0 & 0 & 0 & 0 & 0 & 0 & 1 & 0 & 0 \\
        0 & 0 & 0 & 0 & 0 & 0 & 0 & 1 & 0 \\
        \hline
        0 & 0 & 0 & 0 & 0 & 0 & 0 & 0 & 1
    \end{array}\right) .
\end{equation}
In this case we have \((1 - \sM_{\linebub{}_1^{z}})^2 = 0\).
This is consistent with {\it two} sequential cuts in the $p_3^2$ channel (the long direction) of the 2-loop triangle  vanishing.
Finally, the clockwise monodromy around infinity (where we approach infinity above the real line) is
\begin{align}
    \mathscr{M}_{\linebubUp z}^\infty &=
    \mathscr{M}_{\linebub{}^z_0} \cdot \mathscr{M}_{\linebub{}^z_1} \\
    &=
    \left(\begin{array}{c|cc|ccc|cc|c}
        1 & -2 \pi i & 2 \pi i & (2 \pi i)^2  & 0 & \frac{1}{2} (2 \pi i)^2 & -\frac{1}{2} (2 \pi i)^3 & 0 & 0 \\
        \hline
        0 & 1 & 0 & 0 & 2 \pi i & 0 & 0 & - \frac{1}{2} (2 \pi i)^2 & 0 \\
        0 & 0 & 1 & 2 \pi i & 0 & 2 \pi i & - (2 \pi i)^2 & 0 & 0 \\
        \hline
        0 & 0 & 0 & 1 & 0 & 0 & 2 \pi i & -2 \pi i & \frac{1}{2} (2 \pi i)^2 \\
        0 & 0 & 0 & 0 & 1 & 0 & 0 & -2 \pi i & 0 \\
        0 & 0 & 0 & 0 & 0 & 1 & -2 \pi i & 0 & 0 \\
        \hline
        0 & 0 & 0 & 0 & 0 & 0 & 1 & 0 & 2 \pi i \\
        0 & 0 & 0 & 0 & 0 & 0 & 0 & 1 & 0 \\
        \hline
        0 & 0 & 0 & 0 & 0 & 0 & 0 & 0 & 1
    \end{array}\right) \nonumber
\end{align}
and again we get  \((1 - \sM_{\linebubUp z})^3 = 0\).

To compute the monodromy matrices associated with contours in $\zb$, we can use the fact that $z$ and $\zb$ can be exchanged in the connection from Eq.~\eqref{eq:two_loop_connection} via  conjugation by the matrix 
\begin{equation}
  C = \left(\begin{array}{c|cc|ccc|cc|c}
    1 & & & & & & & & \\
    \hline
    & 1 & 0 & & & & & & \\
    & 0 & 1 & & & & & & \\
    \hline
    & & & 0 & -1 & 0 & & & \\
    & & & -1 & 0 & 0 & & & \\
    & & & 0 & 0 & 1 & & & \\
    \hline
    & & & & & & 0 & -1 & \\
    & & & & & & -1 & 0 & \\
    \hline
    & & & & & & & & -1
  \end{array}\right).
\end{equation}
That is, we have  \(\omega(z, \zb) \to \omega(\zb, z) =  C \omega C^{-1}\). Thus, we also have that
\begin{gather}
  \mathscr{M}_{\linebub{}^{\bar{z}}_0} = C \mathscr{M}_{\linebub{}^z_0} C^{-1}, \qquad
  \mathscr{M}_{\linebub{}^{\bar{z}}_1} = C \mathscr{M}_{\linebub{}^z_1} C^{-1}.
\end{gather}
These are the last monodromy matrices that are needed to construct the discontinuity operators in Eq.~\eqref{sdisc}.

\section{Cuts of the three-loop triangle}
\label{sec:A_3looptri}
In this Appendix, we work out the details of the calculations in Section~\ref{sec:3loopT2cuts}. We start by computing the sum of the two cuts involving $\mc{C}_1$ and write
\begin{multline}
    \left. T_3^{\mc{C}_1,\mc{C}_2} \right\vert_{-i \varepsilon\text{ on r.h.s.}} + T_3^{\mc{C}_1,\mc{C}_3}
    =
    \frac{1}{2}
    \int\!\!\! \frac{d^4 k_1}{i\left(2\pi\right)^4}\left(-2\pi i\right) \delta \left(k_1^2-m^2\right) \Theta (k_1^0)  \left(-2\pi i\right) \delta \left[\left(p_2-k_1\right)^2\right] \Theta (p_2^0-k_1^0)
    \\
    \times
    \frac{1}{\left(p_3+k_1\right)^2}
    \sum \text{Cut}_{k_1^2} T^{2} \left[\left(p_3+k_1\right)^2,k_1^2,p_3^2\right] \, ,
    \label{eq:CutLp1TLp1}
\end{multline}
where $\smash{\sum \text{Cut}_{k_1^2} T_{2} \left[\left(p_3+k_1\right)^2,k_1^2,p_3^2\right]}$ is the sum of cuts in $k_1^2$ through the two-loop triangle $T_2$ with masses $\left(p_3+k_1\right)^2$, $k_1^2$ and $p_3^2$. We take the particle with momentum $k_1$ to have a small mass $m$ to regulate the IR divergences that arise in the cut calculations, and work to leading power in $m^2$. The factor of $\frac{1}{2}$ arises because the mass regulator does not capture the $\frac{1}{L!}$ arising from a product of $L-1$ massless vertices, as worked out in Appendix~\ref{sec:massless3pt}. The sum of the cuts through the two-loop triangle is given by~\cite{Abreu:2014cla},
\begin{multline}
    \sum \text{Cut}_{k_1^2} T^{2} \left[\left(p_3+k_1\right)^2,m^2,p_3^2\right] \\
    =
    \frac{2 \pi i}{256 \pi^4} \frac{1}{ \left(p_3+k_1\right)^4} \frac{1}{\left(1-x\right) \left(1-\bar{x}\right) \left(x-\bar{x}\right)}
    \Big\{
    3 \left[ \text{Li}_3 \left(x\right) - \text{Li}_3 \left(\bar{x}\right) \right] - \ln \left(-x \bar{x} \right)\left[ \text{Li}_2\left(x\right) - \text{Li}_2 \left(\bar{x} \right) \right]
    \Big\} \, ,
\end{multline}
with
\begin{align}
    x \bar{x} & = \frac{m^2}{\left(p_3+k_1\right)^2}\, ,\\
    \left(1-x\right) \left(1-\bar{x}\right) & = 
    \frac{p_3^2}{\left(p_3+k_1\right)^2}\, .
\end{align}
Working to leading power in $k_1^2=m^2$, we can take either $x$ or $\bar{x}$ to be small. The final answer is independent of which one is picked, so we assume that $\bar{x}$ is small, and hence $\bar{x} = \frac{m^2 \left(1-x\right)}{p_3^2 x}$.

Using the delta functions, and performing the integral over the azimuthal angle, the phase space can be written as
\eq{
    \int\!\!\! \frac{d^4 k_1}{i\left(2\pi\right)^4}\left(-2\pi i\right) \delta \left(k_1^2-m^2\right) \Theta (k_1^0)  \left(-2\pi i\right) \delta \left[\left(p_2-k_1\right)^2\right] \Theta (p_2^0-k_1^0)
     =
    \frac{i}{16 \pi} \int_{-1}^1 d \cos \theta \, .
}
In the rest frame of $p_2$, the propagator in $p_3+k_1$ becomes
\eq{
    \left(p_3+k_1\right)^2
    =
    p_3^2 - m_2 \left(\omega_3- p \cos \theta\right) \, ,
}
where $p$ is the magnitude of the three-momentum of the outgoing particles, and where we have dropped power corrections in $m^2$ and hence used that $\omega_{k_1} = \vert \vec{k}_1 \vert = m_2/2$. Changing variables from $\cos \theta$ to $x = 1-\frac{p_3^2}{\left(p_3+k_1\right)^2}$ gives a Jacobian of
\eq{
    d \cos \theta = -\frac{\left(p_3+k_1\right)^4}{m_2 p p_3^2} d x \, .
}
In this frame, the energy of $p_3$ is $\smash{\omega_3=\frac{m_2^2+p_3^2-p_1^2}{2 m_2}}$ and momentum of the outgoing particles is $p = - \frac{p_1^2}{2 m_2} \left(z-\zb\right)$,
which gives
\eq{
    \left(p_3+k_1\right)^2
    \cong
    \frac{1}{2} p_1^2 \left[\left(1-z\right)\left(1-\cos\theta\right) + \left(1-\zb\right) \left(1+\cos\theta\right) \right] \, ,
}
and hence we have
\eq{
    \frac{1}{1-x} = \frac{1}{2} \frac{1+\cos \theta}{1-z} + \frac{1}{2} \frac{1-\cos\theta}{1-\zb} 
}
to leading power in $m^2$. This shows that $x=\zb$ for $\cos \theta=-1$ and $x=z$ for $\cos \theta=1$. 
Putting everything together, phase space along with the propagator in $p_3+k_1$ can be written as
\begin{multline}
    \int \frac{d^4 k_1}{i \left(2\pi\right)^4}\left(-2\pi i\right) \delta \left(k_1^2\right) \Theta (k_1^0)  \left(- 2\pi i\right) \delta \left[\left(p_2-k_1\right)^2\right] \Theta (p_2^0-k_1^0)
    \frac{1}{\left(p_3+k_1\right)^2}
    \\
    = - \frac{i}{16 \pi}  \int_{\zb}^z d x
    \frac{1}{m_2 p \left(1-x\right)} \, .
\end{multline}
Then, dropping polylogarithms in $\bar{x}$ that are subleading in the limit $m^2\to 0$,
\begin{equation}
    \left. T_3^{\mc{C}_1,\mc{C}_2} \right\vert_{-i \varepsilon\text{ on r.h.s. of cut}} + T_3^{\mc{C}_1,\mc{C}_3}
    =
    \frac{1}{16 \left(4\pi\right)^4} \frac{1}{m_2 p \, p_3^4}
    \int_{\zb}^z \frac{d x}{x}
    \Big\{
    3 \text{Li}_3 \left(x\right) - \ln \left(-x \bar{x} \right) \text{Li}_2\left(x\right) 
    \Big\} \, .
\end{equation}
The integration contour from $\zb$ to $z$ in the region $R^2$ can be taken to be a straight line from $\zb<0$ to $z$, with $0<z<1$.
Integrating this expression and using $p=-\frac{p_1^2}{2 m_2} \left(z-\bar{z}\right)$ gives
\begin{multline}
    \left. T_3^{\mc{C}_1,\mc{C}_2} \right\vert_{-i \varepsilon\text{ on r.h.s. of cut}} + T_3^{\mc{C}_1,\mc{C}_3}
    =
    \frac{1}{2048 \pi^4} \frac{1}{p_1^2 p_3^4 \left(z-\zb\right)}
    \Big\{-3 \left[\text{Li}_4\left(z\right) - \text{Li}_4\left(\zb\right) \right] \\
    \hspace{2cm}
    +
    \ln\left(-\frac{m^2}{p_3^2}\right)
    \left[\text{Li}_3\left(z\right) - \text{Li}_3\left(\zb\right) \right]
    -\frac{1}{2} \left[\text{Li}_2^2\left(z\right) - \text{Li}_2^2\left(\zb\right) \right]
    \Big\} \, .
\end{multline}
Next, we calculate the double cut $\mc{C}_2 \mc{C}_3$, with all other propagators having a $+i\varepsilon$. We write the cut as
\begin{multline}
    T_3^{\mc{C}_2,\mc{C}_3} = \frac{1}{2}
    \int \frac{d^4 k_2}{i\left(2\pi\right)^4}\left(-2\pi i\right) \delta \left(k_2^2-m^2\right) \Theta (k_2^0) 
    \text{Cut}_{\left(p_2-k_2\right)^2} B \left(p_2^2,k^2,\left(p_3+k_2\right)^2,p_1^2\right)
    \\ \times
    \frac{1}{\left(p_3+k_2\right)^2}
    \text{Cut}_{k_2^2} T^{2}\left[\left(p_3+k_2\right)^2,k_2^2,p_3^2\right] \, , 
\end{multline}
where $\text{Cut}_{\left(p_2-k_2\right)^2} B$ is an $s-$channel cut through a box with one massive internal line,%
\begin{align}
\text{Cut}_s \text{B} \left(p_2^2,k_2^2,\left(p_3+k_2\right)^2,p_1^2\right) & =
        \begin{gathered}
        \begin{tikzpicture}[baseline=-3.5,scale=2.0]
        \draw[line width=1] (0,0) -- (1,0);
        \draw[line width=1] (1,0) -- (1,1);
        \draw[line width=1] (1,1) -- (0,1);
        \draw[line width=1] (0,1) -- (0,0);
        \draw[line width=1] (0,0) -- (-1,0);
        \draw[line width=1] (0,1) -- (-1,1);
        \draw[line width=1] (1,0) -- (2,0);
        \draw[line width=1] (1,1) -- (2,1);
        \node[text width=0.5cm,color=darkgreen] at (0.7,0.5) {$\mc{C}_s$};
        \draw[dashed,line width=1,color=darkgreen] (0.5,-0.1) -- (0.5,1.1);
        \draw[->,line width=0.2mm] (-0.1,0.35) -- (-0.1,0.65);
        \draw[->,line width=0.2mm] (0.35,1.1) -- (0.65,1.1);
        \draw[->,line width=0.2mm] (1.1,0.65) -- (1.1,0.35);
        \draw[->,line width=0.2mm] (0.65,-0.1) -- (0.35,-0.1);
        \draw[->,line width=0.2mm] (-0.65,-0.1) -- (-0.35,-0.1);
        \draw[->,line width=0.2mm] (-0.35,1.1) -- (-0.65,1.1);
        \draw[->,line width=0.2mm] (1.35,1.1) -- (1.65,1.1);
        \draw[->,line width=0.2mm]  (1.65,-0.1) -- (1.35,-0.1);
        \node[text width=1cm] at (-1.2,0) {$p_2$};
        \node[text width=1cm] at (-1.2,1) {$k_2$};
        \node[text width=2cm] at (2.62,1) {$-p_3-k_2$};
        \node[text width=1cm] at (2.4,0) {$p_1$};
        \node[text width=1cm] at (-0.1,0.5) {$k_1$};
        \node[text width=2cm] at (0.65,1.3) {$k_1-k_2$};
        \node[text width=2cm] at (1.8,0.5) {$k_1+p_3$};
        \node[text width=2cm] at (0.65,-0.3) {$k_1-p_2$};
        \end{tikzpicture}
        \end{gathered}
        \\ &
        =
        \frac{1}{16 \pi} \frac{\log \left[- \frac{m^2 p_1^2 \left(p_3^2-2\omega_{k_2} \left(\omega_3-p\cos\theta\right) \right)}{2 m_2 \left(p_3^2-m_2 \left(\omega_3-p \cos \theta\right) \right)^2 \omega_{k_2}} \right] + 2 \pi i}{m_2 \left(p_3^2 -m_2 \left(\omega_3-p \cos\theta\right) \right) \omega_{k_2}}
\end{align}
where $\theta$ is now the angle between $p_3$ and $k_2$. The cut of the three mass triangle is given by
\eq{
    \text{Cut}_{k_2^2} T^{1} = \frac{i}{8 \pi \left(\xi-\bar{\xi}\right)} \frac{1}{\left(p_3+k_2\right)^2} \ln \left( \frac{1-\xi}{1-\bar{\xi}}\right)
}
with
\eq{
    \xi \bar{\xi} = \frac{k_2^2}{\left(p_3+k_2\right)^2} \, , \hspace{2cm}
    \left(1-\xi\right) \left(1-\bar{\xi}\right) = 
    \frac{p_3^2}{\left(p_3+k_2\right)^2} \, ,
}
where we take $k_2^2>0$, $p_3^2<0$ and it can be shown that for these cuts, $\left(p_3+k_2\right)^2<0$. As before, we assume that $\bar{\xi}$ is small. We make a change of variables from $\omega_{k_2}$ and $\cos \theta$ to $\xi$ and $x$, defined by
\eq{
    \xi = 1 - \frac{p_3^2}{\left(p_3+k_2\right)^2} \, , \hspace{2cm}
    x = 1 - \frac{p_3^2}{p_3^2-m_2 \left(\omega_3-p\cos\theta\right)} \, ,
}
with Jacobian
\begin{equation}
    \frac {\partial (\xi, x)}{\partial (\omega_{k_2}, \cos \theta)} =
    \begin{pmatrix}
        \frac {\partial \xi}{\partial \omega_{k_2}} & \frac {\partial \xi}{\partial \cos \theta} \\
        0 & \frac {\partial x}{\partial \cos \theta}
    \end{pmatrix},
\end{equation}
where
\eq{
    \frac{\partial \xi}{\partial \omega_{k_2}} = \frac{-2 p_3^2 \left(\omega_3-p \cos \theta\right)}{\left(p_3+k_2\right)^4} \, ,
    \hspace{2cm}
    \frac{\partial x}{\partial \cos \theta} = \frac{m_2 p p_3^2}{\left[ p_3^2 - m_2 \left(\omega_3-p \cos \theta\right) \right]^2} \, .
}
The limits of the $\xi$ integrals are at 0 and $x$, while the limits of the $x$ integration are at $z$ and $\zb$. Putting everything together, we get
\eq{
    T_3^{\mc{C}_2,\mc{C}_3} =
    \frac{1}{4096 \pi^4 m_2 p_3^2}  \int_{\zb}^z \frac{d x}{x p p_3^2} \int_0^x d \xi
    \ln \left[\frac{-m^2 \left(1-x\right) x  p_1^2}{p_2^2 p_3^2 \xi}\right]
    \frac{\ln \left(1-\xi\right)}{\xi} \, .
}
Performing the integrals in $\xi$ and $x$, and using that $p=-\frac{p_1^2}{2 m_2} \left(z-\zb\right)$ results in
\begin{multline}
    T_3^{\mc{C}_2,\mc{C}_3} =
    \frac{1}{2048 \pi^4 p_1^2 p_3^4 \left(z-\zb\right)}
    \Big\{ \left[- \ln \left( - \frac{m^2}{p_3^2} \right) + \ln \left(z \zb\right) + 2 \pi i \right] \left[ \text{Li}_3\left(z\right) - \text{Li}_3\left(\zb\right) \right]
    \\
    + \left[\frac{1}{2} \text{Li}^2_2 \left(z\right) - \frac{1}{2}  \text{Li}^2_2 \left(\zb\right) \right]
    - \left[ \text{Li}_4 \left(z\right) - \text{Li}_4 \left(\zb\right) \right]
    \Big\} \, .
\end{multline}

\section{Massless three-point vertices}
\label{sec:massless3pt}
When calculating cut graphs, we sometimes encounter subgraphs with cuts of massless lines on either side of a three-point vertex. This appendix discusses two important subtleties involved in computing these cut subgraphs. The first relates to evaluating the diagrams in dimensional regularization, and the second comes from delta functions evaluated at the endpoints of integration.

When evaluating diagrams with massless three-point vertices in dimensional regularization using the covariant cutting rules, %
one gets a delta function in the angle between the two particles multiplied with its argument raised to a power. For example,
consider the graph
\eq{
    \begin{gathered}
\begin{tikzpicture}[baseline=-3.5]
\node at (0,0) {\parbox{40mm} {\resizebox{40mm}{!}{
 \fmfframe(0,00)(0,0){ \begin{fmfgraph*}(140,70)
     \fmfstraight
	\fmfleft{L1}
	\fmfright{R1,R2}
	\fmf{plain_arrow,label=$p_3$}{L1,v1}
	\fmf{plain,label=$1$,tension=0.5,l.s=right}{v1,v2}
	\fmf{plain,label=$3$,l.s=right,tension=0}{v2,v3}
	\fmf{plain,label=$2$,l.s=right,tension=0.5}{v3,v1}
	\fmf{plain,label=$4$,tension=0.5}{v2,v4}
	\fmf{plain,label=$6$,tension=0.5}{v4,v5}
	\fmf{plain,label=$5$,tension=0.5,l.s=right}{v5,v3}
	\fmf{plain_arrow,tension=1.3}{R1,v4}
	\fmf{plain_arrow,tension=1.3}{R2,v5}
	\fmfv{label=$p_1$}{R2}	
	\fmfv{label=$p_2$}{R1}  \end{fmfgraph*} }}}};
\draw[line width = 1, dashed, darkgreen,smooth] plot [smooth] coordinates {(-0.7,-0.5) (0.3,0.1) (2.2,0.3)};
\draw[line width = 1, dashed, darkgreen,smooth] plot [smooth] coordinates {(0.1,-0.8) (1,-0.4) (2.2,-0.2)};
\draw[darkred, line width = 1] (0.3,-0.3) ellipse (0.45 and 0.25);
\end{tikzpicture}
\end{gathered}
\hspace{6pt} 
    \propto \int_{-1}^1 d \cos \theta \left(1-\cos^2 \theta\right)^{\frac{d-4}{2}} \delta \left(1-\cos\theta\right) \, ,
}
which contributes to $(\disc_{p_2^2})^2 \Phi_2$. The dashed lines correspond to cuts and the circled subgraph is the problematic three-point vertex. Here,
 $\theta$ is the angle between internal particles 1 and 3 in the diagram.
 
The first problem with this expression is that the limit $d\to 4$ is not smooth. For $d>4$ the integral is zero, for $d=4$ it is finite, and for $d<4$ it is divergent. In~\cite{Abreu:2014cla} it was argued that one should use the $d>4$ result and set all such graphs to zero. Indeed, such an approach seems to work in the examples considered in~\cite{Abreu:2014cla}.
 However, it may give results for the cut graphs that are inconsistent with the discontinuities, as discussed below Eqs.~\eqref{sumiszero} and \eqref{twoandthree}.
An alternative to using dimensional regularization is to give the internal lines a small mass $m_\text{reg}$ and take the limit $m_\text{reg}\to 0$. Although masses are not great regulators in general, particularly in gauge theories where they can violate gauge invariance, for the Feynman integrals we consider in this paper they always seems to give results for the cuts consistent with the discontinuities.

The second problem is that, even if a graph or sum of graphs is IR finite in $d = 4$, the delta function of the angle between the two particles may need to be evaluated at one of the endpoints of the limits of integration. Such expressions are not generally well-defined, and more careful analysis is needed. As we will show, this ultimately results in a combinatorial factor of $\frac{1}{L!}$ compared to the na\"ive expectation of setting $\smash{\int_{-1}^1 \delta(1-\cos\theta) d \cos \theta}$ to 1, where $L-1$ is the number of massless three point vertices in the cut diagram.

To see how the combinatorial factor arises, we calculate the $L$-loop triangle:
\begin{equation}
\begin{gathered}
\begin{tikzpicture}[baseline=-3.5,scale=0.8]
        \draw[line width = 1] (0,0) -- (1,0);
        \draw[line width = 1] (1,0) -- (9,2);
        \draw[line width = 1] (3,0.5) -- (9,-1);
        \draw[line width = 1] (5,1) -- (9,0);
        \draw[line width = 1] (7,1.5) -- (9,1);
        \draw[line width = 1] (1,0) -- (9,-2);
        \draw[line width = 1] (10,-2.25) -- (11,-2.5);
        \draw[line width = 1] (10,2.25) -- (11,2.5);
        \draw[line width = 1] (11,-1.5) -- (10,-1.25);
        \draw[line width = 1] (11,-0.5) -- (10,-0.25);
        \draw[line width = 1] (11,0.5) -- (10,0.75);
        \draw[dashed,line width = 1,color=darkgreen] (2,2.4) -- (2,-2.4);
        \draw[dashed,line width = 1,color=darkgreen] (4,2.4) -- (4,-2.4);
        \draw[dashed,line width = 1,color=darkgreen] (10.25,2.4) -- (10.25,-2.4);
        \node[text width=0.5cm] at (0,0.3) {$p$};
        \node[text width=0.5cm,color=darkgreen] at (2.0,3) {$\mathcal{C}_{1}$};
        \node[text width=0.5cm,color=darkgreen] at (4.0,3) {$\mathcal{C}_{2}$};
        \node[text width=0.5cm,color=darkgreen] at (10.25,3) {$\mathcal{C}_{L}$};
        \node[text width=0.5cm] at (1.5,0.5) {$k_1$};
        \node[text width=0.5cm] at (3.5,1.0) {$k_2$};
        \node[text width=0.5cm] at (11.5,2.75) {$k_L$};
        \node[text width=1.5cm] at (1.3,-0.5) {$p-k_1$};
\end{tikzpicture}
\end{gathered}
\end{equation}
The incoming particle is massive with $p^2=m^2$, and we cut the massless propagators with momentum $k_1, \ldots, k_L$, $p-k_1$, and $k_2-k_1, \ldots, k_L-k_{L-1}$.
Na\"{i}vely, using the covariant cutting rules, one would put all the cut particles on-shell and the diagram above would be given by
\begin{multline}
    T = 
    i^L
    \int \frac{d^4 k_1}{\left(2\pi\right)^4} \cdots 
    \frac{d^4 k_L}{\left(2\pi\right)^4}
    \left(2\pi\right) \delta\left(k_1^2\right) 
    \Theta\left(k_1^0\right)
    \cdots
    \left(2\pi\right) \delta\left(k_L^2\right)
    \Theta\left(k_L^0\right)
    \left(2\pi\right) \delta\left[\left(p-k_1\right)^2\right] \Theta\left(p^0-k_1^0\right)
    \\
    \times
    \left(2\pi\right) \delta\left[\left(k_1-k_2\right)^2\right]
    \Theta\left(k_1^0-k_2^0\right)
    \cdots
    \left(2\pi\right)
    \delta\left[\left(k_{L-1}-k_{L}\right)^2\right]
    \Theta\left(k_{L-1}^0-k_L^0\right)\,.
\end{multline}
We label the angle between $k_i$ and $k_j$ by $\theta_{i,j}$, define $\omega_i \equiv k_i^0$, and denote the angle between $k_1$ and the $z$-axis as $\theta$. In the center of mass frame of $p$, the above expression can be written
\begin{multline}
    T = i^L \int \frac{d^3 k_1}{\left(2\pi\right)^3 2 \omega_{1}}
    \cdots \frac{d^3 k_L}{\left(2\pi\right)^3 2 \omega_{L}}
    \left(2\pi\right) \delta \left(m^2-2 m \omega_{1} \right)
    \left(2\pi\right) \delta \left[-2 \omega_{1} \omega_{2} \left(1-\cos\theta_{1,2}\right)\right]
    \\
    \times
    \cdots \left(2\pi\right) \delta \left[-2\omega_{L-1} \omega_{L} \left(1-\cos\theta_{L-1,L}\right)\right]
    \Theta\left(\omega_{1}>\omega_{2}> \cdots > \omega_{L} \right)\,.
\end{multline}
Extracting the Jacobian factors results in
\begin{multline}
    T = \frac{i^L}{\left(8\pi\right)^{L} m} \int_0^\infty d \omega_{1} \int_0^{\omega_{1}} \frac{d \omega_{2}}{\omega_{2}} \cdots \int_0^{\omega_{L-2}} \frac{d \omega_{L-1}}{\omega_{L-1}} \int_0^{\omega_{L-1}} d \omega_{L}
    \delta\left(\omega_{1}-\frac{m}{2}\right)
    \\ \int_{-1}^1 d \cos \theta \int_{-1}^1 d \cos \theta_{1,2} \cdots \int_{-1}^1 d \cos \theta_{k_{L-1},k_L}
    \delta \left(1-\cos\theta_{1,2}\right) \cdots
    \delta \left(1-\cos\theta_{k_{L-1},k_L} \right)\,.
    \label{eq:FeynmLcomb}
\end{multline}
This integral is ambiguous, since the delta functions of the angles are evaluated at the integration endpoints. To evaluate it properly, we must go back to the TOPT expression for the corresponding diagram, where we have a handle on how to make sense of these products of delta functions. Namely, we know that they arise when using the relation
\eq{\lim_{\varepsilon\to 0}\left(\frac{1}{E+i\varepsilon}-\frac{1}{E-i\varepsilon}\right) = - 2 \pi i \delta(E)\,.
\label{eq:deltadef}
}
Thus, when we encounter a delta function that is evaluated at an integration endpoint, this implies we have used the distributional identity in Eq.~\eqref{eq:deltadef} too early. For massless three-point vertices, we should instead use the expression
\eq{
    \frac{1}{E+i\varepsilon}-\frac{1}{E-i\varepsilon}
    =
    -2 i
    \frac{\varepsilon}{E^2+\varepsilon^2} \, ,
    \label{eq:deltaEdef}
}
and only take the limit $\varepsilon \to 0$ after all the integrals have been evaluated. To shorten our equations, we define the expression that appears on the right-hand side of Eq.~\eqref{eq:deltaEdef} as $\delta^\varepsilon \equiv \frac{1}{\pi} \frac{\varepsilon}{x^2+\varepsilon^2}$.

\subsection*{Two loops}
For extra clarity, we now show how the correct combinatoric factor results in the  two-loop case. The $L$-loop case is worked out analogously afterwards; it involves the same ideas but with longer expressions. The two-loop TOPT diagram is given by
\eq{
    T = i^2 \int \frac{d^3 k_1}{\left(2\pi\right)^3 2 \omega_{1}} \int \frac{d^3 k_2}{\left(2\pi\right)^3 2 \omega_{2}} \frac{1}{2 \omega_{1}} \frac{1}{2 \omega_{1-2}}\left(2\pi\right) \delta^{\varepsilon} \left(m-2\omega_{1}\right) \left(2\pi\right) \delta^{\varepsilon} \left(m-\omega_{1}-\omega_{2} -\omega_{1-2} \right)
}
with $\omega_{1-2}=\sqrt{\omega_{1}^2+\omega_{2}^2-2 \omega_{1} \omega_{2} \cos\theta_{1,2}}$. We have already imposed three-momentum conservation. We perform the azimuthal integrals, and change variables from $\cos\theta_{1,2}$ to $\omega_{1-2}$ to get
\eq{
    T = \frac{i^2}{\left(2\pi\right)^2 2^4} 
    \int d \omega_{1} \int d \cos \theta \int d \omega_{2} \int_{\omega_{1}-\omega_{2}}^{\omega_{1}+\omega_{2}} d \omega_{1-2}
    \delta^{\varepsilon} \left(m-2\omega_{1}\right) \delta^{\varepsilon} \left(m-\omega_{1}-\omega_{2} -\omega_{1-2} \right)\,.
    \label{eq:TOPTcomb2loop}
}
We now use that
\eq{
    \int \delta^{\varepsilon} (x) dx = \frac{1}{\pi} \int \frac{\varepsilon  }{x^2+\varepsilon^2} dx
    =
    \frac{1}{\pi}
    \arctan \left( \frac{x}{\varepsilon}\right)
}
to write
\begin{multline}
    T
    =
    \frac{i^2}{\left(2\pi\right)^2 2^4} \int d \cos \theta
    \int d \omega_{2}
    \int \frac{d \omega_{1}}{\omega_{1}}
    \delta^{\varepsilon} \left(m-2\omega_{1}\right)
    \\ \times
    \frac{1}{\pi} \left[
    \arctan \left( \frac{m-2\omega_{1}}{\varepsilon} \right)
    -
    \arctan \left( \frac{m-2\omega_{1}-2\omega_{2}}{\varepsilon} \right)
    \right]\,.
\end{multline}
We can plug in $\omega_{1} = \frac{m}{2}$ everywhere except at singular points, to get
\eq{
    T = \frac{1}{2^6 \pi^3} \int d \cos \theta
    \int d \omega_{2}
    \int d \omega_{1}
    \delta^{\varepsilon} \left(m-2\omega_{1}\right)
    \left[
    \arctan \left( \frac{m-2\omega_{1}}{\varepsilon} \right)
    -
    \arctan \left( \frac{-2\omega_{2}}{\varepsilon} \right)
    \right]\,.
    \label{eq:2loopcomb}
}
Since $\frac{d}{dx} \arctan\left(\frac{x}{\varepsilon}\right) = \pi \delta^{\varepsilon}\left(x\right)$, we get
\begin{multline}
    \int_{0}^\infty d \omega_{1} \, \pi \delta^{\varepsilon} \left(m-2\omega_{1}\right) \left[\arctan \left(\frac{m-2\omega_{1}}{\varepsilon} \right) - \arctan \left(\frac{-2\omega_{2}}{\varepsilon} \right) \right]
    \\
    =
    \frac{1}{2} \left[ \left( \arctan\left(\frac{m-2\omega_{1}}{\varepsilon}\right) - \arctan\left(\frac{-2\omega_{2}}{\varepsilon}\right) \right)^2 \right]_0^\infty= \frac{\pi^2}{2} \, ,
    \label{eq:arctan2loop}
\end{multline}
where we have taken the limit $\varepsilon\to 0^+$ when writing the last equation. The factor of $\frac{1}{2}$ in this equation, arising from the integral over the product of $\arctan$ and a $\delta^{\varepsilon}$ function, has the same origin as the $\frac{1}{L!}$ factor in the $L$-loop case. The combinatorial factor arises because the $\delta^{\varepsilon}$s in Eq.~\eqref{eq:TOPTcomb2loop} only have support on the endpoint of the sequential delta function. We plug this into Eq.~\eqref{eq:2loopcomb} to get
\eq{
    T =
    \frac{i^2}{2^7 \pi^2 m} \int d \cos \theta \int d \omega_{2}\,.
}
Comparing to the $L=2$ case of Eq.~\eqref{eq:FeynmLcomb}, we learn that we must multiply the right hand side of $\smash{\int_{-1}^1\delta \left(1-\cos\theta_{1,2}\right)\stackrel{?}{=}1}$ by a combinatorial factor of $\frac{1}{2}$. Although this factor of $\frac{1}{2}$ could potentially be justified in the two-loop case by claiming that the delta function in Eq.~\eqref{eq:FeynmLcomb} is only integrated up to its endpoint, and hence should be evaluated to give $\frac{1}{2}$, that argument does not generalize to the $L$-loop case, where we will see that we encounter a combinatorial factor of $\frac{1}{L!}$ rather than $\frac{1}{2^L}$.

\subsection*{\texorpdfstring{$L$}{L} loops}
The $L$-loop TOPT diagram is given by
\begin{multline}
    T = i^L \int \frac{d^3 k_1}{\left(2\pi\right)^3 2 \omega_{1}}
    \cdots \frac{d^3 k_L}{\left(2\pi\right)^3 2 \omega_{L}}
    \frac{1}{2 \omega_{1}}
    \frac{1}{2 \omega_{1-2}}
    \cdots
    \frac{1}{2 \omega_{(L-1)-L}}
    \left(2\pi\right)
    \delta^{\varepsilon}\left(m-2\omega_{1}\right)
    \\ \times
    \left(2\pi\right) \delta^{\varepsilon}\left(m-\omega_{1}-\omega_{2} - \omega_{1-2}\right)
    \left(2\pi\right)
    \delta^{\varepsilon}\left(m-\omega_{1}-\omega_{3}-\omega_{1-2} - \omega_{2-3}\right)
    \\ \times
    \cdots
    \left(2\pi\right)
    \delta^{\varepsilon}\left(m-\omega_{1}-\omega_{L} -\omega_{1-2} - \cdots - \omega_{(L-1)-L} \right) \, ,
\end{multline}
where
\eq{
    \omega_{i-j} = \sqrt{\omega_{i}^2+\omega_{j}^2-2 \omega_{i} \omega_{j} \cos\theta_{i,j}}\,.
}
Preforming the azimuthal integrals gives
\begin{multline}
    T =  \frac{i^L}{2^{2L} \left(2\pi\right)^{ L}} \int_0^\infty \omega_{1} d \omega_{1} \int_0^{\omega_{1}} \omega_{2} d \omega_{2} \cdots \int_0^{\omega_{L-1}} \omega_{L} d \omega_{L}
    \frac{1}{\omega_{1}}
    \frac{1}{\omega_{1-2}}
    \cdots
    \frac{1}{\omega_{(L-1)-L}}
    \\ \times
    \int_{-1}^1 d \cos \theta \int_{-1}^1 d \cos \theta_{1,2} \cdots \int_{-1}^1 d \cos \theta_{k_{L-1},k_L}
    \delta^{\varepsilon}\left(m-2\omega_{1}\right)
    \\ \times \delta^{\varepsilon}\left(m-\omega_{1}-\omega_{2} - \omega_{1-2}\right)
    \delta^{\varepsilon}\left(m-\omega_{1}-\omega_{3}-\omega_{1-2} - \omega_{2-3}\right)
    \\ \times
    \cdots
    \delta^{\varepsilon}\left(m-\omega_{1}-\omega_{L} -\omega_{1-2} - \cdots - \omega_{(L-1)-L} \right)\,.
\end{multline}
We change variables from the $\cos \theta_{i,i+1}$ variables to $x_1, \cdots, x_{L-1}$ with $x_i = \omega_{i-(i+1)}$. The Jacobian for each $i$ is given by
\eq{
    J_i = \left( \frac{\partial \omega_{k_i-k_{i+1}}}{\partial \cos \theta_{i,i+1}} \right)^{-1}
    =
    -
    \left( \frac{\omega_{i} \omega_{k_{i+1}}}{ \omega_{k_{i}-{k_{i+1}}}} \right)^{-1} \, ,
}
so
\begin{multline}
    T =  \frac{i^L}{2^{2L} \left(2\pi\right)^L}
    \int_{-1}^1 \cos \theta \int_0^\infty \frac{d \omega_{1}}{\omega_1} \int_0^{\omega_{1}} \frac{d \omega_{2}}{\omega_{2}} \cdots \int_0^{\omega_{L-2}} \frac{d \omega_{L-1}}{\omega_{L-1}} \int_0^{\omega_{L-1}} d \omega_{L}
    \delta^{\varepsilon} \left(m-2 \omega_{1}\right)
    \\ \times
    \int_{\omega_{1}-\omega_{2}}^{\omega_{1}+\omega_{2}} d x_1  \delta^{\varepsilon}\left(m-\omega_{1}-\omega_{2} - x_1 \right)
    \int_{\omega_{2}-\omega_{3}}^{\omega_{2}+\omega_{3}} d x_2
    \delta^{\varepsilon}\left(m-\omega_{1}-\omega_{3} - x_1 - x_2 \right)
    \\ \times
    \cdots
    \int_{\omega_{L-1}-\omega_{L}}^{\omega_{L-1}+\omega_{L}} d x_{L-1}
    \delta^{\varepsilon}\left(m-\omega_{1}-\omega_{L} - x_1 - x_2 - \cdots - x_{L-1} \right)\,.
\end{multline}
Shifting the integrals gives
\begin{multline}
    T =  \frac{i^L}{2^{2L+1} \left(2\pi\right)^{ L}} \int_{-1}^1 d \cos \theta \int_0^\infty \frac{d x_0}{x_0} \int_0^{x_0} \frac{d \omega_{2}}{\omega_{2}} \cdots \int_0^{\omega_{L-2}} \frac{d \omega_{L-1}}{\omega_{L-1}} \int_0^{\omega_{L-1}} d \omega_{L}
    \delta^{\varepsilon} \left(x_0-\frac{m}{2}\right)
    \\ \times
    \int_{-\omega_{2}+x_0}^{\omega_{2}+x_0} d x_1  \delta^{\varepsilon}\left(m-\omega_{2} - x_0 - x_1 \right)
    \int_{\omega_{2}-\omega_{3}+x_1}^{\omega_{2}+\omega_{3}+x_1} d x_{1,2}
    \delta^{\varepsilon}\left(m-\omega_{3} - x_0 - x_{1,2} \right)
    \\ \times
    \cdots
    \int_{\omega_{L-1}-\omega_{L}+x_{1,L-2}}^{\omega_{L-1}+\omega_{L}+x_{1,L-2}} d x_{1,L-1}
    \delta^{\varepsilon}\left(m-\omega_{L} - x_0 - x_{1,L-1} \right) \, ,
    \label{eq:delta_endpoints}
\end{multline}
where $x_{1,i} = x_1 + \cdots + x_i$ and $x_0=\omega_1$. We now have a product of delta functions where each is evaluated at the endpoint of the previous one. To handle this more carefully, we use the $\delta^{\varepsilon}$ distributions. In particular, we investigate the expression
\begin{multline}
\mathcal{I} =
\int_0^\infty d x_0 \delta^{\varepsilon} \left(x_0-\frac{m}{2}\right)
\int_{- \omega_2 + x_0}^{\omega_2 + x_0} d x_1 \delta^\varepsilon(m - \omega_2 - x_0 - x_1)
\int_{\omega_2 - \omega_3 + x_1}^{\omega_2 + \omega_3 + x_1} d x_{1, 2} \delta^\varepsilon(m  - \omega_3 - x_0 - x_{1, 2}) \times \\ \cdots
\int_{\omega_{L - 1} - \omega_L + x_{1, L - 2}}^{\omega_{L - 1} + \omega_L + x_{1, L - 2}} d x_{1, L - 1} \delta^\varepsilon(m - \omega_L - x_0 - x_{1, L - 1}) F(x_0,x_1, x_{1, 2}, \dotsc, x_{1, L - 1}),
\end{multline}
where \(F\) is a test function, which we take to be a smooth function of compact support. We aim to compute the \(\epsilon \to 0^+\) limit of this integral.  We use the fact that if \(x = a + \epsilon y\), then
\[
  \delta^\epsilon(x - a) d x = \frac {d y}{\pi (1 + y^2)}.
\]
Using this formula repeatedly, we find
\begin{multline}
\mathcal{I} =
\int_{-\frac{m}{2\varepsilon}}^\infty \frac{d y_0}{\pi (1+y_0^2)}
\int_{y_0}^{y_0+\frac {2 \omega_2} \epsilon} \frac {d y_1}{\pi (1 + y_1^2)}
\int_{y_1}^{y_1 + \frac {2 \omega_3} \epsilon} \frac {d y_{1, 2}}{\pi (1 + y_{1, 2}^2)} \cdots
\int_{y_{1, L - 2}}^{y_{1, L - 2} + \frac {2 \omega_L} \epsilon} \frac {d y_{1, L - 1}}{\pi (1 + y_{1, L - 1}^2)} \\
\times F(m/2+\epsilon y_0,m/2 - \omega_2 + \epsilon y_1,
  m/2 - \omega_3 + \epsilon y_{1, 2}, \dotsc,
  m/2 - \omega_L + \epsilon y_{1, L - 1}).
\end{multline}
Since the function \(F\) is smooth, we can series expand it around \(\epsilon = 0\).  We keep only the zeroth-order terms in the expansion; the higher-order terms do not contribute in the limit \(\epsilon \to 0^+\).

If \(\omega_i\) vanishes, then the integral over \(y_{1, i - 1}\) vanishes, as the upper and lower integration limits are coincident. If all of the \(\omega_i\) all strictly positive, then the upper integration limits all become \(+ \infty\) in the \(\epsilon \to 0^+\) limit.  Hence, we obtain
\begin{multline}
\lim_{\epsilon \to 0^+} \mathcal{I} =
F(m/2, m/2 - \omega_2, \dotsc, m/2 - \omega_L)
\\
\times
\int_{-\infty}^\infty \frac{d y_0}{\pi (1+y_0^2)}
\int_{y_0}^\infty \frac {d y_1}{\pi (1 + y_1^2)}
\int_{y_1}^\infty \frac {d y_{1, 2}}{\pi (1 + y_{1, 2}^2)} \cdots
\int_{y_{1, L - 2}}^\infty \frac {d y_{1, L - 1}}{\pi (1 + y_{1, L - 1}^2)}.
\end{multline}
Performing the integrals one by one, we get an $\arctan$ function raised to a power each time, just as in Eq.~\eqref{eq:arctan2loop}. The result after performing $L-1$ integrations is 
\begin{multline}
    \lim_{\epsilon \to 0^+} \mathcal{I} =
    F(m/2, m/2 - \omega_2, \dotsc, m/2 - \omega_L)
    \\
    \times
    \left(-1\right)^{L-1}
    \int_{-\infty}^\infty 
    \frac{d y_0}{2^{L-1} \left(L-1\right)! \pi^L \left(1+y_0^2\right)}
    \left(\pi-2\arctan\left(y_0\right) \right)^{L-1}
    .
\end{multline}
The last integral evaluates to
\begin{multline}
    \lim_{\epsilon \to 0^+} \mathcal{I} =
    F(m/2, m/2 - \omega_2, \dotsc, m/2 - \omega_L)
    \\
    \times
    \left(-1\right)^{L}
    \left[
    \frac{1}{2^L L! \pi^L}
    \left(\pi-2\arctan\left(y_0\right) \right)^{L}
    \right]_{-\infty}^{\infty}
    =
    \frac{1}{L!}
    .
\end{multline}
Making use of this in Eq.~\eqref{eq:delta_endpoints}, we get
\eq{
    T = \frac{i^L}{\left(8\pi\right)^{L} L! m} \int_{-1}^1 d \cos \theta \int_0^{m/2} \frac{d \omega_{2}}{\omega_{2}} \cdots \int_0^{\omega_{L-2}} \frac{d \omega_{L-1}}{\omega_{L-1}} \int_0^{\omega_{L-1}} d \omega_{L}\,.
}
In particular, this result has an extra factor of $\frac{1}{L!}$ compared to what one would get by evaluating each of the delta functions to 1. Although we can compute these integrals in TOPT, it is harder to find this combinatorial factor using covariant Feynman rules.

\end{fmffile}

\bibliographystyle{utphys}
\bibliography{steinmann.bib}

\end{document}